%% file: these.tex
\documentclass[11pt,nocut]{report}

\usepackage{latex_style/packages}
\usepackage{latex_style/notations}
\usepackage{psl_cover/psl-cover}

\title{Long-range interactions in the avalanches of elastic interfaces.}
\author{Cl\'ement Le Priol}
\date{}

\institute{l'\'Ecole Normale Supérieure}
\doctoralschool{Physique en \^Ile-de-France}{564}
\specialty{Physique Théorique}
\date{13 novembre 2020}

\jurymember{2}{Léonie Canet}{Université Grenoble Alpes}{Rapporteur}
\jurymember{3}{Stéphane Santucci}{\'Ecole Normale Supérieure de Lyon}{Rapporteur}
\jurymember{4}{Jean-François Joanny}{Collège de France}{Président}
\jurymember{5}{Ludovic Berthier}{Université de Montpellier}{Examinateur}
\jurymember{6}{Laurent Ponson}{Sorbonne Université}{Invité}
\jurymember{7}{Pierre Le Doussal}{\'Ecole Normale Supérieure}{Directeur de thèse}
\jurymember{8}{Alberto Rosso}{Universit{\'e} Paris-Saclay}{Co-directeur de thèse}

\frabstract{
Les syst{\`e}mes d{\'e}sordonn{\'e}s soumis {\`a} une contrainte lentement croissante pr{\'e}sentent 
souvent une dynamique par sauts caract{\'e}ris{\'e}e par des pouss{\'e}es d’activit{\'e}, 
appel{\'e}es avalanches, qui sont la marque d’une transition de phase hors-{\'e}quilibre. 
Cette th{\`e}se traite de la dynamique des interfaces {\'e}lastiques au voisinage de la transition 
de d{\'e}pi{\'e}geage, et plus sp{\'e}cialement de l’effet des interactions {\`a} longue-port{\'e}e.
Je d{\'e}rive la forme d’échelle universelle des fonctions de corr{\'e}lation du champ des vitesses locales. 
Cela permet de tester la classe d’universalit{\'e} de la transition, ainsi que la port{\'e}e des interactions.
J’{\'e}tudie aussi la statistique des clusters qui se forment en pr{\'e}sence d’interactions 
{\`a} longue-port{\'e}e et montre comment celle-ci est reli{\'e}e {\`a} la statistique des avalanches globales.
Enfin je pr{\'e}sente des avanc{\'e}es analytiques vers la compr{\'e}hension de la structure spatiale des avalanches dans un mod{\`e}le de champ-moyen qui est le mod{\`e}le de force brownien.
}

\enabstract{
Disordered systems submitted to a slowly increasing external stress often reacts with a jerky dynamics characterized by bursts of activity, called avalanches, which are the manifestation of an out-of-equilibrium phase transition. This thesis focuses on the dynamics of elastic interfaces near
the depinning transition, and more specially on the effect of long-range interactions.
I derive the universal scaling form of the correlation functions of the local velocity field.
It allows to test the universality class of the transition as well as the range of the interactions.
I also study the statistics of the clusters that form in presence of long-range interactions and show
how it relates to the statistics of the global avalanches. 
Finally I present analytical advances towards the understanding of the spatial structure of avalanches 
within a mean-field model, the Brownian force model. 
}

\frkeywords{D{\'e}sordre, dynamique, transition de phase hors-{\'e}quilibre, avalanches, 
d{\'e}pi{\'e}geage, interactions à longue-port{\'e}e.}
\enkeywords{Disorder, dynamics, out-of-equilibrium phase transition, avalanches, depinning, long-range interactions.}

\begin{document}

\maketitle{}
\pagenumbering{roman}
\input{Intro/Remerciements.tex}
{%
	\hypersetup{linkcolor=black}
	\setcounter{tocdepth}{1}
	\tableofcontents
}

\setcounter{tocdepth}{3}
\input{Intro/intro.tex}


\clearpage
\pagenumbering{arabic}

\input{Chapter_1/Des_and_depinning.tex}
\newpage
\thispagestyle{empty}
\mbox{}
\newpage
\input{Chapter_2/Experiments.tex}
\newpage
\thispagestyle{empty}
\mbox{}
\newpage
\input{Chapter_3/Chapter_3.tex}
\input{Chapter_4/Chapter_4.tex}
\input{Chapter_5/Chapter_5.tex}
\input{Conclusion/Conclusion.tex}
\newpage
\thispagestyle{empty}
\mbox{}
\newpage

\appendix
\input{Appendix/Appendix.tex}
%
%
\bibliography{BiblioThese.bib}
\bibliographystyle{ieeetr}
\end{document}

%% file: Intro/Remerciements.tex


\chapter*{Remerciements}\label{chap: remerciements}
\addcontentsline{toc}{chapter}{Remerciements}%

Arrivé au terme de mes trois années de thèse qui furent enrichissantes tant sur le plan personnel qu'intellectuel, j'aimerai remercier les personnes qui ont rendu cette aventure possible et qui m'ont soutenu.
En premier lieu j'aimerais remercier mes directeurs, Pierre Le Doussal et Alberto Rosso, pour m'avoir fait confiance il y a bientôt quatre ans en acceptant d'encadrer ma thèse. 
J'ai beaucoup appris à leur côté, que ce soit la physique du depinning, des techniques analytiques, comment calibrer des simulations numériques ou comment rédiger un article. 
Je remercie particulièrement Pierre pour les dizaines d'heures passées à m'enseigner le BFM puis la FRG.
Il a su me guider dans des calculs difficiles qu'il prenait le temps de vérifier en détail. 
Je remercie Alberto pour le temps passé à échanger autour des résultats de mes simulations ainsi que pour la convivialité des déjeuners chez lui en période de rédaction d'articles. 
J'ai eu la chance d'avoir des directeurs bienveillants à l'esprit positif qui m'ont toujours encouragé et ont été d'une disponibilité exceptionnelle. 
Je les remercie enfin pour leurs multiples relectures attentives de ce manuscrit.
\vspace{.3cm}

Le travail présenté dans cette thèse doit aussi beaucoup à la collaboration mené avec Laurent Ponson que je remercie. Pour le théoricien que je suis, la confrontation avec des données expérimentales fut un aspect très enrichissant de mon travail. 
Je remercie aussi Mathias Lebihain avec qui j'ai pu échanger sur l'analyse de données expérimentales et Vincent Démery pour m'avoir mis le pied à l'étrier avec les figures de lignes élastiques.
\vspace{.3cm}

Je tiens à remercier les personnes qui ont rendu mon travail possible au LPT, fusionné entre temps au sein du LPENS, notamment Viviane Sebille, Sandrine Patacchini, et Christine Chambon pour leur aide avec toutes les démarches administratives. Un grand merci aussi à Marc-Thierry Jaekel pour m'avoir aidé à maintes reprises à utiliser les ordinateurs du LPT et aux membres du SI qui ont pris sa relève au LPENS.
\vspace{.3cm}

Je remercie Stéphane Santucci et Léonie Canet d'avoir accepté d'être rapporteurs de ce manuscrit. Merci à Ludovic Berthier et à Jean-François Joanny d'avoir accepté de compléter le jury. 
\vspace{.3cm}


J'aimerai aussi remercier les personnes qui ont fait que le laboratoire fut pour moi un endroit accueillant. Merci notamment à Gregory Schehr, Satya Majumdar, Jan Troost et Camille Aron pour les discussions dans les couloirs ou lors du déjeuner. 
Je remercie aussi mes co-thésards avec qui j'ai partagé de nombreux repas, discussions et bières dans le jardin et à la Montagne : mon "grand-frère" de thèse Kraj, qui fut le premier à m'accueillir au LPT, mes premiers camarades de bureaux, Dongsheng et Songyuan, mon "petit frère" Tristan, mes nouveaux camarades Augustin, Arnaud et Cathelijne ainsi que les sympathiques José et Clément.
\vspace{.3cm}

J'ai eu la chance en parallèle de ma thèse de faire des exposés au Palais de la Découverte. J'y ai pris beaucoup de plaisir. Merci aux médiateurs et doctorants de l'unité de physique qui m'y ont accueilli. 
\vspace{.3cm}

J'ai une pensée reconnaissante pour le système éducatif français dans son ensemble et pour tous les enseignants que j'ai pu rencontrer et qui m'ont permis d'arriver jusqu'à cette thèse. 
\vspace{.3cm}

Au-delà de ma thèse, j'aimerai exprimer ma gratitude à tous les amis avec qui j'ai pu échanger et partager des interrogations sur la recherche, avoir des discussions stimulantes sur l'écologie et que faire dans le futur et qui m'ont adressé leurs encouragements durant ces trois années. Je reste proche d'eux qu'ils soient à Paris, en Puisaye, à Grenoble ou en Afrique du Sud. Merci à Ruben, Loucas, Hédi, Malo, Maxime, Élise et toute la team des grimpeurs à Cancun (désormais grimpeurs plus grimpeurs), Simon, Ombline, Gaëlle et Cyril.
Un grand merci aussi à mes amies de la CNV qui m'ont apporté beaucoup de soutien dans cette dernière année. Je pense particulièrement à Julie, Hélène et Christine ainsi qu'à ma chère binôme Aurélie.
\vspace{.3cm}

Merci Gaétane pour les doux moments que nous avons partagé et pour tes encouragements dans ma phase d'écriture.
\vspace{.3cm}

Finalement je tiens à exprimer ma gratitude à mes parents et à mes frères pour leur constant support et leur inestimable présence à mes côtés. 
Merci pour l'ambiance chaleureuse qui m'a soutenu alors que je démarrai la rédaction de ma thèse lors du premier confinement. Je remercie Antonin pour la finalisation de certaines figures du chapitre 1. Et bien sûr merci à mes parents pour tout ce qu'ils m'ont offert depuis mon enfance. Je leur dois notamment le goût de l'effort intellectuel sans lequel je n'aurai pas entrepris cette thèse.



%% file: Intro/intro.tex


\chapter*{Introduction}\label{chap: Intro}
\addcontentsline{toc}{chapter}{Introduction}%

There are many natural systems that, when submitted to a slowly increasing external stress, respond with sudden discrete events called \emph{avalanches}. 
Besides snow avalanches, examples are the earthquakes in tectonic dynamics, the cascades of plastic events in amorphous materials 
or the sudden jumps in the magnetization of ferromagnetic materials.
A common ingredient in these phenomena is the interplay between disorder and the interactions between many elements in the system. 
Most of the time the system is pinned by the disorder in some metastable state and does not move until a first element, the seed of the avalanche, gets destabilized by the slowly increasing external stress, also called the drive.
The response of the seed may in turn trigger the instability of other elements in the system, leading to a collective response, the avalanche.
The sizes of the avalanches are random and, in many case, their statistical distribution is a broad power law. 
A famous example is the Gutenberg-Richter law for the magnitude of earthquakes.

In this thesis I focus on avalanches of elastic interfaces, which propagate in a disordered medium under the action of an external force. 
If the force is low the interface remains pinned by the disorder. 
It frees itself (it depins) when the force exceeds a certain threshold. 
In the 80's, D.S. Fisher proposed to interpret the depinning transition of elastic interfaces as a dynamical critical phenomenon~\cite{fisher1985}.
This opened the way for the application to avalanches of tools developped in the 70's to study equilibrium critical phenomena and especially the functional renormalization group. 
One of the main success of the theory of critical phenomena was 
to show that, near the transition, the large-scale behavior of the system does not depend on the microscopic parameters (like the precise type of disorder) but only on a few relevant parameters. 
Therefore, systems that are a priori very different can share the same large scale behavior. In this case they are said to be in the same universality class. 
The concept of universality allows to describe complex real systems with rather simple models that only incorporate the relevant parameters determining the correct universality class.
In the case of the depinning transition of elastic interfaces, one of the relevant parameters is the range of the elastic interactions, which might be short-range or long-range.
In 1984, Joanny and de Gennes showed that the wetting front of a liquid on a disordered solid substrate can be effectively described by an elastic line with long-range elasticity~\cite{joanny1984}. 
One year later Rice showed that the same model describes the front of a crack in a disordered brittle material~\cite{rice1985}.
My thesis work contributed to a better understanding of the effect of long-range interactions in the avalanches of elastic interfaces. 

In chapter 1 of this thesis I introduce the main elements of the theory of the depinning transition of elastic interfaces. It should allow the reader to understand the following chapters. 
In particular I describe the phase diagram of the transition and give a set of critical exponents
that are associated to this transition.
I also explain that, in presence of long-range interactions, avalanches have a more complex spatial structure than the avalanches in short-range interactions systems. Indeed they are separated in several connected components called clusters.

In chapter 2 I present several experimental phenomena that are described by this theory. 
These phenomena are the motion of domain walls in ferromagnets, the propagation of crack fronts and of wetting fronts in disordered media. 
I also present two systems, earthquakes and avalanches of plastic events in amorphous materials, which cannot be described by elastic interfaces, but for which connections with the depinning have been pointed out. These systems are characterized by long-range interactions. One can hope that a better understanding of the effect of long-range interactions in the depinning framework can shed more light 
on these phenomena.

In chapter 3 I present several methods to characterize the universality class of the depinning transition. These methods rely on the determination of critical exponents or scaling functions. 
I characterize the universal scaling form of the correlation functions of the local velocity field near the depinning transition. 
I show its relevance on numerical simulations and experimental data of crack propagation that I analyzed.
This provides a new method to assess the universality class of the transition. It also gives a direct probe of the range of the interactions. This result, presented in the second part of chapter 3, corresponds to the publication~\cite{lepriol2020}.

In chapter 4 I study the statistics of the clusters inside long-range avalanches with the help of extensive numerical simulations. I find that the number of clusters inside an avalanche seems to obey the statistics of a Bienaym{\'e}-Galton-Watson process. Based on this observation I build a scaling theory that relates the exponents characterizing the statistics of the clusters to the ones of the global avalanches. This chapter corresponds to the publication~\cite{lepriol2020a}.

I start the last chapter by presenting two standard mean-field models of the depinning transition.
These models do not capture the spatial structure of avalanches. 
This structure can be studied, at a mean-field level, in a more recent model, the Brownian force model. I explain how the statistics of different observables can be computed within this model. 
Informations about the spatial structure of long-range avalanches can be gained by studying a non-linear and non-local differential equation. Finally I present the progress that I made in this direction.


%% file: Chapter_1/Des_and_depinning.tex


\chapter{Disordered elastic systems and the depinning transition}\label{chap: DES and depinning}

Many experimental systems can be described by the standard model of \emph{disordered elastic systems} (also called \emph{random manifold}) of internal dimension $d$ embedded in a medium of dimension $d+N$, $N$ being the number of transverse dimensions.
In this thesis I focus on $N=1$ in which case the manifold is usually called an elastic \emph{interface}.
The competition between elasticity and disorder is at the root of a complicated energy landscape and a very rich physics~: on the one hand the elastic energy acts to align all the points and flatten the interface; on the other hand each point of the interface is looking for a local minimum of the disorder potential.
To minimize the total energy the interface must distort itself in order to find a balance between the elastic and the disorder energy. This makes the interface rough.
In my thesis I am interested in what happens to the interface when it is pulled out-of-equilibrium
by an external driving force. 
In this chapter I introduce some of the theoretical concepts used to discuss the problem of the elastic interface driven out of equilibrium when there is only one direction of motion ($N=1$).

\section{Phenomenology of the depinning transition}\label{sec: Ch1 Phase diagram}

\begin{figure}
	\centering
	\includegraphics[scale=0.45]{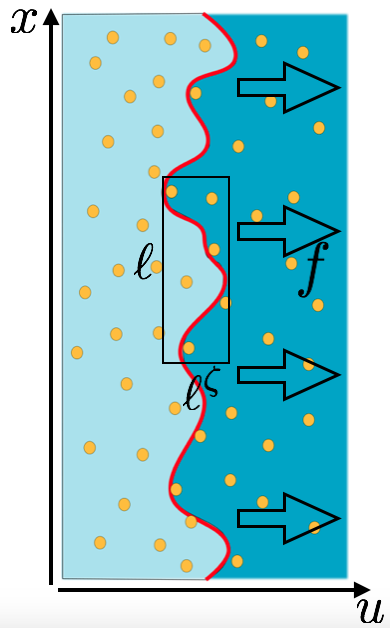}
	\caption{Sketch of an elastic line in a disordered medium. The line is pinned by the disorder. The competition between disorder and elasticity makes the line rough : A portion of internal extension $\ell$ has a typical width $\ell^{\zeta}$. The line can move under the application of an external force $f$. \label{fig: Line_1}} 
\end{figure}
\begin{figure}
	\centering
	\includegraphics[scale=0.3]{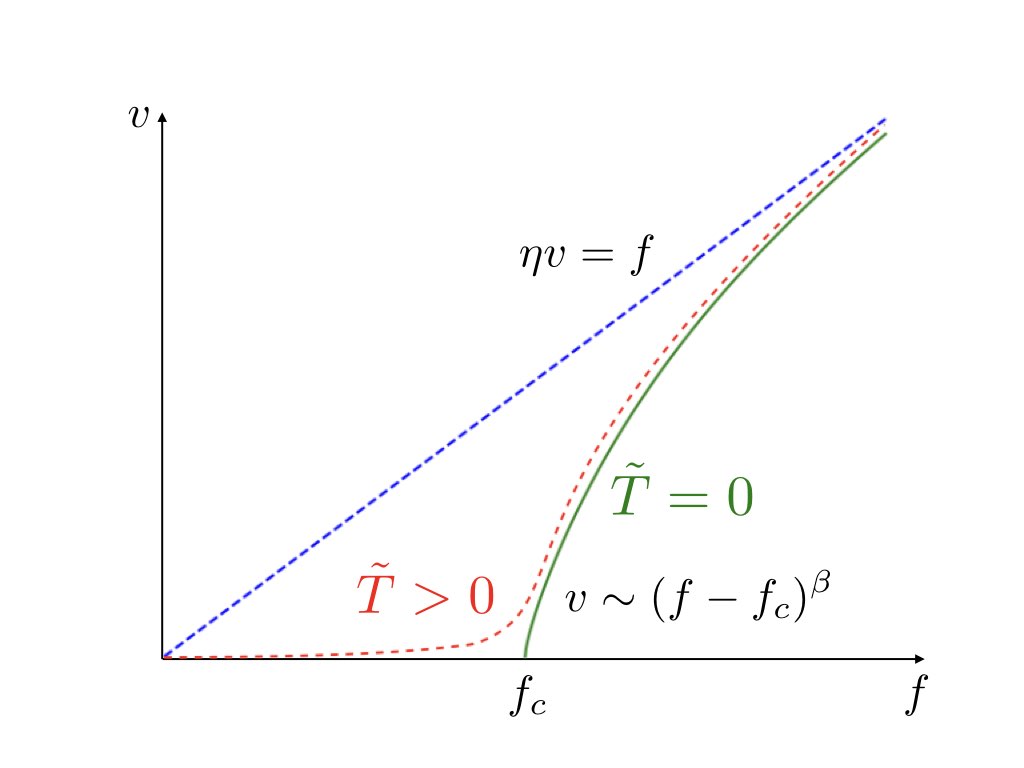}
	\caption{Velocity-force characteristics. In the absence of disorder the velocity is proportional to the applied force (blue dashed line). In the presence of disorder, at zero temperature ($\tilde{T}=0$), the line polarizes under the application of a small external force but it stays pinned in some local minimum of the energy landscape. There is a finite critical force $f_c$ above which the velocity becomes non-zero (green line). At finite temperature ($\tilde{T}>0$) the interface moves slowly for $f$ below $f_c$ (red dashed line). The effect of a finite temperature is explained in section~\ref{sec: ch1 effect of temperature}. \label{fig: Phase diagram}} 
\end{figure}

In figure~\ref{fig: Line_1} I show a sketch of an elastic interface in a $2$ dimensional disordered medium. 
(Whenever the internal dimension is $d=1$ as in this case, I will also use the term \textit{line} in an equivalent manner to \emph{interface}.) 
$x$ denotes the internal coordinate of the interface and  $u$ the transverse direction, along which the motion occurs. 
We assume that the position of the interface is described by a univalued function $u(x)$.
The yellow dots represent the \emph{quenched} (or \emph{frozen}) disorder that pins the line at some favored locations. 
It counters the elastic energy and makes the line random and rough~: the interface is not flat but has transversal displacements. So a portion of the interface has a \emph{width} that grows with the size of the portion. 
Portions of a given size $\ell$ have an average width $W(\ell)$ that scales as a power of $\ell$.
Thus we can define a \emph{roughness exponent} $\zeta$ by the scaling relation~: 
\begin{equation}
W(\ell) \sim \ell^{\zeta} \, . \label{eq: def1 zeta}
\end{equation}
This scaling is illustrated in figure~\ref{fig: Line_1}.

In many experimental systems the effect of temperature can often be neglected.
So we assume that the system is at zero temperature $\tilde{T}=0$\footnote{In this thesis the temperature is denoted by $\tilde{T}$ to avoid confusion with the avalanche duration $T$ that will be introduced below.}.
In absence of an applied force, $f=0$, because of the competition between elasticity and disorder, there are many local minima of the energy.
These are called \emph{metastable states}, while the global energy minimum is called the \emph{ground state}.
Let us look at what happens when we apply an external force $f$ on the line, starting from $f=0$ and very slowly increasing the force along the transverse direction. We assume that the line starts in some randomly chosen metastable configuration which is a local minimum of the energy for $f=0$. At the beginning the line will stay pinned in the same configuration because the force $f$ is weaker than the local disorder pinning force. Then it will reach a point where $f$ is strong enough to overcome the local disorder. A portion of the line will move forward and eventually lock into a new more stable configuration with lower energy. Because the duration of this move is usually short in comparison to the time the line stays in a configuration, we say that the line \emph{jumps} from one configuration to another. Such a jump is also called an \emph{avalanche}. As we continue to increase $f$ there will be a succession of jumps between metastable configurations. Eventually we will reach a critical force $f_c$ beyond which the line will move away from a metastable configuration but will never find a new one~: it acquires a finite velocity.  As $f$ continues to increase the motion of the line becomes faster and it eventually reaches a fast flow regime where the velocity is proportional to the applied force.

With these considerations in mind we can draw the schematic velocity-force characteristics presented in figure~\ref{fig: Phase diagram}.
Here $v$ is the mean velocity of the line, averaged both over space and time. 
We are interested in the long-time steady-state regime where inertia can be neglected.
The blue dashed line is the characteristics of a pure system, i.e. with no disorder. In this case the velocity is proportional to the force.
In the presence of disorder and at zero temperature we can distinguish three regimes:
\begin{itemize}
\item $f < f_c$. The interface polarizes~: it might jump from one metastable state to an other more stable state but it stops at some point. Hence in the steady-state regime the average velocity is zero. 
\item At $f_c$ the interface acquires a finite velocity that grows as $v \sim (f-f_c)^{\beta}$ for $f>f_c$.
\item $f \gg f_c$. The interface moves fast with a mean velocity that tends to become proportional to the force, as in the pure case, with the disorder playing the role of an effective thermal noise.
\end{itemize}

The transition from $v=0$ to $v>0$ that occurs at $f_c$ is called the \emph{depinning transition}.
The  velocity-force characteristics at $T=0$ resembles the curve of an order parameter in a second-order phase transition, like the magnetization in the Ising spin model~\cite{kardar2007}. 
By analogy with equilibrium critical phenomena it has been suggested that the depinning transition could be interpreted as a dynamical critical phenomenon~\cite{fisher1985} (also called \emph{out-of-equilibrium phase transition}).
We can distinguish two dynamical phases~:
a moving phase above $f_c$ and a non-moving phase below.
The depinning transition is also characterized by a correlation length that diverges at the critical point and universal critical exponents that do not depend on the details of the system but only on a few relevant parameters such as the internal dimension $d$ of the system or the range of the elastic interactions.
But since the system is out-of-equilibrium one cannot compute the free-energy of the system. 
Hence the usual way of computing the order parameter as the derivative of the free-energy 
with respect to the external field cannot be applied.
Nevertheless the analogy has proven to be very fruitful~: a functional renormalization group (FRG) theory has been developped for the depinning transition enabling the computation of critical
exponents and scaling forms near the transition~\cite{nattermann1992a, narayan1992, narayan1993, chauve2000, chauve2001, ledoussal2002, wiese2006}.
In the two following sections I introduce the main critical exponents characterizing the depinning transition.

\subsection{Below the threshold : quasi-static avalanches}
When $f$ is below $f_c$ there are well defined metastable configurations where the line is motionless. 
This allows to define without any ambiguity an \emph{avalanche} as being the jump between 
two successive metastable configurations.
A portion of the line of length $\ell$ moves
defining the \emph{extension} of the avalanche.
The area swept by the line during this jump is called the \emph{size} of the avalanche and is denoted by $S$
(see left panel of figure \ref{fig: Line SRE vs LRE}). 
The average width of an avalanche is given by equation \eqref{eq: def1 zeta}.
Hence the \emph{roughness exponent} gives the scaling of the size of an avalanche with its extension~:
\begin{equation}
S \sim \ell^{d} W(\ell) \sim \ell^{d+\zeta} \, . \label{eq: S sim ell^{d+zeta}}
\end{equation} 
Avalanches are the collective motion of many points of the interface.
Their extension has an upper bound that depends on a \emph{correlation length} $\xi$ that diverges as $f$ approaches $f_c$.
This divergence is characterized by the \emph{correlation length exponent} $\nu$~:
\begin{equation}
\xi \sim |f-f_c|^{-\nu} \, . \label{eq: def nu below threshold}
\end{equation}
Near the critical point the avalanche sizes have a broad power law distribution with a cutoff 
$S_{\text{cut}}$ that scales as $\xi^{d+\zeta}$~: 
\begin{equation}
P(S) \sim S^{-\tau} g_S\left(S/S_{\text{cut}}\right) \, , \label{eq:P(S) sim S^{-tau} }
\end{equation}
where $g_S(s)$ is a scaling function that is constant for $s \ll 1$ and presents a fast decay (e.g. exponential) for $s > 1$ and we introduced the \emph{avalanche size exponent} $\tau$.
$P(S)$ is sometimes simply refered to as the avalanche distribution.
At the critical point the cutoff diverges and the distribution becomes scale-free.
\newline
Finally the avalanches have a random \emph{duration} $T$. It is also power-law distributed near the critical point. Introducing an \emph{avalanche duration exponent} $\tilde{\alpha}$ we can write~:
\begin{equation}
P(T) \sim T^{-\tilde{\alpha}} g_T\left(T/T_{\text{cut}}\right) \, , \label{eq:P(T) sim T^{-talpha} }
\end{equation}
where $g_T$ is some scaling function of form similar as $g_S$.
As for the size, the duration of an avalanche scales as a power law of its extension. We define 
the \emph{dynamical exponent} $z$ by the relation~:
\begin{equation}
T \sim \ell^{z} \, . \label{eq: def exponent $z$}
\end{equation}
Combining \eqref{eq: def exponent $z$} and \eqref{eq: S sim ell^{d+zeta}} gives a scaling relation between the avalanche size and duration~:
\begin{equation}
S\sim T^{\gamma} \quad , \quad \gamma=\frac{d+\zeta}{z} \, . \label{eq: S sim T^{gamma}}
\end{equation}

\begin{figure}[t!]
	\begin{subfigure}[t]{0.48\linewidth}%
	\centering
	\includegraphics[width=1\linewidth]{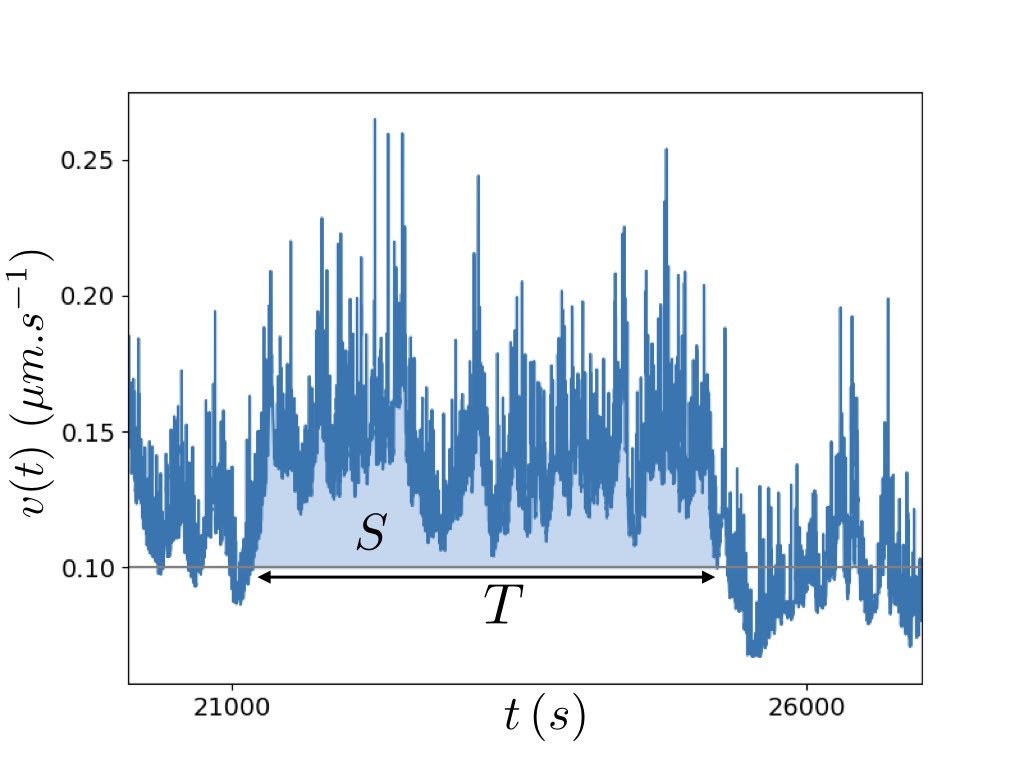}
	\caption{Instantaneous velocity signal of the center-of-mass of an interface. An avalanche is defined as a continuous period during which the center-of-mass velocity signal $v(t)$ exceeds a threshold $v_{th}$ (thin grey line). The duration $T$ and size $S$ of the avalanche are illustrated.\label{Fig: cdm velocity d'Alembert}}
	\end{subfigure}
	\hspace{.5cm}
    \begin{subfigure}[t]{0.48\linewidth}%
    \centering
	\includegraphics[width=1\linewidth]{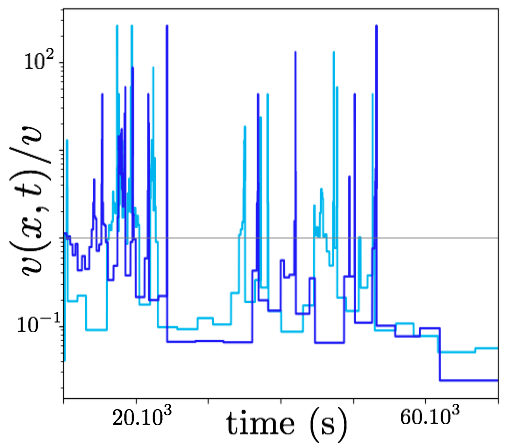}
	\caption{Velocity of two points close to each other along the line. The velocity is rescaled by the mean velocity of the line $v$. Each point has a very intermittent velocity and they are correlated. \label{Fig: 2-point velocity}}
	\end{subfigure}
	\caption{Experimental velocity signals. The data are issued from the d'Alembert crack experiment presented in section~\ref{sec: Ch2 Crack propagation}. The method to reconstruct the velocity signal is presented in appendix~\ref{sec: appendix experiment analysis}}
\end{figure}\label{Fig: ch1 Velocity signals}

\subsection{Above the threshold : intermittency and spatial correlations}
In experiments a mean velocity $v>0$ is often imposed to the interface. This is the most realistic way to trigger several avalanches in a finite amount of time. Hence the system is above the critical point and the transition is approached by making $v$ as small as possible.
In this case the line never completely stops and subsequents avalanches are triggered before the previous ones stop. This complicates the task of defining avalanches without ambiguity.

For a finite system, when $v$ is very small, $v\to 0^+$, the instantaneous velocity $v(t)$ of the center of mass is very \emph{intermittent}~:
it is very large for short period and very low for longer periods.
The mean velocity $v$ is the time average of the instantaneous velocity $v(t)$.
A standard procedure to define the avalanches is to threshold the velocity signal of the center-of-mass of the interface~:
avalanches are defined as the periods where $v(t)$ exceeds a threshold $v_{th}$. 
The duration $T$ of the avalanche is the time period where $v(t)$ stays above $v_{th}$ and the avalanche size $S$ is the integral of the difference between the signal and the threshold over this period (see figure \ref{Fig: cdm velocity d'Alembert}).
However this kind of analysis raises important issues :
if the threshold is too large, a single avalanche may be interpreted as a series of seemingly distinct events,  
while if it is too small subsequent avalanches can be merged into a single event~\cite{janicevic2016, bares2013}. 
Furthermore it does not take into account the spatial fluctuations of the velocity (a portion of the interface can be moving fast while other portions stay pinned) and local events can be skipped.
Hence disentangling avalanches and accurately measuring critical exponents are a major challenge.
In chapter~\ref{chap: chap3} I will propose an alternative method to determine the critical exponents, which is based on the correlations of the local velocity signal. 

In figure~\ref{Fig: 2-point velocity} I show the velocities of two points that are close to each other along the line in a crack propagation experiment~\cite{lepriol2020}. 
We see that they are intermittent, as is the center-of-mass velocity. We also see that the two points move fast or slow roughly at the same time~: there are spatial correlations in the system. 
The motion of the interface occurs through the motion of correlated domains. 
These domains have a maximal size that diverges as $f$ approaches $f_c$ from above as~:
\begin{equation}
\xi \sim (f-f_c)^{-\nu} \, . \label{eq: def nu above threshold}
\end{equation}
The exponent $\nu$ is the same in equations \eqref{eq: def nu above threshold}
and \eqref{eq: def nu below threshold}~\cite{narayan1993}. Hence the correlation length scales in the 
same way below and above threshold.

When the threshold force $f_c$ is approached from above the mean velocity $v$ vanishes.
By analogy with equilibrium critical phenomena we expect that it does so in a singular manner.
Thus we define the \emph{depinning exponent} $\beta$ that characterizes the onset of positive velocity at $f_c$~:
\begin{equation}
v \sim (f-f_c)^{\beta} \, . \label{eq: def beta}
\end{equation}

We have defined a set of six exponents $\zeta$, $\beta$, $\nu$, $z$, $\tau$ and $\tilde{\alpha}$. It is customary in equilibrium critical phenomena that the exponents characterizing the phase transition are not independent but linked by scaling relations. I will derive the scaling relations for the depinning transition in section \ref{sec: Ch1 STS and scaling relations}. 

\subsection{Effect of the temperature}\label{sec: ch1 effect of temperature}
Although within this thesis I focus exclusively on an interface at zero temperature, let me briefly mention the effect of a finite temperature $\tilde{T}$. 

When $\tilde{T}>0$, the thermal fluctuations allow the interface to explore the energy landscape around a metastable configuration. Some portion of the line can pass energy barriers
and reach states of lower energy. A small applied force induces an asymmetry in this exploration and
thus gives rise to a finite steady velocity $v>0$ (red dashed line in figure~\ref{fig: Phase diagram}). This phenomenon is called \emph{creep}. For 
$f \ll f_c$ the creep law for the velocity is~:
\begin{equation}
v \sim \exp \left( -\frac{C f^{-\mu}}{\tilde{T}} \right) \, , \label{eq: creep law}
\end{equation}
where $C$ is some non-universal constant and $\mu \simeq 1/4$ in $d=1$ \cite{ioffe1987, feigelman1989, nattermann1990, chauve2000}.
This formula has been experimentally confirmed in a magnetic experiment where the authors successfully fitted the velocity-force characteristics of a domain wall with a stretched exponential $v\sim \exp\left( f^{-0.25}\right)$ over eleven decades in velocity~\cite{lemerle1998} (I present magnetic domain walls experiment in section {\red \ref{sec: Ch2 Barkhausen noise}}).
Recent progress in numerical simulations enabled to explore in details the dynamics of the creep. 
This topic has been recently reviewed in Ref.~\cite{ferrero2020}. 

At $f_c$ the temperature induces a \emph{thermal rounding} of the transition. The velocity 
vanishes with the temperature as $v \sim \tilde{T}^{\psi}$.
Above $f_c$ a small temperature is irrelevant at scales below $\xi\sim (f-f_c)^{-\nu}$. At scales larger than $\xi$ the combination of the temperature and a finite velocity $v$ supersedes the pinning 
effect of the disorder~: the line advances as if there were only thermal noise and no quenched disorder~\cite{chauve2000, giamarchi2009}.

\subsection{Generalization to more complex interfaces}\label{sec: generalization overhangs bubbles}

\begin{figure}
	\centering
	\includegraphics[scale=0.45]{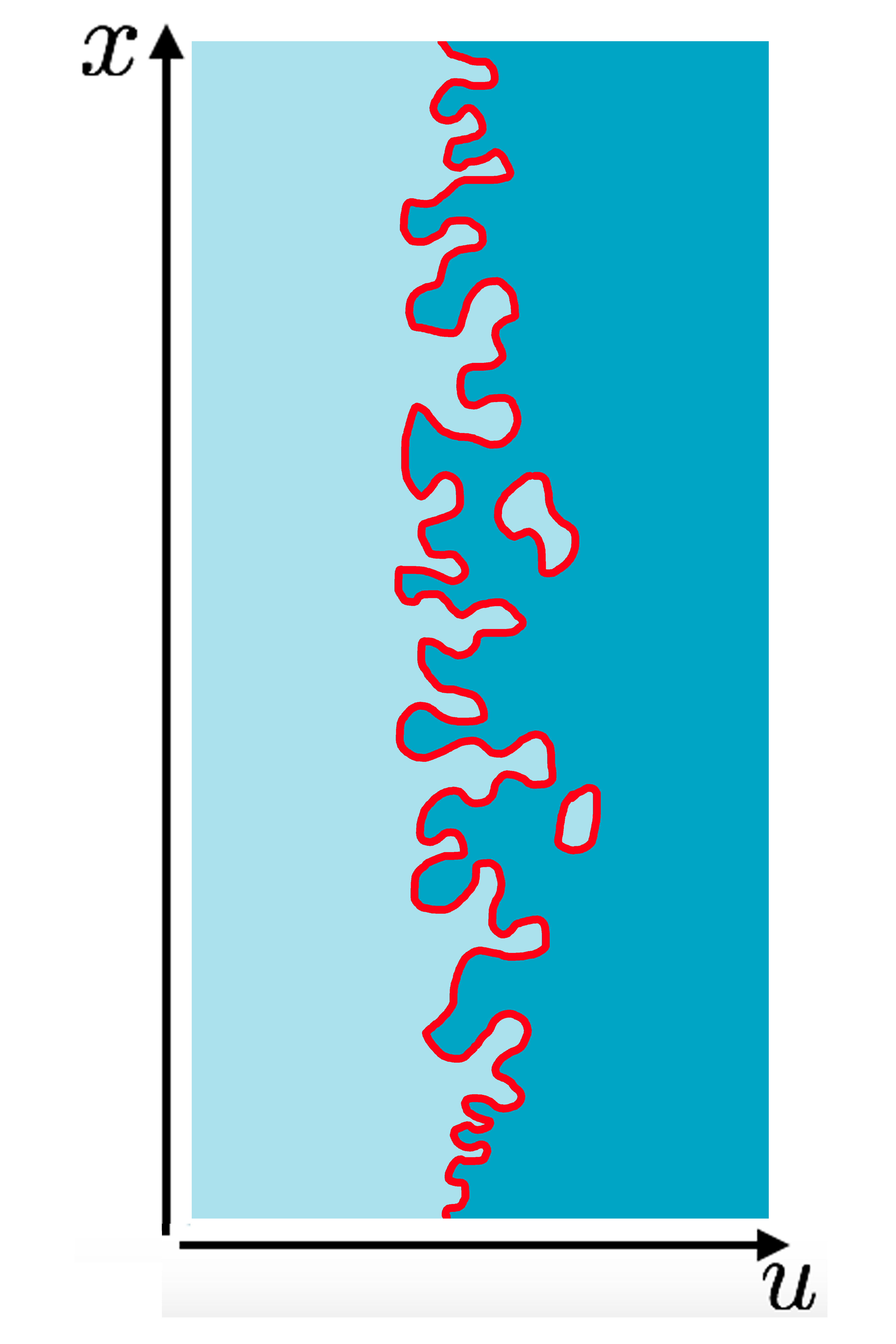}
	\caption{Sketch of an interface with overhangs and bubbles. These elements are not accounted for by the model studied in this thesis. \label{fig: Bad_line}} 
\end{figure}

An important assumption made in the description above is that the position of the interface 
is described by a univalued function $u(x)$. 
Figure~\ref{fig: Bad_line} shows a sketch of an interface with overhangs and bubbles. 
These elements are not accounted for by our model, although they are generally present in real interfaces.
Indeed the presence of such elements at small scales do not change the large-scale physics of the interface.
Thus the description can be simplified in order to make the model analytically tractable.
Recent works have focused on the accurate description of an interface like the one in figure~\ref{fig: Bad_line} with numerical simulations using a phase-field model based on a Ginzburg-Landau Hamiltonian and built a connection between this model and the theory of single-valued elastic interfaces~\cite{caballero2018, guruciaga2019, caballero2020}.
However in this thesis we shall only consider univalued interfaces.

\section{Equation of motion}\label{sec: Ch1 Eq of motion}

In this section I introduce the equation of motion for an elastic interface driven in a disordered medium. I consider the general case of a $d$ dimensional interface moving along a single transverse dimension. The internal coordinate of the interface is denoted by a vector $x\in \mathbb{R}^d$.
The position of the interface is a single valued time-dependent function $u(x,t)$.

We consider an overdamped regime where the inertia term is small in comparison to the friction term and can be neglected ($\mu\partial^2_{t} u(x,t) \ll \eta \partial_{t} u(x,t)$ where
$\mu$ is the mass density of the interface and $\eta$ is the friction coefficient). 
The local velocity, $v(x,t)=\partial_t u(x,t)$, is determined by the sum of the three forces mentioned above~:
\begin{equation}
\eta v(x,t) = \eta \partial_t u(x,t) = f(x,t) + F_{\el}(x,t) + F_{\dis} \left( x, u(x,t) \right) \label{eq: Motion 1}
\end{equation}
where $f(x,t)$ is the external driving force, $F_{\el}(x,t)$ is the elastic force and $F_{\dis}\left(x, u(x,t)\right)$ is the force resulting from the \emph{quenched} disorder.
Now I discuss the different forms for each force.

\subsection{The external force}
In section \ref{sec: Ch1 Phase diagram} we considered the case of an homogeneous external force f. 
This was convenient in order to draw the velocity-force characteristics in figure~\ref{fig: Phase diagram}. 
However, as we will see in chapter~\ref{chap: experiments}, in most experimental systems, the external force is better described by a spring pulling the interface. 
This motivates the use of a different convention~:
\begin{equation}
f(x,t) = m^2 \left( w(t) - u(x,t) \right) \, . \label{eq: F ext}
\end{equation}
The term $m^{2}$ corresponds to the stiffness of the spring. But in all this thesis we shall call $m$ the \emph{mass}.
The reason is that this term also arises naturally in the field theory approaches of this problem where it acts as an infrared cutoff control parameter, the critical point being reached when $m \to 0$. 
The name \emph{mass} originates from quantum field theory where, when dealing with particles, it is the mass of particles that set the infrared cutoff.
Concerning the disordered elastic interfaces, it was introduced by M{\'e}zard and Parisi in the replica field theory~\cite{mezard1991}
and by Le Doussal, Wiese and Chauve in the FRG, first for the depinning transition~\cite{ledoussal2002} and then for the study of an interface at equilibrium~\cite{ledoussal2004}.
The effect of the mass is to confine the interface in a harmonic well of curvature $m^2$ centered at the position $w(t)$ and implies an upper bound on the fluctuations of the line at large scales. 
We can associate to this force a time dependent confining Hamiltonian~:
\begin{align}
f(x,t) &= -\frac{\delta}{\delta u(x,t)} \mathcal{H}_{w(t)}[u] \, , \\
\mathcal{H}_{w}[u] &= \frac{m^2}{2} \int \left(w-u(x) \right)^2 d^d x \label{eq: confining Hamiltonian} \, .
\end{align}
The confining Hamiltonian~\eqref{eq: confining Hamiltonian} was introduced in~\cite{ledoussal2006a, ledoussal2010a} to put the FRG on a more rigorous setting 
and it allowed the numerical determination of the FRG fixed point functions~\cite{middleton2007, rosso2007}.
Indeed the driving \eqref{eq: F ext} is also very convenient for numerical simulations for two reasons.
The first one is that the interface can be driven at finite mean velocity $v$ by setting $w(t)=vt$.
The second is that we can simulate as many avalanches as desired (which is necessary for computing probability distributions) in a quasi-static setting by slowly
incrementing $w$ in order to destabilize only one point of the interface and then let the system
freely evolve.

\begin{figure}
	\centering
	\includegraphics[scale=0.5]{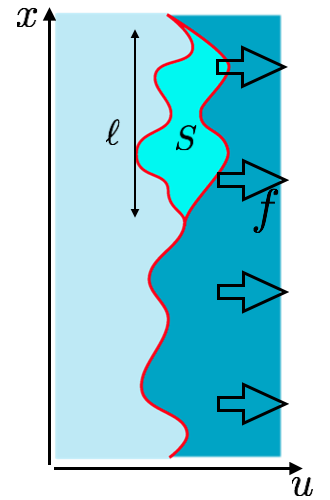}\includegraphics[scale=0.5]{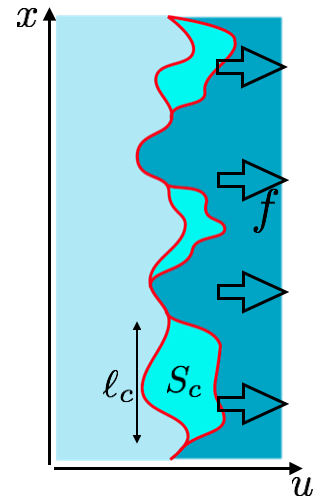}
	\caption{Illustration of the difference between short-range (SR) and long-range (LR) elasticity. 
	\textit{Left : SR elasticity.} The avalanche occurs in a single connected component. 
	\textit{Right : LR elasciticy.} The avalanche can be made of several spatially separated connected components, called \emph{clusters}. The moving portions were marginally stable before the avalanche, while the portions in between remain pinned by stronger disorder forces. 
	\label{fig: Line SRE vs LRE}} 
\end{figure} 

\subsection{The elastic force}\label{sec: ch1 elastic force}

In the case of an elastic line with short-range elasticity a portion of the line is only pulled by its nearest neighbors.
The elastic force thus only depends on the local curvature of the line and it can be expressed as a power series in the gradient $\nabla_x u(x)$. 
Since the elasticity must preserve the central symmetry $u(x) \rightarrow u(-x)$ only even power are allowed.
One can show that in the large scale limit only the lowest order term is relevant (see e.g.~\cite{rosso2002}). The minimal version of the short-range (SR) elastic force and the associated hamiltonian
are thus~:
\begin{align}
F_{\el}^{\SR}(x) &= c\nabla_x^{2} u(x) \label{eq: Fel SR} \, , \\
\mathcal{H}_{\el}^{\SR}[u] &= \frac{c}{2} \int \left(\nabla_x u(x)\right)^2 d^d x  \label{eq: Hamiltonian el SR} \, ,
\end{align}
where $c$ is a dimensionless coefficient that sets the strength of the elastic interactions and we have the relation
$F_{\el}^{\SR}(x) = -\frac{\delta}{\delta u(x)}\mathcal{H}_{\el}^{\SR}[u]$.
Equation \eqref{eq: Motion 1} with the elastic force \eqref{eq: Fel SR} is called the
quenched Edwards-Wilkinson equation.

When considering crack fronts, wetting fronts and some magnetic domain walls, the elastic interface is an effective description for forces that can be mediated through the embedding medium (these systems are presented in chapter~\ref{chap: experiments}).
This leads to long-range (LR) interactions and to non-local elasticity.
In the most general case, long-range elastic forces can be written with an arbitrary kernel~$\mathcal{C}$~:
\begin{align}
F_{\el}^{\LR} (x,t) &=  \int \mathcal{C}(x-y)\left( u(y,t)-u(x,t)\right) d^d y  \, ,\label{eq: F el LR generalized}\\
\mathcal{H}_{\el}^{\LR}[u] &= \frac{1}{2} \int \mathcal{C}(x-y)\left(u(y)-u(x)\right)^2 d^dxd^dy \label{eq: Hamiltonian el LR} \, .
\end{align}
For an elastic interaction the kernel must be positive\footnote{Here the word kernel must not be understood in its mathematical sense, mathematically it is actually a function of the variable $x$.}~:
$\mathcal{C}(x) \geq 0$ for all $x\neq 0$. Also there must be some $x$ where $\mathcal{C}(x) > 0$ as otherwise there would be no interaction. Note that $\mathcal{C}(0)$ does not need to be defined as it multiplies $0$ in \eqref{eq: F el LR generalized}.
We shall only consider isotropic interfaces where the strength of the interaction only depends on the distance between points~:
$\mathcal{C}(x) = \mathcal{C}\left(|x|\right)$.
In the case of the experimental systems just mentioned above the kernel and the elastic force take the forms~:
\begin{align}
&\mathcal{C}(y-x)=\frac{c}{|y-x|^{d+\alpha}}\, , \text{ with }  \alpha=1 \, , \label{eq: LR elasticity kernel} \\
&F_{\el}^{\LR}(x) = c\int \frac{u(y)-u(x)}{|y-x|^{d+\alpha}} dy \, . \label{eq: F el LR}
\end{align} 
As $|y-x|^{-(d+\alpha)} \to +\infty$ when $y\to x$ one could worry that the integral in~\eqref{eq: F el LR} diverges. But the stronger the interaction is, the closer $u(y)$ and $u(x)$ are attracted. So we can expect that $u(y)-u(x)$ vanishes sufficiently fast when $y\to x$ so that the integral is convergent. Another reason to expect the convergence is that physicaly the elastic force must be finite. 
Only $\alpha=1$ corresponds to experimental systems. But we will nevertheless study the effect of the long-range kernel~\eqref{eq: LR elasticity kernel} for different values of $\alpha$. 
Indeed it allows to better understand what happens for $\alpha=1$ by observing wich quantities depend on $\alpha$ and which do not.

A salient difference between short-range and long-range elasticity is illustrated in figure~\ref{fig: Line SRE vs LRE}. When the elasticity is short-ranged a point of the interface starts to move only if a neighboring point did so. Hence avalanches occur in a single connected component. 
On the other hand with a long-range kernel the motion of a portion of the interface can trigger the 
instability of points at finite and even large distances. 
So the avalanche can be made of several connected 
components, spatially separated by portions of the interface that are more stable. 
These components are called \emph{clusters}. A major part of the present work was to study the statistics of the clusters and their relation with the statistics of the global avalanche. This will be presented in chapter~\ref{chap: cluster statistics}.

\paragraph{Fourier transform of the elastic force}
It will be useful to write the elastic force in Fourier space.
For the sake of simplicity, I restrain here to $d=1$.
The Fourier transform and inverse Fourier transform of a function $h$ are defined as~:
\begin{equation}\label{eq: def Fourier Transform.}
\hat{h}(q) = \int_{-\infty}^{\infty} e^{-iqx} h(x) dx \, , \quad
h(x) = \int_{-\infty}^{\infty} e^{iqx} \hat{h}(q)  \frac{dq}{2\pi}
\end{equation}
With this definition the Fourier transforms of the elastic forces \eqref{eq: Fel SR} and \eqref{eq: F el LR generalized} read~:
\begin{align}
\hat{F}_{\el}^{\SR} (q,t) &= - c q^{2} \hat{u}(q,t)  \, , \label{eq: F el SR Fourier} \\
\hat{F}_{\el}^{\LR} (q,t) &= \left( \hat{\mathcal{C}}(q) - \hat{\mathcal{C}}(0)\right)\hat{u}(q,t) \label{eq: F el LR Fourier generalized} 
\end{align}
and the equation of motion \eqref{eq: Motion 1} becomes~:
\begin{equation}\label{eq: Motion Fourier}
\eta \partial_t \hat{u}(q,t) = \hat{f}(q,t) + \hat{F}_{\el}(q,t) + \hat{F}_{\dis} \left( q, g[u] \right) 
\end{equation}
where $g[u]$ is some functional that depends on the configuration of the interface.

If we specify the LR kernel to be of the form~\eqref{eq: LR elasticity kernel} we can compute its Fourier transform.
I focus on its scaling here and give the details of the computation in the appendix~\ref{sec:Appendix Fourier Transform elastic force}. 
For $0<\alpha<2$, we have\footnote{Note that when computing the Fourier transform there is a divergence for $x\to 0$ so that $\hat{\mathcal{C}}(q) = +\infty$.
However we are interested in the term $\left( \hat{\mathcal{C}}(q) - \hat{\mathcal{C}}(0)\right)$ which remains finite for $0< \alpha < 2$.}~:
\begin{equation}\label{eq: C(q)-C(0) chap 1}
\hat{\mathcal{C}}(q) - \hat{\mathcal{C}}(0) = c\int \frac{\cos(qx) - 1}{\abs{x}^{1+\alpha}} dx \sim -|q|^{\alpha} \, .
\end{equation} 
If $\alpha>2$ there are even power of $q$ up to $\alpha$ that comes out.
The large scale behavior we are interested in corresponds to the small $q$ regime in Fourier space.
For $\alpha > 2$ the elastic force is dominated by $q^2 \hat{u}(q,t)$ while it is proportional to 
$|q|^{\alpha} \hat{u}(q,t)$ for $\alpha <2$.
It is thus usual to generically write the elasticity force in Fourier space as 
\begin{equation}
\hat{F}_{\el} (q,t) = - c_{\alpha} |q|^{\alpha} \hat{u}(q,t) \, , \label{eq: F el LR Fourier}
\end{equation}
with the understanding that $\alpha=2$ corresponds to short-range elasticity. 
This statement may sound surprising. 
Indeed when considering the elasticity in real space the kernel
$\mathcal{C}(y-x) = \frac{c}{|y-x|^{d+2}}$
is very different from the short-range elasticity $c\nabla_x^2$.
There should always be differences since the possibility of having several clusters in an avalanche can not be removed with the long-range kernel, no matter the value of $\alpha$.
The statement "$\alpha=2$ corresponds to short-range elasticity" can be made more precise in the following way~: 
For a long-range kernel with $\alpha \geq 2$ the critical exponents defined in section \ref{sec: Ch1 Phase diagram} are expected to take the same values as for the interface with short-range elasticity.
Or in other words the interface with LR elasticity and $\alpha \geq 2$  belongs to the same universality class as the one with SR elasticity.
So we expect to have very few clusters and the avalanche size to be overwhelmingly dominated by a single cluster.

\paragraph{A novel characteristic length}
Taking the Fourier transform of equation \eqref{eq: Motion 1} we see that the external force 
\eqref{eq: F ext} and the elastic force \eqref{eq: F el LR Fourier} combine into a total term
equal to $\left( m^2 + c_{\alpha}|q|^{\alpha} \right) \hu(q,t)$. Since $c_{\alpha}$ is dimensionless, $m^{2/\alpha}$ has the dimension of an inverse length. So the mass induces a characteristic length~:
\begin{equation}
\ell_m \sim m^{-2/\alpha} \, . \label{eq: def xi_m }
\end{equation}
Because the mass sets the distance to the critical point we can identify this length with the one 
in equation \eqref{eq: def nu below threshold}. This leads to the scaling relation~:
\begin{equation}
m^2 \sim |f-f_c|^{\alpha \nu} \, . \label{eq: scaling m2 |f-fc|}
\end{equation}

\subsection{The disorder force}\label{sec: Ch1 disorder force}
The disorder force derives from a disorder potential $F_{\dis}(x,u) = -\partial_u V(x,u)$. We define the correlations of the disorder using the covariance functions : 
\begin{align}
\overline{V(x,u) V(x',u')}^c &= \delta^{d}(x-x') R_0(u-u') \, , \label{eq: def correl potential} \\
\overline{F_{\dis}(x,u) F_{\dis}(x',u')}^c &= \delta^{d}(x-x') \Delta_0(u-u') \, , \label{eq: def correl disorder force} \\
\Delta_0 (u) &= -R_0''(u) \, , \label{eq: Delta_0 (u) = -R_0''(u)}
\end{align}
where $R_0''$ is the second derivative of the single-variable function $R_0$, $\overline{\mbox{...\rule{0pt}{2.5mm}}}$
denotes the average over disorder and the superscript $...^c$ stands for the connected correlations  (i.e. cumulants, or covariances): 
$\overline{V(x,u) V(x',u')}^c = \overline{V(x,u) V(x',u')} - \overline{V(x,u)^2}$.

\begin{figure}
	\centering
	\includegraphics[scale=0.5]{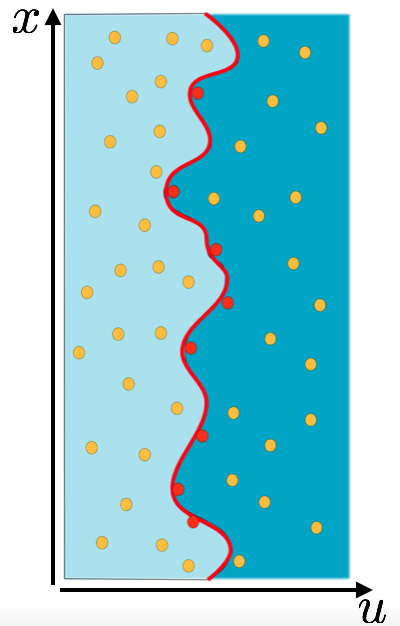}\includegraphics[scale=0.5]{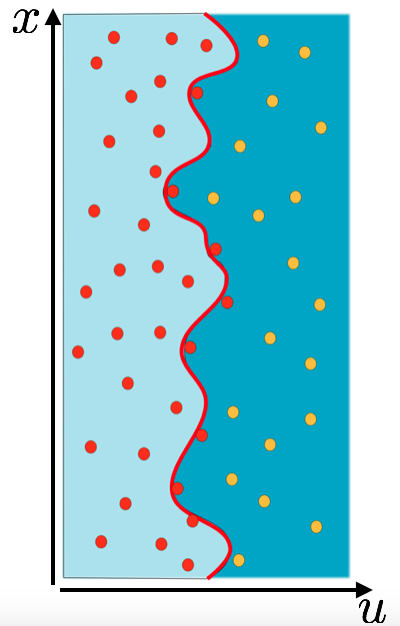}
	\caption{There are two possibilities for the disorder. \textit{Left :} Random Bond disorder. Only the impurities which are located on the line (in red) contribute to the pinning. \textit{Right :} Random Field disorder. The impurities favor one of the two phases separated by the interface (for instance they can favor an upward magnetization in a disordered magnet). Hence all the impurities located on one side of the interface contribute to the pinning. \label{fig: Line RB vs RF}} 
\end{figure} 

The impurities can couple to the interface in two very different ways leading to two universality classes of the disorder called Random Bond and Random Field. These two kind of couplings are illustrated in figure~\ref{fig: Line RB vs RF}. 

\paragraph{Random Bond (RB)}
The impurities act effectively only on the interface. In this case we can associate to the disorder a Hamiltonian
\begin{equation}
\mathcal{H}_{\dis}^{\RB}[u] = \int V \left(x,u(x) \right) dx \label{eq: Hamiltonian disorder RB}
\end{equation}
The potential has short-range correlations. They typically decay as 
$R_0(u) \sim e^{-u/r_f}$ where $r_f$ is the range of the correlations and is typically given by
the size of the impurities in the system or the thickness of the interface (in the model we assume 
for simplicity that the interface is infinitely thin but a real interface necessary has a finite thickness). 

\paragraph{Random Field (RF)}
The impurities couple differently to both sides of the interface. For instance in an uniaxial ferromagnet the impurities can favor an upward magnetization.
Thus the disorder energy involves an integral not only at the interface but over one whole phase.
We can take into account this coupling by writing~:
\begin{equation}
\mathcal{H}_{\dis}^{\RF}[u] = \int V^{\eff} \left(x,u(x) \right) dx \quad ; \quad
V^{\eff} \left(x,u \right) = \int_0^{u} V(x,z) dz \, .
\label{eq: Hamiltonian disorder RF}
\end{equation}
The effective potential has long-range correlations and 
$F_{\dis}(x,u) = -\partial_u V^{\eff} (x,u) = -V(x,u)$ so the disorder force is short-range correlated with the same correlations as the potential for the Random Bond disorder. 

\vspace{5pt}
In both cases, RB and RF, the force has short-range correlations that decay over a typical scale $r_f$ as
\begin{equation}
\Delta_0 (u) \sim e^{-u/r_f} \, . \label{eq: disorder correlation decay}
\end{equation}

$R_0$ and $\Delta_0$ are the correlator of the microscopic disorder, also called \emph{bare disorder}.
Under the renormalization procedure these correlator change form and flow to fix points that we denote $R$ and $\Delta$ respectively. These fix points correspond to the effective disorder felt at large scale by the interface. 
At equilibrium all nonperiodic microscopic disorder flow either to a RB or a RF fix point. Thus the properties of the interface, such as its roughness, depend on the universality class, RB or RF, of the disorder. 
But at the depinning transition all (nonperiodic) disorder flow to the RF fix point, hence giving rise to the same large-scale physics~\cite{chauve2000, ledoussal2002, rosso2007}.

\section{The Larkin length and the upper critical dimension}\label{sec: Ch1 Larkin length and d_uc}

We consider the balance between the disorder and the elastic force on a portion of interface of extension $L$ and width $W(L)$. The typical elastic force on such a portion estimated from 
\eqref{eq: F el LR} is 
\begin{equation}
f_{\el} \sim L^d \frac{c W(L)}{L^{\alpha}} \, .
\end{equation}
On the other hand the typical pinning force estimated from \eqref{eq: def correl disorder force} is 
\begin{equation}
f_{\text{pin}} \sim \sqrt{\frac{\Delta(0)}{L^d}} \, .
\end{equation}
The balance of the two forces yields 
\begin{equation}
L^{2\alpha-d} = \frac{ \left( c W(L)\right)^2}{\Delta(0)} \, .
\end{equation}
A characteristic length $L_c$ appears when $W(L)$ reaches the correlation length of the disorder $r_f$, $W(L_c) = r_f$, leading to
\begin{equation}\label{eq: Larkin length}
L_c = \left( \frac{c^2 r_f^2}{\Delta(0)} \right) ^{\frac{1}{2\alpha-d}} \, .
\end{equation}
$L_c$ appeared in a model by Larkin \cite{larkin1970} and is known as the \emph{Larkin length}.
Below the Larkin length, for $L<L_c$, the balance of disorder and elasticity tells us that the interface wanders as
\begin{equation}\label{eq: Zeta Larkin}
W(L) \sim L^{\zeta_L} \quad ; \quad \zeta_L = \frac{2\alpha-d}{2} \, ,
\end{equation} 
where $\zeta_L$ is the roughness exponent from the Larkin model. 
Within this model the interface width grows unboundedly with $L$ for
$d< d_{\text{uc}}=2 \alpha$, i.e. $\zeta_L>0$, but remains bounded for $d > d_{\text{uc}}$. 
Here $d_{\text{uc}}=2 \alpha$ is called the \emph{upper critical dimension} and plays a special role. 

The Larkin model assumes a single energy minimum and breaks down for $L>L_c$. 
Indeed above the Larkin length one must take into account the existence of many metastable states and the scaling~\eqref{eq: Zeta Larkin} breaks down.
The interface can be viewed as a collection of domains of size $L_c$ that are independently pinned.
A more detailed analysis for $L>L_c$ shows that~:
\begin{itemize}
\item if $d > d_{\text{uc}}$ the elastic energy dominates over the disorder. The interface appears to be flat at large scales which means that the roughness exponent is $\zeta=0$.
In this case the system falls into the \emph{mean-field} universality class. Mean-field models will be described in chapter~\ref{chap: chap5}.
\item if $d < d_{\text{uc}}$ the elastic and the disorder energy are of the same order
and their competition results in a non trivial roughness exponent $\zeta>0$. Hence at large scales the interface is rough and the transversal displacement can grow unboundedly. 
\end{itemize}
For $d < d_{\text{uc}}$ if the system size is $L<L_c$ there is no avalanche. The system has a purely elastic response with no intermittence.
On the other hand when $L$ becomes larger than $L_c$ the intermittent avalanche dynamics appear. 
The crossover between these two behaviors has been observed in Ref. \cite{tanguy2004}.

\section{Scaling relations between exponents}\label{sec: Ch1 STS and scaling relations}

In section \ref{sec: Ch1 Phase diagram} I introduced exponents characterizing the depinning transition. As for equilibrium phase transitions we can establish relations between the exponents using scaling arguments. In this section I derive four relations, leaving only two independents exponents out of six.

From equations \eqref{eq: def nu above threshold}, \eqref{eq: def beta} $v$ scales as $\xi^{-\beta/\nu}$. 
From \eqref{eq: def1 zeta} and \eqref{eq: def exponent $z$} the mean velocity inside an avalanche of extension $\ell$ scales as $\ell^{\zeta-z}$. 
Making the assumption that the mean velocity inside the largest avalanches (of extension $\xi$)
is proportional to the mean velocity of the interface we can say that $v\sim \xi^{\zeta-z}$ and
we have the first relation~:
\begin{equation}
\beta = \nu (z-\zeta) \, . \label{eq: scaling relation beta nu z zeta}
\end{equation}

Sizes and durations are defined for each avalanche. Because they are defined for the same events and scale together the probability laws are equal~: $P(S)dS=P(T)dT$. 
Inserting the scaling relation \eqref{eq: S sim T^{gamma}} in the power-regime of the equality we can write 
$T^{-\gamma \tau} T^{\gamma-1} dT \sim T^{-\tilde{\alpha}}dT$ with $\gamma=(d+\zeta)/z$. This yields the second relation~:
\begin{equation}
\gamma = \frac{d+\zeta}{z} = \frac{\tilde{\alpha}-1}{\tau-1} \, .\label{eq: scaling relation ta tau gamma}
\end{equation}

\vspace{0.5cm}
The last two relations derive from a symmetry of the problem called \emph{Statistical Tilt Symmetry}.
 
\paragraph{Statistical Tilt Symmetry}
We consider an interface with a general long-range elastic kernel $\mathcal{C}$ subject to an
homogeneous external force $f$. 
We tilt the disorder by adding to the right-hand side of equation \eqref{eq: Motion 1} a field 
$F_{\tilt}(x)$ that only depends on $x$ and has zero spatial average~: $\int F_{\tilt}(x) dx =0$. 
What is the response of the interface to such a tilt ?
The tilt can be absorbed in a new displacement field that is conveniently defined in Fourier as~:
\begin{equation}
\hat{\tu}(q,t) = \hat{u}(q,t) + \frac{1}{\hat{\mathcal{C}}(q)-\hat{\mathcal{C}}(0)}\hat{F}_{\tilt}(q) \, . \label{eq: STS def new displacement field}
\end{equation}
We write the temporal evolution of $\tu$ in Fourier using equations \eqref{eq: F el LR Fourier generalized} and \eqref{eq: Motion Fourier}~:
\begin{align}
\eta \partial_t \hat{\tu}(q,t) &= \eta \partial_t \hat{u}(q,t)
    = \hat{f}(q,t)+ \left( \hat{\mathcal{C}}(q)-\hat{\mathcal{C}}(0) \right) \hat{u}(q,t) +\hat{F}_{\dis}\left(q, g[u] \right) +
    \hat{F}_{\tilt}(q) \notag \\
    &= \hat{f}(q,t) + \left( \hat{\mathcal{C}}(q)-\hat{\mathcal{C}}(0) \right) \left( \hat{\tu}(q,t) - \frac{1}{\hat{\mathcal{C}}(q)-\hat{\mathcal{C}}(0)}\hat{F}_{\tilt}(q) \right) +\hat{F}_{\dis}\left(q, g[u] \right) +\hat{F}_{\tilt}(q) \notag \\
    &= \hat{f}(q,t) + \left( \hat{\mathcal{C}}(q)-\hat{\mathcal{C}}(0) \right) \hat{\tu}(q,t) +\hat{F}_{\dis}\left(q, g[u] \right) \label{eq: STS new eq of motion Fourier}
\end{align}
Coming back to real space we see that the new displacement field obeys the following equation of motion~:
\begin{equation}
\eta \partial_t \tu(x,t) = f(x,t) + \int \mathcal{C}(y-x)\left(\tu(y)-\tu(x)\right) dy + 
F_{\dis}\left(x, \tu(x,t) - \mathcal{S}(x)\right) \, 
\label{eq: STS def new eq of motion}
\end{equation}
where $\mathcal{S}(x)$ is the inverse Fourier Transform of $\frac{1}{\hat{\mathcal{C}}(q)-\hat{\mathcal{C}}(0)}\hat{F}_{\tilt}(q)$.
The effect of the tilt has been absorbed into the new displacement field  with a shift in the disorder field. This is the so-called Statistical Tilt Symmetry.

We are interested in the statistical response of the interface to the tilt, which means we have to take the average over disorder. Since the correlation function \eqref{eq: def correl disorder force} 
$\overline{F_{\dis}(x, u)F_{\dis}(x', u')}$ only depends on the difference $|u-u'|$ the field 
$F_{\tilt}$ disappears from the average $\overline{\tu}$. 
Hence the response function of the original field to the tilt is~:
\begin{equation}
\chi_{q} = \frac{\partial \overline{\hat{u}}(q)}{\partial \hat{F}_{\tilt}(q)} = 
\frac{\partial \left(\overline{\hat{\tu}}(q) - \left( \hat{\mathcal{C}}(q)-\hat{\mathcal{C}}(0) \right)^{-1} \hat{F}_{\tilt}(q)\right)}{\partial \hat{F}_{\tilt}(q)} = 
-\left( \hat{\mathcal{C}}(q)-\hat{\mathcal{C}}(0) \right)^{-1} \sim |q|^{-\alpha} \, . \label{eq: STS response function}
\end{equation}
In the last scaling I assumed a kernel $\left(\hat{\mathcal{C}}(q) - \hat{\mathcal{C}}(0) \right) \sim - |q|^{\alpha}$.
A dimensional analysis of equation \eqref{eq: STS response function} allows us to establish new scaling relations between the exponents of the depinning transition.
I use the square brackets $[..]$ to denote the dimension of the quantities. $[\xi]$ is the dimension
of a length along the interface. 
The displacement field has the dimension $[u] = [\xi]^{\zeta}$ and the tilt has the dimension of a force $[F] = [\xi]^{-1/\nu}$. So the response function to the tilt has the dimension
$[\chi] = [\xi]^{\zeta + 1/\nu}$ which must be equal to $[\xi]^{\alpha}$ from \eqref{eq: STS response function}. This yields the second scaling relation~:
\begin{equation}
\nu = \frac{1}{\alpha - \zeta} \, . 
\end{equation}\label{eq: scaling relation nu zeta alpha}

To derive the third relation we consider the mean polarizability of the interface below $f_c$ i.e.
the mean displacement of the interface upon an infinitesimal increment $df$ in the applied force~:
\begin{equation}
\chi_{\uparrow} = \frac{d\langle u \rangle}{df}
\end{equation}
The advance of the interface is the product of
the average size of the avalanches triggered when $f \rightarrow f+df$ times
the number of avalanches triggered 
divided by the interface size $L^d$.
The avalanche size distribution is a power law of exponent $\tau>1$ with a small scale cutoff $S_0$ that depends on the microscopic details of the problem and a large scale cutoff 
$S_{\text{cut}} \sim \xi^{d+\zeta}$. Hence the mean avalanche size scales as 
$\langle S \rangle \sim S_0^{\tau-1} S_{\text{cut}}^{2-\tau}\sim \xi^{(d+\zeta)(2-\tau)}$.
The number of avalanches triggered is proportional to the system size and to $df$. Introducing
the density of avalanches $L^d \rho_f$ we can write 
$\chi_{\uparrow} =\rho_f \langle S \rangle$.
We assume that the density of avalanches remains finite as $f$ approaches $f_c$ so we have a 
divergence in the mean polarizability that is only induced by the divergent length $\xi$~:
$\chi_{\uparrow} \sim\xi^{(d+\zeta)(2-\tau)}$.
On the other hand since $\chi_{\uparrow}$ has the same dimension as $\chi_q$ in \eqref{eq: STS response function} we can expect that it diverges  as 
$\chi_{\uparrow} \sim\xi^{\alpha}$.
This yields the last scaling relation~\cite{narayan1993}~:
\begin{equation}
\tau = 2 - \frac{\alpha}{d+\zeta} \, . \label{eq: scaling relation tau alpha zeta}
\end{equation}
This relation, and its generalization to other observables, was also 
obtained with the FRG field theory in~\cite{dobrinevski2014}.

\section{Middleton theorems}\label{sec: Ch1 Middleton thm}

Our quenched problem has a glassy energy landscape with many metastable states. 
This can render the analysis of the dynamics very difficult. For instance one could fear the dynamics to be chaotic with divergent solutions or multiple attractors. Three theorems first stated by Middleton \cite{middleton1992a} in 1992 and then rigorously proven in Ref.~\cite{baesens1998}
\footnote{although only in the SR elasticity framework}
simplify the analysis of the dynamics by showing that there is a unique attractor for the dynamics when the driving is monotonous (i.e. $\partial_t f(x,t) \geq 0$ for all $x$ and $t$)~: the solutions of the problem for a given monotonous driving starting in different initial configurations will reach the same configuration after some transient time. So after a transient time the evolution of the system does not depend on its previous history. In this section I state the three theorems and give some proofs but not all. 
Especially I prove the first theorem and deduce a consequence which is the uniqueness of the mean velocity $v$ for a given force $f$. 
For the second theorem I prove only a weaker version and I do not attempt to prove the third one which is more difficult to prove rigorously.
In this section I use dots to denote time derivative : $\du(x,t) = \partial_t u(x,t)$.

The first theorem states that if two configurations do not cross at a given time then they will never cross.
\begin{theorem}[No Passing Rule]
If $u^{(1)}$ and $u^{(2)}$ are two solutions of the dynamics given by equation \eqref{eq: Motion 1} such that $u^{(1)}$ is behind $u^{(2)}$ at some time $t_0$, i.e. $\forall x$, $u^{(1)}(x,t_0) \leq u^{(2)}(x,t_0)$, then $u^{(1)}$ will stay behind $u^{(2)}$ for all time $t$ greater than $t_0$~:
$\forall x$, $\forall t \geq t_0$, $u^{(1)}(x,t) \leq u^{(2)}(x,t)$.
\end{theorem}

To prove this theorem let us assume that $u^{(1)}$ catches up $u^{(2)}$ and let us take the time $t_1$ when this happens for the first 
time (note that with the hypothesis of the theorem we can have $t_1=t_0$) and a point $x_1$ where 
$u^{(1)}(x_1,t_1) = u^{(2)}(x_1,t_1)$. At this point both solutions feel the same external force $f(x_1,t_1)$ and the same disorder force $F_{\dis}\left(u^{(2)}(x_1,t_1)\right)$. Only the elastic forces differ. We take an 
elastic force with an arbitrary elastic kernel
$F_{\el}(x,t) = \int \mathcal{C}(x-y)\left(u(y,t)-u(x,t)\right) dy$. The difference of the velocities at position $x_1$ and time $t_1$ is~:
\begin{equation}
\eta \left( \du^{(2)}(x_1,t_1) - \du^{(1)}(x_1,t_1)\right) = 
\int \mathcal{C}(x-y)\left(u^{(2)}(y,t_1)-u^{(1)}(y,t_1)\right) dy \geq 0
\end{equation}
where the inequality arises from the positivity of $\mathcal{C}(x)$ and the fact that $u^{(2)}$ is still in front of $u^{(1)}$ at $t_1$. Hence if $u^{(1)}$ cathes up $u^{(2)}$ at some point, the velocity of $u^{(2)}$ at this point is greater or equal than the one of $u^{(1)}$ so $u^{(1)}$ cannot pass $u^{(2)}$.
\vspace{4pt}

The \emph{no passing rule} has two important consequences that are useful to study the dynamics of the system. 

First if for a given constant and homogeneous external force $f$ there is one stable configuration in the system then all solutions coming from the left of this configuration must stop before (or at) this configuration. Hence the existence of one solution with $v=0$ implies that the steady-state of all solutions coming from the left is also $0$.
This implies in turn that there is a unique critical force $f_c$ at which the steady-state mean velocity becomes non zero.

A second consequence is the uniqueness of the mean velocity for a given $f$ above $f_c$. To see this point let us consider two solutions $u^{(1)}$ and $u^{(2)}$ of the dynamics \eqref{eq: Motion 1} with $f(x,t)=f$ constant and homogeneous such that $u^{(1)}$ is behind $u^{(2)}$ at some initial time $t_0$ and let us note $v^{(1)}$ and $v^{(2)}$ their mean steady-state velocity that we assume to be strictly positive. Since $u^1$ can not pass $u^2$ we necessarily have $v^{(1)} \leq v^{(2)}$. Now I argue that for a finite system of size $L^d$ there must be some time $t_1 > t_0$ such that $u^{(1)}(t_1)$ has completely passed in front of $u^{(2)}(t_0)$. Indeed for a finite system the interface at any time is bounded in a band of width 
proportional to $L^{\zeta}$ around its center-of-mass position. Since $v^{(1)} >0$ there exists some time $t_1$ at which the center of mass of $u^{(1)}(t_1)$ will have passed the one of $u^{(2)}(t_0)$ by more than the few $L^{\zeta}$ necessary for having $u(x,t_1) > u(x,t_0)$ for any point $x$. 
The time translation of $u^{(1)}$, $\tu^{(1)}(t) = u^{(1)}(t + t_1)$ is in front of $u^{(2)}$ at time $t_0$ and has mean velocity $v^{(1)}$. Repeating the same argument as above we hence have $v^{(2)} \leq v^{(1)}$. This proves that $v^{(1)}=v^{(2)}$ and thus that the steady state velocity is unique for a given force and that the velocity force characteristics drawn in figure~\ref{fig: Phase diagram} is defined without any ambiguity.

\vspace{7pt} 
I now present the second theorem.
\begin{theorem}[Monotonicity]
Let $u$ be a solution of the dynamics given by equation \eqref{eq: Motion 1} with a monotonous driving
($\dot{f}(x,t) \geq 0$ or $\dot{w}(t) \geq 0$ if the driving is of the form \eqref{eq: F ext}).
If there is a time $t_0$ such that (i) the velocity of all points is positive $(\forall x$, $\du(x,t_0) \geq 0)$ and (ii) at least one point has a strictly positive velocity  $(\exists x_0$ such that $\du(x_0, t_0) >0)$ ; 
then the motion of the interface is strictly monotonous~: $\forall x$, $\forall t > t_0$,
$\du(x,t) > 0$.
\end{theorem}

This theorem is actually anterior to Middleton's paper. It was mathematically proven in 1985 in Ref.~\cite{hirsch1985}. Rigorously proving the theorem is not obvious and requires caution. Especially the proof given by Middleton is false\footnote{It uses the fact that if a function $\phi(t)$ satisifies 
$\phi(t^*)=0$ and $\partial_t \phi(t^*)\geq 0$ then $\phi(t)\leq 0$ for some $t$ preceding $t^*$. However this is false as one can see by considering $\phi(t)=t^2$ and $t^*=0$.}. 
Here I only prove a weaker version of the theorem, namely that the motion of the interface is monotonous $\du(x,t) \geq 0$, $\forall x$ $\forall t \geq t_0$
(instead of \emph{strictly} monotonous).
 
I prove the weak version for the case of the driving \eqref{eq: F ext} $f(x,t)=m^2\left(w(t)-u(x,t)\right)$ with $\dot{w}(t) \geq 0$.
Let us assume that the velocity of some point vanishes and let us take the first 
time $t_1 \geq t_0$ when this happens and a point $x_1$ where $\du(x_1,t_1) = 0$.
Taking a time derivative of \eqref{eq: Motion 1} at $(x_1,t_1)$ gives~:
\begin{align}
\eta \ddot{u}(x_1,t_1) &= m^2\left(\dot{w}(t_1) - \du(x_1,t_1)\right) 
+ \int \mathcal{C}(y-x)\left( \du(y,t_1) - \du(x_1,t_1)\right) dy +\du(x_1,t_1)\partial_u F_{\dis}\left(x_1, u(x_1,t_1) \right) \, \notag \\
	&= m^2\dot{w}(t_1) + \int \mathcal{C}(y-x)\left( \du(y,t_1) - \du(x_1,t_1)\right) dy \, \notag \\
	&\geq m^2\dot{w}(t_1) \geq 0 \, .
\end{align}
The last inequality shows that the velocity cannot become negative. 

Note that if one rules out by hand the (a priori very unlikely) possibility that the velocities of all points vanish at the exact same time then the strict monotonicity of the solution ensues~\cite{rosso2002a}. 

\vspace{7pt} 
Finally the third theorem states 
\begin{theorem}[Uniqueness of asymptotic solution]
For a given monotonous driving there is a unique asymptotic (long-time) solution of the dynamics
\eqref{eq: Motion 1} up to time-translations.
\end{theorem}
This theorem is more difficult to show rigorously and I do not attempt to prove it here.

\section{Conclusion}

The energy landscape of an elastic interface embedded in a disordered medium presents many metastable states. 
Due to the interplay between disorder and elasticity, an interface submitted to an external force stays pinned if the force is lower than a finite critical force $f_c$. 
Above the critical force, the line acquires a finite velocity.
This transition between a non-moving phase below $f_c$ and a moving phase above $f_c$ has been interpreted as an out-of-equilibrium phase transition.
Several critical exponents can be associated to this transition. 
They are linked by scaling relations, leaving only two independent exponents.
The elastic interactions along the interface can be short-range (SR) or long-range (LR).
In the latter case the range of the interactions is controlled by a continuous parameter $\alpha$.
While SR avalanches are made of a single connected component, 
LR avalanches have a more complex spatial structure and are made of several connected components
called \emph{clusters}.
In experiments where the interface is driven at a finite velocity, this complex spatial structure
makes the reconstruction of the avalanches a difficult problem.
Motivated by this problem, I worked on the effect of a finite driving velocity and the scaling of the velocity field above $f_c$.
I understood the scaling of the correlation functions of the velocity field. This
result, presented in chapter~\ref{chap: chap3}, provides a simple, robust and efficient method to identify the universality class of a critical phenomenon. 
The clusters also have their own statistical distribution, characterized by new exponents. 
With the help of numerical simulations I have built a "scaling theory" for these objects and related the new exponents to the ones defined for the global avalanches in section~\ref{sec: Ch1 Phase diagram}. This scaling theory is presented in chapter~\ref{chap: cluster statistics}.


%% file: Chapter_2/Experiments.tex


\chapter{Experimental realisations}\label{chap: experiments}

In the previous chapter I introduced the basic concepts of the physics of disordered elastic interfaces and the depinning transition. 
In this chapter I present experimental phenomena whose physical understanding has made progress thanks to the depinning framework.  
I first present three systems where experimental observations are well described by the depinning theory, 
namely the Barkhausen effect, the crack propagation and the propagation of a wetting front on a rough substrate.
Depinning is also relevant for other experimental systems that I do not discuss here such as Charge Density Waves and pinned vortex lattices in type-II superconductors.
The last two sections of the chapter are devoted to earthquakes and plastic deformations in amorphous materials. Although these phenomena are not explicitly described by an elastic interface, connections with the depinning have been pointed out in several works.

A common feature of these systems is that their collective avalanche behaviors arise from interactions that are long-range. Studying the effects of the long-range interactions on the local dynamics and spatial structure of the avalanches is therefore a topic of importance. The results presented in chapters~\ref{chap: chap3} and~\ref{chap: cluster statistics} improve the understanding of these effects for the systems characterized by long-range elasticity. I hope that they will also contribute to a better understanding of earthquakes and plasticity avalanches. 

\section{The Barkhausen noise}\label{sec: Ch2 Barkhausen noise}

\begin{figure}[tb]
	\begin{subfigure}[t]{0.48\linewidth}%
	\centering
	\includegraphics[width=1\linewidth]{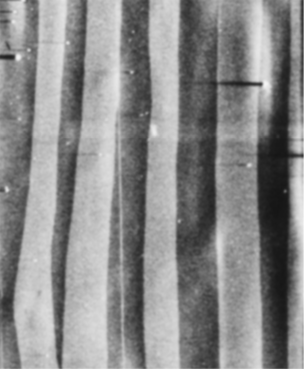}
	\caption{The domain structure of $\text{Fe}_{21}\text{Co}_{64}\text{B}_{15}$ amorphous alloy observed by scanning electron microscope. The domains are separated by walls parallel to the magnetization. This is the typical structure observed in soft ferromagnetic materials. Image taken from~\cite{zapperi1998}.\label{Fig: BK Domains image}}
	\end{subfigure}
	\hspace{.5cm}
    \begin{subfigure}[t]{0.48\linewidth}%
    \centering
	\includegraphics[width=.65\linewidth]{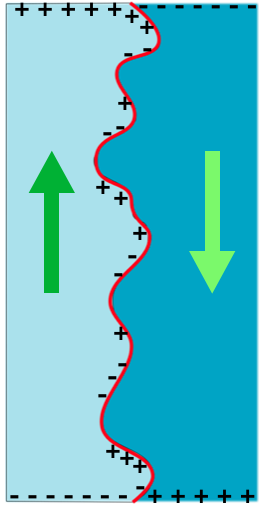}
	\caption{Exemple of surface magnetic charges at the boundary of the sample and on the Domain Wall between two domains of opposite magnetization. On the DW the dipolar interactions between magnetic charges induce a long-range elasticity.\label{Fig: BK free magnetic charges}}
	\end{subfigure}
	\caption{Domains in magnetic materials. \label{Fig: BK 2}}
\end{figure}

When slowly varying the magnetic field through a piece of ferromagnetic material such as iron, the magnetization in the material changes and generates "irregular induction pulses in a coil wound around the sample, that can then be heard as a noise in a telephone"~\cite{barkhausen1919} (and~\cite{durin2006} for a translation in english of the original article).
This noise, recorded for the first time by Heinrich Barkhausen in 1919~\cite{barkhausen1919}, is nowadays known as 
the \emph{Barkhausen noise} or the \emph{Barkhausen effect}.
Barkhausen originally assumed that the noise was created by flipping magnets.
It was the first evidence of the existence of the magnetic domains postulated by Pierre-Ernest Weiss in 1906. 
However in 1949 it was evidenced that the jumps in the magnetization were actually related to the irregular motion of the domain walls (DWs) between domains of opposite magnetization~\cite{durin2006, williams1949}\footnote{Depending on the magnetization of the sample, there can be other contributions. The DW motion is the essential contribution to the magnetization in the central part of the hysteresis loop, around the coercive field~\cite{bertotti1998}}. Domain walls are visible in figure \ref{Fig: BK Domains image} which shows an image of magnetic domains in a ferromagnetic alloy.

\begin{figure}[tb]
	\begin{subfigure}[t]{0.48\linewidth}%
	\centering
	\includegraphics[width=1\linewidth]{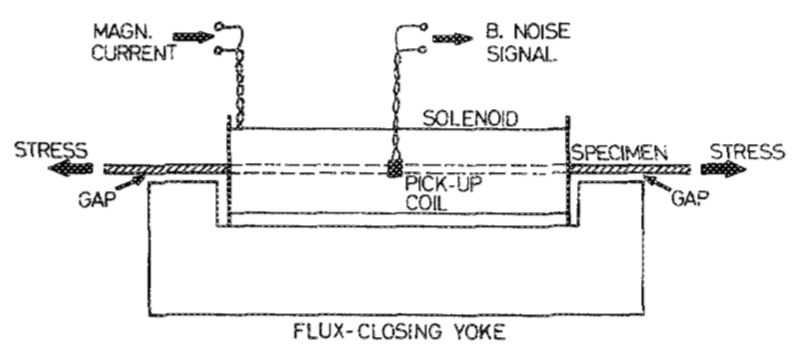}
	\caption{Experimental setup for recording Barkhausen noise. Reprinted from \cite{alessandro1990a}. \label{Fig: BK setup}}
	\end{subfigure}
	\hspace{.5cm}
    \begin{subfigure}[t]{0.48\linewidth}%
    \centering
	\includegraphics[width=1\linewidth]{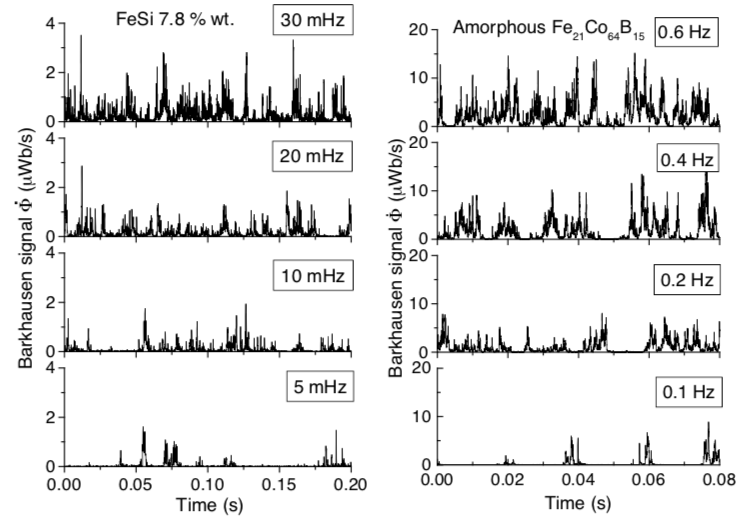}
	\caption{Examples of Barkhausen signals in two different samples under varying external field. The labels indicate the frequency of the applied magnetic field (note that the period of the signal is much larger than the duration of the measurement). When the frequency is lowered the pulses become well-separated. Reprinted from \cite{durin2006}. \label{Fig: BK signal}}
	\end{subfigure}
	\caption{Barkhausen noise experiment}
\end{figure}\label{Fig: BK 1}

Figure \ref{Fig: BK setup} shows a sketch of an experimental setup used to record the Barkhausen noise. A ferromagnetic material is put in the center of a solenoid. A current in the solenoid generates in its center a uniform magnetic field,
hereafter called \emph{external field}. 
This field induces a magnetization of the sample which in turn induces a response magnetic field.  
A pickup coil wounded around the sample detects the induced magnetic flux $\Phi$.
Figure \ref{Fig: BK signal} shows the magnetic flux change $\dot{\Phi} = \partial_t \Phi$ in two different samples and for different oscillation frequencies of the applied field. The applied signal is triangular in order to have a constant growth rate of the external field during the measurement. 
When the driving rate is high the signal is constituted of intertwinned overlapping pulses. 
At low driving rate the pulses become well separated in time and one can define Barkhausen jumps as the events when the signal exceeds some small threshold. This jumps have a given size $S$ and duration $T$ (defined in figure~\ref{Fig: cdm velocity d'Alembert})
which are power law distributed with exponents 
$\tau$ and $\ta$~:
\begin{equation}
P(S) \sim S^{-\tau} \quad , \quad P(T) \sim T^{-\ta} \, . \label{eq: BK exponents size and duration}
\end{equation}

\paragraph{The ABBM model}
In 1990 Alessandro, Beatrice, Bertotti and Montorsi proposed a simple model, now called ABBM model, 
which is consistent with experimental data~\cite{alessandro1990, alessandro1990a}.

When a sample is magnetized  it induces a \emph{demagnetizing field} that opposes the magnetization. It arises from the discontinuity in the normal component of the magnetization at the boundary of the sample (or at the interface between two domains of different magnetization). By analogy with electrostatics the demagnetizing field $H_{\text{dm}}$ can be seen as deriving from a potential created by surface \emph{magnetic charges}~\cite{durin2006, colaiori2008}~:
\begin{align}
H_{\text{dm}} &= -\nabla_{\vec{r}} \int \frac{\sigma dS'}{|\vec{r} - \vec{r}'|} \, , \label{eq: demagnetizing field 1} \\
\sigma &= \vec{n} \cdot \left( \overrightarrow{M}_1 - \overrightarrow{M}_2 \right) \, , \label{eq: surface magnetic charge}
\end{align}
where $\sigma$ is the surface charge density, $\vec{n}$ the vector normal to the domain interface
and $\overrightarrow{M}_1$, $\overrightarrow{M}_2$ are the magnetization in two different domains.
Surface magnetic charges can be visualized in figure \ref{Fig: BK free magnetic charges}.
In the most simple model the demagnetizing field is approximated as uniform over the sample and proportional to its mean magnetization $m$ 
and can be written as $H_{\text{dm}} = -km$ where $k$ is a geometry dependent demagnetizing factor~\cite{alessandro1990, urbach1995}. 
The magnetization rate is then equal to the total field $H_{\text{ext}} + H_{\text{dm}}$ plus the disorder pinning force. 
ABBM chose this last term to be a Brownian process $W(m)$.
Assuming that the external field is increased at a constant rate $c$ the equation for the magnetization rate reads~:
\begin{equation}
\frac{dm}{dt} = ct - km + W(m) = k(vt-m) + W(m)\, , \label{eq: ABBM model}
\end{equation}
with $v=c/k$ the mean velocity of the interface.
If the magnetization is induced by the motion of a single domain wall the mean magnetization $m$ can be identified with the average position of the wall and
equation \eqref{eq: ABBM model} is also the equation of motion for the center of mass of the DW with 
the driving force~\eqref{eq: F ext} deriving from a harmonic confining potential..
As one can see in figure \ref{Fig: BK Domains image} there are actually many domain walls in one sample. In this case the ABBM model still describes correctly the magnetization of the material~\cite{alessandro1990, colaiori2008}.

As we will see below the ABBM model notably succeeds in predicting the exponents for the power law distribution of $P(S)$ and $P(T)$.
Solving the ABBM model yields the following exponents~\cite{bertotti1994, ledoussal2009c}~:
\begin{align}
\tau = \frac{3}{2} - \frac{c}{2\sigma} \quad , \quad \ta = 2 - \frac{c}{\sigma} \label{eq: ABBM predictions}
\end{align}
where $2\sigma$ is the variance of the Brownian process $W$~: $\overline{\left(W(m')-W(m)\right)^2} = 2\sigma|m-m'|$. The power laws have large scale cutoffs that depend on $k$.
These exponents are valids for $c/ \sigma < 1$. This corresponds to a driving velocity regime where the interface stops infinitely often. When $c/ \sigma > 1$ the interface never stops and avalanches cannot be defined. This point will be discussed further in section~\ref{sec: statistics of the ABBM ch5}.

\paragraph{Critical exponents and range of the interactions}
In the litterature the exponents $\tau$ and $\ta$ experimentally measured span a broad range of values. 
However when restricting to a careful analysis at low driving rate around the coercive field (i.e. at the center of the hysteresis loop) they appear to split up in two classes~\cite{durin2006}. 
In Ref.~\cite{durin2000} the size and duration distributions were measured for polycristalline SiFe and amorphous alloys like $\text{Fe}_{21}\text{Co}_{64}\text{B}_{15}$
(the corresponding signals are shown in figure \ref{Fig: BK signal}).
The authors found $\tau = 1.5 \pm 0.05$, $\ta = 2 \pm 0.2$ for polycristalline SiFe and 
$\tau = 1.27 \pm 0.03$, $\ta = 1.5 \pm 0.1$ for amorphous alloys. 
This suggests that the domain walls in ferromagnets belong to two different universality classes.
The first set of values is consistent with the predictions of the ABBM model \eqref{eq: ABBM predictions} in the limit of a vanishing driving rate ($c \to 0$) which are also the predictions of the mean-field depinning model~\cite{durin2000, narayan1993}. The elasticity of the DW is dominated by the long-range dipolar interaction between magnetic charges (see figure \ref{Fig: BK free magnetic charges}) which yields the long-range elasticity kernel \eqref{eq: LR elasticity kernel} $\mathcal{C}(|x-y|) \sim |x-y|^{-(2+1)}$ where $2$ is the internal dimension of the DW~\cite{durin2000}. Thus the system is at the upper critical dimension where the mean-field theory is valid.
The second set of values is consistent with numerical simulations of the depinning of an interface with short-range elasticity~\cite{durin2000}. 
In this case the elasticity of the DW is dominated by the surface tension $\gamma \nabla_x^2 u(x)$\footnote{During the experiment the samples were subject to a tensile stress (as depicted in figure \ref{Fig: BK setup}) which increases the surface tension of the DWs~\cite{durin2000}.}.
These experiments were reanalyzed recently and the avalanche shapes were
compared in great details with the predictions from the FRG field theory, with
an excellent agreement, confirming the above scenario~\cite{durin2016}.

\section{Crack front propagation}\label{sec: Ch2 Crack propagation}

Crackling noise can also be recorded during the failure of brittle heterogeneous materials under slow external loading. 
Here the crackling signal refers to the mean crack speed and its variation with time.
In this section I briefly describe two experiments that provide a direct observation
of an in-plane crack propagation. The first one, originally performed by J. Schmittbuhl and K.J. M{\aa}l{\o}y in 1997, is a reference experiment in the crack field and is coloquially known as "the Oslo experiment"~\cite{schmittbuhl1997}. 
The second one, designed by J. Chopin and L. Ponson, is directly inspired by the former but slightly different. To differentiate it from the first one I refer to it as "d'Alembert experiment", because it was performed at Institut d'Alembert in Paris.
I analyzed the data issued from this experiment and the results I obtained are presented in chapter~\ref{chap: chap3} and in the publication~\cite{lepriol2020}.

\begin{figure}[t!]
	\begin{subfigure}[t]{0.48\linewidth}%
	\centering
	\includegraphics[width=1\linewidth]{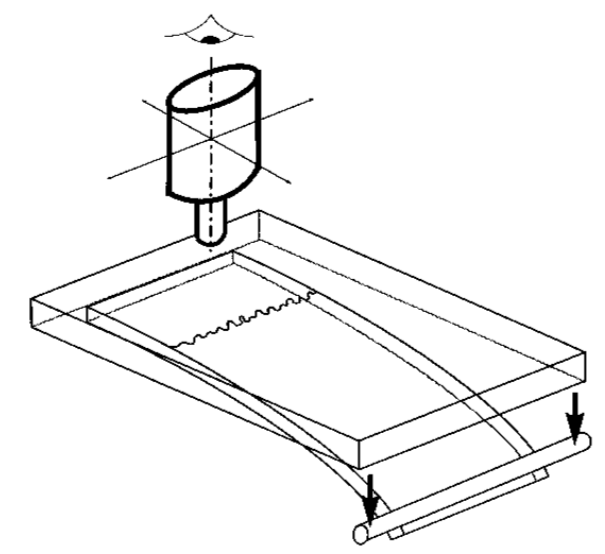}
	\caption{Sketch of the Oslo experiment. The thin bottom plate is pushed down with a cylindrical rod. The crack front can be directly observed with a camera. Reprinted from \cite{delaplace1999}. \label{Fig: sketch crack Oslo}}
	\end{subfigure}
	\hspace{.5cm}
    \begin{subfigure}[t]{0.48\linewidth}%
    \centering
	\includegraphics[width=1\linewidth]{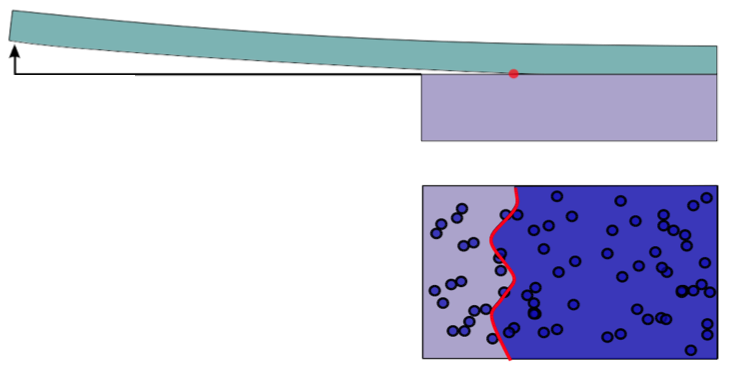}
	\caption{Sketch of the d'Alembert experiment. The thin plate is pulled up. The crack front is locally pinned by ink dots. See text for details. Courtesy L. Ponson. \label{Fig: sketch crack d'Alembert}}
	\end{subfigure}
	\caption{Schemes of crack propagation experiments.}
\end{figure}
\begin{figure}[t!]
	\centering
	\includegraphics[width=.8\linewidth]{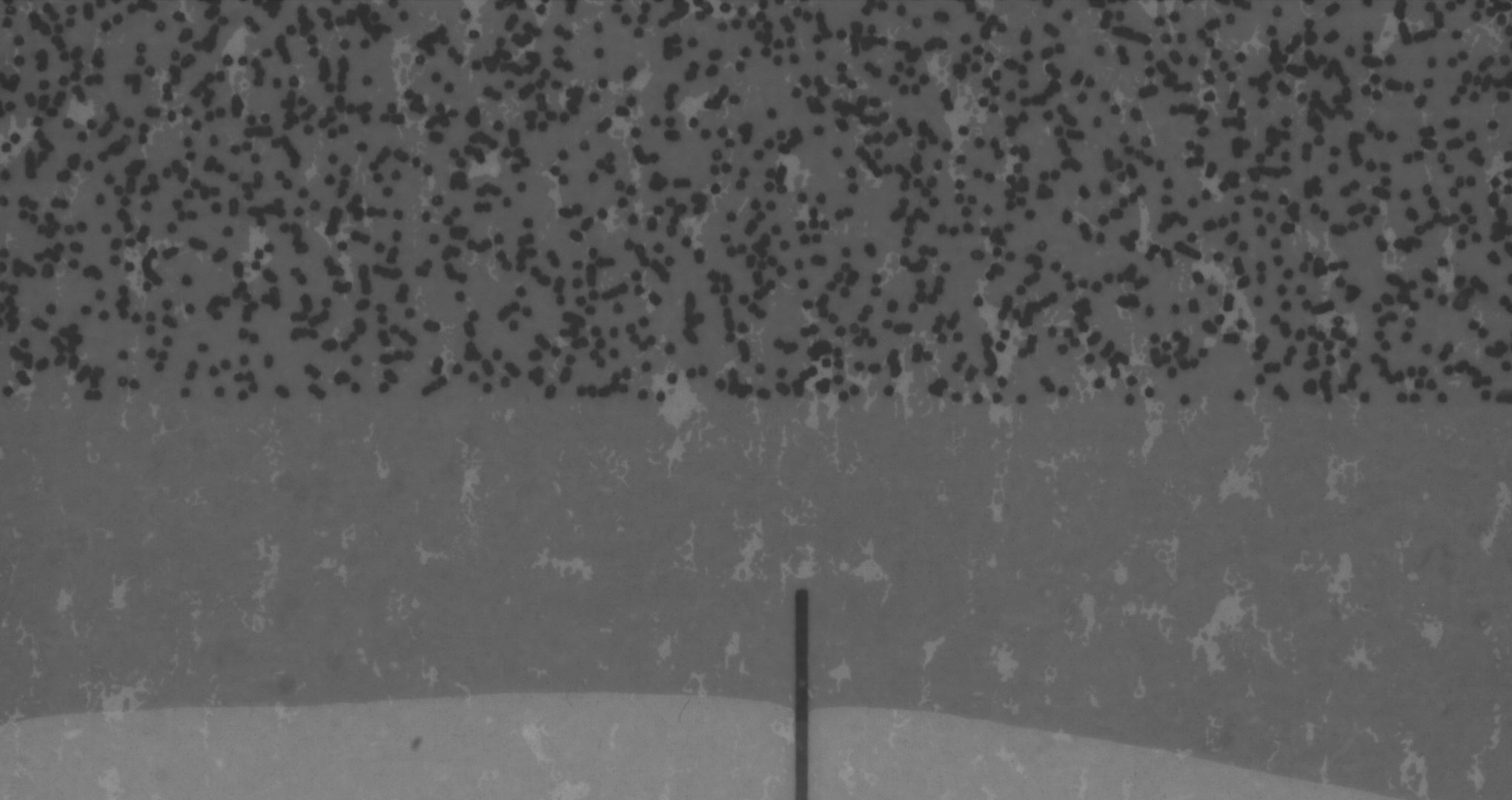}
	\newline
	\vspace{0.3cm}
	\includegraphics[width=1\linewidth]{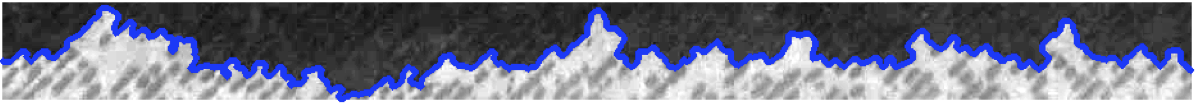}
	\caption{D'Alembert experiment. \emph{Top :} Picture of the disorder. The ink dots of diameter $100 \mu$m have a stronger adhesion energy than the background transprency ($6.1$ J/m$^2$ versus $2.2$ J/m$^2$) and thus pin the crack front.
	\emph{Bottom :} The crack front (blue line) appears as the interface between a dark and a clear phase on the image.
	Courtesy L. Ponson.
\label{Fig:Crack experiment d'Alembert}}
\end{figure}

\paragraph{Oslo experiment}
The setup consists of two transparent plates of Plexiglas annealed together at 
$200^{\circ} C$ and several bars of pressure.
The annealed surface is a plane of weaker toughness.
Before annealing defects are introduced by sandblasting the surfaces with steel particles of 
$50 \mu m$ diameter~\cite{schmittbuhl1997}. This introduces random flucutations in the toughness of the weak plane. 
Then a crack is opened by pulling apart the two plates (so called Mode I crack, see figure \ref{Fig: sketch crack Oslo}).
It propagates along the weak plane of the annealed surface.
Pictures of the system are taken with a digital camera mounted on a microscope. 
The crack front appears as the interface between a clear and a dark region on the image 
(as can be seen on the bottom picture of figure \ref{Fig:Crack experiment d'Alembert} taken on the d'Alembert experiment).

The direct observation of the front allowed to measure the roughness exponent $\zeta$. 
The first measurements yield values in the range $\zeta = \zeta^- \approx 0.50 - 0.66$~\cite{schmittbuhl1997, delaplace1999}. 
This is at variance with the predictions of the model. The crack front is expected to be described by an elastic string with the long-range kernel \eqref{eq: LR elasticity kernel} with $\alpha=1$~\cite{gao1989}. For this model the roughness exponent has been estimated very precisely by a numerical method which give the value $\zeta = 0.388\pm 0.002$~\cite{rosso2002}.
The discrepancy was resolved in 2010 when it was found that the value $\zeta^-$ actually describes the roughness of the front at short scales and that $\zeta$ crossovers to $\zeta^+ \approx 0.35 - 0.37$ at large scales~\cite{santucci2010}.
The crossover is governed by the Larkin length~\cite{laurson2010} below which the Larkin model predicts a roughness exponent $\zeta_L = \alpha - d/2 = 0.5$ (cf. equation~\eqref{eq: Zeta Larkin}) consistent with the experimental results.

The recording of the front position with a fast camera also gives access to the local velocity signal using the Waiting Time Matrix method~\cite{maloy2006} that is presented in appendix~\ref{sec: appendix experiment analysis}. 
Despite the success of the depinning model to explain both the roughness exponent and the crackling noise statistics~\cite{bonamy2008}, many questions related to the local crack dynamics remain open. These questions concern the cluster structure of the avalanches 
and the distribution of the local velocity along the front and its correlations~\cite{tallakstad2011}.
I will tackle these issues in the following chapters. 
Especially I present new results regarding the correlation functions of the local velocity in chapter~\ref{chap: chap3}
and I discuss the statistics of the clusters in chapter~\ref{chap: cluster statistics}.

\label{Fig: Crack 1}

\paragraph{D'Alembert experiment}
In this experiment a 5 mm thick plexiglas plate of PMMA (polymethyl methacrylate) is detached from a 20 mm thick PDMS (polydimethylsiloxane) substrate using a beam cantilever geometry (see sketch in figure \ref{Fig: sketch crack d'Alembert}).
The disorder is introduced by printing ink dots of diameter $100 \mu$m with a density of $20\%$ on
a commercial transparency which is then bonded to the plexiglas plate before annealing with the substrate (see figure \ref{Fig: sketch crack d'Alembert} and top picture in figure \ref{Fig:Crack experiment d'Alembert}). The adhesion energy to the substrate is stronger for the ink dots than for the transparency ($6.1$ J/m$^2$ versus $2.2$ J/m$^2$) so the ink dots are pinning centers for the interface.
This experiment has a controlled disorder where both the size of the defects as well as their pinning force are precisely known. As in the Oslo experiment the crack front propagates along the weak plane between the two plates and its successive positions are recorded with a digital camera. 

\vspace{0.5cm}
In the chapter~\ref{chap: chap3} I present results regarding the correlation functions of the local velocity obtained with the data issued from the d'Alembert experiment. 
In the following section I give a derivation of the equation of motion for the in-plane propagation 
of a crack front.

\subsection{Derivation of the equation of motion of a crack in a disordered material}\label{sec: derivation crack propagation equation ch2}

Since the work of Gao and Rice~\cite{gao1989} it is known that a crack front in an elastic material can be effectively described by an elastic line with a long-range kernel
$\mathcal{C}(x-y) = \pi^{-1}|x-y|^{-(1+\alpha)}$ with $\alpha=1$. 
In Ref.~\cite{ponson2017} it is shown that the driving is homogeneous of the form 
$f(t) = k\left(v_d t - u(x,t)\right)$ where 
$\bar{u}(t)$ is the mean position of the front, $v_d$ is the driving velocity
and $k$ is a stiffness (it corresponds to $m^2$ in equation \eqref{eq: F ext}).
This driving, corresponding to effectively pulling the line with a spring, poises the system near the critical point.
The whole equation of motion reads~:
\begin{equation}
\frac{1}{\mu} \partial_t u(x,t) = k\left(v_d t - u(x,t)\right) + 
	\frac{1}{\pi} \int \frac{u(y,t)-u(x,t)}{|y-x|^{1+\alpha}} dy + F_{\dis}\left(x, u(x,t)\right)
\end{equation}\label{eq: Motion crack experiment}
where the mobility $\mu$ is the inverse of the friction coefficient $\eta$ in the equation \eqref{eq: Motion 1}.
Two parameters, $k$ and $v_d$, set the distance to the critical point. A priori the critical point can be approached experimentally by taking $k$ and $v_d$ as small as possible. It turns out that this is not the case in the d'Alembert experiment. This is because the PDMS substrate is not perfectly elastic but viscoelastic~\cite{chopin2018}. 


In the following subsections I derive the equation of motion for the crack, first in an homogeneous elastic material and then in a disordered one.

\subsubsection{Crack propagation in homogeneous elastic material}

\begin{figure}[t!]
	\centering
	\includegraphics[width=.7\linewidth]{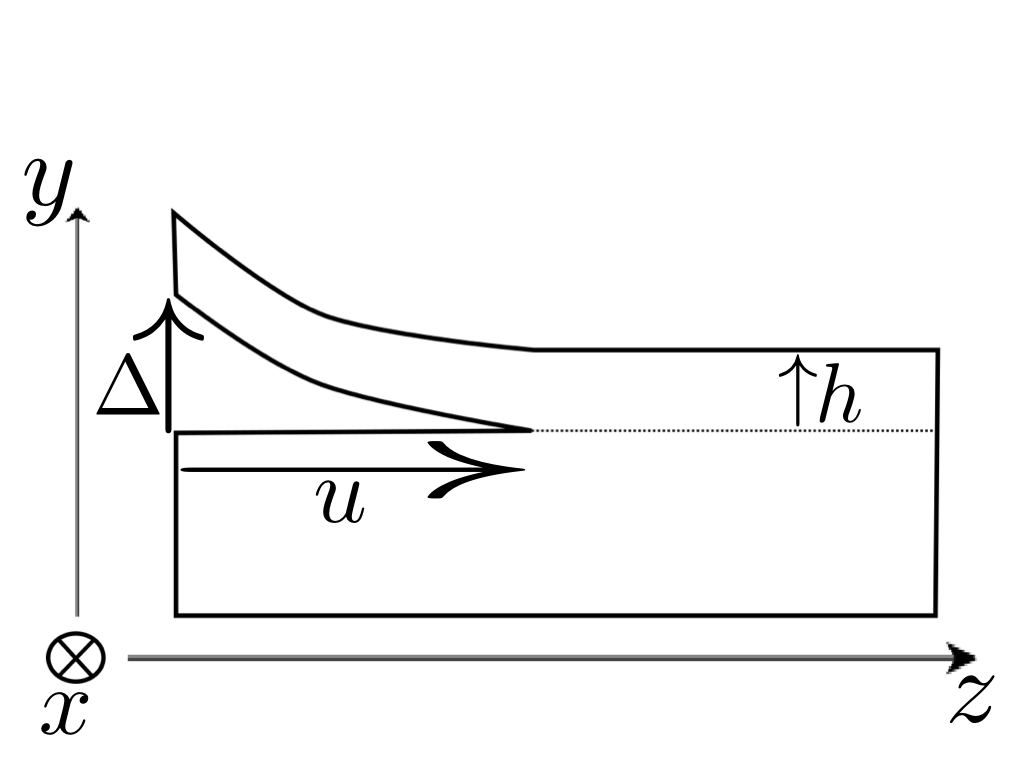}
	\caption{Sketch of a crack of length $u$ for an homogeneous elastic material of unit width (along
	the $x$ direction).
	The upper plate has height $h$ and a vertical displacement $\Delta$ is imposed at its end.
	This plate can be described as an Euler-Bernoulli cantilever beam.
\label{Fig:Sketch crack}}
\end{figure}

It is convenient to start with the homogeneous elastic material. Here the front is prefectly flat and is characterized by its position $u$ and  speed $\dot u$ (see Fig. \ref{Fig:Sketch crack}).
Note that a priori the crack front position is a vector $\vec u = (u_y, u_z)$. However in our experiment the crack propagates at the interface between two materials hence it is natural to assume in-plane propagation where $u_y$ stays constant and only $u_z:=u$ evolves. 
Following  Griffith's idea~\cite{griffith1921}
the  evolution in time of the crack  is determined by the energy balance (per unit surface) between the energy released when the material is fractured and the energy needed to create new fracture surfaces :
 \begin{equation}\label{eq:Griffith criterion}
G^{\text{dyn}}(u, \dot u) = G_c \, .
\end{equation}
The fracture energy per unit surface $G_c$ is constant for homogeneous elastic materials. 
$G^{\text{dyn}}$ is the energy release rate which accounts for the release of potential elastic energy minus the kinetic term.
It displays a simple velocity dependence~\cite{freund1990}~:
\begin{equation}\label{eq:Freund relation}
G^{\text{dyn}}(u, \dot u) = \left( 1- \frac{\dot u}{c_R} \right) G^{\text{el}}(u) \, ,
\end{equation} 
where $c_R$ is the Rayleigh wave\footnote{Rayleigh wave are acoustic waves that travel along the surface of solids}
velocity and $G^{\text{el}}(u)$ is the elastic energy release rate. 

In the experimental setup sketched in Fig. \ref{Fig:Sketch crack} a displacement $\Delta$ is imposed at the end of the upper plate. The elastic energy associated with the deformation of the plate is a function of the imposed
displacement $\Delta$ and of the crack length $u$. 
In particular if one describes the plate as an Euler-Bernoulli cantilever beam of unit width and height $h$, the 
elastic energy writes~\cite{freund1990}~:
\begin{equation}\label{eq:Beam elastic energy}
E^{\text{el}}(u, \Delta) = \frac{E h^3 \Delta ^2}{8 u^3} \, ,
\end{equation}
with $E$ the Young modulus of the plate. 
The elastic energy release rate is then 
\begin{equation}
G^{\text{el}} (u, \Delta) = - \frac{d E^{\text{el}}}{du} (u, \Delta) = \frac{3}{8} \frac{E h^3 \Delta ^2}{u^4} \, .
\end{equation}

The experiment starts by imposing an initial displacement $\Delta_0$ which opens the crack up to a length $u_0$ 
such that $G^{\text{el}} (u_0, \Delta_0) = G_c$.
Then the displacement is increased as $\Delta(t) = \Delta_0+v_0 t$ and 
the crack  moves from $u_0$ to $u_0 + u(t)$. Keeping  $v_0 t \ll \Delta_0$, $u(t) \ll u_0$ one can write the first order expansion of the elastic energy release rate :
\begin{equation}\label{eq:expansion loading}
 G^{\text{el}}(u_0 + u(t), \Delta_0 + v_0 t) = G_c   +  \dot G_0 t + G_0' u(t) \, ,
\end{equation} 
where $\dot G_0 = v_0 \partial_{\Delta} G^{\text{el}}(u_0,\Delta_0)$ and 
$G_0' = \partial_u G^{\text{el}}(u_0,\Delta_0)$.
When $\dot u \ll c_R$, combining equation \eqref{eq:expansion loading} with equations \eqref{eq:Griffith criterion} and \eqref{eq:Freund relation} to first order yields the following equation of motion :
\begin{equation}
\frac{1}{\mu} \dot u =  k \,\left( v_d t - u \right) \, , 
\end{equation}
with $\mu = c_R$, $v_d=-\frac{\dot G_0}{ G'_0} = \frac{u_0}{2\Delta_0} v_0$ and $k= -\frac{G'_0}{G_c} = \frac{4}{u_0}$.
Thus by varying $v_0$ one can control the steady velocity $v_d$ of the crack propagation.

\subsubsection{Crack propagation in disordered elastic material}

When the material is heterogeneous the fracture energy displays local fluctuations around its mean value $G_c$ :
\begin{equation}\label{eq:local fluctuations}
G_c(x,u) = G _c + \delta G_c (x, u) \, .
\end{equation}
As a consequence the crack front $u(x,t)$ becomes rough. This non trivial shape  introduces a correction in the elastic energy release rate, which was computed to first order in perturbation by Rice~\cite{rice1985}~:
 \begin{equation}\label{eq:Rice1985}
G^{\text{el}}\left(x, u(x,t), \Delta \right) = G^{\text{el}}\left(u(x,t), \Delta\right)  \left(1 + \frac{1}{\pi} \int \frac{u(y,t)-u(x,t)}{|y-x|^2}dy \right)
\end{equation}
The balance between the energy release and the fracture energy still holds but must now be written at the local level :
\begin{equation}\label{eq:Griffith criterion_local}
G^{\text{dyn}}(x,u(x,t), \dot u(x,t)) =  \left( 1- \frac{\dot u(x,t)}{c_R} \right) 
		G^{\text{el}}\left(x, u(x,t), \Delta \right) = G_c(x,u(x,t)) \, .
\end{equation}
In presence of impurities the first order expansion of $G^{\text{el}}$ becomes :
\begin{equation}\label{eq:expansion loading2}
G^{\text{el}}(x, u(x,t),t)=   G_c   +  \frac{G_c}{\pi} \int \frac{u(x',t)-u(x,t)}{|x'-x|^2}dx' + \dot G_0 t+G_0'  u(x,t) \, .
\end{equation} 
By combining together equations \eqref{eq:expansion loading2}, \eqref{eq:Griffith criterion_local}, \eqref{eq:local fluctuations} and \eqref{eq:Freund relation} one obtains :

\begin{equation}\label{eq:Movement equation appendix}
\frac{1}{\mu} \dot u =  k \,\left( v_d t - u(x,t) \right) + \frac{1}{\pi} \int \frac{u(x',t)-u(x,t)}{|x'-x|^2}dx'  - \frac{\delta G_c}{G _c} \, ,
\end{equation}
where $\mu=c_R$.
This equation is equivalent to equation \eqref{eq: Motion crack experiment} with the disorder arising 
from the toughness fluctuations $- \frac{\delta G_c(x,u)}{G _c}$.
Note that $k=\frac{4}{u_0}$ hence the critical point can be approached by starting with a large $u_0$.
In the d'Alembert experiment $u_0$ is of the order of a few centimeters which is much larger than the fluctuations of the front which are of the order of the millimeter. 

\section{Wetting of disordered susbtrates}\label{sec: ch2 wetting}

\begin{figure}[t]
	\centering
	\includegraphics[width=1\linewidth]{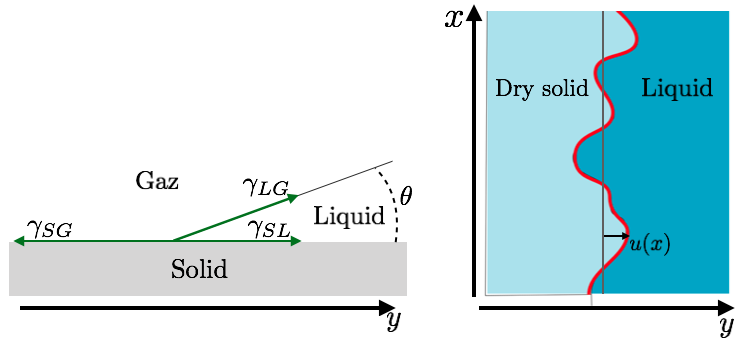}
	\caption{\emph{Left :} Sketch of partial wetting. The contact angle $\theta$ is such that the interfacial tensions $\gamma_{SG}$, $\gamma_{SL}$ and $\gamma_{LG}$ verify the Young-Dupr{\'e} relation~\eqref{eq: Young-Dupre}.
	\emph{Right :} $u(x)$ is the deviation of the contact line from a flat configuration.  \label{Fig: Young-Dupre}}
\end{figure}

When a liquid is dropped on a solid substrate it can either spread and form a film that covers the solid (\emph{total wetting}) or form a meniscus (\emph{partial wetting}).
In the latter case, the line where the liquid surface meets the solid is called 
\emph{contact line}, \emph{wetting front} or \emph{triple line}.
It has been shown independently by Joanny and de Gennes~\cite{joanny1984} and by Pomeau and Vannimenus~\cite{pomeau1985} that the contact line can be described as an elastic line with long-range elasticity arising from the surface tension of the meniscus.
In this section I reproduce the computation from Joanny and de Gennes of the elastic energy associated to a small deformation of the line, as it is useful to understand the assumptions under which the form of the elastic energy holds (this has also been done in the PhD Thesis~\cite{rosso2002a}).
Then I present an experiment and discuss results about the roughness exponent.

\subsubsection{Derivation of the elastic energy of a contact line}

For partial wetting the meniscus is characterized by the contact angle $\theta$ between the surface of the solid and the liquid/gaz interface (see figure~\ref{Fig: Young-Dupre} left). 
The contact line is submitted to forces coming from the surface tensions 
$\gamma_{SG}$, $\gamma_{SL}$ and $\gamma_{LG}$ of the solid/gaz, solid/liquid and liquid/gaz interfaces respectively.
At equilibrium the resulting component of the total force parallel to the solid surface must be nul.
This gives the Young-Dupr{\'e} relation~:
\begin{equation}\label{eq: Young-Dupre}
\gamma_{LG} \cos (\theta) = \gamma_{SG} - \gamma_{SL} \, .
\end{equation}
If the solid's surface is perfect, $\cos (\theta)$ is constant and the contact line is flat. 
The presence of disorder, that can be either chemical heterogeneities or a rough surface, 
induce fluctuations in $\gamma_{SG}-\gamma_{SL}$ 
which cause a distortion of the contact line. 

We want to compute the energy cost of the distortion.
Let $(x,y)$ be the coordinate on the solid surface and $z(x,y)$ be the height of the liquid-gaz interface. Let $u(x)$ be the local deformation of the contact line from its mean position (see figure~\ref{Fig: Young-Dupre} rigth).
We assume small deformations ($|u(x)| << 1$) and do a perturbative computation at the lowest order in $u$.

A condition on the curvature of the interface can be obtained from the Laplace-Young equation for the capillary pressure. It states that the pressure difference between the liquid and the gaz at the interface is 
$\Delta P = P_{L} - P_{G} = 2 \gamma_{LG} H$ where $H$ is the curvature of the interface. 
Far from the contact line the interface is flat hence $\Delta P=0$. 
The pressure difference being constant everywhere, we obtain the Laplace condition~:
\begin{equation}\label{eq: Lapace condition}
\frac{\partial ^2 z}{\partial x^2} + \frac{\partial ^2 z}{\partial y^2} = 0 \, . 
\end{equation}
We look for a solution under the form~:
\begin{equation}\label{eq: Ansatz z(x,y)}
z(x,y) = \sin (\theta) y + \sin (\theta) \int \frac{dk}{2\pi} \beta(k) e^{ikx -|k|y} \, .
\end{equation}
The first term is the solution for a flat contact line. 
The presence of $e^{-|k|y}$ in the second term ensures that the correction vanishes when $y\to +\infty$.
We can express $\beta(k)$ as a function of the distortion $u(x)$ by using the boundary condition~:
\begin{equation}\label{eq: Boundary Condition wetting}
z \left(x,u(x) \right) = 0 \, .
\end{equation}
Introducing the Fourier components of the distortion $u(x)$
\begin{equation}
\hat{u}(k) = \int_{-\infty}^{+\infty} e^{-ikx} u(x) dx \, ,
\end{equation}
we have perturbatively~:
\begin{equation}
\beta(k) =  \hat{u}(k) + O(u^2) \, .
\end{equation}
The deformation of the interface height $z$ induces a correction to the capillary energy. It reads~:
\begin{equation}
U_{\text{cap}} = \frac{\gamma_{LG}}{2} \int_{-\infty}^{+\infty} dx \int_{u(x)}^{+\infty} 
\left( \left( \nabla z \right)^2 - \sin^2 \theta \right) dy \, .
\end{equation}
Integrating over $x$ and $y$ we have~:
\begin{equation}
U_{\text{cap}} = \gamma_{LG} \sin^2 (\theta) \int_{-\infty}^{+\infty} |k| |\hat{u}(k)|^2 \frac{dk}{2\pi} \, .
\end{equation}
This result can be Fourier transformed and read in real space~:
\begin{equation}\label{eq: Capillary energy final}
U_{\text{cap}} = \gamma_{LG} \sin^2 (\theta) \int \frac{\left( u(x)-u(x') \right)^2}{2\pi(x-x')^2} dx dx' \, .
\end{equation}
This result is exactly the elastic energy~\eqref{eq: Hamiltonian el LR} with the LR kernel~\eqref{eq: LR elasticity kernel}
with $\alpha=1$.

\subsubsection{Experimental situation}

\begin{figure}[t]
	\centering
	\includegraphics[width=.7\linewidth]{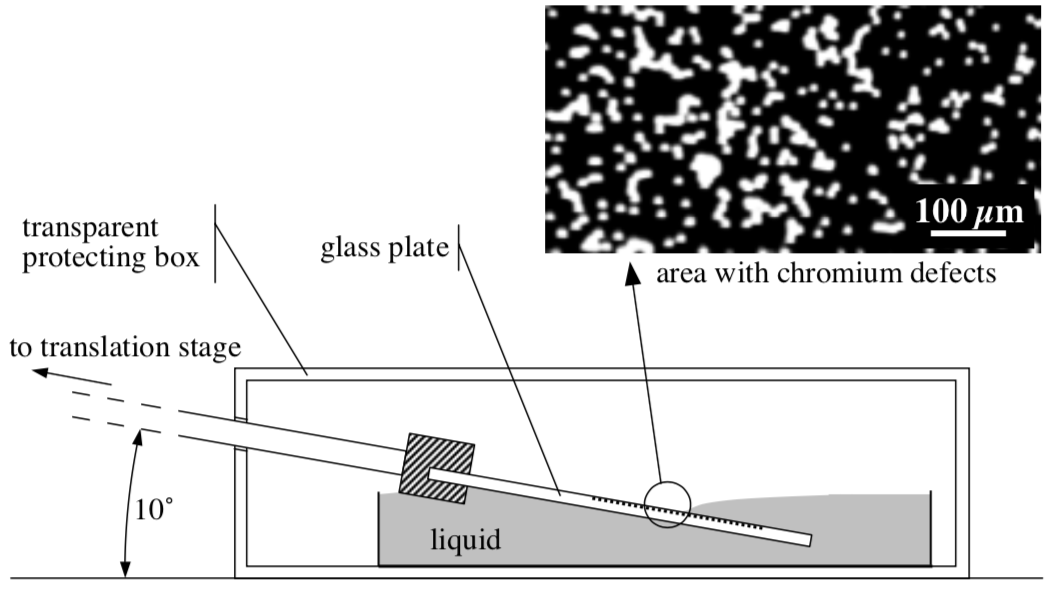}
	\newline
	\vspace{0.3cm}
	\includegraphics[width=.7\linewidth]{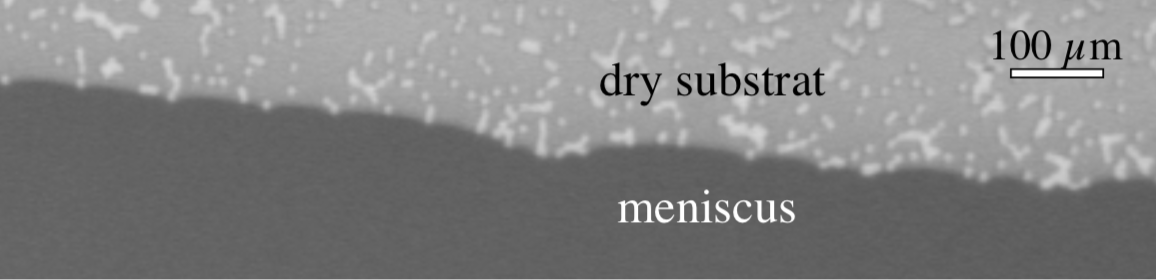}
	\caption{Wetting experiment. 
	\emph{Top :} Sketch of the experimental setup.
	\emph{Bottom :} Picture of the \emph{contact line} which separates the meniscus from the dry substrate.
	Images taken from \cite{moulinet2002}. \label{Fig: wetting front}}
\end{figure}

Figure \ref{Fig: wetting front} shows a sketch of an experiment, studied in Refs.~\cite{moulinet2002, moulinet2004, ledoussal2009a}, that allows to observe a disordered wetting front and follow its motion.
A glass plate is dipped into a liquid (here water or an aqueous solution of glycerol).
The front is observed from above using a camera.
Disorder has been introduced on the plate by the presence of squares of chromium (of size $10\times 10$ $\mu m^2$). 
When the plate is removed from the liquid at a constant velocity (ranging from $0.2$ $\mu m/s$ to $20$ $\mu m/s$ in \cite{moulinet2002}) the contact line recedes on the substrate.
The precise form of the elastic kernel taking into account
gravity, as well as arbitrary contact angle and plate inclination was computed in~\cite{ledoussal2010}.
For our purpose it is sufficient to approximate it as follows~\cite{ertas1994, ledoussal2009a, ledoussal2010}, using the notations defined in chapter~\ref{chap: DES and depinning}~:
\begin{equation}
\eta \partial_t u(x,t) = m^2\left(vt - u(x,t)\right) + \int \frac{u(x',t)-u(x,t)}{(x-x')^2} dx'  + F_{\dis}\left(x, u(x,t)\right) \, , \label{eq: ch2 motion wetting line}
\end{equation}
The LR force is given by the Joanny-de Gennes kernel. The driving force
contains a confining mass term with $m^2 \sim \gamma_{LG}/L_{\rm cap}$ where $L_{\rm cap}$ is
the capillary length, the large scale cutoff length which arises in wetting because of gravity~\cite{joanny1984, pomeau1985, nikolayev2003, ledoussal2010}.
Note that the driving force takes the same form as in equation~\eqref{eq: F ext}.

This experiment allows in particular to measure the roughness exponent $\zeta$.
However while the best numerical estimate ot the roughness exponent for a line with the elastic kernel
$\mathcal{C}(x-x') \sim |x-x'|^{-2}$ is $\zeta = 0.388\pm 0.002$~\cite{rosso2002}, experiments consistently yield value close to $\zeta=0.5$ ($\zeta=0.51\pm0.03$ in~\cite{moulinet2002}, $\zeta=0.505$ in~\cite{moulinet2004} and $\zeta$ in the range $0.45-0.57$ for different samples in~\cite{bormashenko2015}).
This discrepancy shed doubts on the completeness of the model \eqref{eq: ch2 motion wetting line}.
Possible additional effects were unveiled in~\cite{golestanian2003}
such as more complex dissipation mechanisms, a coating transition, and the existence
of additional non linear terms breaking the $u \to -u$ symmetry which are allowed at depinning. 
In~\cite{ledoussal2006} it was shown that for strong enough disorder these non linear
terms become relevant, leading to an exponent $\zeta \approx 0.5$. 
However, in~\cite{ledoussal2009a}
the renormalized disorder correlator $\Delta$ was measured in the same experiment and found to be in good agreement with the prediction from FRG, valid for quasistatic depinning with irrelevant non linear terms (although it is unclear
if the experimental results are precise enough to distinguish between the two fixed points). 
It is quite possible that the situation is the same as for cracks (which are supposed to be described by the same LR elasticity kernel, see section~\ref{sec: Ch2 Crack propagation})~: originally the roughness exponent was systematically measured in the range $0.5-0.6$. It is only in 2010 that a crossover to $\zeta$ in the range $0.35-0.37$ at larger scale was detected. 
The roughness exponent has not been measured at large scale in wetting experiments since then.
There is one study posterior to this date (Ref.~\cite{bormashenko2015} cited above) where 
$\zeta$ is measured but only at small scales. 

This debate on the value of the roughness exponent calls for alternative methods to identify the universality class of the system. In chapter~\ref{chap: chap3} I present a new method based on the correlations of the local front velocity that I expect to be simple and robust.

\section{Earthquakes}\label{sec: Ch2 Earthquakes}

\begin{figure}[t!]
	\begin{subfigure}[t]{0.48\linewidth}%
	\centering
	\includegraphics[width=1\linewidth]{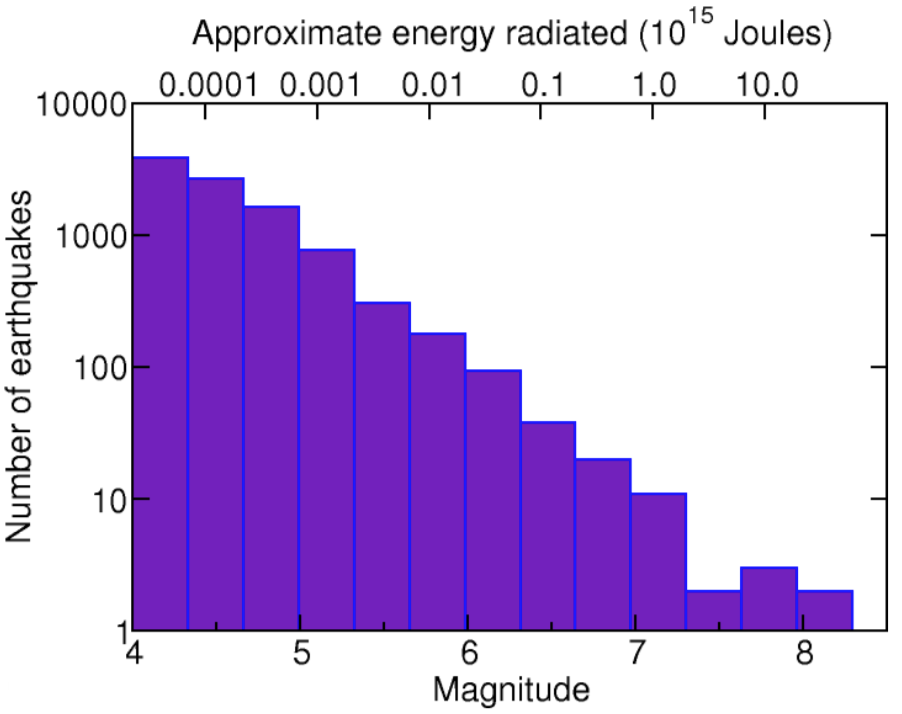}
	\caption{Histogram of the number of earthquakes recorded on earth in the year 1995 as a function of their magnitude. It is a power law distribution in term of their seismic moment or energy radiated. Reprinted from \cite{sethna2001}. \label{Fig: GR law 1995}}
	\end{subfigure}
	\hspace{.5cm}
    \begin{subfigure}[t]{0.48\linewidth}%
    \centering
	\includegraphics[width=1\linewidth]{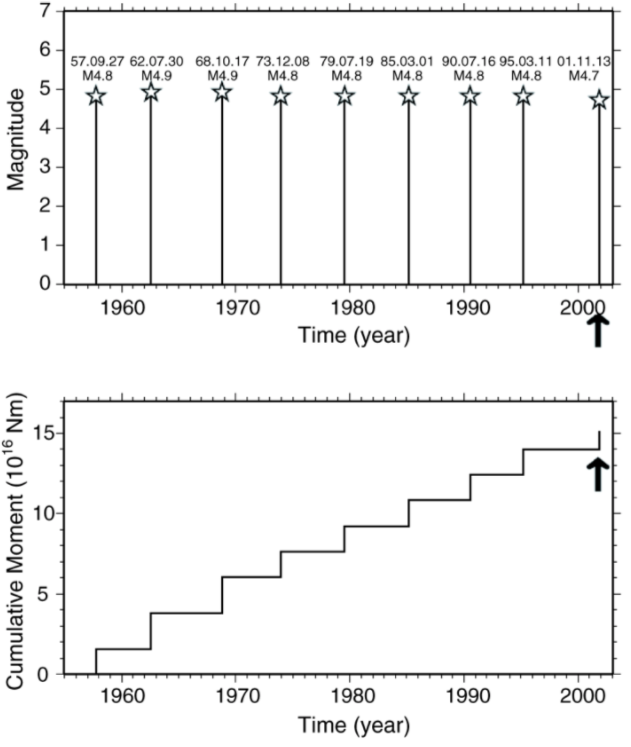}
	\caption{\emph{Top :} Large earthquakes in Kamaishi with nearly the same magnitude $m\simeq 4.8$ and a quasi-periodic occurence. \emph{Bottom :} Cumulative seismic moment of the Kamaishi Earthquakes. It is dominated by the large earthquakes identified in the upper panel. Originally from~\cite{okada2003}, taken from \cite{landes2014}. \label{Fig: Characteristic EQ law}}
	\end{subfigure}
	\caption{Gutenberg-Richter (left) and characteristic (right) earthquakes distributions. \label{Fig: EQ distributions}}
\end{figure}

Earthquakes are a natural large-scale phenomena that display three characteristic features intrisic to avalanche dynamics~:
(i) a slow driving rate, (ii) abrupt events separated by long quiescence periods  and
(iii) power law distribution of the energy released during these events.

Earthquakes are a consequence of \emph{plate tectonics.} 
The \emph{lithosphere} (the rigid outermost shell of the earth) is broken into many plates which are in motion relatively to each other. 
This motion is mainly driven by the movements of the underlying \emph{asthenosphere} (the ductile part of the mantel below the lithosphere)
at velocities ranging from millimeters to a few centimeters per year.
At the boundary between plates the earth crust is cut up by many \emph{faults}. 
Along the faults friction forces oppose the relative motion between plates. Stress and elastic energy are accumulated and the plates are blocked as long as the accumulated stress remains below the \emph{static friction threshold}.
When this threshold is met at some place, the plates slip locally. A local slip can nucleate a larger slip zone and slipping regions have sizes ranging from meters to hundreds of kilometers while the slip $\Delta u$ (the relative displacement between the two plates) ranges from millimeters to meters.
During the slip elastic energy is released through acoustic waves which make the earth quake. Depending on the size of the rupture, earthquakes durations range from milliseconds to a few minutes~\cite{ben-zion2008}.

\subsubsection{Characteristics and statistics of earthquakes}
For an indivivual earthquake a scalar seismic moment is defined as 
$M_0 = \mu_0 A \overline{\Delta u}$ with $\mu_0$ being the shear modulus of the lithosphere, 
$A$ the fault area slipping and $\overline{\Delta u}$ the average slip over this area.
The energy released during the earthquake is approximately proportional to the seismic moment~: 
$E \propto M_0$~\cite{dearcangelis2016}. 
Non specialists are more used to hear about the \emph{magnitude} $m$ of earthquakes. 
The magnitude and the seismic moment are linked via the relation~\cite{dearcangelis2016}~: 
\begin{equation}
\log (M_0) = \frac{3}{2} m +16.1 \, . \label{eq: moment magnitude relation}
\end{equation}
The statistics of the earthquakes is given by the famous \emph{Gutenberg-Richter law}~: in a given region during a large enough time period the number $N(m)$ of earthquakes with magnitude larger than $m$ exhibits an exponential decay. The law is usually stated under the form~\cite{Gutenberg1944}~:
\begin{equation}
\log \left(N(m) \right) = a - b m \, . \label{eq: G-R law}
\end{equation}
where $b$ is the Gutenberg-Richter (G-R) exponent and $a$ is a constant such that $10^{a}$ is proportional to the period over which the statistics is recorded and to the seismicity of the region (i.e. the frequency at which earthquakes occur).
The G-R exponent $b$ is about $1$ but varies in the range $0.8-1.2$ depending on the region of the world.
From equations \eqref{eq: moment magnitude relation} and \eqref{eq: G-R law} it follows that the seismic moments are power law distributed~:
\begin{equation}
P(M_0) \sim M_0^{-(1+2b/3)} \, , \label{eq: power law seismic moment}
\end{equation}

While the regional earthquake distribution are power law distributed, for a single fault the situation is sometimes different. The small events are power law distributed but there is a peak of large events with an almost constant magnitude. 
These \emph{characteristic earthquakes}~\cite{wesnousky1994} occur quasi-periodically~\cite{ben-zion2003}.
The Gutenberg-Richter and characteristic earthquakes distributions are illustrated in figure \ref{Fig: EQ distributions}.
Other faults exhibit a true G-R distribution. 
A possible reason for this difference is that the interaction between neighboring faults induce a different emergent behavior than the one of the isolated system. 

\subsubsection{Link with the depinning model}
Earthquake physics is very complicated and can of course not be fully captured by the depinning model.
Especially there are many friction effects in fault that are not captured by the depinning model 
(see the section 3.4.2 of the PhD Thesis~\cite{landes2014} for more details on this point).
Nevertheless the elastic interface is a starting point for simple fault models where friction ingredients are added by hand and which can help to gain insights on faults and earthquakes physics.
Here I do not intend to perform an exhaustive review of the results obtained by adding ingrediend to the elastic interface but only to mention a few exemples.

A fault is modeled as a $2d$ interface. The direction of displacement is in the plane of the fault. Strictly speaking there are two possible directions of displacement but the tectonic driving imposes one main direction. 
The relative displacement along the fault can thus be represented by a function 
$u:  \mathbb{R}^2 \rightarrow \mathbb{R}$.
The elasticity has a long-range kernel 
$\mathcal{C}(x-y) \sim |x-y|^{-(d+\alpha)}$ with $d=2$ the dimension of the interface and 
$\alpha=1$~(\cite{fisher1997} and references therein).

The first ingredients that can be added are \emph{frictional weakening} and its counterpart \emph{frictional strengthening}. Frictional weakening describes the fact that the dynamic frictional forces are weaker than the static frictional forces. It is relevant in the context of solid friction. Once the static threshold is met and the motion starts it is immediately accelerated.
Frictional strengthening is just the opposite~: the dynamic frictional forces are higher than the static ones. It is relevant for liquid friction. As the velocity increases the motion become more and more inhibited. 
In Ref.~\cite{fisher1997} it was found that when adding frictional weakening to the elastic depinning model the avalanches exhibit the characteristic earthquake distribution. 
When the interface presents regions of frictional weakening surrounded by regions of frictional strengthening an initial large slip, corresponding to a large earthquake, is followed by afterslips which trigger \emph{aftershocks}~\cite{perfettini2018}.

One can also add \emph{viscoelastic relaxation}~: when a material is subject to a stress, molecular rearrangements relax the internal stress.
In Ref.~\cite{jagla2014} a viscoelastic relaxation is added along the elastic interaction. The consequence is that coherent oscillations of the stress emerge. These coherent oscillations generate quasi-periodic global events (the whole system moves) which are analogous to the quasi-periodic characteristic earthquakes.

\section{Elastoplasticity and the yielding transition}\label{sec: Ch2 Plasticity}

\begin{figure}[tb!]
	\centering
	\includegraphics[width=.8\linewidth]{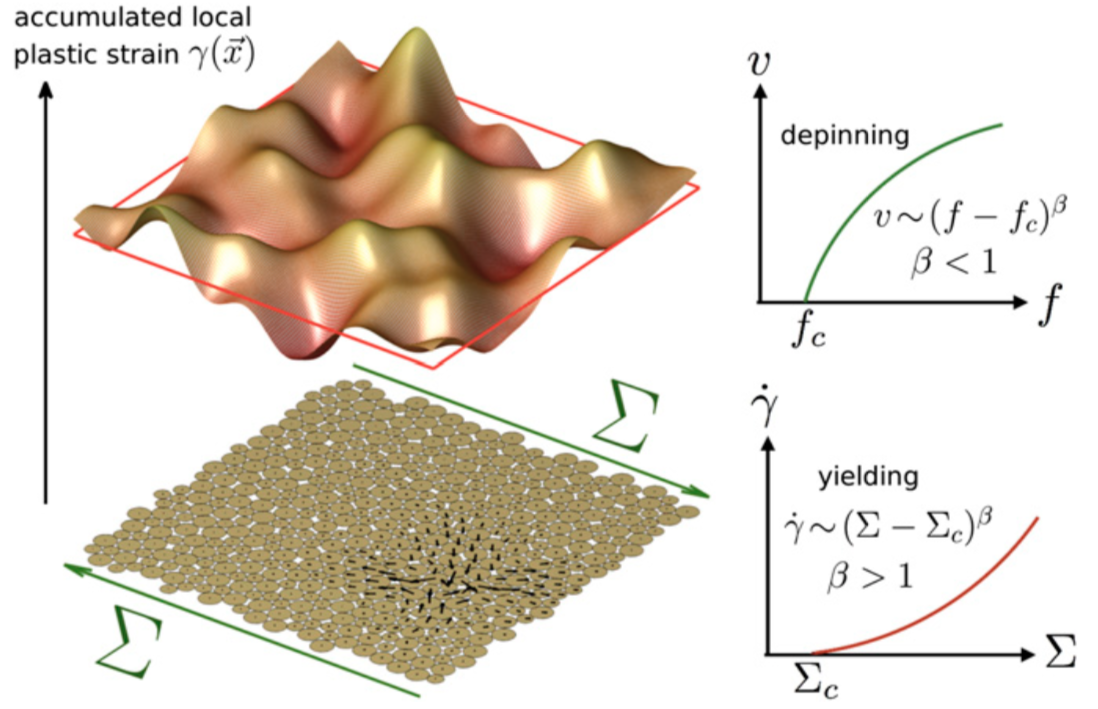}
	\caption{Analogy between the yielding and the depinning transitions.
	\emph{Left :} The accumulated local plastic strain under an external shear stress $\sigma$ (bottom) can be seen as a $d$-dimensional interface in $d+1$ dimensions (top). 
	\emph{Right :} The departure of the global strain rate $\dot{\gamma}$ from zero above the critical stress $\Sigma_c$ is characterized by a critical exponent $\beta$, a feature shared with the depinning.
	Image taken from~\cite{lin2014}.
\label{Fig: yielding-depinning}}
\end{figure}

In the depinning transition, an elastic interface starts to flow above a finite force threshold $f_c$. 
A behavior that is similar in many respects is shared by some amorphous solids. In contrast to crystalline solids, amorphous solids have no internal structure.
Typical exemples of such materials are foams, mayonnaise or whipped cream. When left at rest they are solids~:
they do not flow and keep their initial shape.
But one can make them flow easily by applying a sufficiently large shear stress, in which case they behave more like liquids.
See~\cite{nicolas2018} for a recent review on the deformation and flow of amorphous solids.
As for the depinning the onset of the flow arises at a finite value of the stress, called \emph{yield stress} because the solid yields. These materials are named \emph{yield stress solids} and the transition from solid to liquid behaviour is called the \emph{yielding transition}.

The deformation of the solid under the shear stress occurs through local \emph{shear transformations}
(also named \emph{plastic events}) where the particles rearrange locally.
This rearrangement reduces the local stress which is redistributed elastically in the rest of the material. The redistributed stress can trigger new plastic events, possibly leading to 
an \emph{avalanche} of plastic events and a macroscopic deformation of the material. 

Formally the local deformation from the initial configuration is measured by the local plastic strain
$\gamma (x)$, where $x \in \mathbb{R}^d$ is the internal coordinate of the solid. 
It can be seen as a $d$-dimensional interface in $d+1$ dimension, as illustrated in figure \ref{Fig: yielding-depinning}. 
As the depinning, the yielding transition is characterized by a set of critical exponents characterizing a diverging length scale, the avalanche size distribution or the strain rate versus stress curve (see right of figure \ref{Fig: yielding-depinning} for the latter).
The dynamics of the plastic strain is governed by~:
\begin{equation}
\partial_t \gamma(x) = \Sigma + \int \mathcal{G}(x-y)\gamma(y) + \sigma_{\dis}\left(x, \gamma(x)\right) \, . \label{eq: plasticity}
\end{equation}
This equation is very similar to the equation~\eqref{eq: Motion 1} governing the dynamics of an elastic interface. The plastic strain $\gamma$ plays the role of the displacement $u$, the external shear stress $\Sigma$ corresponds to the external force $f$ and there is a quenched disorder $\sigma_{\dis}$.
However there is one crucial difference~: the interaction kernel $\mathcal{G}(x)$ is non-positive but changes sign.
For an infinite two-dimensional system it reads~\cite{picard2004}~:
\begin{equation}
\mathcal{G}(x) = \mathcal{G}(|x|, \phi) = \frac{\cos (4\phi)}{\pi |x|^2} \, , \label{eq: eshelby kernel}
\end{equation}
where $\phi$ is the angle between the shear direction and the vector $x$.
In three dimensions it also displays a four-fold symmetry and decay with the distance as $1/|x|^d$. 

Let us define $\delta(x) = -\Sigma - \int \mathcal{G}(x-y)\gamma(y) - \sigma_{\dis}\left(x, \gamma(x)\right)$. A similar quantity can be defined for elastic interfaces. 
A plastic event is triggered when $\delta$ becomes negative.
Between plastic events $\delta$ is positive and is a measure of the distance to instability.
In Ref.~\cite{lin2014a} the authors pointed out that the distribution $P(\delta)$ is a key difference between the yielding and the depinning transitions. In elastic interfaces submitted to a monotonous driving, $\delta$ can only decrease until the instability is reached and a local motion occurs. This leads to a constant density of $\delta$ close to instability~: 
$P(\delta) \simeq P(\delta=0)$ for $\delta \ll 1$~\cite{fisher1998}.
For plasticity however the stability of a point can be increased by a neighbouring event. 
Because the points passing down $\delta=0$ get a higher stability after the plastic event, the zone of positive $\delta$ close to zero is depleted. In \cite{lin2014a} it is argued that when $\delta$  approaches zero $P(\delta)$ vanishes as~:
\begin{equation}
P(\delta) \sim \delta^{\theta} \, , \label{eq: P(delta) plasticity}
\end{equation}
thus defining a new critical exponent $\theta>0$ (whereas for the depinning $\theta=0$).
Taking into account this difference the authors took advantage of the analogy between the yielding and the depinning to establish scaling relations for the critical exponents of the yielding transition~\cite{lin2014}.

In chapter~\ref{chap: chap3} I establish the universal scaling function of the correlation functions of the local velocity field  in elastic interface avalanches. Based on this result, I propose a conjecture for the correlation functions of the local plastic strain rate.
In chapter~\ref{chap: cluster statistics} I explain that, in presence of long-range interactions, avalanches of elastic interfaces organizes into clusters for which we can define new critical exponents. I derive scaling relations linking these new exponents to the ones of the global avalanches. 
I hope that this approach, focusing on clusters of local events, can be extended to plasticity avalanches. 


\section{Conclusion}\label{sec: Ch2 Conclusion}

Crack and wetting fronts as well as domain walls in ferromagnets are well described by the model of an elastic interface in a disordered medium. 
For the two former systems, not only the mean signal but also the spatial evolution of the front can be observed. The elastic interaction in these systems is long-range. Understanding the effect of long-range elasticity is crucial to correctly describe the spatial progression of the front. 
In chapters~\ref{chap: chap3} and~\ref{chap: cluster statistics} I present new results regarding the statistics of the local velocity and the distribution of the avalanche clusters.

Earthquakes and the plasticity of amorphous solids are phenomena that are not directly described by the elastic interface model but still connection have been made that helped a better understanding of these phenomena. They are also characterized by long-range interactions, although not strictly elastic. 
Thus we can hope that a better comprehension of the long-range interaction in the depinning framework could be of some help in understanding these phenomena too.


%% file: Chapter_3/Chapter_3.tex


\chapter{Assessing the Universality Class of the transition}\label{chap: chap3}

In presence of an experimental critical phenomenon, or of a new model, assessing the universality class of the transition is a crucial task. 
As a universality class can be defined by a fixed set of critical exponents, assessing the universality class of a system initially went through determining some critical exponents and confronting experimental measurement with theoretical predictions.
Since the exponents are not independent, it is sufficient to determine a small number of exponents, two in the case of the depinning transition, to characterize the universality class.
Besides the critical exponents, other characterization of the transition were developped, based on the avalanche dynamics.

In section~\ref{sec: Ch3 Existing methods} I  present some methods to characterize the universality class of the depinning transition. I start with the standard methods to determine the roughness exponent $\zeta$, the correlation length exponent $\nu$, the velocity exponent $\beta$ and the dynamical exponent $z$.
These methods are commonly used in numerical simulations but, except for $\zeta$, are less suited for real experiments.
The remainder of the section focuses on the methods to characterize the avalanches, namely determining the avalanche size exponent $\tau$ and the avalanche duration exponent $\tilde{\alpha}$, or characterizing the avalanche shape. 
These methods require to define avalanches, a task that in real experiments raises important issues.

In section~\ref{sec: Ch3 mon travail} I discuss the new method that we proposed in~\cite{lepriol2020}.  We found the scaling forms of the two-point space and time correlation functions of the local velocity field and show that they are related to two independent exponents. 
This new method does not require to define avalanches and can be easily applied to study systems for which the velocity field is accessible.



\section{Existing methods to characterize the transition}\label{sec: Ch3 Existing methods}

\subsection{Roughness exponent}\label{sec: Ch3 Roughness exponent}

The determination of the roughness exponent $\zeta$ has been the focus of many works. 
It is easily accessible in experiments where a direct observation of the front is possible such as 
the crack and wetting front experiments presented in sections~\ref{sec: Ch2 Crack propagation} and \ref{sec: ch2 wetting}.
Perturbative dimensional expansion within the FRG provided analytical predictions~\cite{chauve2001, ledoussal2002}. 
It was determined with high accuracy for different models using numerical simulations~\cite{rosso2002, rosso2003}.
I discussed in the previous chapter the evolution of the measured values of $\zeta$ in the crack and wetting front experiments and the agreement with the theoretical values. In this section I present the different methods to determine $\zeta$.
There are three widely used definition of $\zeta$. Two of them are equivalent while the third one does not allow to measure an exponent $\zeta > 1$.

\paragraph{Average width of the interface}
The first definition, used in simulations~\cite{rosso2002}, wetting front~\cite{moulinet2002} and crack front experiments~\cite{schmittbuhl1997, delaplace1999}, relies on the scaling of the width of portions of the interface and was already  
mentioned in section~\ref{sec: Ch1 Phase diagram}. 
Considering a portion of the interface of size $\ell$ we can formally define its width as the mean quadratic displacement over this portion~:
\begin{equation}
W^2_u (\ell) = \left\langle \left( u - \langle u \rangle \right)^2 \right\rangle 
	= \langle u^2 \rangle - \langle u \rangle^2
\end{equation}
where $\langle ... \rangle$ denotes the spatial average over the portion considered. For a line we 
have $\langle u \rangle = \ell^{-1} \int_0^{\ell} u(x) dx$.
$W^2_u (\ell)$ can fluctuate depending on the portion considered and the realization of the disorder. 
To get rid of the fluctuations we must take the average over the disorder~:
\begin{equation}
W^2(\ell) = \overline{W^2_u (\ell)}
\end{equation}
where $\overline{\mbox{...\rule{0pt}{2.5mm}}}$ denotes the average over the disorder.
$W(\ell)$ represents the average transversal displacement, or average width, of a portion of the interface of size $\ell$.
The first definition of the roughness exponent, already given in equation~\eqref{eq: def1 zeta}, is the scaling relation~:
\begin{equation}
W(\ell) \sim \ell^{\zeta} \, . \label{eq: def11 zeta}
\end{equation}
Besides taking the average over disorder, the whole probability distribution of 
$W^2_u (\ell)$ has also been studied and used to determine the value of $\zeta$~\cite{moulinet2004, rosso2003a}.




\paragraph{Structure factor}
The second definition is used in analytical works and is also well-suited for the analysis of experimental or numerical data. 
Within this definition, $\zeta$ characterizes the power law decay of the structure factor~:
\begin{equation}
S_q := \overline{u_q u_{-q}} = \overline{|u_q|^2} \sim q^{-(1+2\zeta)}\, , \label{eq: def2 zeta}
\end{equation}
where $u_q$ is the Fourier Transform of the interface position $u(x)$. 
It was for instance used in~\cite{schmittbuhl1997, delaplace1999, maloy2001} and
I also use it in chapter~\ref{chap: cluster statistics} to measure the roughness exponents for several values of the range parameter $\alpha$.

\paragraph{Spatial correlations of the displacement field}
A third definition uses the correlation function of the interface displacement~:
\begin{equation}\label{eq: def3 zeta}
\tilde{C}(x) := \overline{\langle \left(u(x)-u(0) \right)^2 \rangle} \sim x^{2\tilde{\zeta}} \, .
\end{equation}
This definition is not completely equivalent to the previous ones. Especially 
it cannot yield 
a roughness exponent greater than one.
To see this point we use that $u(x)-u(0)=\sum_{y=0}^{x-1}\Delta_y$ with $\Delta_y=u(y+1)-u(y)$ and write
\begin{align}
\tilde{C}(x) = \sum_{y_1=0}^{x-1} \sum_{y_2=0}^{x-1} 
	\overline{\langle \Delta_{y_1} \Delta_{y_2} \rangle}
\end{align}
and notice that on average 
\begin{align}
\overline{\langle \Delta_{y_1} \Delta_{y_2} \rangle} \leq \overline{\langle \Delta_{y_1}^2 \rangle} = \tilde{C}(1) \, .
\end{align}
This shows that $\tilde{C}(x)$ is bounded by $x^2 \tilde{C}(1)$ which prevents from measuring 
a roughness exponent greater than one.
Nevertheless numerical results suggest that $\tilde{\zeta} = \zeta$ if $\zeta < 1$~(\cite{rosso2002a} and references therein).
This definition was used for instance in~\cite{duemmer2007}.

\subsection{Correlation length exponent, velocity exponent and dynamical exponent}\label{sec: Ch3 Duemmer&Krauth}

The correlation length exponent $\nu$, velocity exponent $\beta$ and dynamical exponent $z$ are linked together with the roughness exponent $\zeta$ via two scaling relations derived in section~\ref{sec: Ch1 STS and scaling relations} 
and that I recall here~:
\begin{equation}\label{eq:ch3 scaling relations recall}
\nu=1/(\alpha-\zeta) \quad , \quad 
\beta=\nu(z-\zeta) \, .
\end{equation}
The measurement of a single  exponent, or of a single function of them (e.g. $\zeta/z$), along with the knowledge of the exponent $\zeta$ is therefore sufficient to determine all values. 
If an additional exponent is measured, it allows to prove the consistency of the measurements. 
In this section I present different methods that give access to these exponents. 
The first one relies on the spatial correlations of the velocity field and directly inspired the new method that I describe in section \ref{sec: Ch3 mon travail}.

\begin{figure}[t!]
	\centering
	\includegraphics[width=0.6\linewidth]{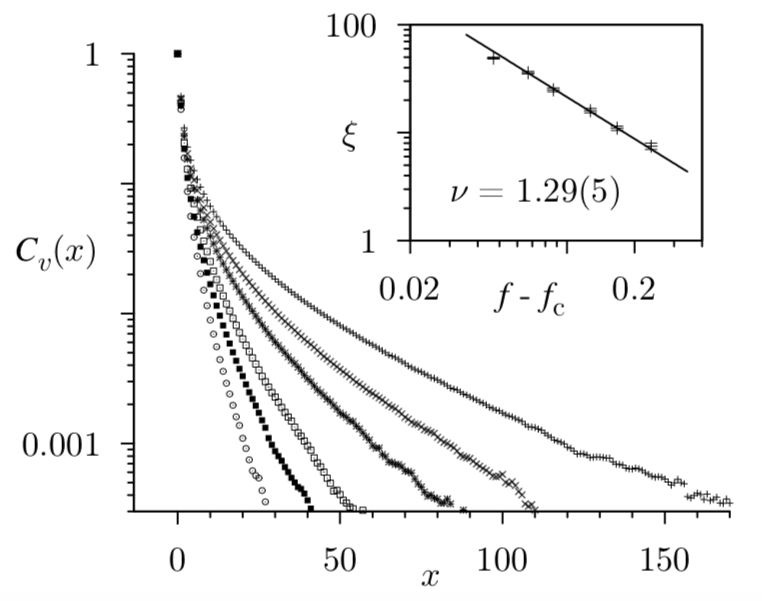}
	\caption{Image from~\cite{duemmer2005}. Connected spatial correlation function of the velocity $C_v(x)$ \eqref{eq: def spatial correlation function} for different values of $f-f_c$. The inset shows the correlation length $\xi$ extracted from fits with the functional form \eqref{eq: fit spatial correlation function duemmer}. The correlation length exponent $\nu$ is determined by a fit of the formula $\xi \sim (f-f_c)^{-\nu}$. \label{Fig: Cv(x) Duemmer&Krauth}}
\end{figure}

\paragraph{Correlation length of the velocity field}
One of the methods to measure the correlation length exponent $\nu$ relies on the connected spatial correlation function of the local velocity field~:
\begin{equation}
C_v(x) := \langle v(x,t)v(0,t) \rangle^c = \langle \left(v(x,t)-v\right)\left(v(0,t)-v\right)\rangle
	= \langle v(x,t)v(0,t)\rangle -v^2 \, , \label{eq: def spatial correlation function}
\end{equation}
where $v=\langle v(x,t) \rangle$ is the mean velocity of the line, averaged over both space and time.
In~\cite{duemmer2005} the authors performed a numerical simulation of the dynamics of a line with SR elasticity and computed the function $C_v(x)$.
Their plot of this function is shown in figure~\ref{Fig: Cv(x) Duemmer&Krauth}.
Since the motion of the line is correlated on domains of size up to $\xi \sim |f-f_c|^{-\nu}$ one can expect that the correlation function $C_v(x)$ decays fast for $x > \xi$.
The authors thus fitted the correlation function with the functional form
\begin{equation}
C_v(x) \sim x^{-\kappa} e^{-x/\xi} \, . \label{eq: fit spatial correlation function duemmer}
\end{equation}
This fit gives a value of $\xi$ that depends on the external force $f$. Repeating the operation for several values of $f-f_c$ the authors could fit the power law $\xi \sim |f-f_c|^{-\nu}$ which gave a value of the exponent $\nu=1.29\pm0.05$ for SR elasticity (see inset in figure \ref{Fig: Cv(x) Duemmer&Krauth}, this value is a bit smaller, but coherent within the range of uncertainty, than two other determinations that will be given below).
In this work the value of the exponent $\kappa$ did not converge and it was assumed to be non-universal. In \ref{sec: Ch3 mon travail} I will present the universal scaling form of the correlation functions of the velocity field and show that $\kappa$ is actually a universal exponent that is related to the exponents $\beta$ and $\nu$.

\paragraph{Initial growth of the interface width}
The scaling relation~\eqref{eq: def11 zeta} tells us that in the steady-state regime, the width of a
portion of interface of lateral extension $\ell$ scales as $W(\ell)\sim \ell^{\zeta}$.
If a simulation, or an experiment, starts from an initial flat configuration (where $W(\ell)=0$) at $t=0$ there must be a transient regime during which the width of the interface grows.
The dynamical exponent, along with the roughness exponent characterizes this growth at early time~: 
$W(t,\ell) \sim t^{\zeta/z}$~\cite{barabasi1995}.
The growth of the interface eventually saturates at $\ell^{\zeta}$ at a time $t^*$ which is such that
$t^{*\zeta/z} = \ell^{\zeta}$, i.e. $t^*=\ell^z$ .
The growth of the interface width has been experimentally measured on the Oslo experiment
yielding a \emph{growth exponent} $\zeta/z \simeq 0.5$~\cite{schmittbuhl1995a}.
This value is in agreement with a later numerical simulation of an elastic line with LR elasticity, 
with $\alpha=1$, where a value $\zeta/z=0.495\pm0.005$ was measured~\cite{duemmer2007}.
Altogether with the value of the roughness exponent $\zeta=0.388\pm0.002$~\cite{rosso2002}, this gives a value for the dynamical exponent  $z=0.770\pm0.005$.

\paragraph{Short-time dynamics}
In~\cite{ferrero2013a} the authors achieved a precise numerical determination of the SR elasticity exponents by studying the non-steady short-time dynamics of the line.
Drawing on the knowledge about equilibrium phase transition, they show that, near the critical point $f=f_c$, the velocity of an initially flat interface with infinite velocity initial condition relaxes 
as $v(t) \sim t^{-\beta/(\nu z)}$.
The exponent characterizing this relaxation is related to the growth exponent via the scaling relation \eqref{eq:ch3 scaling relations recall}~: $\beta/(\nu z) = 1-\zeta/z$.
In their simulations the authors measured both exponents and found
$\beta/(\nu z) = 1-\zeta/z = 0.128 \pm 0.003$.
They also measured $\zeta=1.250 \pm 0.005$ using the structure factor~\eqref{eq: def2 zeta}.
It enabled them to give a full set of precise values~:
$z=1.433\pm0.007$, $\nu=1.333\pm0.007$, $\beta = 0.245\pm0.006$.

\paragraph{Finite-size scaling}
The critical force $f_c$ at which the depinning transition occurs is unambiguously defined in the thermodynamics limit. However, for finite-size samples $f_c$ can fluctuate between different samples.
The fluctuations decrease  when the size $L$ of the sample is increased and are characterized by a finite-size scaling exponent $\nu_{FS}$~\cite{chayes1986} which satisfies
the exact bound $\nu_{FS} \geq 2/(d+\zeta)$~:
\begin{equation}\label{eq: def nu finite-size scaling}
\sigma_{f_c}^2 = \overline{\left( f_c^{smpl} - \overline{f_c^{smpl}}\right)^2} \sim L^{-2/\nu_{FS}}
\end{equation} 
where $f_c^{smpl}$ is the critical force of a given sample and 
$\overline{\mbox{...\rule{0pt}{2.5mm}}}$ denotes the average over many samples.
It has been theoretically expected that $\nu_{FS} = \nu$~\cite{narayan1993, ledoussal2002, fedorenko2006}.
In~\cite{bolech2004} the finite size scaling exponent was numerically measured  $\nu_{FS}=1.33\pm0.01$. This value is in agreement, within the range of uncertainty, with the values measured later in~\cite{duemmer2005} and in~\cite{ferrero2013a}.

\paragraph{Direct fit of the velocity-force characteristics}
Finally the velocity exponent $\beta$ can be obtained by fitting the velocity-force characteristics near the critical point with the power law $v\sim (f-f_c)^{\beta}$.
It has been implemented in numerical simulations, giving values
$\beta=0.33\pm0.02$ for SR elasticity~\cite{duemmer2005} and $\beta=0.625\pm0.005$ for LR elasticity
for $\alpha=1$~\cite{duemmer2007}.
This method requires to know the mean velocity $v$ of the interface, the applied external force $f$ and the critical force $f_c$. The knowledge of at least one of these quantities is often lacking in experiments, so that applying this method requires dedicated work.
This has been achieved in Ref.~\cite{ponson2009} where all quantities have been measured during the failure of Botucatu sandstone specimens. The author obtained the velocity-force characteristics both in the thermal regime below $f_c$ and in the depinning regime above $f_c$ where a fit led to an exponent 
$\beta \simeq 0.80$.

\subsection{Characterizing the avalanches}

Besides measuring the standard critical exponents of the depinning transition, other works dealt
with characterizing the avalanches. 
The initial and natural method was to measure the avalanche size and avalanche duration exponents, $\tau$ and $\ta$. 
Then new methods were proposed based on the universality of the temporal avalanche shape.

The most common way to define avalanches rely on the study of the center-of-mass velocity signal 
$v(t)$ as a function of time.
Avalanches are defined as the periods where the center-of-mass velocity signal $v(t)$ exceeds a threshold $v_{th}$. 
The duration $T$ of the avalanche is the time period where $v(t)$ stays above $v_{th}$ and the avalanche size $S$ is the integral of the difference between the signal and the threshold over this period, $S=\int_{t_0}^{t_0+T} \left(v(t)-v_{th}\right)$, where $t_0$ is the starting time of the avalanche. The definition of $S$ and $T$ is illustrated in the left panel of figure~\ref{Fig:ch3_v(t)_d'Alembert}.
The use of a finite threshold $v_{th}>0$ is necessary because in experiments the interface is driven at a finite mean velocity.

\subsubsection{Avalanche size and duration exponents}\label{sec: Ch3 Avalanche size and duration exponents}

Once avalanches are defined, the statistics of the size and duration can be computed over a large number of avalanches and the probability distributions $P(S)$ and $P(T)$ can be experimentally measured.
The critical exponents $\tau$ and $\tilde{\alpha}$ are then determined by fitting the power law part of the distributions.
This method has been applied with success in Barkhausen noise experiments~\cite{durin2000} (already discussed in section~\ref{sec: Ch2 Barkhausen noise}). 
The experiments yield two distinct sets of values, depending on the material used.
The experimental values were found to be in agreement with numerical values obtained from simulations of two-dimensional ($d=2$) interfaces with different elastic interactions.
The first set of values ($\tau=1.5 \pm 0.05.$, $\ta=2\pm 0.2$) correspond to LR elasticity with $\alpha=1$ while the second one ($\tau=1.27\pm 0.03$, $\ta=1.5\pm 0.1$) is consistent with SR elasticity.
This showed that there are two distinct universality classes within the Barkhausen noise experiments.


\subsubsection{Avalanche shape}\label{sec: Ch3 Avalanche shape}

Besides the power law distributions of $S$ and $T$ and the exponents defined in chapter~\ref{chap: DES and depinning}, an observable that has been the focus of many works is the temporal shape of the global velocity signal
$v(t)$ averaged over many avalanches~(\cite{papanikolaou2011, laurson2013} and references therein). 
In order to compare between many avalanches the time must be properly rescaled.
It is natural to rescale by the avalanche duration $T$, which gives the average avalanche shape at fixed duration $\langle v(t|T) \rangle$~\cite{papanikolaou2011, laurson2013}. More recently the average avalanche shape at fixed avalanche size was also introduced and characterized \cite{dobrinevski2014, durin2016}.

In Ref.~\cite{papanikolaou2011} it has been shown by a renormalization group argument that upon proper rescaling of a few quantities (namely the avalanche duration $T$, the curvature of the harmonic confinement $m^2$ (see equation \eqref{eq: F ext}) and the mean velocity $v$) the average avalanche shape at fixed duration should have a universal scaling form.
The authors solved the mean-field ABBM model in the limit $m^2\to 0$, $v\to 0$ and found that the average shape is an inverted parabola $\langle v(t|T) \rangle \sim t(1-t/T)$.
Their prediction is in very good agreement with a Barkhausen experiment on a thin film of permalloy
for which the mean-field description is expected to be valid (the interface has long-range elasticity $\alpha=1$ and internal dimension $d=2$ hence $d=2\alpha=d_{\text{uc}}$).

\begin{figure}[tb]
	\centering
	\includegraphics[width=0.6\linewidth]{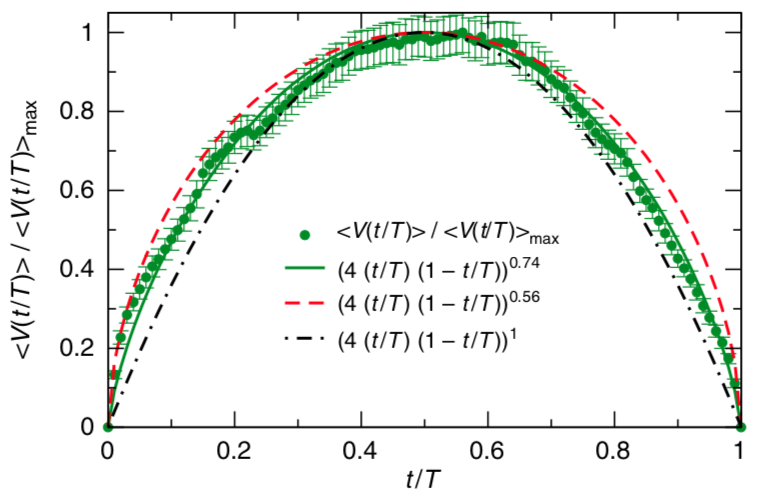}
	\caption{Average avalanche shape in the Oslo crack front experiment. Image from~\cite{laurson2013}. The green dots with statistical error bars are the experimental measurements. The green line is a best fit of equation \eqref{eq: avalanche shape asymmetric scaling} with $\gamma=1.74$ and $a=0$. The scaling form for mean-field (dash-dotted black line, $\gamma=2$) and the symmetrized  form ($a=0$) for SRE (dashed red line, $\gamma=1.56$) are also shown.}
\end{figure}\label{Fig: AvalancheShape Laurson}

A general empirical scaling form beyond mean-field has been then proposed in Ref.~\cite{laurson2013}.
The authors showed that it 
essentially depends only on the exponent $\gamma$ that characterizes the scaling between the size and duration of an avalanche ($\langle S \rangle_T  \sim T^{\gamma}$). 
Scaling arguments led the author to propose the following scaling form~:
\begin{equation}\label{eq: avalanche shape asymmetric scaling}
\langle v(t|T) \rangle = T^{\gamma-1} \left[\frac{t}{T} \left(1-\frac{t}{T}\right) \right]^{\gamma-1}
	\left[ 1-a\left(\frac{t}{T}-\frac{1}{2}\right) \right] \, ,
\end{equation}
where $a$ is a parameter that quantifies the asymmetry of the shape. If $a=0$ the shape is symmetric 
(i.e. $\langle v(t|T) \rangle = \langle v(T-t|T) \rangle$).
The introduction of an asymmetry parameter was justified because in general, and in particular in non mean-field models, the internal dynamics of the avalanche violates time-reversal symmetry.
For a mean-field model $\gamma=2$ and, if the shape is symmetric ($a=0$), the scaling form \eqref{eq: avalanche shape asymmetric scaling} becomes an inverted parabola, in accordance with the prediction from \cite{papanikolaou2011}.
The authors from~\cite{laurson2013} found an excellent agreement between the scaling form \eqref{eq: avalanche shape asymmetric scaling} and shapes obtained from simulations for different values of the elasticity. In particular for $\alpha=1$ the best fit gives a value of $\gamma=1.79\pm0.01$ and an asymmetry factor $a=0.021$. 
The value of $\gamma$ is in agreement with the value
$\gamma=1.80$ given by the scaling relation $\gamma=(d+\zeta)/z$ with $\zeta=0.39$ and $z=0.77$.
They applied the method on the Oslo crack front experiment. The best fit gives a value 
$\gamma=1.74\pm0.08$ which is more precise and closer to the theoretical prediction than the value
obtained from a direct fit $\langle S \rangle_T \sim T^{\gamma}$ that yields $\gamma=1.67\pm0.15$.

More recently the functional renormalization group allowed to obtain more precise analytical predictions that go beyond mean-field theory and empirical scaling forms.
The full scaling functions of the avalanche size versus avalanche duration ($\langle S \rangle_T = g(T)$) and of the avalanche shape at fixed duration or at fixed avalanche size have been predicted~\cite{dobrinevski2014, ledoussal2012b, dobrinevski2012, ledoussal2013, dobrinevski2013a}.
It shows that the form~\eqref{eq: avalanche shape asymmetric scaling} is good but not exact,
as the factor $\left(1-a(t/T-1/2)\right)$ is replaced by a universal function obtained in~\cite{dobrinevski2014}. 
These quantitative predictions were accurately tested on Barkhausen noise experiments~\cite{durin2016}
with LR and SR elasticity by reanalyzing the experiments from~\cite{durin2000}.
The very good fits with the short-range samples provided a reliable test of the theory beyond mean field.

\subsection{Experimental difficulties}\label{sec: Ch3 Difficulties}

Several difficulties, intrisic to experiments, can complicate the analysis of experimental signals and are susceptible to affect the measurements of critical exponents. 
These difficulties can concern the global velocity signal $v(t)$ as well as the local velocity signal $v(x,t)$.

\subsubsection{Difficulties with the global signal}

\paragraph{Thresholding the signal}\label{sec: Ch3 thresholding peril}

\begin{figure}[t!]
	\centering
	\includegraphics[width=0.47\linewidth]{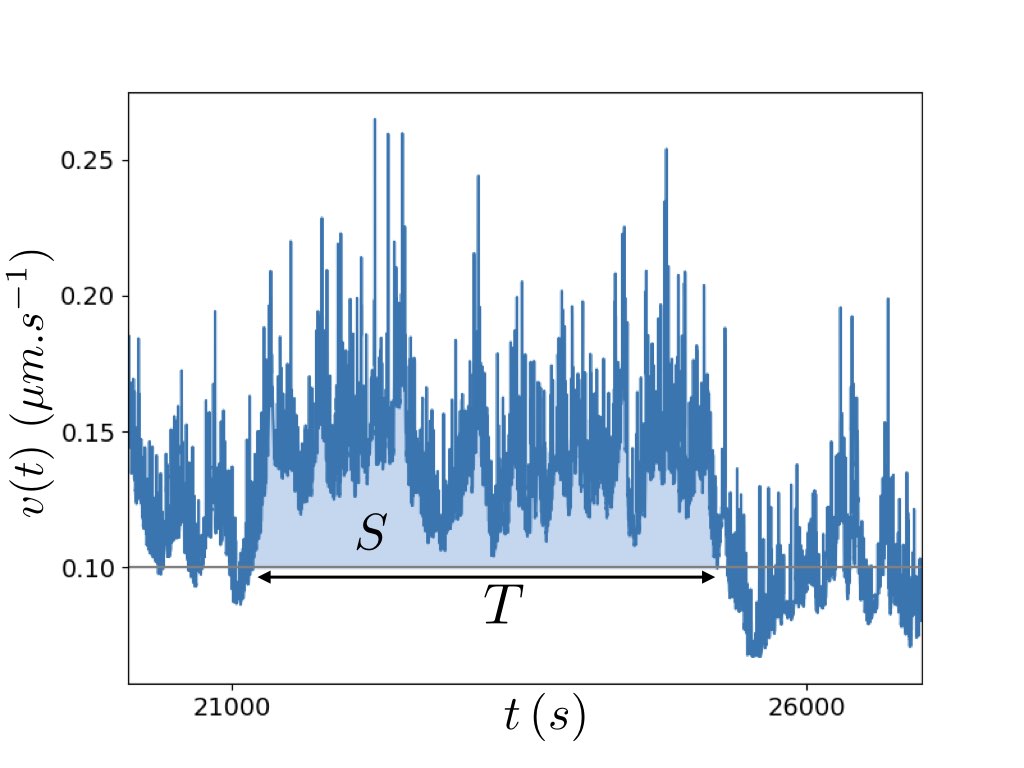}\includegraphics[width=0.47\linewidth]{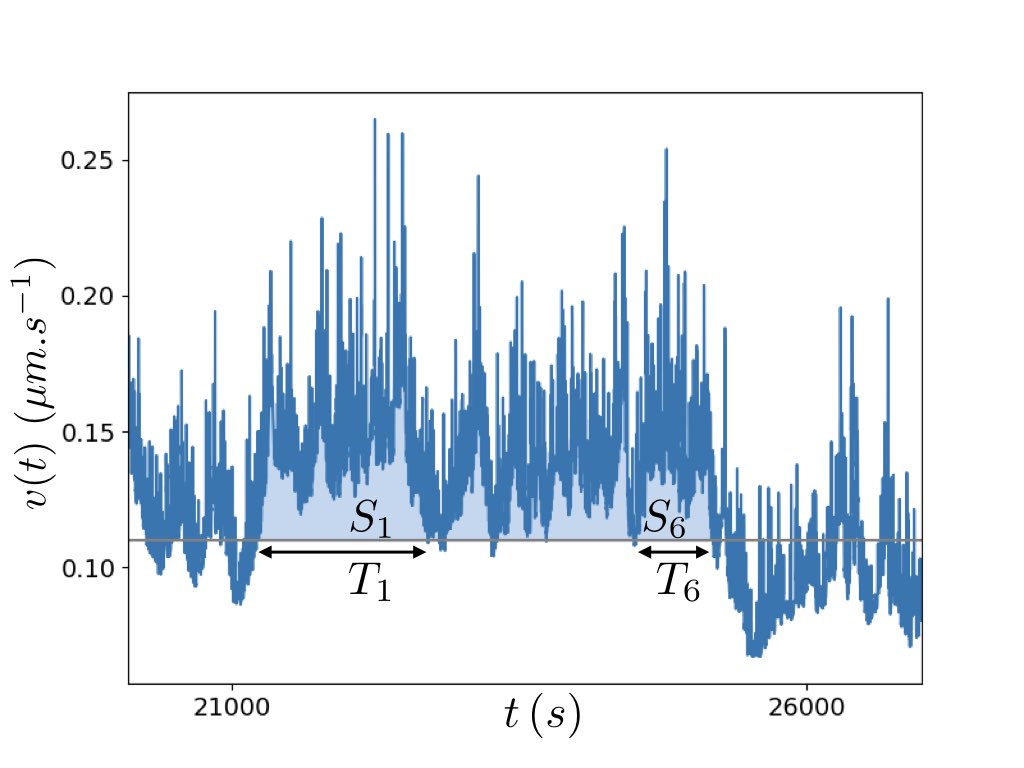}
	\caption{\emph{Left :} An avalanche is defined as a continuous period during which the center-of-mass velocity signal $v(t)$ exceeds a threshold $v_{th}$ (thin grey line). Here $v_{th}=0.10 \,\mathrm{\mu m.s}^{-1}$. 
	\emph{Right :} With the use of a threshold $v'_{th}=0.11\,\mathrm{\mu m.s}^{-1}$ instead of $v_{th}=0.10 \,\mathrm{\mu m.s}^{-1}$ one obtains six consecutives avalanches instead of a single one.
	The velocity signal shown here corresponds to the d'Alembert experiment in the regime of mean velocity $v = 0.13\, \mathrm{\mu m.s}^{-1}$.}
\end{figure}\label{Fig:ch3_v(t)_d'Alembert}

Experimental signals are affected by finite threshold that have different origins. 
First, experimental apparatus often have a finite sensitivity and do not record any signal below a certain threshold. 
Second, the numerization of an analogic signal induces an intrisic threshold. Finally a larger threshold is often imposed to substract noise signal or a finite background activity level. 
The latter case arises in all avalanche experiments where the system is driven at a finite velocity 
and avalanches are defined with respect to a threshold value of the signal, $v_{th}>0$.
This causes an important issue: if the threshold is too large, a single avalanche may be interpreted as a series of seemingly distinct events, while if it is too small subsequent avalanches can be merged into a single event. The sensitivity of the definition of avalanches with respect to the threshold is illustrated in figure~\ref{Fig:ch3_v(t)_d'Alembert}.
This raises the question of the robustness of the measurement of avalanche exponents or avalanche shapes with respect to the threshold.

In Ref.~\cite{janicevic2016} the authors
identify two main perils of a too high thresholding, based on numerical simulations and data from the Oslo crack front experiment.
The first one is that events appear to be time-clustered with a power law distribution of the waiting times between two consecutive events whereas if the events where uncorrelated, as it is expected, 
the waiting times should be distributed exponentially.
The second one is that the scaling exponent of observables might be affected by the threshold (this was already observed in~\cite{laurson2009} for the duration exponent $\tilde{\alpha}$). Especially the exponent $\gamma$ relating the avalanche size and duration ($\langle S \rangle_T \sim T^{\gamma}$)
decreases when the threshold is increased.
The exponential distribution of waiting times and the correct value of $\gamma$ are recovered when the threshold value is much smaller than the mean velocity of the line ($v_{th} \ll v$) while the number of events is maximized when $v_{th} \approx v$. 


\paragraph{Choosing the right interface size}\label{sec: Ch3 Choosing the right interface size}

\begin{figure}
	\centering
	\includegraphics[width=0.6\linewidth]{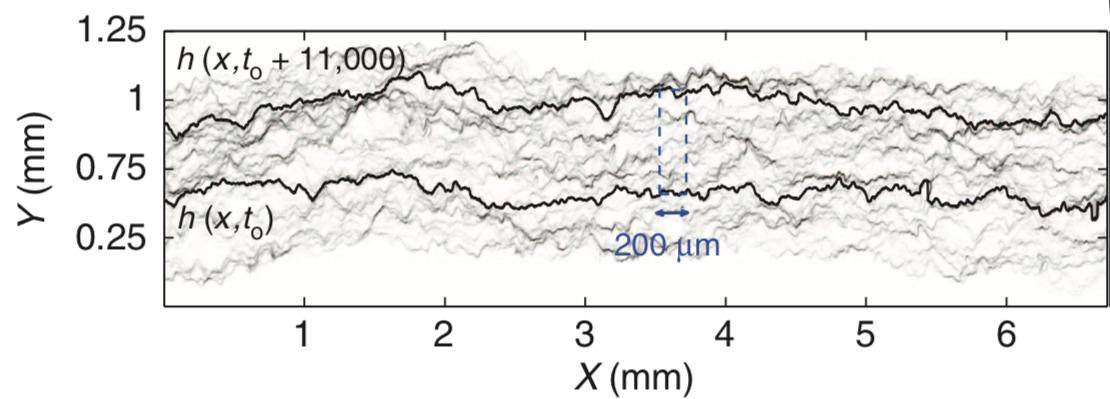}
	\caption{The signal studied in~\cite{laurson2013} is the mean velocity of segments of the interface of lenght $200 \mu$m . Image from~\cite{laurson2013}.}
\end{figure}\label{Fig: Finite Portion Interface Laurson}

In the thermodynamic limit the center-of-mass velocity $v(t)$ of an interface driven at finite velocity should be constant $v(t) \equiv v$. The bursts in $v(t)$ observed in experiments or numerical simulations are a consequence of the finite size of the system.
In a large system the drive can trigger in a very short time interval two avalanches at different spatial locations. These avalanches are distincts but overlap in the center-of-mass signal $v(t)$ leading to a seemingly larger avalanche.
Defining avalanches therefore requires to have an interface that has the "right size". For instance in~\cite{laurson2013} the signal $v(t)$ is defined for portions of the front of length $200~\mu$m
(see figure~\ref{Fig: Finite Portion Interface Laurson}).
This value is two times higher than the velocity correlation length which was previously estimated to be $\xi_v \simeq 100~\mu$m by fitting the spatial velocity correlation function in~\cite{tallakstad2011}. 
The scale chosen is thus larger than the expected avalanche maximal extension but not by much, 
so it minimizes the risk of avalanches overlapping. 
This process also allows to increase the statistics.
Since the observed length of the crack front is 6.72 mm, 33 independent velocity signals can be extracted from one experiment ($6.72/0.2 = 33.6$).
Note that this process requires to know the local velocity field  in order to determine the velocity correlation length

\subsubsection{Difficulty with the local signal $v(x,t)$ : Reconstructing avalanches}\label{sec: Ch3 Reconstructing avalanches}

In wetting front or crack front propagation, such as the Oslo experiment, one has access not only to the global but also to the local velocity signal $v(x,t)$. 
Local events can therefore be defined by thresholding the local velocity signal.
This, of course, raises the same issues as the thresholding from the global signal discussed above.
A second problem is that the events defined by thresholding the local signal are differents from the avalanches obtained by thresholding the global signal.
In Ref.~\cite{maloy2006} the distribution of the size of local events was found to follow a power law with an exponent 
$\tau_c=1.7\pm0.1$ wich is much higher than the scaling prediction for the avalanche size exponent~\eqref{eq: scaling relation tau alpha zeta} $\tau=2-\alpha/(d+\zeta) = 1.28$ for the elastic line with long-range elasticity and $\alpha=1$. 
Indeed the events defined by thresholding the local velocity signal are not avalanches but clusters of activity that belong to larger avalanches. The exponent measured is thus the exponent of the cluster size distribution~\cite{laurson2010}.
Measuring the avalanche size would require to reconstruct the avalanches, that is to determine which events belong to the same avalanche. This is a difficult task that has not been achieved yet.
This issue of reconstructing avalanches can nevertheless be bypassed if one is able to relate the cluster exponent $\tau_c$ to the global avalanche $\tau$. This enterprise has been initiated in~\cite{laurson2010} and 
we complemented it in a recent publication that will be the subject of chapter~\ref{chap: cluster statistics}~\cite{lepriol2020a}.


\section{A novel method based on the universal scaling of the local velocity field}\label{sec: Ch3 mon travail}


In the publication~\cite{lepriol2020} we proposed a new method to determine determine the universality class of a system in simulations and experiments where the local velocity field is accessible. 
This new method does not require to define avalanches thus bypassing the problems just mentionned in section~\ref{sec: Ch3 Difficulties}. Especially no thresholding of the acquired signal is needed (but an intrisic threshold when acquiring the data remains). 
It is based on the two-point spatial and temporal correlations of the velocity field, two quantities that are easy to compute once the local velocity field is known. 
The spatial correlation function has already been studied for the SRE line in Ref.~\cite{duemmer2005}(see section \ref{sec: Ch3 Duemmer&Krauth}). The authors showed that a correlation length $\xi$ and the exponent $\nu$ can be extracted from this function. They also found that the exponent $\kappa$ characterizing the short scale decay of the correlation function is non-universal. 
However if the depinning transition is really a critical phenomenon, we expect the correlation functions of the order parameter, i.e. the velocity, to have a universal scaling form, as it it the case for the correlation function of the magnetization in the Ising model~\cite{kardar2007}.

In this section I establish the universal scaling form of the spatial and temporal correlations functions of the velocity and show the agreement with both numerical simulations and data from the d'Alembert crack experiment.

\subsection{Scaling forms of the correlation functions}

I start by establishing the scaling forms of the correlation functions using scaling arguments for the short-scale behaviors and an effective Langevin equation for the large-scale behaviors.
Since it is often the mean velocity and not the force that is imposed in experiments we choose to work with a driving velocity $v_d$. It is implemented analytically by taking the external force \eqref{eq: F ext} to be 
$f(x,t)=m^2\left(v_d t - u(x,t)\right)$. In the stationary regime the global velocity of the line averaged over time and space is equal to the driving velocity~: $v=v_d$.
Let us write explicitly the equation of motion that we study in this section~:
\begin{equation}
v(x,t) = \partial_t u(x,t) = m^2\left(v_d t - u(x,t)\right) + \int \frac{u(y)-u(x)}{|y-x|^{1+\alpha}} + \eta \left(x, u(x,t)\right) \, , \label{eq: eq of motion ch3}
\end{equation}
where $\eta(x,u)$ denotes the quenched disorder force and we use the LR elastic kernel \eqref{eq: LR elasticity kernel} for the elastic force.
The elastic coefficient and the friction coefficient are set to equal to one for the sake of simplicity.

We expect the spatial correlation function $C_v(x)$ and the temporal correlation function $G_v(\tau)$ to have the following scaling forms~:
\begin{align}
C_v(x) &:= \langle \left(v(0,t)-v_d\right) \left(v(x,t)-v_d\right) \rangle = \langle v(0,t) \; v(x,t) \rangle - v_d^2 = v_d^2 \; \mathcal{F} \left( \frac{x}{\xi_v} \right) \, , \label{eq:Cv(x)} \\
G_v(\tau) &:= \langle \left(v(x,t)-v_d\right) \left(v(x,t+\tau)-v_d\right) \rangle = \langle v(x,t)\; v(x,t+\tau) \rangle - v_d^2 = v_d^2 \; \mathcal{G} \left( \frac{\tau}{t^*} \right)\, . \label{eq:Gv(tau)}
\end{align}
where $\mathcal{F}$ and $\mathcal{G}$ are universal functions. 
The brackets $\langle ... \rangle$ denote an average over both time and space which, by ergodicity, is equivalent to the average over disorder. 
The presence of the factor $v_d^2$ in front of the universal functions ensures the homogeneity of the equations. 
The proposed scaling forms rely on the existence of a single characteristic scale~: 
\begin{align}
\xi_v \sim v_d^{-\nu/\beta} \, , \label{eq: scaling xi_v} 
\end{align}
called the correlation length at finite velocity. Its scaling arises from the combination of the scalings of the velocity $v_d=v\sim \left(f-f_c \right)^{\beta}$ and of the correlation length $\xi \sim (f-f_c)^{-\nu}$.
The dynamical exponent $z$ then gives a naturally associated correlation time~: 
\begin{equation}
t^* \sim \xi_v^z \sim v_d^{-\nu z/\beta}\, . \label{eq: scaling t^*}
\end{equation} 

Besides $\xi_v$ there are three other lengths in the system~:
\begin{itemize}
\item $a$ the size of a pixel,
\item $L$ the total length of the line,
\item $\ell_m \sim m^{-2/\alpha}$ the characteristic length induced by the harmonic confinement and beyond which the interface is flattened.
\end{itemize}
In order for our scaling assumptions to hold $\xi_v$ must be the limiting length scale and so must be smaller than $\ell_m$ and $L$. It must also be larger than the pixel size $a$ if we want to detect the small scale behavior. 
Finally critical behaviors are visible only in systems whose size can be considered to be infinite. This means that $L$ must be larger than $\ell_m$. Otherwise the scaling form would be blurred by finite size effects. 
Therefore the four lengths must be ordered as follows~:
\begin{equation}
a \ll \xi_v \ll \ell_m \ll L \, .
\end{equation}
\smallskip

In the following subsections I derive the asymptotic forms of $\mathcal{F}(y)$ and $\mathcal{G}(y)$ via a scaling analysis based on the existence of a unique correlation length and a unique correlation time when $v_d$ is small. 
Below the scales $\xi_v$ and $t^*$ we expect to find a critical behavior governed by the exponents of the depinning transition. Above these scales we should recover the large force (equivalent to high velocity) behavior.

\subsubsection{Scaling argument for the critical behaviors}
\begin{figure}[t!]
	\centering
	\includegraphics[width=0.6\linewidth]{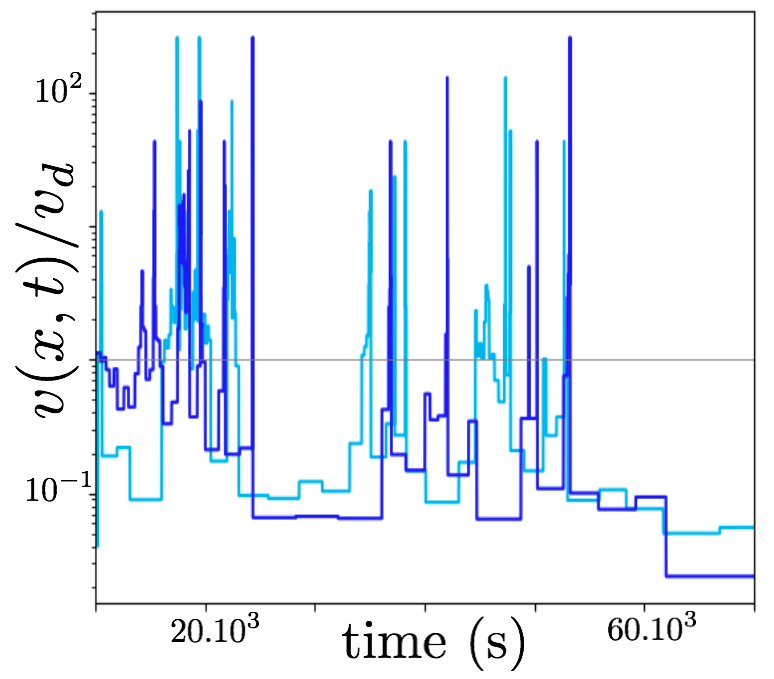}
	\caption{Velocity of two points close to each other along the line. The velocity is rescaled by the driving velocity $v_d$. Each point has a very intermittent velocity and they are correlated. The data are issued from the d'Alembert crack experiment. \label{Fig: 2-point velocity ch3}}
\end{figure}
The figure \ref{Fig: 2-point velocity ch3} shows the velocities of two points close to each other along the line in the d'Alembert crack experiment. When the drive is slow, $v_d \ll 1$, the local velocity is very intermittent~: during avalanches it takes values of order $v^{\max}$ independent of $v_d$ and it is almost zero during quiescence periods. In order for the mean velocity to be equal to $v_d$, the realizations of values of order $v^{\max}$ arise with a probability proportional to $v_d$. 
The main contribution to the spatial correlations $C_v(x)$ comes from the realizations for which both $v(0,t)$ and $v(x,t)$ are of order $v^{\max}$. The expectation of $v(x,t)$ conditioned by $v(0,t)$ being of order $v^{\max}$ has a universal scaling form~:
$ \langle v(x,t)|v(0,t) \sim v^{\max} \rangle = v_d \mathcal{H}(x/\xi_v )$.
Once again the factor $v_d$ ensures the homogeneity of the equation.
In order to have a finite expectation when $v_d \to 0$ we must have $ \mathcal{H}(y) \sim y^{-\beta/\nu}$ when $y \to 0$. Hence for $x \ll \xi_v$ we have~:
\begin{equation}
C_v(x) \sim v_d \times \langle v(x,t)|v(0,t) \sim v^{\max} \rangle \underrel{x \ll \xi_v}{\sim} v_d^2 \left(\frac{x}{\xi_v}\right)^{-\beta/\nu} \, . \label{eq: asymptote small scales Cv(x)}
\end{equation}
The same reasoning gives the small scale asymptote of the temporal correlation function. 
We have the universal scaling form of the conditional expectation
$\langle v(x,t+\tau)|v(x,t) \sim v^{\max} \rangle = v_d \tilde{\mathcal{H}}(\tau/t^* )$ and the requirement of this expectation to be finite when $v_d \to 0$ implies that
$ \tilde{\mathcal{H}}(y) \sim y^{-\beta/(\nu z)}$ when $y \to 0$. 
Thus at small time, $\tau \ll t^*$, the temporal correlation function reads~:
\begin{equation}
G_v(\tau) \sim v_d \times \langle v(x,t+\tau)|v(x,t) \sim v^{\max} \rangle \underrel{\tau \ll t^*}{\sim}v_d^2 \left(\frac{\tau}{t^*}\right)^{-\beta/(\nu z)} \, . \label{eq: asymptote small scales Gv(tau)}
\end{equation}
Note that the last terms in \eqref{eq: asymptote small scales Cv(x)} and \eqref{eq: asymptote small scales Gv(tau)} scale as $v_d x^{-\beta/\nu}$ and $v_d \tau^{-\beta/(\nu z)}$ respectively.
Hence the correlation functions $C_v(x)$ and $G_v(\tau)$ are of order $v_d$ in this regime.

\subsubsection{Thermal approximation for the large scale behaviors}\label{sec: thermal approximation ch3}

For the purpose of computing the large scale behaviors of the correlation functions 
it is convenient to rewrite equation \eqref{eq: eq of motion ch3} in the 
comoving frame : $u(x,t) \rightarrow v_d t + u(x,t)$. The disorder term then becomes $\eta\left(x, v_d t + u(x,t) \right)$.
To describe large scales $x > \xi_v$ or $\tau > t^*$, a reasonable approximation is to use an effective model
where the disorder is replaced by a white noise. 
Its correlations read :
\begin{equation}\label{eq:Correlations disorder}
\langle \eta\left(0, u(0,0)\right) \eta\left(x, v_d \tau + u(x,\tau)\right)\rangle = \delta(x)\delta\left(v_d\tau +\Delta u (x,\tau)\right) \, 
\end{equation}
where $\Delta u (x,\tau) := u(x,\tau) - u(0,0)$.
For simplicity the coefficient in front of the delta functions in \eqref{eq:Correlations disorder} is set to unity (which can always be done by a choice of units). 
At large length scale $x > \xi_v$ the correlation is zero because of the term $\delta(x)$.
At large time scale, we see by dimensional analysis that\footnote{This comes from the combination of $\tau \sim x^z$ and  $\Delta u \sim x^{\zeta}$.}
$\Delta u \sim \tau^{\zeta/z}$.
Hence $v_d \tau / \Delta u \sim (t^*)^{-\beta/(\nu z)} \tau^{1-\zeta/z} \sim \left(\tau / t^* \right)^{\beta/(\nu z)}$.
So at large time scale, $\tau > t^*$, $\Delta u (x,\tau)$ is subdominant compared to $v_d \, \tau$. 
On large scales the disorder can thus be replaced
by an effective thermal noise $\eta (x, v_d t)$ and 
the equation of motion \eqref{eq: eq of motion ch3} becomes a Langevin equation~:
\begin{align}
\partial_t u(x,t) &= -m^{2}u(x,t) + \eta\left(x, v_d t \right) + \int \frac{u(x',t) - u(x,t)}{|x'-x|^{1+\alpha}} dx' \, ,  \label{eq: Langevin eq. real space} \\
\partial_t \hu(q,t) &= -m^{2}\hu(q,t) + \hat{\eta}\left(q, v_d t \right) -|q|^{\alpha} \hu(q,t) 
 		= -b(q)\hu(q,t) + \hat{\eta}\left(q, v_d t \right) \, . \label{eq: Langevin eq. fourier space}
\end{align}
where $b(q) = m^2 + |q|^{\alpha}$ and $\hat{\eta}\left(q, v_d t \right)$ is the Fourier Transform of 
$\eta(x,v_d t)$ with respect to the first variable only.
The Fourier transformation from equation \eqref{eq: Langevin eq. real space} to \eqref{eq: Langevin eq. fourier space} holds for $0 < \alpha < 2$. 
Note that equation \eqref{eq: Langevin eq. fourier space} with $\alpha = 2$ corresponds to short-range (SR) elasticity (see section \ref{sec: ch1 elastic force}).
It is a first order linear differential equation and its solution reads :
\begin{equation}\label{eq: solution Langevin equation}
\hu(q, t) = \int_{-\infty}^{t} e^{-b(q) (t-t')}  \hat{\eta}(q,v_d t') dt' \, .
\end{equation}

With this solution we can compute the general two-point correlation function, which we denote $C(x,\tau)$, that depends on both time and space. It can be written as a Fourier transform~:
\begin{equation}\label{eq:General 2-points correlations}
C(x,\tau) := \langle \partial_t u(y,t) \partial_t u(y+x,t + \tau) \rangle = \int \frac{dq_1 dq_2}{(2\pi)^2} e^{iq_1 y} e^{iq_2(x+y)}
			 \langle \partial_t \hu(q_1,t) \partial_t \hu(q_2,t + \tau) \rangle \, .
\end{equation}
We can compute the correlation term in Fourier space by plugging in 
\begin{equation}
\partial_t \hu(q,t) = \hat{\eta}(q,v_d t) - b(q) \int_{-\infty}^{t} e^{-b(q) (t-t')}  \hat{\eta}(q,v_d t') dt'
\end{equation} 
with the noise correlations 
\begin{equation}
\langle \hat{\eta}(q_1,v_d t_1) \hat{\eta}(q_2,v_d t_2) \rangle =  2 \pi \delta(q_1+q_2) \delta(t_1-t_2) / v_d \, .
\end{equation}
The result is~:
\begin{align}
\langle \partial_t \hu(q_1,t) \partial_t \hu(q_2,t + \tau) \rangle
	&= \frac{ 2 \pi \delta(q_1+q_2)}{v_d} \left( \delta(\tau) - b(q_2) e^{-b(q_2)\tau} + b(q_1)b(q_2)
	\frac{e^{-b(q_2)\tau}}{b(q_1)+b(q_2)} \right) \, .
\end{align}
Performing one integral over $q$, equation \eqref{eq:General 2-points correlations} now reads :
\begin{equation}\label{eq: General 2-points correlations simplified ch3}
C(x,\tau) =  \frac{1}{v_d} \delta(x) \delta(\tau) 
		- \frac{1}{2 v_d} \int \frac{dq}{2\pi} e^{iqx} b(q) e^{-b(q)\tau} \, .
\end{equation}
The first term is local and originates from the delta function approximation to the noise,
and represents the correlations at short scales, not described accurately by the present effective model.
The large scale behavior is described by the second term, on which we now focus.

The integral in \eqref{eq: General 2-points correlations simplified ch3} is easily computed for 
$\alpha=1$ which is the case of most interest since it describes physical systems. Setting the mass to zero we find,
at large scales $x > \xi_v$ or $\tau > t^*_v$,
\begin{equation}\label{eq: C(x,tau) alpha=1 ch3} 
C(x,\tau) \simeq  \frac{1}{2 v_d} \partial_\tau \int \frac{dq}{2\pi} e^{iqx} e^{-|q| \tau} 
= \frac{1}{2 \pi v_d} \partial_\tau \frac{\tau}{x^2 + \tau^2} = \frac{1}{2 \pi v_d}  \frac{x^2-\tau^2}{(x^2 + \tau^2)^2}
\end{equation}

\paragraph{Spatial correlations}
To obtain the spatial correlations, we set $\tau=0$. For $\alpha=1$ we find the decay
\begin{equation}\label{eq: tail spatial correlation alpha=1 ch3}
C_v(x) 
	    \underrel{x\gg\xi_v}{\simeq} \frac{1}{2 \pi v_d}  \frac{1}{x^2} 
\end{equation}
To extend this result to general $0<\alpha \leq 2$ we start from
the second term of \eqref{eq: General 2-points correlations simplified ch3} where $\tau$ is set to 0. 
The Fourier transform of $|q|^{\alpha}$ is regularized by introducing a factor $e^{-\epsilon |q|}$
and then 
taking the limit $\epsilon \to 0$ at the end. Using
\begin{align}
\int \frac{dq}{2\pi} e^{iqx-\epsilon |q|} |q|^{\alpha} &= \frac{\Gamma(1+\alpha)}{\pi (\epsilon^2+x^2)^{\frac{1+\alpha}{2}}}
   \cos \left( (1+\alpha) \arctan \left(\frac{x}{\epsilon} \right) \right) \, ,
\end{align}
we find the general result~:
\begin{equation}\label{eq: Cv(x)_appendix}
C_v(x) \underrel{x\gg\xi_v}{\simeq} \, \frac{\Gamma(1+\alpha) \sin \left(\frac{\pi \alpha}{2} \right)}{2\pi v_d} \,  \frac{1}{x^{1+\alpha}} \quad \text{for} \quad \alpha < 2 \, .
\end{equation}

\paragraph{Short-range elasticity}
For short-range elasticity ($\alpha=2$) the prefactor in \eqref{eq: Cv(x)_appendix}
vanishes, and the large distance decay is not a power law anymore within this model,
but is much faster.
One can obtain some qualitative idea by considering 
a model with a finite correlation length along $x$, e.g. replacing $\delta(x)\rightarrow e^{-\frac{|x|}{\ell}}$.
The disorder correlator then becomes~:
\begin{equation}
\langle \hat{\eta}(q_1, v_d t_1) \hat{\eta}(q_2, v_d t_2) \rangle = \frac{2 \ell}{1+(q_2 \ell)^2} \, (2 \pi) \delta(q_1+q_2) \, \frac{\delta(t_1-t_2)}{v_d} \, . 
\end{equation}
When computing the spatial correlation function in the limit $m\to 0$, we obtain, discarding all $\delta(\tau)$ and
$\delta(x)$ terms (i.e. assuming $\ell$ is the largest length)~:
\begin{equation}
C_v(x) \underrel{x\gg\ell}{\simeq} - \frac{1}{v_d \ell} \int \frac{dq}{2 \pi} e^{i q x} \frac{(\ell q)^2}{1+ (\ell q)^2}
\simeq \frac{1}{2 v_d \ell^2} e^{- |x|/\ell} \, , \label{eq: tail Cv(x) SRE ch3}
\end{equation}
which is a rationale for the exponential decay observed in Ref.~\cite{duemmer2005}. It is then
reasonable to expect that $\ell$ will be of order $\xi_v$. We emphasize that this is not an accurate calculation which would require to account for more details about
the renormalized disorder.

\paragraph{Temporal correlations}
We now turn to the temporal correlations. For $\alpha=1$
we set $x=0$ and $\tau > 0$ in \eqref{eq: C(x,tau) alpha=1 ch3}  and obtain~:
\begin{equation}\label{eq: tail Gv(tau) alpha=1}
G_v(\tau) \underrel{\tau \gg t^*}{\simeq}  - \frac{1}{2 \pi v_d}  \frac{1}{\tau^2} \, .
\end{equation}
For general $\alpha$ we set $x=0$ and $\tau>0$ in the second term of \eqref{eq: General 2-points correlations simplified ch3}~:
\begin{align}
G_v(\tau) &\underrel{\tau \gg t^*}{\simeq} -\, \frac{1}{2v_d} \int \frac{dq}{2\pi} b(q)e^{-b(q)\tau} 
\underrel{\tau \gg t^*}{\simeq} -\, \frac{e^{-m^2\tau}}{2\pi v_d} \int_0^{\infty}  dq (m^2 + |q|^{\alpha}) e^{-|q|^{\alpha} \tau} \notag \\
G_v(\tau) &\underrel{\tau \gg t^*}{\simeq} -\, \frac{e^{-m^2\tau} (1+m^2\alpha\tau)\Gamma\left(\frac{1}{\alpha}\right)}{2\pi v_d \alpha^2} \; \frac{1}{\tau^{1+\frac{1}{\alpha}}} \, .
\end{align}
In the limit $m \to 0$ this reduces to :
\begin{equation}\label{eq: tail Gv(tau) generic alpha}
G_v(\tau) \underrel{\tau \gg t^*}{\simeq} -\,\frac{\Gamma\left(\frac{1}{\alpha}\right)}{2\pi v_d \alpha^2} \; \frac{1}{\tau^{1+\frac{1}{\alpha}}} \, .
\end{equation}
Before ending this section let us mention that the approach we used, based on replacing the quenched noise with a velocity dependent thermal noise, does not allow to recover the correct dependence on $v_d$ but only the large scale 
dependence on $\tau$ and $x$. 
Indeed, in the replacement \eqref{eq:Correlations disorder}
we have not tried to be accurate~: one could refine the model by multiplying 
by a prefactor with the correct dimension, and appropriate dependence in velocity (which 
could in principle be predicted by the functional renormalization group~\cite{chauve2000}.) 

\subsubsection{Complete scaling forms}
Collecting the results \eqref{eq: asymptote small scales Cv(x)}, \eqref{eq: asymptote small scales Gv(tau)} for the short scale asymptotes and \eqref{eq: Cv(x)_appendix}, \eqref{eq: tail Gv(tau) generic alpha} for the large scale tails we can write the full scaling forms
(we recall that $C_v(x)=v_d^2 \mathcal{F}\left(x/\xi_v\right)$ and 
$G_v(\tau)=v_d^2 \mathcal{G}\left(\tau/t^*\right)$)~:
\begin{align}
\mathcal{F}(y) &\sim \left\lbrace 
\begin{array}{l l l}
y^{-\frac{\beta}{\nu}} \; &  \text{ if } \; y \ll 1 \, , & \\
y^{-(1+\alpha)} \; &  \text{ if } \; y \gg 1 \, , & \text{for } \; \alpha < 2 \, , \\
e^{-y} \; &  \text{ if } \; y \gg 1 \, , & \text{for SR elasticity} \; (\alpha = 2) \, ,
\end{array} \right. \label{eq: full scaling form Cv(x) generic alpha} \\
\mathcal{G}(y) &\sim \left\lbrace 
\begin{array}{l l l}
y^{-\frac{\beta}{\nu z}} \; & \text{ if } \; y \ll 1 \, , & \\
-y^{-\left(1+\frac{1}{\alpha}\right)} \; &  \text{ if } \; y \gg 1 \, , & \text{ for } \; \alpha \leq 2 \,\, .
\end{array} \right. \label{eq: full scaling form Gv(tau) generic alpha}
\end{align}
In particular for $\alpha=1$ we have~:
\begin{align}
\mathcal{F}(y) &\sim \left\lbrace 
\begin{array}{c}
y^{-\frac{\beta}{\nu}} \quad  \text{ if } \; y \ll 1 \, , \\
y^{-2} \quad \text{ if } \; y \gg 1 \, ,
\end{array} \right. \label{eq: full scaling form Cv(x) alpha=1} \\
\mathcal{G}(y) &\sim \left\lbrace 
\begin{array}{c}
y^{-\frac{\beta}{\nu z}} \quad \text{ if } \; y \ll 1 \, , \\
-y^{-2} \quad \text{ if } \; y \gg 1 \, .
\end{array} \right. \label{eq: full scaling form Gv(tau) alpha=1}
\end{align}

Note that at large distance, the decay of the spatial correlation function provides exactly the range $1 + \alpha$ of the elastic interactions.
The negative sign in front of the large scale decay of the temporal correlation function is the mark of anticorrelations~: when the time delay $\tau > t^*$ the local velocity $v(x,t+\tau)$ is likely to be low if $v(x,t)$ was high, and to be high if $v(x,t)$ was low.

The simplest explanation for this anticorrelation predicted in
$G_v(\tau)$ is that it arises from the alternance of avalanches and quiescience
periods. Note that an anticorrelation was also predicted between the sizes
(and the velocities) of successive avalanches, and measured in numerics~\cite{ledoussal2019}.
This prediction however was obtained in the quasistatic limit $v_d \to 0$ 
using the mass as a cutoff and expressed as a function of the scaling
variable $W m^{\zeta}$, where $W=w_2-w_1$, the first avalanche being
triggered at $w_1$ and the second at $w_2$. To make the connection
to the present observation more precise, and confirm whether it is part of the same phenomenon, one would need to extend the result of~\cite{ledoussal2019} to zero mass and 
a finite $v_d$, with $W= v_d \tau$ corresponding to a time separation $\tau$.

\subsection{Numerical model : a cellular automaton}\label{sec: ch3 numerics}

\begin{figure}[!p]
	\begin{subfigure}[t]{1\linewidth}%
	\centering
	\includegraphics[width=.65\linewidth]{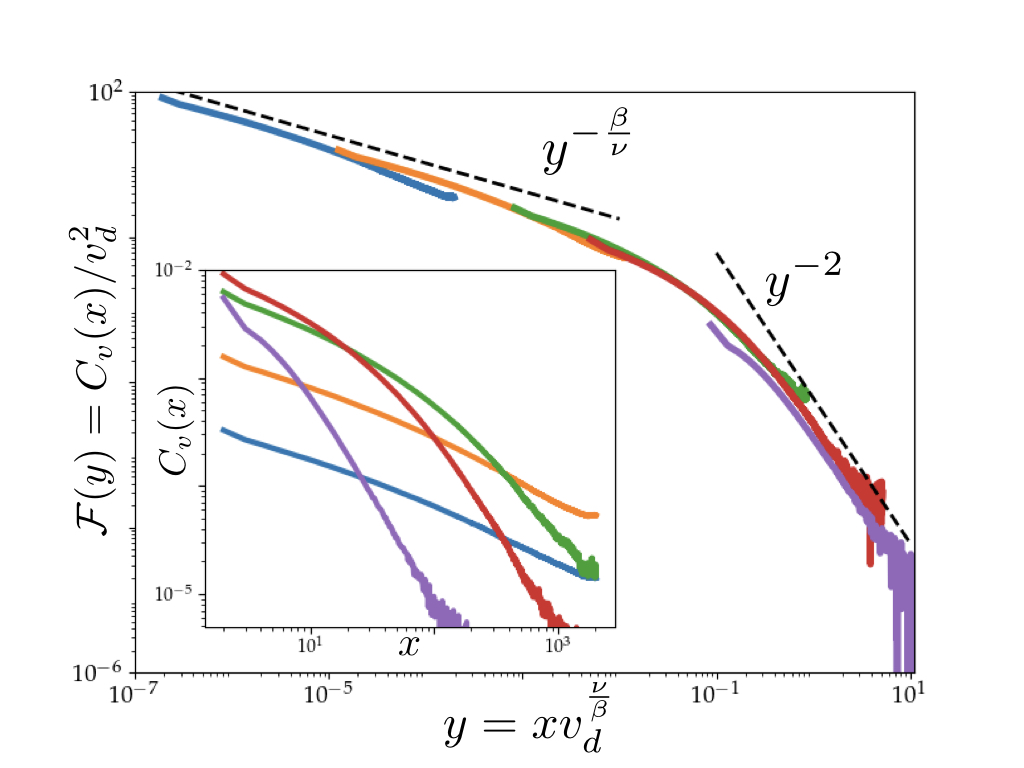}
	\caption{Spatial correlations in the cellular automaton for driving velocities 
$v_d=0.002$ (blue), $0.01$ (yellow), $0.05$ (green), $0.1$ (red) and $0.3$ (purple). 
\textbf{Main panel :} A perfect collapse is observed using the scaling form \eqref{eq:Cv(x)}.
The asymptotic behaviors \eqref{eq: full scaling form Cv(x) alpha=1} are verified. In particular, going at large distances the decay $y^{-2}$ of the elastic interaction is recovered while at small distances the critical behavior 
$\beta/\nu \simeq 0.385$ is captured.
From the crossover the length $\xi_v$ is estimated to be $\xi_v \simeq 0.07 \, v_d^{-\nu/\beta}$.
\textbf{Inset :} Nonrescaled correlation function $C_v(x)$.
System size : $L = 4096$, mass: $m^2 = 10^{-3}$.
\label{Fig:Cv(x)_simus}}
	\end{subfigure}
	\vspace{1.5cm}
    \begin{subfigure}[t]{1\linewidth}%
    \centering
	\includegraphics[width=.65\linewidth]{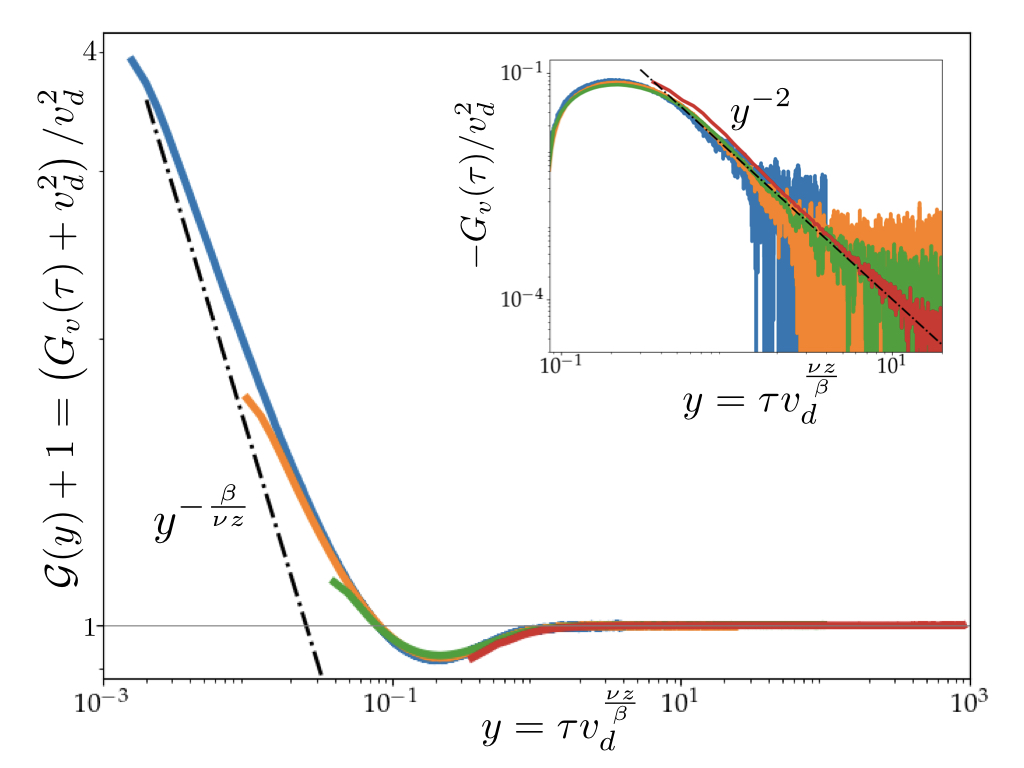}
	\caption{Temporal correlations in the cellular automaton for driving velocities $v_d=0.02$ (blue), $0.05$ (yellow), $0.1$ (green) and $0.3$ (red). 
\textbf{Main panel :} The scaling form \eqref{eq:Gv(tau)} and the asymptotic behaviors \eqref{eq: full scaling form Gv(tau) alpha=1} are verified.
In particular at large time we observe anticorrelations
while at small time the depinning behaviour $y^{-\beta/(\nu z)}$ is recovered ($\beta/(\nu z) \simeq 0.500$).
\textbf{Inset :} Zoom on the anticorrelation. For $\alpha=1$ the expected decay is $1/y^2$.
System size : $L$ and $m$ range from $L=2048$, $m^2=10^{-3}$ ($v_d=0.3$) to $L=32768$, $m^2=10^{-4}$ ($v_d=0.02$). \label{Fig:G(tau)_simus}}
	\end{subfigure}
	\caption{Correlation functions in the simulation. \label{Fig:Correlations_simus}}
\end{figure}

In order to test the scaling predictions 
I implemented a cellular automaton simulation of equation \eqref{eq: eq of motion ch3}.
The three variables $u$, $x$ and $t$ are integer and we assume periodic boundary conditions (PBC) along $x$ which takes values ranging from $0$ to $L-1$.
The local velocity is defined as :
\begin{equation}\label{eq: def v(x,t) automaton}
v(x,t) = \theta \left( m^2 \left( w_0 + v_d t - u(x,t) \right) + \sum_{x'} \frac{u(x',t) - u(x,t)}{|x'-x|^{2}} +  \eta\left(x, u(x,t)\right) \right) \, ,
\end{equation}
$\theta$ being the Heaviside function.
Here the quenched disorder  pinning force $\eta$ should be negative and uncorrelated. In practice, I take identical and independent variables whose distribution is the negative part of the normal law. At each time step all the points feeling a positive total force jump one step forward,
$u(x,t+1) = u(x,t)+1$,
while the other points - which feel a negative force - stay pinned.
Then the time is incremented, $t\rightarrow t+1$, and the forces are recomputed~: new pinning forces are drawn for the jumping points, the elastic force is updated by using a Fast Fourier Transform (FFT) algorithm and the driving force is incremented by $m^2 v_d$. I give more details about the computation of the elastic force in appendix~\ref{sec: appendix computation elasticity FFT}.

The simulation start from a flat line with $f(x,t)=0$. The driving force is first incremented by an amount $m^2 w_0$ such that the line moves enough to forget the initial flat configuration 
(as assured by the third Middleton theorem) and reaches a random critical configuration characterized by the depinning roughness exponent $\zeta$. 
Then I turn on the dynamics (the time $t$ starts to be incremented and $w_0$ is replaced by $w_0+v_d t$) and wait until reaching the steady state (characterized by $v=v_d$) before computing the correlation functions.

The results are shown on figure \ref{Fig:Correlations_simus}. 
Figure \ref{Fig:Cv(x)_simus} shows the spatial correlations for driving velocity ranging from $v_d=0.002$ to $v_d=0.3$. The inset shows the correlation functions before rescaling. 
In the main panel the correlation functions have been rescaled using the scaling assumption \eqref{eq:Cv(x)}. They perfectly collapse on a main curve corresponding to the scaling form 
$\mathcal{F}(y)$ given in equation \eqref{eq: full scaling form Cv(x) alpha=1}, confirming the 
predicted asymptotes. In particular the decay in $1/x^2$ is the fingerprint of the long-range nature of the elasticity. It also confirms that a unique correlation length $\xi_v$ controls the dynamics.
The value of the small scale exponent is $\beta/\nu = 0.385$.

The temporal correlations are shown on figure \ref{Fig:G(tau)_simus}. 
Driving velocities range from $v_d=0.02$ to $v_d=0.3$.
In order to visualize the anticorrelations on a log-log scale I do not plot $G_v(\tau)$ but $\left( G_v(\tau) + v_d^2 \right)/v_d^2$. The parts of the curves below $1$ correspond to anticorrelation.
Once again the curves rescaled using equation \eqref{eq:Gv(tau)} collapse on a unique main curve
corresponding to the scaling form \eqref{eq: full scaling form Gv(tau) alpha=1}.
The exponent $\beta/(\nu z) \simeq 0.50$ at small scale is recovered.
It is remarkable that the power law behavior $y^{-\beta/(\nu z)}$ holds for the non connected function $\mathcal{G}(y)+1$ until the time when anticorrelation appears.
The inset shows the power law decay $1/\tau^{2}$ in the anticorrelation regime.

\subsection{Experiments}

I also analyzed data issued from the d'Alembert experiment presented in section~\ref{sec: Ch2 Crack propagation}.
Here I first present the results for the correlation functions $C_v(x)$ and $G_v(\tau)$. The curves corresponding to different driving velocities are collapsed using a different rescalings than the ones predicted in equations~\eqref{eq:Cv(x)}, \eqref{eq:Gv(tau)}. 
In a following subsection I explain how this difference arises from the visco-elastic behaviour of the silicone substrate used in the d'Alembert experiment.

\begin{figure}[!p]
	\begin{subfigure}[t]{1\linewidth}%
	\centering
	\includegraphics[width=.65\linewidth]{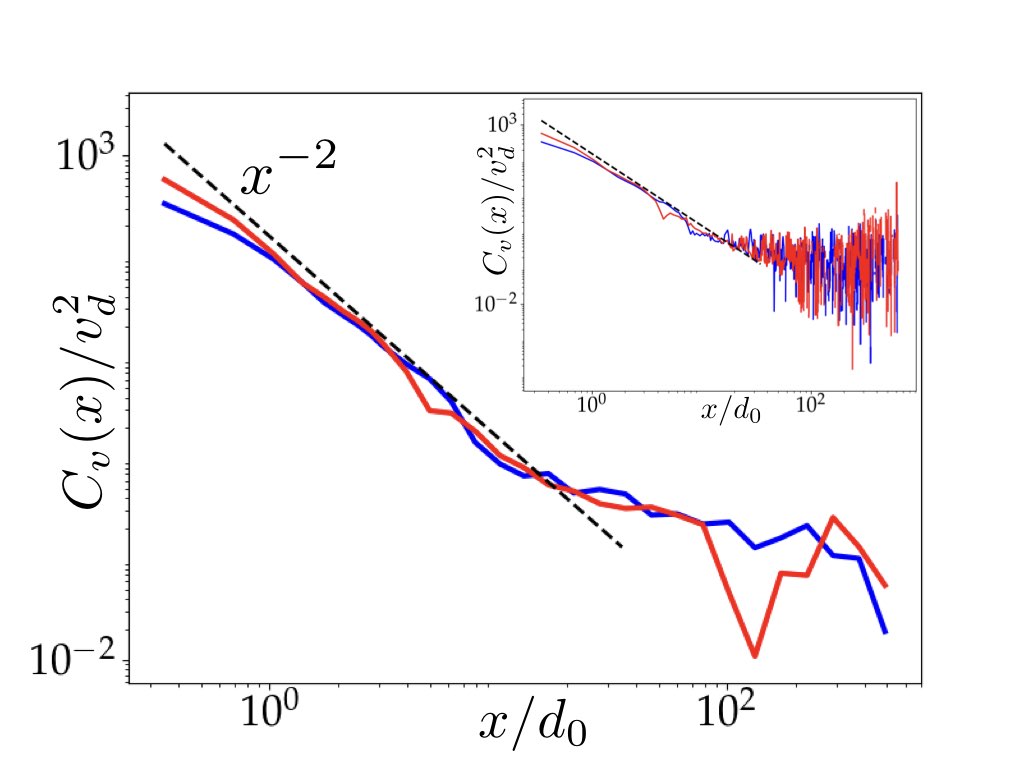}
	\caption{Spatial correlations in the experiment for driving velocities 
	$v_1 = 132$ nm/s (blue) and $v_2=31$ nm/s (red). 
	\emph{Inset :} Raw signal.
	\emph{Main panel :} 	The signal has been smoothened at large $x$ by averaging over windows of equal logarithmic size.
	The large distance decay $x^{-2}$ of the elastic 	interactions is observed.
	Note that no rescaling of the x-axis was performed.
	\label{Fig:Cv(x)_experiment}}
	\end{subfigure}
	\vspace{1.5cm}
    \begin{subfigure}[t]{1\linewidth}%
    \centering
	\includegraphics[width=.65\linewidth]{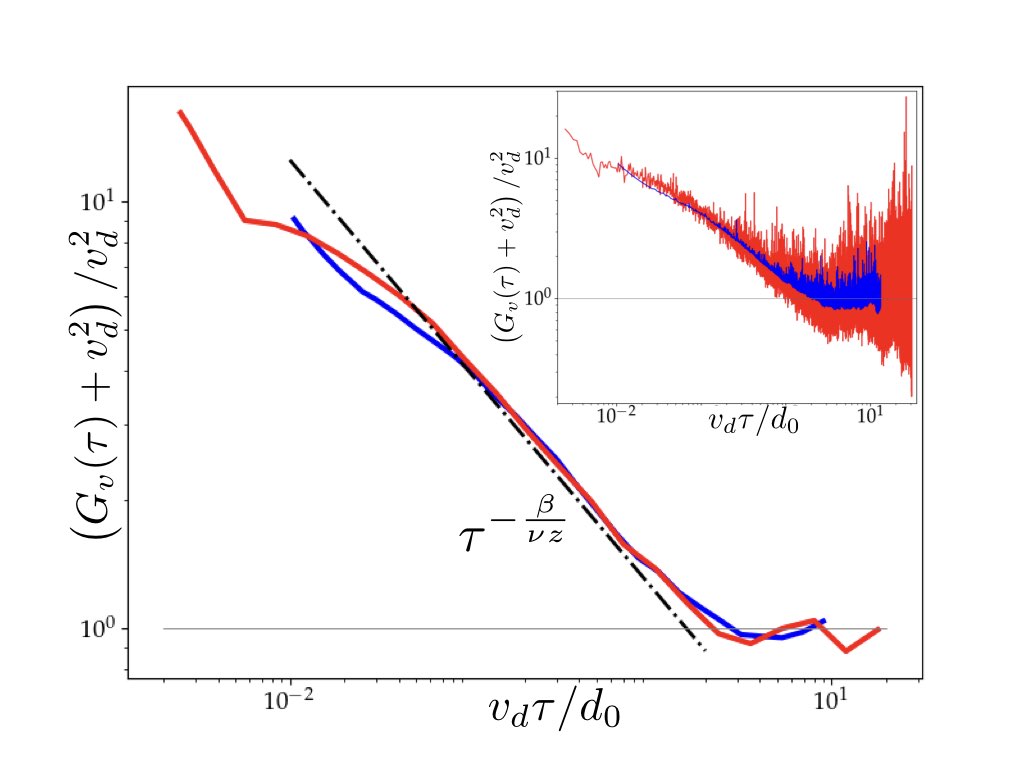}
	\caption{Temporal correlations in the experiment for driving velocities 
	$v_1 = 132$ nm/s (blue) and $v_2=31$ nm/s (red) .
	\emph{Inset :} Raw signal.
	\emph{Main panel :} 	The signal has been smoothened by averaging over windows of equal logarithmic size.
	The asymptotic predictions of~\eqref{eq: full scaling form Gv(tau) alpha=1} are verified :
	anticorrelation is observed at large time and the depinning power law decay is recovered at
	short time.
	The time axis has been rescaled by $v_d$ and thus corresponds to a distance. \label{Fig:G(tau)_experiment}}
	\end{subfigure}
	\caption{Correlation functions in the experiment. \label{Fig:Correlations_experiment}}
\end{figure}

The local crack velocity is computed using the methodology of the Waiting Time Matrix (WTM) introduced in Ref.~\cite{maloy2006}.
We count the number of frames during which the front stays inside each pixel which provides the waiting time in this pixel. Doing so for every pixel we obtain the WTM of the image.
The local crack speed in a pixel of coordinate $(x,y)$ is
$v(x,y) = p/WTM(x,y)$ where $p=34.7~\mu$m is the pixel size.
The position of the front of an abscisse $x$ at time $t$ is obtained by summing the waiting times along $x$ until the sum reaches $t$. I give more details about the method in appendix~\ref{sec: appendix experiment analysis}.
Two different driving velocity regimes are tested : $v_1= 132 \pm 3~\mathrm{nm.s}^{-1}$ and $v_2 = 31 \pm 1~\mathrm{nm.s}^{-1}$. 
The resulting correlation functions are shown on figure~\ref{Fig:Correlations_experiment}.

The plot of the spatial correlation function confirms the large distance decay as $1/x^2$ corresponding to the range of the elasticity. This is an experimental confirmation that the elastic kernel of the crack front is long-range in this experiment as predicted in Ref.~\cite{chopin2018}. 
The $x$-axis has been rescaled by the defect size $d_0=100~\mu$m. 
At variance with the numerical simulation, rescaling with a power of $v_d$ was not necessary to obtain a collapse of the two curves. 
This rather counter-intuitive behavior is explained by the fact that the silicone substrate is not perfectly elastic but visco-elastic. This induces a velocity dependence of the material toughness (or equivalently of the disorder pinning force)~\cite{kolvin2015} and varying the driving velocity does not change the distance to the critical point. I return to this point at the end of this section.
For both velocities the large scale behavior breaks down below $d_0$ for distances of $2$-$3$ pixels 
but the expected small scale behavior is not visible due to a lack of spatial resolution.

Regarding the temporal correlation function, a similar behavior as in the numerics is observed, with a crossover from a power law decay to anticorrelation.
The power law decay is consistent with the depinning prediction $G_v(\tau) \sim \tau^{-\beta/\nu z}$. 
The two curves corresponding to the crack speeds $v_1$ and $v_2$ are collapsed by rescaling the time axis by $v_d$ instead of $t^* = v_d^{\beta/(\nu z)}$. As for the spatial correlation this is explained by the velocity dependence of the pinning force that I discuss below. 
Rescaling the time by $v_d$ allows to compare the distance travelled by the drive instead of the time.

\begin{figure}[b!]
	\begin{subfigure}[t]{0.48\linewidth}%
	\centering
	\includegraphics[width=1\linewidth]{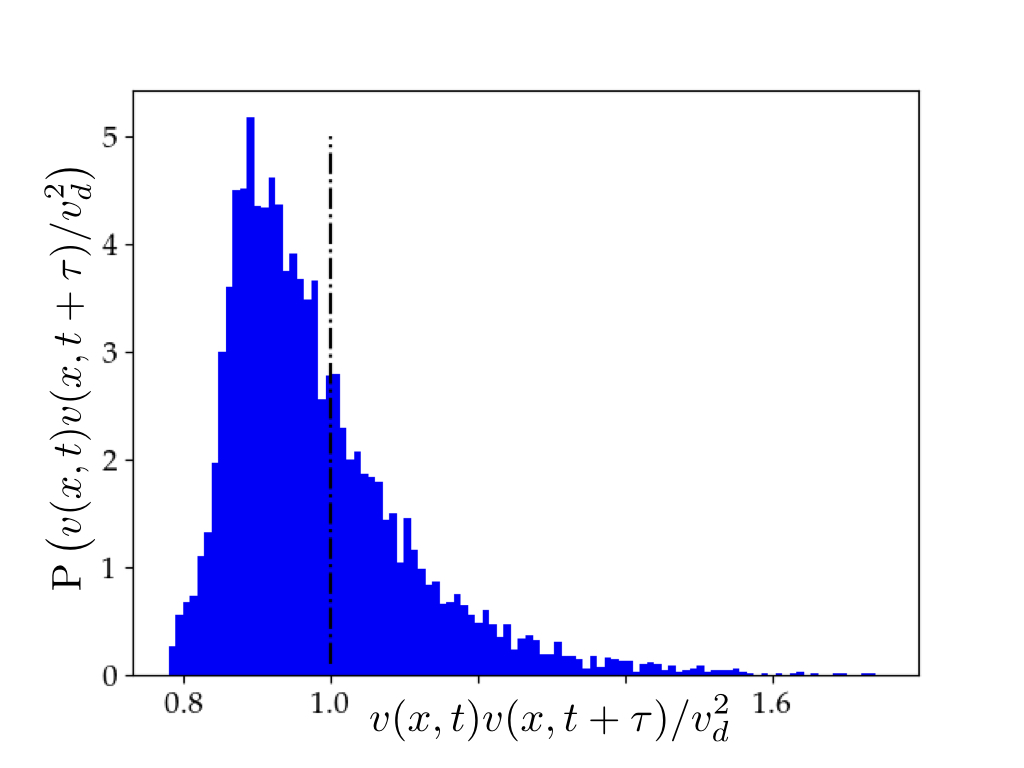}
	\caption{\label{Fig:PGtau_v1_experiment}}
	\end{subfigure}
	\hspace{.5cm}
    \begin{subfigure}[t]{0.48\linewidth}%
    \centering
	\includegraphics[width=1\linewidth]{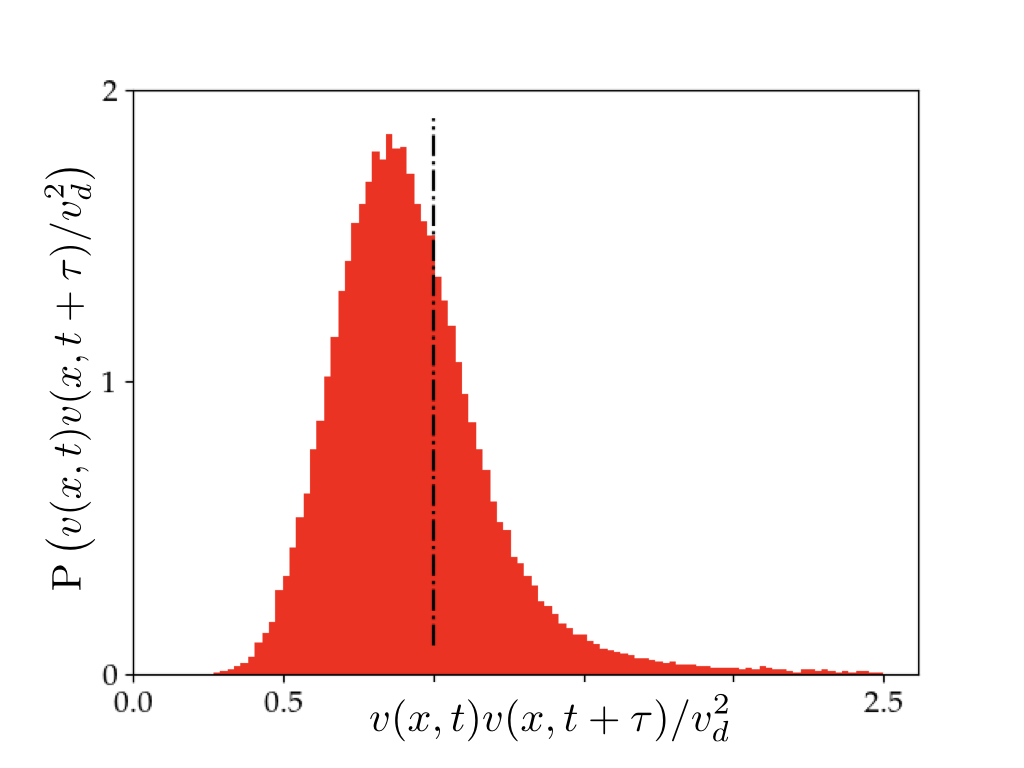}
	\caption{\label{Fig:PGtau_v2_experiment}}
	\end{subfigure}
	\caption{\textit{Left :} Histogram of $v(x,t)v(x,t+\tau)$ for driving velocity $v_1 = 132$ nm/s and for $\tau > 3$ in rescaled units (this corresponds to the point where we start to have some signal below $1$ on the right panel)
\textit{Right :} Same histogram for driving velocity $v_2 = 31$ nm/s and for $\tau > 2$ in rescaled units. \label{Fig: Histograms Gtau experiment}}
\end{figure}

\paragraph{Statistical significance of the anticorrelation.}
When we compute the correlation functions $C_v(x)$ and $G_v(\tau)$ from the experimental data we first obtain a very noisy curve, visible in the insets of figure~\ref{Fig:Correlations_experiment}. In the main panels the curves are smooth because I took the average values over bins of equal logarithmic size. 
Looking at the raw signal for the temporal correlations one might wonder if the anticorrelation visible on the smoothened curve in the main panel is real or is an artifact due to statitical noise.
To discriminate I plot on figure~\ref{Fig: Histograms Gtau experiment} the histogram of  $v(x,t)v(x,t+\tau)/v_d^2$ for all $\tau$ large enough so that the function $G_v(\tau)$ potentially has anticorrelation. The histograms are clearly peaked below $1$ for both driving velocities $v_1$ and $v_2$. The mean and median are respectively 0.983 and 0.951 for $v_1$ and 0.944 and 0.896  for $v_2$.
This shows that the anticorrelation is real and that the signal above 1 is due to the noise.

\subsubsection{Impact of the visco-elasticity on the correlation length $\xi_v$}\label{sec:Impact of viscoelasticity}

In the d'Alembert experiment a plexiglas plate of PMMA (polymethyl methacrylate) is detached from a substrate made of PDMS (polydimethylsiloxane).
The PDMS is not perfectly elastic but displays a visco-elastic behavior. 
In this subsection I show how this behavior modifies the equation of motion~\eqref{eq:Movement equation appendix} derived for a crack front in section~\ref{sec: derivation crack propagation equation ch2} and its consequence for the correlation length $\xi_v$. 

The main impact of the visco-elastic behavior is that the fracture energy $G_\mathrm{c}$ defined in equation \eqref{eq:Griffith criterion} is not independent of the velocity anymore.
Instead it shows a rather strong dependence with crack speed~\cite{chopin2018, kolvin2015}~:
\begin{equation}
G_\mathrm{c} = G_\mathrm{c}(\du) \simeq \left( 1 + \frac{\du}{v_c} \right)^{\gamma} \, ,
\end{equation} 
where $v_c$ and $\gamma$ are material dependent parameters characterizing the energy dissipation process occuring when the material is broken.
In our experiment we have $v_c \ll v_2 < v_1$ and $\gamma \simeq 1/3$~\cite{chopin2018}, 
so that $G_c(\du) \sim \du ^{\gamma}$ in the range of crack speeds investigated. Hence the expansion for the fracture energy in presence of impurities (see equation \eqref{eq:local fluctuations}) should be modified as follow~:
\begin{equation}\label{eq:G_c velocity expansion}
\frac{G_c(x,u,\dot u)}{G _c(v_d)} = 1 + \frac{\delta G_c(x,u, v_d)}{G _c(v_d)} + \frac{\gamma}{v_d} (\dot u-v_d) \, .
\end{equation}
The last term in \eqref{eq:G_c velocity expansion} modifies the equation of motion~\eqref{eq:Movement equation appendix} which becomes~:
\begin{equation}\label{eq:Movement equation appendix 2}
\left( \frac{1}{c_R} + \frac{\gamma}{v_d} \right) v = k \,\left( v_d t - u(x,t) \right) - \frac{\delta G_c}{G_c} + \frac{1}{\pi} \int \frac{u(y)-u(x)}{|y-x|^2}dy 
\end{equation}
where the constant $\gamma$ has been absorbed in the loading. 
Equations \eqref{eq:Movement equation appendix} and \eqref{eq:Movement equation appendix 2} have the same
form but the mobility has been renormalized and is now
$\mu = \left( \frac{1}{c_R} + \frac{\gamma}{v_d} \right)^{-1}$ instead of $\mu=c_R$ in \eqref{eq:Movement equation appendix}.
In the experiment the driving velocity satisfies $v_c \ll v_d \ll c_R$ so that $ \mu \simeq \frac{v_d}{\gamma}$.

Functional renormalization group calculations allow a more precise determination of the dynamical correlation length  $\xi_v$ than the
scaling $\xi_v\sim v_d^{-\nu/\beta}$ that I used in this chapter. 
It has been shown that $\xi_v$ depends not only on the driving velocity $v_d$ but also on the mobility $\mu$~\cite{chauve2000}~:
\begin{equation}\label{eq: xi_v exact from FRG}
\xi_{v} = L_c \left( \frac{\mu f_c}{ v_d} \right)^{\frac{\nu}{\beta}} \, ,
\end{equation}
where $L_c$ is the Larkin length and $f_c$ the critical force. 
For our experimental conditions the ratio $\frac{\mu}{v_d}\simeq \gamma^{-1}$ does not depend on $v_d$ so neither does $\xi_v$.
This explains why rescaling the $x$-axis by $v_d^{-\nu/\beta}$ was not necessary to collapse the spatial correlation functions. 
Tuning $v_d$ does not change $\xi_v$ and we cannot come closer to the 
depinning critical point. 

The formula \eqref{eq: xi_v exact from FRG} being exact it allows to make a numerical estimation of the correlation length. 
In Ref.~\cite{Demery2014} $L_c$ and $f_c$ have been estimated to be $L_c = d_0/\sigma^2$ and $f_c = \sigma^2$~
where $\sigma^2 = \left\langle \delta G_c^2 \right\rangle$ is the variance of the toughness spatial fluctuations.
In the d'Alembert experiment the disorder is controlled and we have $\sigma \simeq 1/2$. A numerical application hence yields~:
\begin{equation}
\xi_v \simeq 4 d_0 \left( \frac{3}{4}\right)^{2.6} \simeq 2 d_0 \, .
\end{equation}
This value is consistent with the breaking of the large scale behavior on figure~\ref{Fig:Cv(x)_experiment} that occurs at a distance of order $d_0$.
A priori we would expect the depinning prediction to hold when $\xi_v$ is large compared to the microscopic scale of the disorder. It is therefore remarkable that the power law decay observed for the temporal correlation function in the experiment is consistent with the depinning prediction $\tau^{-\beta/\nu z}$
even if $\xi_v$ is of the order of $d_0$.

\subsection{Comparison with previous results}

The local velocity correlation functions $C_v(x)$ and $G_v(\tau)$ have already been computed in previous simulations and experiments. However the connection with the scaling forms~\eqref{eq: full scaling form Cv(x) generic alpha} and \eqref{eq: full scaling form Gv(tau) generic alpha} was missing.
In section~\ref{sec: Ch3 Duemmer&Krauth} I already explain how the correlation function $C_v(x)$ was used to extract the scale $\xi_v$ for an interface with short-range elasticity in Ref.~\cite{duemmer2005}. The authors observed an exponential cutoff but the small scale exponent $\beta/\nu$ was not predicted.
On the experimental side both the spatial and temporal correlation functions have been computed for the Oslo crack experiment~\cite{tallakstad2011}.
As for the d'Alembert experiment, the collapse of $C_v(x)$ was obtained without rescaling of the $x$-axis and the collapse of $G_v(\tau)$ was obtained by rescaling the time axis by $v_d$.
The functions were found to scale as $C_v(x) \sim x^{-\tau_x}$ and $G_v(\tau) \sim \tau^{-\tau_t}$
with exponents $\tau_x = 0.53 \pm 0.12$ and $\tau_t = 0.43$ a bit away from the depinning 
predictions $\beta / \nu \simeq 0.38$ and $\beta / (\nu z) \simeq 0.50$.
However the fits were performed using exponential cutoffs  at large distance and time which might affect the effective exponents. 
Furthermore the anticorrelation in time was not observed.
Regarding this point one must note that standard log-log plot routines discard negative values.
Anticorrelation is thus invisible on a direct log-log plot of the connected correlation function\footnote{This costed me three weeks searching for the anticorrelation in the experiment although I was expecting it.} $G_v(\tau)$. 
In order to see the anticorrelation one must use alternative plots. 
This is why I plotted the non-connected correlation function 
$G_v(\tau)/v_d^2 + 1 = \left\langle v(x,t) v(x,t+\tau) \right\rangle/v_d^2$.
It would be interesting to test how far the behavior predicted in this chapter could capture the Oslo experiments, especially the small $x$ regime of the spatial correlations which was resolved in~\cite{tallakstad2011}.
Finally let us note that Gjerden \textit{et al}.~\cite{gjerden2014} computed the same correlation functions in simulations
of a fiber bundle model that mimics the presence of damages in front of the crack.
Their model is expected to fall into the depinning universality class with long-range
elasticity~\cite{gjerden2014}. They measured small scale exponents $\tau_x=\tau_t = 0.43$ with cutoffs faster than exponential.

\section{Conjecture for the correlation functions in plasticity}

In section~\ref{sec: Ch2 Plasticity} we saw that the avalanches of plastic events in amorphous materials near the yielding transition share some common feature with the avalanches near of elastic interfaces near the depinning transition.
In amorphous materials the interactions are mediated by a kernel of the form
$\mathcal{G}(x) = \mathcal{G}(|x|, \phi) = \frac{\cos (4\phi)}{\pi |x|^d}$.
By analogy with~\eqref{eq: full scaling form Cv(x) generic alpha} I propose the following scaling form for the spatial correlation function of the local plastic strain rate, $\dot{\gamma}(x,t)$~:
\begin{equation}\label{eq: conjecture spatial correlations plasticity}
C(|x|, \phi) := \langle \dot{\gamma}(x,t) \dot{\gamma}(0,t) \rangle^c = \left\lbrace
\begin{array}{l l}
|x|^{-\beta/\nu} \cos(4\phi) & \text{ if } |x| < \xi_{\dot{\gamma}} \sim \dot{\gamma}^{-\nu/\beta} \, ,\\
|x|^{-d} \cos(4\phi) & \text{ if } |x| > \xi_{\dot{\gamma}}  \, ,
\end{array} \right.
\end{equation}
where $\dot{\gamma}$ is the average global strain rate.

\section{Conclusion}\label{sec: Ch3 Conclusion}

In this chapter I computed the scaling forms of the spatial and temporal correlation functions of the velocity field for an elastic interface driven at finite velocity. Numerical as well as experimental results confirm the predicted scaling forms. 
This is the first time that anticorrelation is predicted and observed in depinning systems at finite drive.
These findings open new perspectives for the experimental study of disordered elastic interfaces. As the correlations of the local velocity display universal features of the depinning even when the driving speed is finite, the critical behavior can be investigated at intermediate distance from the critical point.  
This provides a robust and efficient method to identify the universality class of the transition and to test the relevance of specific depinning models. 
In particular the difficulty of correctly defining avalanches before accessing the critical exponent is bypassed by this new method. 

Finally it is important to remark that the scaling forms~\eqref{eq:Cv(x)} and \eqref{eq:Gv(tau)} are very general and can be expected to be valid for all out-of-equilibrium transitions with avalanche dynamics. The asymptotic forms~\eqref{eq: full scaling form Cv(x) generic alpha} and \eqref{eq: full scaling form Gv(tau) generic alpha} are also very general, as beyond $\xi_v$ 
the spatial correlations decay as $1/x^{d+\alpha}$ for a long-range model ($d$ being the spatial dimension) and 
exponentially fast for short-range elasticity. It would certainly be insightful to probe this behavior in various problems, including those where the nature of the elastic interactions still needs to be deciphered, 
and especially to test the conjecture~\eqref{eq: conjecture spatial correlations plasticity} in the context of the yielding transition where avalanches of plastic events are observed \cite{lin2014}.


%% file: Chapter_4/Chapter_4.tex


\chapter{Cluster Statistics}\label{chap: cluster statistics}

Crack fronts, wetting lines and some domain walls are described by an elastic line with long-range (LR) elasticity. A key feature of LR elasticity is that the motion of a portion of the interface can trigger the instability of points at finite and even large distances while points in between remain stable. 
Therefore avalanches can be made of multiple, spatially disconnected components called \emph{clusters}~\cite{laurson2010}, even in the quasi-static limit.
Clusters have been experimentally studied in the Oslo experiment where the local velocity is accessible~\cite{maloy2006, tallakstad2011}.
The local velocity is known thanks to the Waiting Time Matrix (see appendix~\ref{sec: appendix experiment analysis}). It is defined in one pixel as 
\begin{equation}\label{eq: ch4 local velocity matrix}
V(x,y) = \frac{p}{WTM(x,y)} \, ,
\end{equation}
where $p$ is the pixel size and $WTM(x,y)$ is the time spent by the front in the pixel at location $(x,y)$.
A binary matrix $V_C$ is then obtained by applying a treshold to the spatial velocity matrix~:
\begin{equation}\label{eq: ch4 thresholded matrix for clusters}
V_C(x,y) = \left\lbrace
\begin{array}{l}
1 \text{ if } V(x,y) \geq C v \, ,\\
0 \text{ if } V(x,y) < C v \, ,
\end{array}\right.
\end{equation}
where $v$ is the mean velocity of the front and $C$ is an arbitrary constant of the order of a few unities. 
\begin{figure}[t]
	\centering
	\includegraphics[width=.95\linewidth]{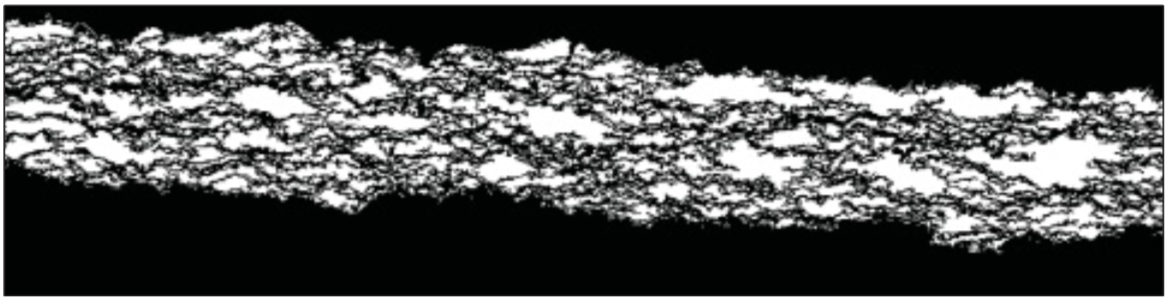}
	\includegraphics[width=.95\linewidth]{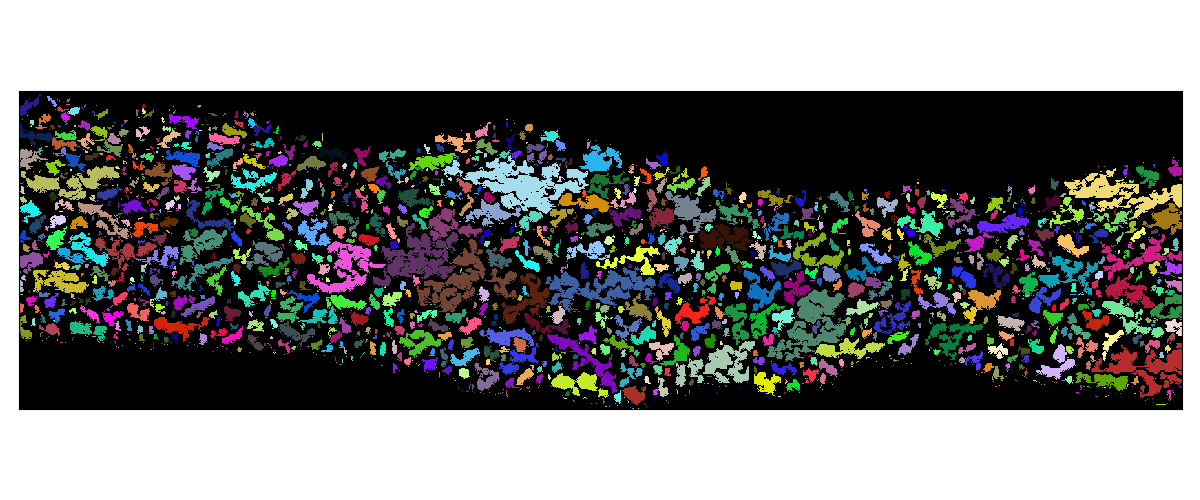}
	\caption{Images of thresholded matrix $V_C$ in the Oslo \emph{(top, taken from~\cite{tallakstad2011})} and d'Alembert \emph{(bottom)} experiments. 
	The front is moving upward.
	White and colored clusters correspond to the local velocities being $C$ times larger than the mean velocity $v$ with $C=2$ (top) and $C=10$ (bottom). In the bottom image, each cluster has a different color. \label{ClusterMaps}}
\end{figure}
In the studies~\cite{maloy2006, tallakstad2011} values of $C$ ranging from $2$ to $22$ have been tested. 
In the first study, the authors found that the cluster sizes follow a power law distribution with an exponent $\tau_c = 1.7 \pm 0.1$~\cite{maloy2006}.
This value is much higher than the exponent for the global avalanche size $\tau=1.28$.
It was pointed out by Laurson et al.~\cite{laurson2010} that the events studied in~\cite{maloy2006} were clusters and not global avalanches.
Laurson et al. performed the first theoretical study of the cluster statistics and
proposed a scaling relation between the clusters and avalanches exponents~:
\begin{equation}\label{eq: ch4 tau_c = 2tau-1}
\tau_c =  2\tau-1 \, .
\end{equation}
This relation was argued based on scaling arguments and was supported by numerical observations. Indeed they run an automation simulation of the model and measured exponents $\tau = 1.25\pm 0.05$ for the global avalanche sizes and $\tau_c = 1.53 \pm 0.05$ for the cluster sizes. 
A neat experimental confirmation was brought one year later by Tallakstad et al.~\cite{tallakstad2011} when they measured $\tau_c=1.56\pm 0.04$ on the Oslo experiment, a value in perfect agreement with the conjecture~\eqref{eq: ch4 tau_c = 2tau-1} and the theoretical value $\tau=1.28$
(which comes from the scaling relation $\tau=2-\alpha/(d+\zeta)$ with the best numerical determination $\zeta=0.39$~\cite{rosso2002}).

We completed this study and critized the argument in a recent publication~\cite{lepriol2020a} which is the subject of this chapter.
Using an automaton model, I performed extensive simulations of the quasi-static dynamics of a line with long-range elasticity.
Drawing on Ref.~\cite{tanguy1998} I studied various values of the elasticity parameter $\alpha$ in order to check if scaling relations like~\ref{eq: ch4 tau_c = 2tau-1} depend on $\alpha$ or not.
My results confirm the relation~\eqref{eq: ch4 tau_c = 2tau-1}.
However it turns out that the assumptions on which the derivation proposed in~\cite{laurson2010} relies are false. 
In this chapter I review the derivation given in~\cite{laurson2010} and show why it is incorrect. 
Then I give a new derivation of this relation. 
Laurson et al. had the idea to study the number of clusters inside one avalanche $N_c$ in order to link the cluster statistics to the global avalanche one.
We build on this idea and study the whole probability distribution of $N_c$ which gives the key to the derivation of~\eqref{eq: ch4 tau_c = 2tau-1}.

Recently, Planet et al. proposed a different scaling relation linking $\tau_c$ and $\tau$~\cite{planet2018}.
This scaling relation relies on the scaling of the size of the events with their duration which is assumed to be different for the clusters and for the global avalanches. 
The scaling relations are tested on an  imbibition front propagation experiment for which the elasticity kernel is different than for crack front.
I will also present and discuss this study. 
 
The chapter is organized as follows.
In the first section I define the quantities studied in this chapter.
In section~\ref{sec: recall of previous results} I recall some previously known results about the avalanche statistics.
In section~\ref{sec: clusters statistics} I review the argument of Ref.~\cite{laurson2010} and show why it is false. Then I study the statistics of the number of clusters inside one avalanche and deduce several scaling relations (some being new) linking the cluster and avalanche exponents. Among others the relation~\eqref{eq: ch4 tau_c = 2tau-1} is recovered.
I conclude this section by discussing the study~\cite{planet2018}.
In section~\ref{sec: statistics gaps and diameter} I extend the study to the statistics of gaps and of the avalanche diameter.
Finally all the predictions regarding critical exponents are summarized and compared with numerical measurements in a table at the end of the chapter (section~\ref{sec: table of exponents}).

\section{Introduction of the observables and various critical exponents}\label{sec: introduction of observables}

\begin{figure}
	\centering
	\includegraphics[width=.8\linewidth]{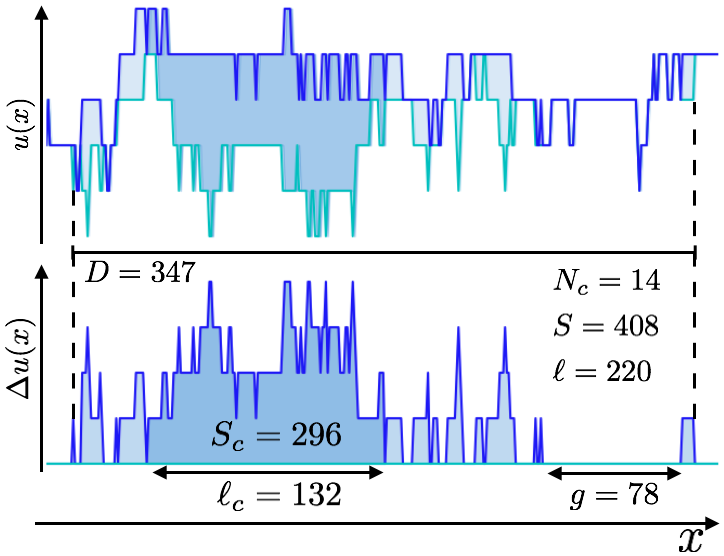}
	\caption{Example of a simulated avalanche with $\alpha=1$. The upper image shows two successice metastable configurations of the line. The avalanche is the advance from the lower configuration to the upper one. The lower image shows the relative displacement of the final configuration with respect to the initial. The observables are defined in the text.  \label{fig: Ch4 SeeAvalanche}} 
\end{figure}

Many observables can be associated to an avalanche. Figure~\ref{fig: Ch4 SeeAvalanche} introduces the ones that I study in this chapter.
The upper image shows two successive metastable configurations of the line. The area in between has been swept by the line during the avalanche and corresponds to the avalanche total size $S$. 
The clusters are more distiguishable on the lower image where the relative displacement of the final configuration with respect to the initial one is shown. 
The avalanche can be decomposed into $N_c$ disconnected clusters of size $S_c$ and \emph{extension} $\ell_c$.
Between two consecutive clusters there is a \emph{gap} of length $g$. 
We define two more global quantities for the avalanche : the diameter $D$ is the distance between the extremal points of the avalanche and the linear size $\ell$ is the total length of all portions of the line wich moved. 
In the case of SR elasticity, where avalanches are made of one single component, these quantities are the same and are usually referred to as \emph{avalanche extension} $\ell$~\cite{delorme2016}.
The distinction in presence of LR elasticity has been first made by D.S. Fisher~\cite{fisher1998}.
We have the relations $S =  \sum_{i=1}^{N_c} S_{c,i}$, $\ell = \sum_{i=1}^{N_c} \ell_{c,i}$ and $D = \ell + G$ where $G=\sum_{j=1}^{N_c-1} g_j$ is the total gap length.
We are interested in the statistics of all these quantities which display power law distributions at small scales and faster decays at large scales.
We define the following critical exponents~:
\begin{align}
P(S) &\sim S^{-\tau} \quad \text{for} \quad S<S_{m} \quad \text{(total size)} \label{eq: def tau} \\
P(S_c) &\sim S_c^{-\tau_c} \quad \text{for} \quad S_c<S_{c, m} \quad \text{(cluster size)}  \label{eq: def tau_c} \\
P(N_c) &\sim N_c^{-\mu} \quad \text{for} \quad N_c< N_{m} \quad \text{(number of clusters)}  \label{eq: def tau_Nc} \\
P(\ell) &\sim \ell^{-\kappa} \quad \text{for} \quad \ell<\ell_{m} \quad \text{(linear size)}  \label{eq: def kappa} \\
P(\ell_c) &\sim \ell_c^{-\kappa_{_c}} \quad \text{for} \quad \ell_c<\ell_{c, m}  \quad \text{(cluster extension)} \label{eq: def kappa_c} \\
P(g) &\sim g^{-\kappa_g} \quad \text{for} \quad g < g_{\text{cross}}  \quad \text{(gap)} \label{eq: def kappa_g} \\
P(G) &\sim G^{-\kappa_G} \quad \text{for} \quad G < G_{\text{cross}}  \quad \text{(gaps sum)} \label{eq: def kappa_G} \\
P(D) &\sim D^{-\kappa_D} \quad \text{for} \quad D < D_{\text{cross}}  \quad \text{(diameter)} \label{eq: def kappa_D}
\end{align}
In the above expressions the subscript '$m$' indicates that the quantities decay faster than a power law at large scales while the subscript '$\text{cross}$' indicates a crossover to a faster (and $\alpha$ dependant) power law.

I investigated the distribution of these quantities using numerical simulations
of the equation of motion~: 
\begin{equation}
v(x,t) = \partial_t u(x,t) = m^2\left(w(t) - u(x,t)\right) + \int \frac{u(y,t)-u(x,t)}{|y-x|^{1+\alpha}} + \eta \left(x, u(x,t)\right) \, , \label{eq: eq of motion ch4}
\end{equation}
where $\eta$ denotes the quenched disorder force and we use the LR elastic kernel \eqref{eq: LR elasticity kernel} for the elastic force.
I implemented an automaton simulation analogous to the one described in section \ref{sec: ch3 numerics}. In particular $u$ and $x$ take integer values and I assume periodic boundary conditions (PBC) along $x$ which takes values ranging from $0$ to $L-1$.
The local velocity is defined as :
\begin{align}\label{eq: def automaton model ch4}
v(x,t) &= \theta \left( \,  m^2 \left( w(t) - u(x,t) \right) + \sum_{y} \frac{u(y,t) - u(x,t)}{|y-x|^{1+\alpha}} + \eta\left(x, u(x,t)\right) \right) \, , 
\end{align}
$\theta$ being the Heaviside function.
The random pinning forces $\eta(x,u)$ are drawn from the negative half of a normal law.
At variance with the simulation used in chapter 3, here $w(t)$ is not increased at a constant rate but 
step by step in order to simulate a quasi-static dynamics.
Starting from an initial stable critical configuration $w$ is incremented just enough so that exactly one point becomes unstable. 
Then the dynamical rules are applied while keeping $w$ fixed. At each time step
all the points feeling a positive total force move one step forward, $u(x,t+1)=u(x,t)+1$, while the other points stay pinned. Then new pinning forces are drawn for the jumping point and the elastic force is updated. 
The dynamics evolve until the line reach a new stable configuration where all points feel a negative total force. The difference between the initial and final configurations defines an avalanche as illustrated on figure \ref{fig: Ch4 SeeAvalanche}. The values of the observables associated to the avalanche are computed and saved. Then $w$ is incremented again to make one point unstable and thus trigger a new avalanche. 
The process is repeated in order to have a statistics of at least $10^{5}$ avalanches. 

\begin{figure}
	\centering
	\includegraphics[scale=0.7]{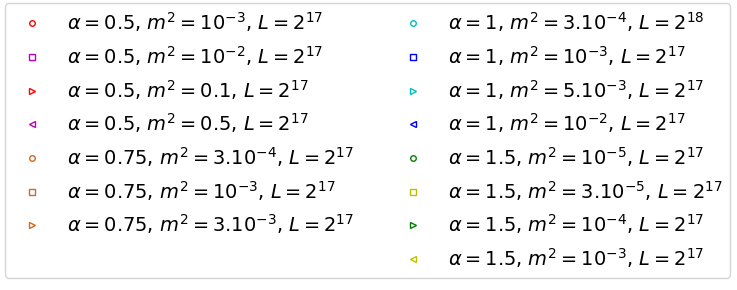}
	\caption{Color code for all figures in this chapter except figures~\ref{fig: P(S_c|S)}, \ref{fig: P(ell_c|ell)}, \ref{fig: P(G), P(D)} and \ref{fig: P(g) and P(g|G)}.
	Red and magenta are used for $\alpha=0.5$, brown for $\alpha=0.75$, cyan and blue for $\alpha=1$ and yellow and green for $\alpha=1.5$.  \label{fig: ColorCode}} 
\end{figure}

Although only $\alpha=1$ corresponds to experimental systems, a better understanding can be gained by studying how the exponents change when $\alpha$ is varied, as it has been done in~\cite{tanguy1998}. In this chapter I study in details four values of $\alpha$~: 
\begin{itemize}
\item $\alpha=1$ which corresponds to the long-range elasticity of fracture fronts and liquid contact lines.
\item $\alpha=0.5$ for which the internal dimension of the line is the upper critical dimension $d_{uc}=2\alpha=1$, thus the mean-field prediction starts to apply with logarithmic corrections.
\item $\alpha=0.75$ which is in the middle between the physical case $\alpha=1$ and the mean-field case $\alpha=0.5$.
\item $\alpha=1.5$ which is in the middle between $\alpha=1$ and $\alpha=2$ which corresponds to short-range elasticity.
\end{itemize}
For each $\alpha$ I performed automaton simulations with $3$ or $4$ different values of the mass and system size $L=2^{17}$ or $2^{18}$. In almost all figures presented in this chapter I use the color code detailed in figure \ref{fig: ColorCode} which also summarizes the parameter values.

\section{Recall of previous results}\label{sec: recall of previous results}

There are already some known results about the global behavior of avalanches. In particular we have seen in chapter 1 that there is a scaling relation linking $\tau$ to the roughness exponent $\zeta$.
It holds for $\frac{d}{2} \leq \alpha \leq 2$ and reads~:
\begin{align}
&\tau = 2 - \frac{\alpha}{d+\zeta} \, .  \label{eq: tau NarayanFisher} 
\end{align}
This relation saturates to the mean-field exponent $\tau=3/2$ for $\alpha \leq \frac{d}{2}$ and to the SR exponent $\tau = 2-2/(d+\zeta)$ for $\alpha \geq 2$.
The mass induces a characteristic length $\ell_{m} = m^{-2/\alpha}$ beyond which the line is flattened by the harmonic confinement. This length directly sets the cutoff for the avalanche total size distribution~:
\begin{align}
&S_{m} \sim \ell_{m}^{d+\zeta} = m^{-2(d+\zeta)/\alpha} \, . \label{eq: relation Smax ellmax} 
\end{align}
For SR elasticity the relation \eqref{eq: relation Smax ellmax} is true not only for the cutoff but links the size and extension of all avalanches~:
\begin{equation}
S \sim \ell^{d+\zeta} \, . \label{eq: S sim ell to the power d+zeta}
\end{equation}
This allows to derive a scaling relation between the size and extension exponents
$\kappa-1 = (\tau-1)(d+\zeta)$. Using equation \eqref{eq: tau NarayanFisher} with $\alpha=2$, $\kappa$ can be reexpressed as~:
\begin{align}
\kappa &= d + \zeta - 1 \, . \label{eq: relation kappa zeta alpha 1} 
\end{align}
We see that for SR avalanches the scaling exponents $\tau$ and $\kappa$ 
depend only on the roughness exponent $\zeta$. 
One can wonder if this is still true for the exponents defined in equations \eqref{eq: def tau} to \eqref{eq: def kappa_D} and valid for LR avalanches. 
It is also not clear whether the relations \eqref{eq: S sim ell to the power d+zeta} still holds when the elasticity is long-range. 
Does it apply to clusters ? Or to the total avalanche ? And in this case should one consider the avalanche diameter or its linear size ?


\begin{figure}
	\centering
	\includegraphics[scale=0.35]{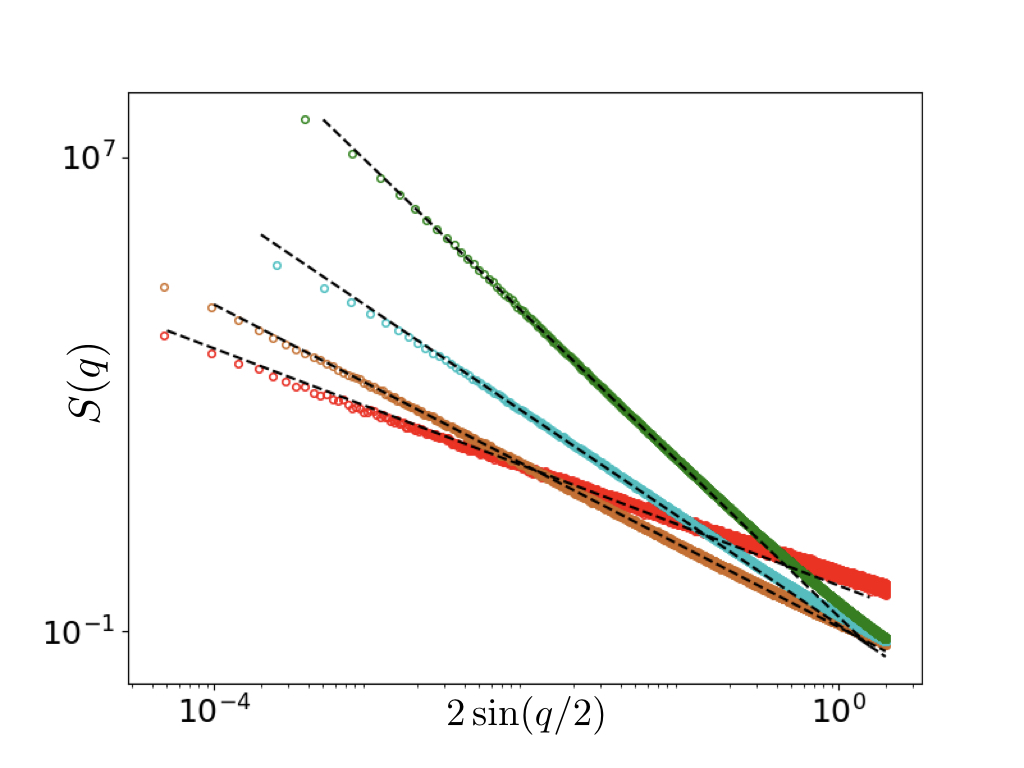}
	\caption{Structure factor. 
	The dashed lines correspond to fit $q^{-(1+2\zeta)}$ with $\zeta=0$ ($\alpha=0.5$), $0.18$ ($\alpha=0.75$), $0.39$ ($\alpha=1$) and $0.77$ ($\alpha=1.5$).
	The color code is the one given in figure \ref{fig: ColorCode} except that $L=25.10^{3}$ for $\alpha=1$ and $L=2^{14}$ for $\alpha=1.5$.\label{fig: Roughness}} 
\end{figure}

\paragraph{Roughness exponent}
If we are to establish a link between the exponents defined above and the roughness exponent our first task must be to determine $\zeta$ which depends on $\alpha$. 
I measured the values of $\zeta$ using the structure factor $S(q) = \overline{u_q u_{-q}} \sim q^{-(d+2\zeta)}$.
The data are shown in figure \ref{fig: Roughness} and the numerical values are summarized in table~\ref{tab: zeta}.
Two values are already known. The mean-field prediction which should hold, up to logarithmic corrections, for $\alpha=0.5$ is $\zeta=0$.
For $\alpha=1$ the value $\zeta=0.39$ has been determined numerically in \cite{rosso2002}. 
There is no exact result to date for $\alpha=0.75$ and $\alpha=1.5$.
Nevertheless $\zeta$ has been computed up to two-loop in an expansion in $\epsilon=2\alpha-d$ using functional renormalization group \cite{ledoussal2002}. The result is (setting $d=1$)~:
\begin{equation}\label{eq: zeta two loop approximation}
\zeta^{(2)} = \frac{2\alpha-1}{3} \left(1 +(2\alpha-1) \frac{\psi(1)+\psi(\alpha)-2\psi(\alpha/2)}{9\sqrt{2}\gamma} \right) \, 
\end{equation}
where $\psi$ is the digamma function\footnote{The digamma function is the logarithmic derivative of the gamma function~: $\psi(x) = \Gamma'(x)/\Gamma(x)$.
The gamma function itself is defined for any complex number with strictly positive real part as~:
$\Gamma(z) = \int_0^{\infty} t^{z-1}e^{-t} dt$}
and $\gamma=\int_0^1 dy \sqrt{y - \log y -1}=0.5482228893$.
The first term into the braces corresponds to the one loop expansion, $\zeta^{(1)}=(2\alpha-1)/3$ (already computed by Ertas and Kardar~\cite{ertas1994}), and the second term to the two loop expansion.
This relation gives $\zeta^{(1)}=1/6$ to one loop and $\zeta^{(2)}=0.21$ to two loop for $\alpha=0.75$ 
and $\zeta^{(1)}=2/3$, $\zeta^{(2)}=0.96$ for $\alpha=1.5$.
It was noticed in \cite{ledoussal2002} that the correct value seems
to be around the midpoint between 1 loop and 2 loop\footnote{For $\alpha=1$, $\zeta^{(1)}=1/3$ and $\zeta^{(2)}=0.46$ while the exact result is $\zeta=0.39$}.
This midpoint is $0.19$ for $\alpha=0.75$ and $0.81$ for $\alpha=1.5$.

\begin{center}
	\begin{tabular}{|c|c|c|c|c|}
		\hline
		$\alpha$ & 0.5 & 0.75 & 1 & 1.5 \\ \hline
		$\zeta$ & 0 (MF) & 0.18 & 0.39~\cite{rosso2002} & 0.77 \\
		\hline
	\end{tabular}
	\captionof{table}{Table of roughness exponents. We use the mean-field value for $\alpha=0.5$ and the determination from~\cite{rosso2002} for $\alpha=1$. The values for $\alpha=0.75$ and 1.5 were obtained by fitting the structure factor in figure~\ref{fig: Roughness}. \label{tab: zeta}}
\end{center}

\begin{figure}
	\centering
	\includegraphics[scale=0.3]{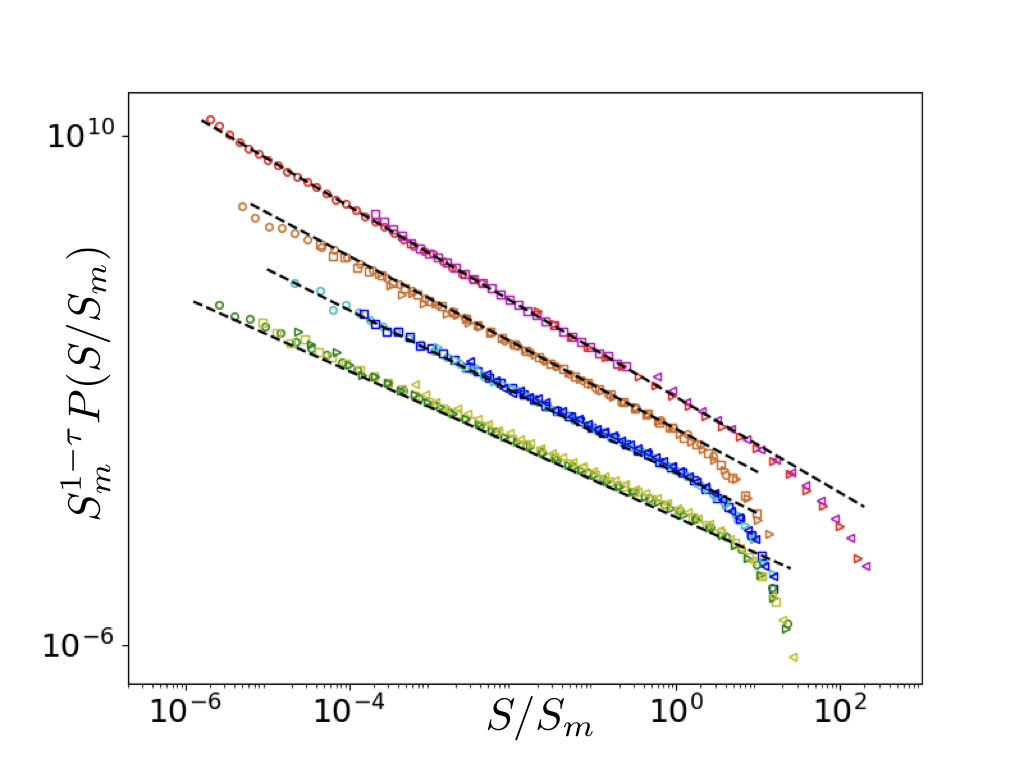}
	\caption{Collapse of the avalanche size distribution. 
	$P(S/S_m)$ is the probability distribution of the variable $S/S_m$. Using equation \eqref{eq: relation Smax ellmax} we define $S_m = \ell_m^{d+\zeta} = m^{-2(d+\zeta)/\alpha}$. Vertical axis is rescaled by $S_m^{1-\tau}$ to ensure the collapse of the distributions for different values of the mass. 	
	The perfect collapse obtained is a test of the validity of equation \eqref{eq: relation Smax ellmax}. The dashed line are fits of the exponent $\tau$. The numerical values are $\tau=1.5$, $1.36$, $1.26$ and $1.14$ (top to bottom).
	For visibility purpose, data have been shifted vertically by a factor $100$ for $\alpha=0.5$, $10$ for $\alpha=0.75$ and $0.1$ for $\alpha=1.5$. \label{fig: P(S)}} 
\end{figure}

\paragraph{Avalanche size}
Figure \ref{fig: P(S)} shows the total size distribution $P(S)$ of avalanches. 
The x-axis has been rescaled using $S_m = \ell_m^{d+\zeta} = m^{-2(d+\zeta)/\alpha}$.
The perfect collapse of the distributions for different masses shows the validity of relation \eqref{eq: relation Smax ellmax} for $\frac{1}{2} \leq \alpha \leq 2$.
The fits of the power law exponents are in agreement with the scaling relation \eqref{eq: tau NarayanFisher} using the values of $\zeta$ listed in table~\ref{tab: zeta}.


\section{Cluster statistics}\label{sec: clusters statistics}

In this section I review the argument of Laurson et al.~\cite{laurson2010} that justified the scaling relation $\tau_c=2\tau-1$~\eqref{eq: ch4 tau_c = 2tau-1} and show why it is incorrect.
Then I study the probability distribution of the number of clusters $N_c$. 
Finally I use this distribution to deduce several scaling relations. 
Among others the relation~\eqref{eq: ch4 tau_c = 2tau-1} is recovered.

\subsection{Review of the derivation by Laurson et al.}

The first theoretical study of the cluster statistics is the paper by Laurson, Santucci and Zapperi~\cite{laurson2010}.
This paper focuses on the case $\alpha=1$ for which they found the scaling relation \eqref{eq: ch4 tau_c = 2tau-1} linking $\tau$ and $\tau_c$.
This relation was supported for $\alpha=1$ by numerical simulations and experiments.
In this section I reproduce their derivation. Then I show that the assumptions on which it relies are false.

\subsubsection{Laurson et al. argument}

The derivation proposed in~\cite{laurson2010} is based on two assumptions : 
\begin{itemize}
\item During the avalanche spreading the evolution of the number of clusters is interpreted as a random walk with the time being replaced by the current size of the avalanche. 
The walks starts at $N_c=1$ at time $1$ and evolves up to time $S$ where $S$ is the avalanche size. At each time step
$(S \rightarrow S+1)$ the number of clusters can increase by $1$ (a new cluster is triggered), remain constant or decrease by $1$ (two clusters merge). The walk has a reflective boundary condition at $N_c=1$.
This implies that the average number of clusters at the end of avalanches of size $S$ scales as~:
\begin{equation}
\langle N_c \rangle_S \sim S^{\gamma_S} \label{eq: Nc sim S to gamma_S}
\end{equation} 
where $\gamma_S$ is the (anomalous) diffusion exponent of the random walk.
The assumption that the process is Markovian leads to $\gamma_S=1/2$.
\item  The clusters belonging to an avalanche of size $S$ have a typical size~:
\begin{equation}
S_c \sim \frac{S}{\langle N_c \rangle}_S \sim S^{1 - \gamma_S} \, . 
\label{eq: Zapperi scaling assumption between Sc and S}
\end{equation}
\end{itemize}

Then the authors apply the relation $P(S_c) dS_c = P(S)dS$ using \eqref{eq: Zapperi scaling assumption between Sc and S} and find~:
\begin{equation}
\tau_c = \frac{\tau - \gamma_S}{1 - \gamma_S} \label{eq: Zapperi conj. generalized}
\end{equation}

The authors have performed automaton simulations of the dynamics for the case $\alpha=1$.
They found a numerical value $\gamma_S=0.47$ and interpreted it as $\gamma_S=1/2$ 
which yields the relation \eqref{eq: ch4 tau_c = 2tau-1} \cite{laurson2010}. 

\subsubsection{Flaws of the arguments}

\begin{figure}
	\centering
	\includegraphics[width=0.7\linewidth]{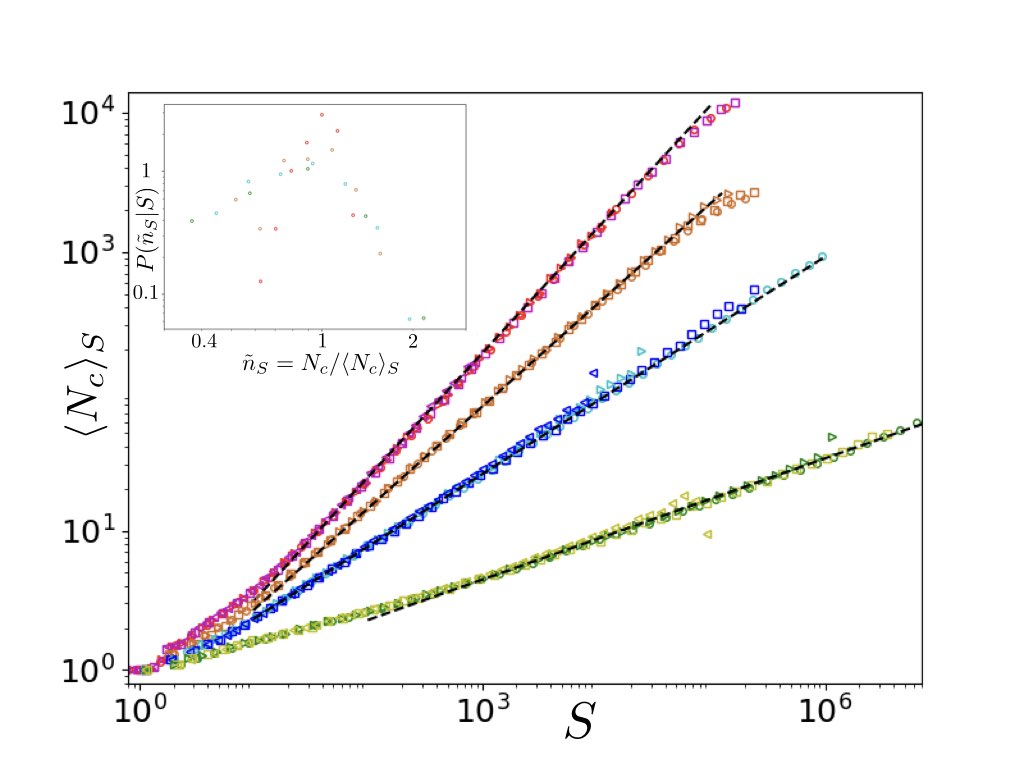}
	\caption{\emph{Main panel :} $\langle N_c \rangle_S$ versus $S$. The dashed lines are fit of the exponent $\gamma_S$.
	From top to bottom the values are $\gamma_S=0.89$ ($\alpha=0.5$), $\gamma_S=0.73$ ($\alpha=0.75$), $\gamma_S=0.52$ ($\alpha=1$) and $\gamma_S=0.29$ ($\alpha=1.5$). 
	\emph{Inset :} Probability of $\tilde{n}_{S} = N_c/\langle N_c \rangle_{S}$ conditioned by the avalanche size $S$. The conditional probability has been computed in small intervals centered around 
	$S=10^3$, $10^3$, $10^4$ and $10^5$ for $\alpha=0.5$, 0.75, 1 and 1.5 respectively.
	The fact that the conditional probabilities are well-peaked justifies the application of the relation $P(N_c)dN_c = P(S)dS$.
	\label{fig: Nc vs S}} 
\end{figure}

Numerical simulations allow to test the two assumptions above for various value of $\alpha$.
The argument that leads to the first relation \eqref{eq: Nc sim S to gamma_S} is supposed to hold for any long-range kernel. In particular the value of $\alpha$ does not play a role in the argument. 
Figure~\ref{fig: Nc vs S} shows that the relation $\langle N_c \rangle_S \sim S^{\gamma_S}$ indeed holds but that the exponent $\gamma_S$ is in general different from $1/2$
and decreases continuously when $\alpha$ increases. 
In particular there is no reason for $\gamma_S$ to be equal to $1/2$ just at $\alpha=1$ (although it is close as I measured $\gamma_S=0.52$ for $\alpha=1$).
As a consequence the particular relation $\tau_c=2\tau-1$ that was established using a value of $\gamma_S=1/2$ should be replaced by the more general relation~\eqref{eq: Zapperi conj. generalized}. 
This relation involves the exponent $\gamma_S$ which must then be determined either numerically or thanks to another prediction. 

\begin{figure}[!p]
	\begin{subfigure}[t]{1\linewidth}%
	\centering
	\includegraphics[width=.72\linewidth]{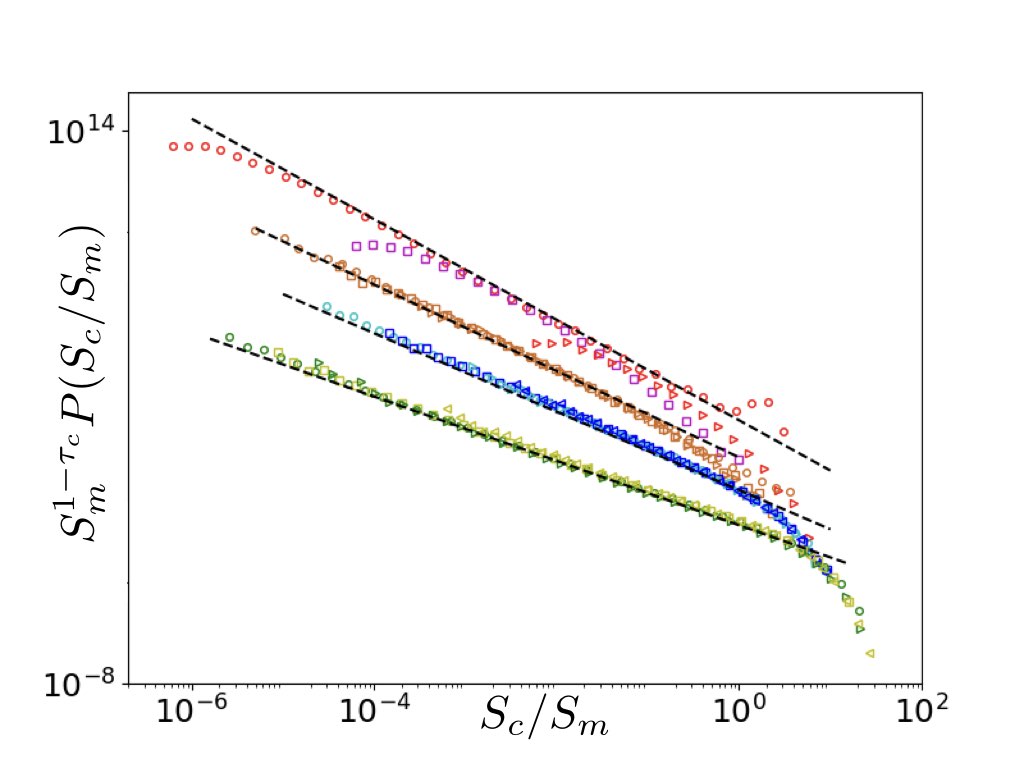}
	\caption{Collapse of the cluster size distributions. The x-axis has been rescaled by $S_m$ defined in equation \eqref{eq: relation Smax ellmax} and the y-axis by $S_m^{1-\tau_c}$ The values of $\tau_c$ are obtained by fitting the power law part of the distributions (dashed lines). The collapse of the distributions for each $\alpha$ shows that the cutoff of the distributions is controlled by $S_m$. It is also a verification of the values of $\tau_c$.
	The fitted values are (from top to bottom, i.e. $\alpha=0.5$ to 1.5) $\tau_c=2$, 1.72, 1.56 and 1.28.
	Data have been shifted vertically for visibility purpose.
	\label{fig: P(S_c)}}
	\end{subfigure}
	\vspace{1cm}
    \begin{subfigure}[t]{1\linewidth}%
    \centering
	\includegraphics[width=.72\linewidth]{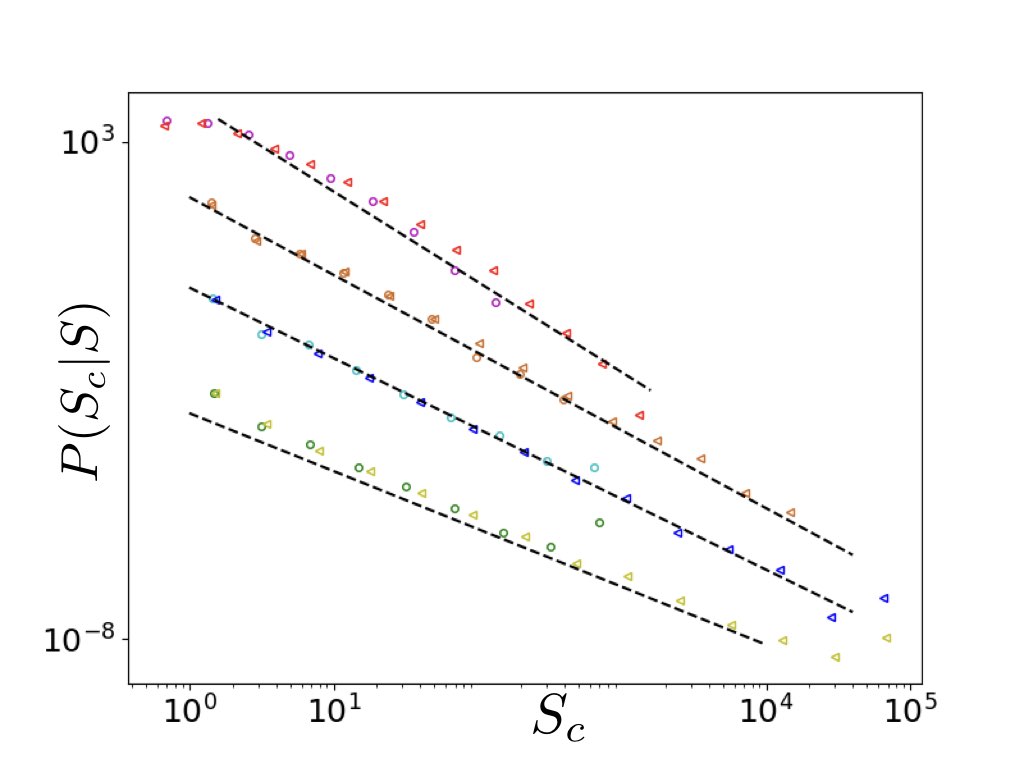}
	\caption{Distribution of cluster sizes $S_c$ conditioned by the total size $S$ of the avalanche to which the clusters belong. Two values of $S$ are compared~: $S=10^3$ (circles) and $S=10^5$ (triangles). We observe that the conditional distribution $P(S_c|S)$ is a power law cut at $S$.
	The dashed lines are fit using the exponents $\tau_c$ determined in the figure above.
	Parameters values are $\alpha=0.5$, $m^2=10^{-3}$, $L=2^{17}$ ; 
	$\alpha=0.75$, $m^2=3.10^{-4}$, $L=2^{17}$ ; $\alpha=1$, $m^2=3.10^{-4}$, $L=2^{18}$ ; 
	$\alpha=1.5$, $m^2=3.10^{-5}$, $L=2^{17}$.
	Data have been shifted vertically for visibility purpose. \label{fig: P(S_c|S)}}
	\end{subfigure}
	\vspace{-1cm}
	\caption{Distributions of cluster sizes. \label{fig: P(S_c) and P(S_c|S)}}
\end{figure}

The second assumption~\eqref{eq: Zapperi scaling assumption between Sc and S} is surprising.
There are many clusters inside one avalanche and there is a priori no reason to expect them to all have the same typical size. 
Note however that a similar equality holds when averaging over many clusters belonging to avalanches of a given fixed size $S$~:
\begin{equation}
\left\langle S_c \right\rangle_S = \frac{S}{\langle N_c \rangle_S} \, . \label{eq: average relation S_c S N_c}
\end{equation}
To see this let's take $M$ avalanches of size $S$ each with a different number of clusters $N_c^{i}$. 
Let $S_c^{i, j}$ be the $j$th cluster of avalanche $i$. The average of the cluster sizes over all the avalanches is~:
\begin{equation*}
\langle S_c \rangle_S = \frac{1}{\sum_{i=1}^{M} N_c^{i}} \sum_{i=1}^{M} \sum_{j=1}^{N_c^{i}}S_c^{i,j}
  = \frac{M\times S}{\sum_{i=1}^{M} N_c^{i}} = \frac{S}{\langle N_c \rangle_S}
\end{equation*}
Although we have an equality for the average size of clusters belonging to avalanches of size $S$, these clusters do not have a typical size.
This is evidenced by figure~\ref{fig: P(S_c|S)} which shows the distribution $P(S_c|S)$ of the cluster sizes conditioned by the total size of the avalanche to which they belong. It appears to be power law distributed until the cutoff that is imposed at $S$. 
In consequence the relation~\eqref{eq: Zapperi scaling assumption between Sc and S} is incorrect.

The last step of the derivation relies on the application of the relation $P(S_c)dS_c = P(S)dS$.
However there are many realizations of $S_c$ associated to one realization of $S$. This renders the relation $P(S_c)dS_c = P(S)dS$ ill-defined. 
Relations of the type $P(X)dX=P(Y)dY$ require a one-to-one correspondance between the variables $X$ and $Y$ in order to be applied. 
This is not the case here as we have seen that there is a broad distribution of $S_c$'s associated to one value of $S$. 
So this relation cannot be applied and the last step of the derivation is not grounded. 

\begin{figure}
	\centering
	\includegraphics[width=0.7\linewidth]{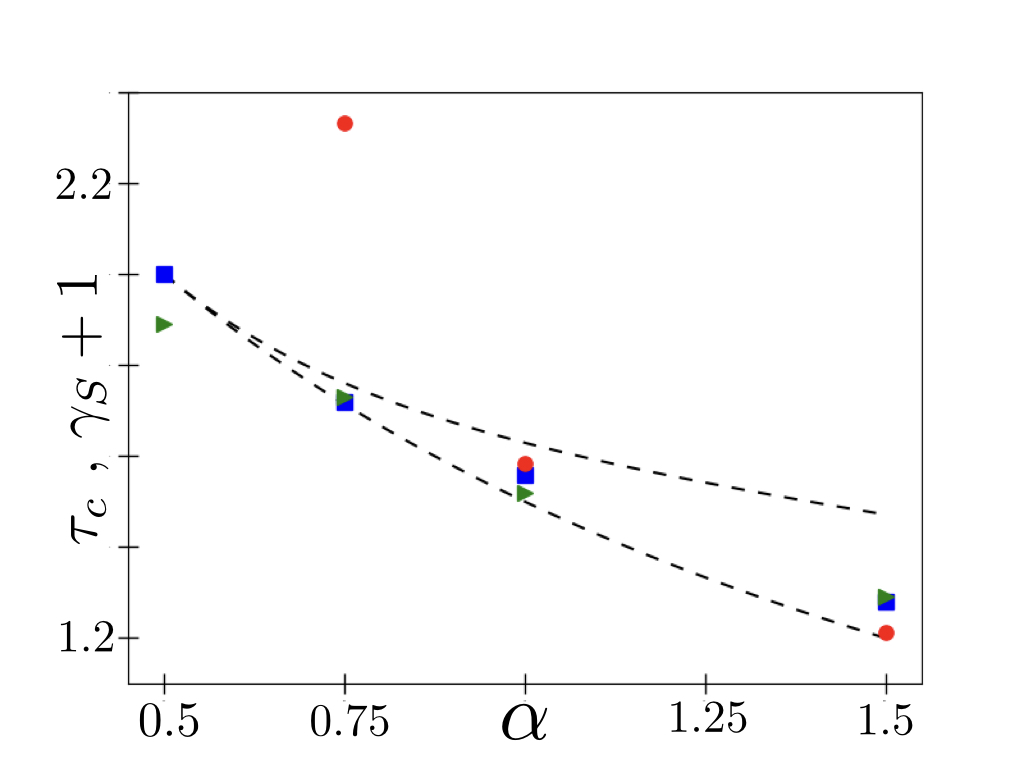}
	\caption{Green triangles: $\gamma_S+1$ as measured from figure \ref{fig: Nc vs S}. 
    The values of $\gamma_S$ are fed into the prediction~\eqref{eq: Zapperi conj. generalized} $\tau_c=(\tau-\gamma_S)/(1-\gamma_S)$ (red cirles) which can be compared with the direct measurements	 of $\tau_c$ issued from figure \ref{fig: P(S_c)} (blue squares).
    Errorbars are smaller than the symbols size.
	The prediction is close from the measurement for $\alpha=1$ but becomes too large at low values of $\alpha$ (the prediction for $\alpha=0.5$ is $\tau_c=5.55$ and lies outside the figure). 
	The dashed lines test our prediction of $\tau_c$ as a function of $\zeta$ (see table~\ref{tab: measurement vs predictions}) with $\zeta^{(1)} = \frac{2\alpha-1}{3}$ (bottom line) and 
	$\zeta^{(2)}$ given in equation~\eqref{eq: zeta two loop approximation} (upper line).
	\label{fig: tau_c and gamma_S vs alpha}} 
\end{figure}

As a final check I plot the cluster size distribution in figure~\ref{fig: P(S_c)}, measure the exponent $\tau_c$ and compare its values with the prediction ~\eqref{eq: Zapperi conj. generalized}.
In order to perform a collapse the data must be rescaled by the cutoff scale. 
There was no prior analytical knowledge of the dependence of the cutoff scale $S_{c,m}$ to the mass.
I used $S_{c,m} = S_m = \ell_m^{(d+\zeta)}$ because it was the only scale homogneous to a size I had at hand.
It appeared to work and it is quite remarkable that $S_m$ controls the cutoffs of both $P(S)$ and $P(S_c)$.
For the purpose of evaluating the prediction~\eqref{eq: Zapperi conj. generalized}
the exponent $\tau$ is determined via the relation~\eqref{eq: tau NarayanFisher} and the values of $\zeta$ in table \ref{tab: zeta} and I use the value of $\gamma_S$ measured in figure~\ref{fig: Nc vs S}. 
For $\alpha$ ranging from $0.5$ to $1.5$ the direct measurements give $\tau_c=2$, 1.72, 1.56 and 1.28 while the prediction~\eqref{eq: Zapperi conj. generalized} yields $\tau_c=5.55$, 2.33, 1.58, 1.21. This values are plotted in figure~\ref{fig: tau_c and gamma_S vs alpha} as blue squares and red circles respectively. 
We see that while the prediction~\eqref{eq: Zapperi conj. generalized} is close from the measurement for $\alpha=1$, it becomes in strong disagreement at low values of $\alpha$.
\vspace{.5cm}

Athough the general prediction~\eqref{eq: Zapperi conj. generalized} is false the main relation \eqref{eq: ch4 tau_c = 2tau-1} proposed in the abstract of Ref.~\cite{laurson2010} seems to be true for any $\alpha$. Indeed inserting \eqref{eq: tau NarayanFisher} and the values of $\zeta$ listed in table \ref{tab: zeta} in this relation we obtain the values
$\tau_c = 2$, $1.73$, 1.56, 1.30 (for $\alpha$ increasing from $0.5$ to $1.5$) which are in excellent agreement with the numerical determinations of $\tau_c$ listed above.
In the following I will present a new derivation of this relation. 
It relies on the whole probability distribution of the number of clusters $N_c$.

\subsection{Statistics of the number of clusters}\label{sec: stat of Nc}

In Ref.~\cite{laurson2010} the number of clusters inside an avalanche, $N_c$, was introduced to link the cluster statistics to the global avalanche one.
Here we follow and deepen this idea by studying the whole probability distribution, $P(N_c)$, of the number of clusters.
The first observations indicated a power law distribution with an exponent $\mu\simeq 1.5$ independent of $\alpha$ and a cutoff $N_{m}$ that increases when the mass is decreased.
From \eqref{eq: Nc sim S to gamma_S} we can expect this cutoff to scale as
$N_{m} \sim S_{m}^{\gamma_S} \sim m^{-2\gamma_S (d+\zeta)\alpha}$ but this makes appear the unknown
exponent $\gamma_S$.
We could have used the numerical values of $\gamma_S$ in order to produce a data collapse but we choose instead to use a different procedure, described in~\cite{rosso2009}, and which does not require
any knowledge of the cutoff scale. 
This procedure makes appear a connection with a famous stochastic process. 
I start by introducing this process before presenting the study of $P(N_c)$.

\subsubsection{The Bienaym{\'e}-Galton-Watson process}

The Bienaym{\'e}-Galton-Watson (BGW) process was introduced independently by Ir{\'e}n{\'e}e-Jules Bienaym{\'e}~\cite{bienayme1845} in 1845 and by Francis Galton and Henry William Watson~\cite{watson1875} in 1875.
They were interested in the distribution of surnames in an idealized population and in the probability that family names die out. They came out with a simple model that nowadays bear their names.
Family names are assumed to be patrilineal (passed from father to son) so the model focuses on male individuals.
Each individual has a random number of offsprings. The assumption of the model is that the number of offsprings per individual is an independent identically distributed (i.i.d)
random variable.
Let us denote $p_k$ the probability that an individual has $k$ offsprings. The fate of the process depends on the average reproduction rate, here denoted by $R$~:
\begin{equation}\label{eq: reproduction rate BGW}
R = \sum_{k=0}^{\infty} k p_k \, .
\end{equation}
If $R > 1$ there is a finite probability that the family line never gets extinct,
while for $R \leq 1$ the line gets extinct with probability one.
In this case, one can define the family size, $S_F$, as the total number of descendants of a single ancestor. It is a stochastic variable that displays a power-law distribution 
$P(S_F) \sim S_F^{-\tau_F}$ with exponent $\tau_F=3/2$.
If $R<1$ this power law is truncated exponentially above a maximal size $S_{F,m} \sim (1-R)^{-2}$.
If $R=1$ the power law is not truncated.

It turns out that the extinction of family names and their statistical distribution in a population are actually governed by many other factors than the single extinction of a family line accounted for in this model (according to the wikipedia page on the BGW process~\cite{wikipedia2020}).
Nevertheless it fairly describes the transmission of Y-chromosomes.
This simple model also describes the initiation of a nuclear chain reaction or an epidemic outbreaks in its early stage.
Closer from our concern it also describes the unfolding of an elastic interface avalanche in the fully-connected model. In this model all sites interact equally with each other and the elastic kernel is a constant, $\mathcal{C}(x-y)=\frac{c}{L}$ where $L$ is the system size, if $x\neq y$. Therefore when one point moves forward it can trigger the instability of a number of other points (its offsprings) anywhere in the system and its number of offsprings can be assumed to be an in i.i.d. random variable. The fully-connected model, which will be presented in more details in section~\ref{sec: FC and ABBM ch5}, belongs to the mean-field universality class.
Therefore the mean-field distribution of avalanche size is the same as the distribution of family size in the BGW model.
We will see in the following subsection that the BGW process is also connected to the distribution of the number of clusters.

\subsubsection{Normalization procedure}

\begin{figure}
	\centering
	\includegraphics[width=0.7\linewidth]{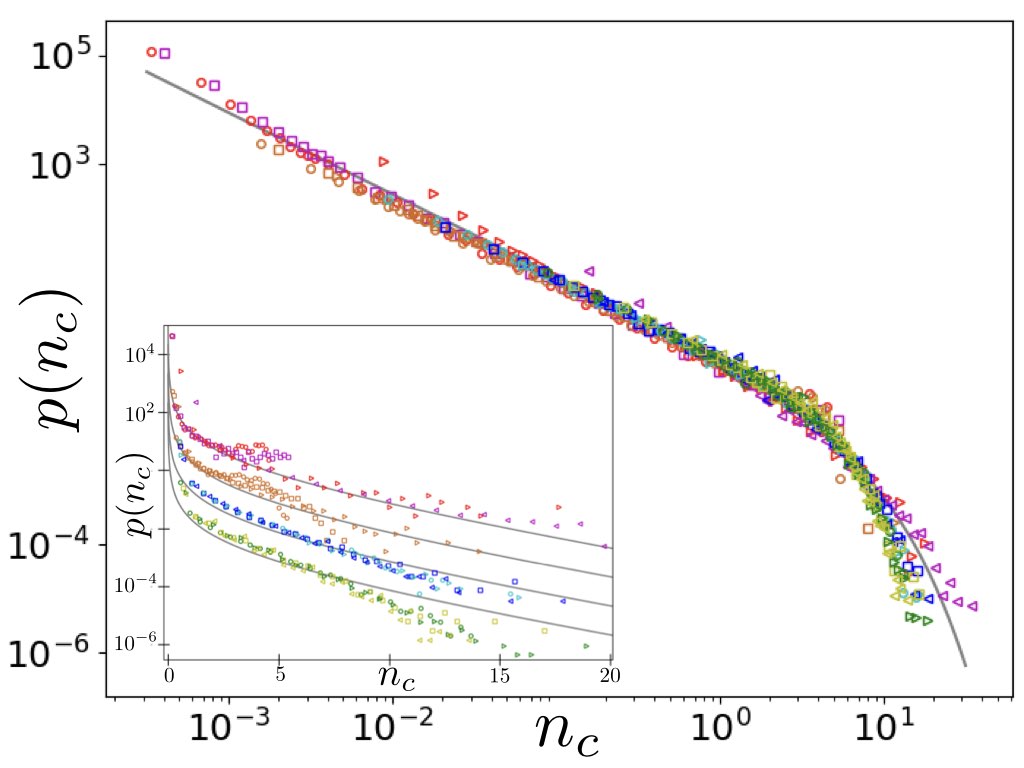}
	\caption{\emph{Main panel :} $p(n_c)$ (as defined in equation \eqref{eq: def p(n_c) 1}) in log-log (left) and log-lin scales (right). The thin gray line corresponds to the BGW function $p_{BGW}$ \eqref{eq: Galton-Watson nc}. \emph{Inset~:} Same quantity in log-lin scale and with a vertical shift of the data to test the exponential decay ($\alpha=0.5$ to 1.5 from top to bottom). \label{fig: p(n_c)}} 
\end{figure}

The number of clusters $N_c$ is a random variable whose distribution is a power law between 
a small scale cutoff $N_{min}=1$ and a large scale cutoff $N_m$. 
$N_c$ must takes integer values but for the sake of simplicity we will assume the distribution of $N_c$ to be continuous.
Its exponent $\mu$ is comprised between 1 and 2. Thanks to the large scale cutoff all of its moments are defined. For any integer $p\geq 1$ we have~:
\begin{equation}
\langle N_c^{p} \rangle \simeq \int_{N_{min}}^{N_{m}} N_c^{p-\mu} dN_c \simeq \frac{N_m^{p-\mu+1}}{p-\mu+1} \, . \label{eq: scaling moments N_c}
\end{equation} 
From this relation we see that $N_m$ scales as the ratio of two consecutive moments. In particular we have $N_m \sim \langle N_c^2 \rangle / \langle N_c \rangle$.
This is a rationale for defining a rescaled variable
\begin{equation}\label{eq: def n_c}
n_c = \frac{2\langle N_c \rangle}{\langle N_c^2 \rangle} N_c \, .
\end{equation}
The FRG predicts that when $m\to 0$ the statistics of $n_c$ is encoded in a universal function $p(n_c)$ which is defined by the relation \cite{rosso2009}\footnote{This prediction was originally made for the avalanche size, nevertheless since $N_c$ scales as a power of $S$ we can expect this result to remain true for the number of clusters.}~:
\begin{equation}\label{eq: def p(n_c) 1}
p(n_c)dn_c = \frac{\langle N_c^2 \rangle}{2\langle N_c \rangle^2} P\left(N_c\right) dN_c \, .
\end{equation}
Note that the function $p(n_c)$ is not a probability distribution. Indeed it is not normalized to unity. It instead obeys the two following normalization conditions~\cite{rosso2009}~:
\begin{align}
\langle n_c \rangle_p &= \int n_c p(n_c) dn_c = 1 \, , \\
\langle n_c^2 \rangle_p &= \int n_c^2 p(n_c) dn_c = 2 \, .
\end{align}
The analytical form of $p(n_c)$ has been computed for the BGW process (corresponding to mean-field avalanches size) and reads~\cite{ledoussal2009, ledoussal2009b}~:
\begin{equation}\label{eq: Galton-Watson nc}
p_{\text{BGW}}(n_c) = \frac{n_c^{-3/2}}{2\sqrt{\pi}} \exp(-n_c/4) \, .
\end{equation}
I computed $p(n_c)$ for $\alpha=0.5$, 0.75, $1$ and $\alpha=1.5$. The results are shown in figure~\ref{fig: p(n_c)}. 
In all cases we see a power law $P(N_c)\sim N_c^{-\mu}$ where the exponent $\mu$ is consistent with the BGW value $\mu=3/2$. 
Moreover the data seem to collapse on the same universal curve and cannot be clearly distinguished 
from the BGW function of equation \eqref{eq: Galton-Watson nc}.
The case $\alpha=0.5$ is mean-field. There we expect the number of clusters to be proportional to the 
avalanche size (up to logarithmic corrections). Thus it is not too surprising to recover the mean-field function in this case. 
There is however no obvious reason to justify that this behavior also holds for $\alpha > 0.5$.
The inset of figure~\ref{fig: p(n_c)} allows to compare the large scale decay of $p(n_c)$ with the mean-field function for each $\alpha$. 
Numerically we can neither confirm nor exclude that $p(n_c)$ is BGW (in the limit $m \to 0$, $L\to \infty$) also for $\alpha=0.75$, $\alpha=1$ and $\alpha=1.5$.
This constitute an intriguing question that remains open.
For what follows we need only to assume that $\mu=3/2$ and we do not mind about the exact decay of the distribution at large $N_c$.

\subsection{New scaling relations}

\begin{figure}[p]
	\centering
	\includegraphics[width=0.7\linewidth]{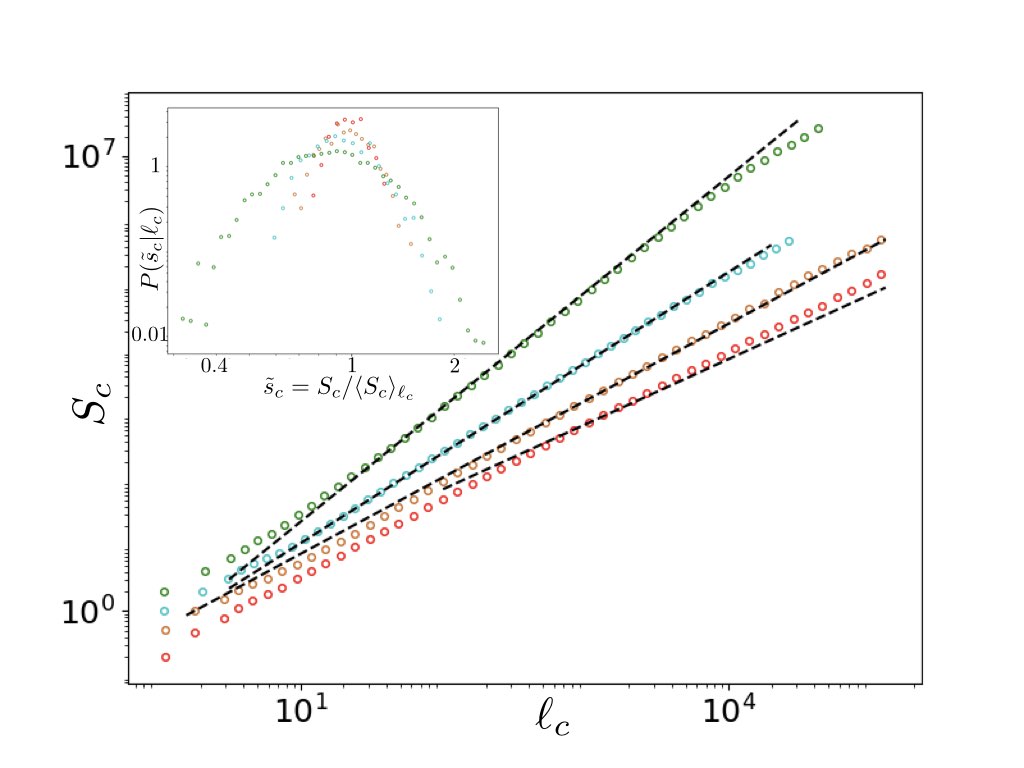}
	\caption{\emph{Main panel :} Averaged cluster size $\langle S_c \rangle_{\ell_c}$ versus cluster extension $\ell_c$.
	 Dashed lines correspond to fits $\langle S_c \rangle_{\ell_c} \sim \ell_c^{1+\zeta}$. \emph{Inset :} Conditional probability $P(S_c|\ell_c)$ computed in a small interval around $\ell_c = 1300$. The probability is well peaked. This justifies the derivation of equation \eqref{eq: relation kappa_c tau_c zeta}. \label{fig: Sc vs ell_c}} 
\end{figure}
\begin{figure}[p]
	\centering
	\includegraphics[width=0.7\linewidth]{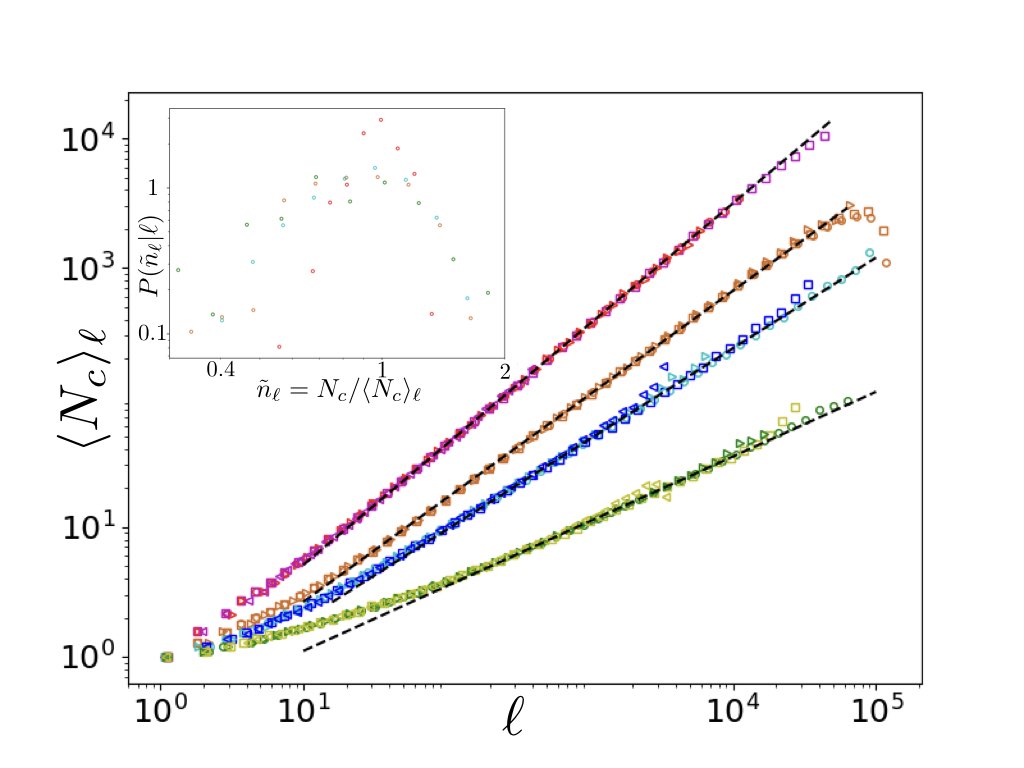}
	\caption{\emph{Main panel :} $\langle N_c \rangle_{\ell}$ versus $\ell$. The dashed lines are fit of the exponent $\gamma_{\ell}$ (defined by the relation $\langle N_c \rangle_{\ell} \sim \ell^{\gamma_{\ell}}$) .
	From top to bottom the values are $\gamma_{\ell}=0.93$, 0.80, 0.70 and 0.50 for $\alpha=0.5$, 0.75, 1 and 1.5 respectively. 
	\emph{Inset :} Probability of $\tilde{n}_{\ell} = N_c/\langle N_c \rangle_{\ell}$ conditioned by the avalanche size $\ell$. The conditional probability has been computed in small intervals centered around 
	$\ell=10^2$, $10^2$, $10^3$ and $10^4$ for $\alpha=0.5$, 0.75, 1 and 1.5 respectively.
	The fact that the conditional probabilities are well-peaked justifies the application of the relation $P(N_c)dN_c = P(\ell)d\ell$.
	\label{fig: Nc vs ell}} 
\end{figure}

\begin{figure}[!p]
	\begin{subfigure}[t]{1\linewidth}%
	\centering
	\includegraphics[width=.72\linewidth]{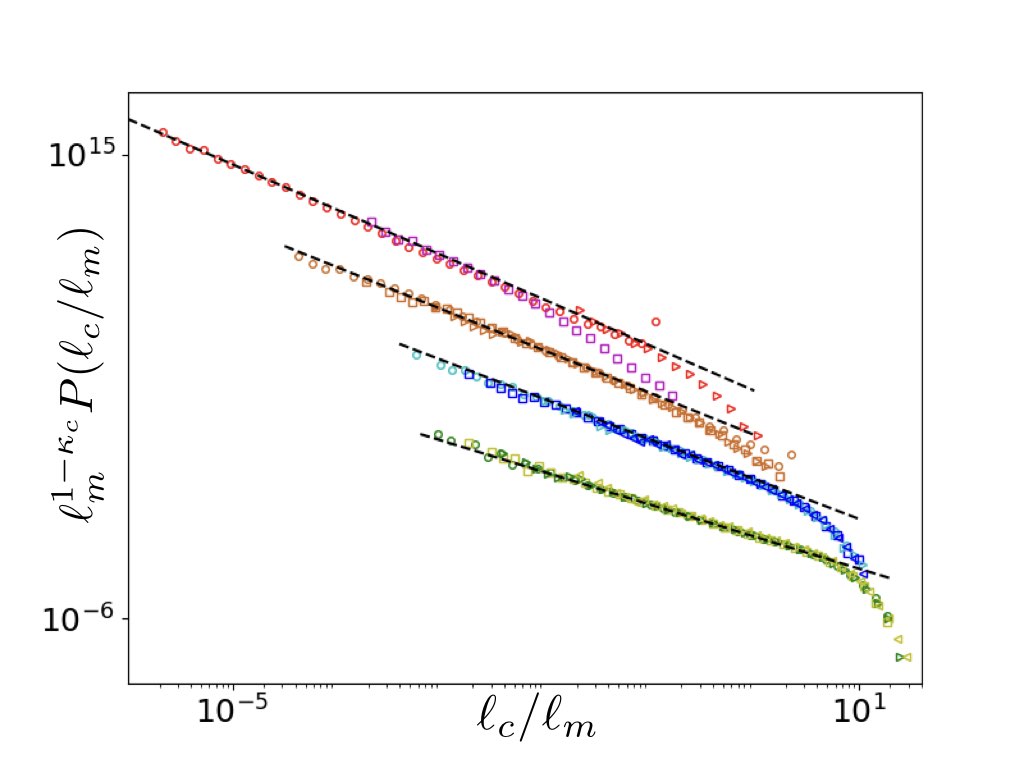}
	\caption{Collapse of the cluster extension distributions. The x-axis has been rescaled by $\ell_m$ defined in equation \eqref{eq: relation Smax ellmax} and the y-axis by $\ell_m^{1-\kappa_c}$ The values of $\kappa_c$ are obtained by fitting the power law part of the distributions (dashed lines). The collapse of the distributions for each $\alpha$ shows that the cutoff of the distributions is controlled by $\ell_m$. It is also a verification of the values of $\kappa_c$.
	The fitted values are (from top to bottom, i.e. $\alpha=0.5$ to 1.5) $\kappa_c=2.05$, 1.90, 1.80 and 1.45.
	Data have been shifted vertically for visibility purpose.
	\label{fig: P(ell_c)}}
	\end{subfigure}
	\vspace{1cm}
    \begin{subfigure}[t]{1\linewidth}%
    \centering
	\includegraphics[width=.72\linewidth]{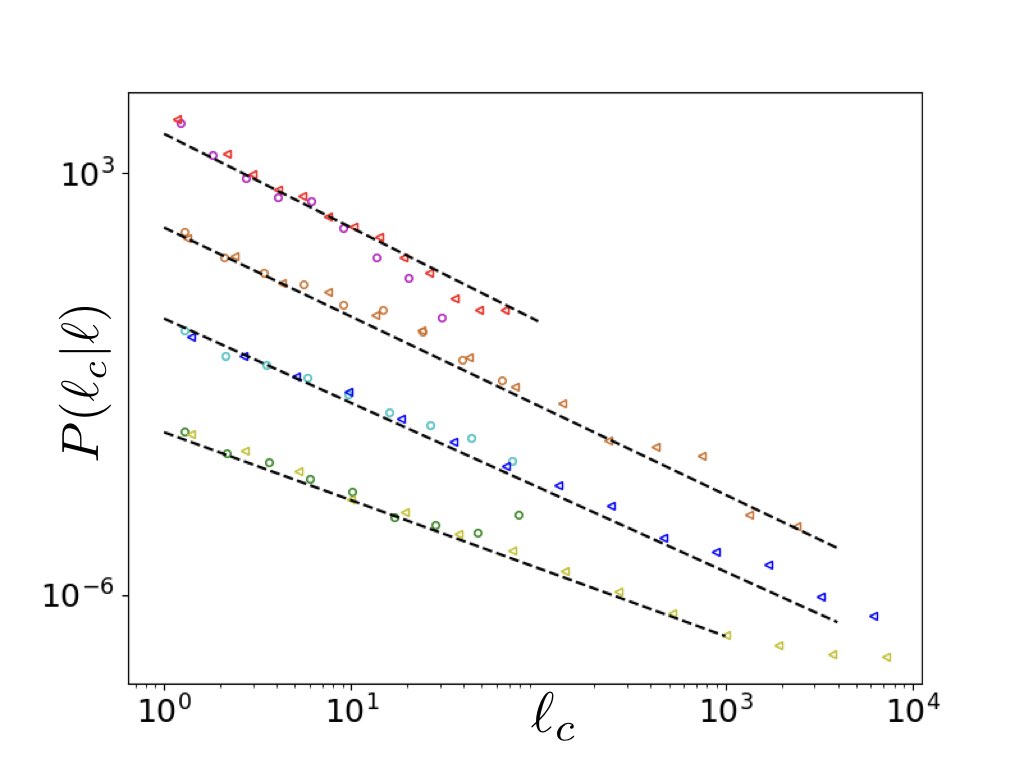}
	\caption{Distribution of cluster extensions $\ell_c$ conditioned by the linear size $\ell$ of the avalanche to which the clusters belong. Two values of $\ell$ are compared~: $\ell=10^2$ (circles) and $\ell=10^4$ (triangles). We observe that the conditional distribution $P(\ell_c|\ell)$ is a power law cut at $\ell$.
	The dashed lines are fits using the exponents $\kappa_c$ determined in the figure above.
	Parameters values are $\alpha=0.5$, $m^2=10^{-3}$, $L=2^{17}$ ; 
	$\alpha=0.75$, $m^2=3.10^{-4}$, $L=2^{17}$ ; $\alpha=1$, $m^2=3.10^{-4}$, $L=2^{18}$ ; 
	$\alpha=1.5$, $m^2=3.10^{-5}$, $L=2^{17}$.
	Data have been shifted vertically for visibility purpose. \label{fig: P(ell_c|ell)}}
	\end{subfigure}
	\vspace{-1cm}
	\caption{Distributions of cluster extensions. \label{fig: P(ell_c) and P(ell_c|ell)}}
\end{figure}

The mean-field BGW exponent $\mu=3/2$ is a key element to propose a new derivation of scaling relations linking the exponents of the clusters to the ones of the global avalanche. 
The reasoning is based on four hypotheses that are all verified numerically~:
\begin{itemize}
\item[(i)] 
Similarly to short-range avalanches the size of a cluster is related to its extension via~:
\begin{equation}\label{eq: Sc vs ell_c}
S_c \sim \ell_c^{d+\zeta} \, .
\end{equation}
This relation is confirmed by figure~\ref{fig: Sc vs ell_c} where $\langle S_c \rangle_{\ell_c}$ is plotted as a function of $\ell_c$. The inset shows that the conditional distribution $P(S_c|\ell_c)$ is peaked around $\langle S_c \rangle_{\ell_c}$ and so the scaling holds in general. 
This allows to apply a relation of the type $P(S_c)dS_c = P(\ell_c) d\ell_c$ and derive
a first scaling relation~:
\begin{equation}\label{eq: relation kappa_c tau_c zeta}
\kappa_c -1 = (\tau_c-1)(d+\zeta) \, .
\end{equation}

\item[(ii)] The distribution of $N_c$ conditioned by a given $S$ is peaked around a typical value which is close from the average $\langle N_c \rangle_S$ (see inset of figure \ref{fig: Nc vs S}). This also holds for the distribution of $N_c$ conditioned by $\ell$ (inset of figure \ref{fig: Nc vs ell}). Thus we can write~:
\begin{equation}
P(S) dS =  P(N_c) dN_c = P(\ell) d\ell \label{eq: P(S)dS=P(N)dN=P(ell)dell}
\end{equation}

\item[(iii)] The distribution $P(S_c|S)$ of cluster sizes conditioned by the total avalanche size is a power law with the same exponent $\tau_c$ as the global cluster distribution and a sharp cutoff at $S$. This is visible on figure \ref{fig: P(S_c) and P(S_c|S)}.
For $\tau_c<2$ it follows that~:
\begin{align}
\langle S_c \rangle_S &\sim S^{2-\tau_c} \, , \label{eq: relation Sc sim S to 2-tau_c} \\
\langle N_c \rangle_S &= \frac{S}{\langle S_c \rangle_S } \sim S ^{\gamma_{S}} \;\text{ with }\; \gamma_{S} = \tau_c-1\, . \label{eq: relation Nc vs S}
\end{align}
The first equality in the second line has been proven above \eqref{eq: average relation S_c S N_c}.
For $\alpha=0.5$ the cluster exponent is $\tau_c=2$. In this case the first relation should be replaced by $\langle S_c \rangle_S \sim \ln(S)$ and the second becomes
$\langle N_c \rangle_S \sim S/\ln(S)$. This can explain that I measured an effective exponent $\gamma_S=0.89$ instead of $1$ in this case. 

The same argument holds for the cluster extension. In this case the avalanche linear size $\ell$ plays the role of the total size $S$. One can see on figure~\ref{fig: P(ell_c) and P(ell_c|ell)} that
$P(\ell_c|\ell)$ has the same exponent $\kappa_c$ as $P(\ell_c)$.
So we also have the relations~:
\begin{align}
\langle \ell_c \rangle_{\ell} &\sim \ell^{2-\kappa_c}\, , \label{eq: relation ell_c sim ell to 2-kappa_c}\\
\langle N_c \rangle_{\ell} &= \frac{\ell}{\langle \ell_c \rangle_{\ell} } \sim \ell ^{\gamma_{\ell}} \;\text{ with }\; \gamma_{\ell} = \kappa_c-1 \, .\label{eq: relation Nc vs ell} 
\end{align}
Once again a logarithmic correction applies for $\alpha=0.5$. 

\item[(iv)] The last hypothesis is the one that we have seen in the former section~\ref{sec: stat of Nc}. 
The number of clusters $N_c$ follows a power law with an exponent $3/2$ which is independent of $\alpha$~:
\begin{equation}
P(N_c) \sim N_c^{-3/2} \, . \label{eq: P(N_c) sim N_c to -3/2}
\end{equation}
\end{itemize}

We can now link the cluster exponents to the global avalanche ones. Inserting the scaling relations
\eqref{eq: relation Nc vs S}, \eqref{eq: relation Nc vs ell} and \eqref{eq: P(N_c) sim N_c to -3/2} into the equality \eqref{eq: P(S)dS=P(N)dN=P(ell)dell} we obtain~:
\begin{align}
\tau_c  = 2\tau -1 \, , \label{eq: relation tau_c tau} \\
\kappa_c = 2 \kappa - 1 \, .\label{eq: relation kappa_c kappa}
\end{align}
We have recovered the relation \eqref{eq: relation tau_c tau} which was proposed in~\cite{laurson2010}.
We see that the same relation holds regarding the cluster extension and the avalanche linear size.
Note that we have not yet used the first hypothesis (i).
This hypothesis allows us to express the cluster exponents in term of only three parameters~:
$d$, $\alpha$ and $\zeta$. 
From equations  \eqref{eq: relation tau_c tau}, \eqref{eq: relation kappa_c tau_c zeta} and
\eqref{eq: tau NarayanFisher} we deduce~:
\begin{align} 
\tau_c &= 3 - \frac{2\alpha}{d+\zeta} \, , \label{eq: relation tau_c zeta alpha} \\
\kappa_c &= 1 + 2(d+\zeta) - 2\alpha \, . \label{eq: relation kappa_c zeta alpha}
\end{align}

The exponents $\tau_c$, $\kappa_c$, $\gamma_S$ and $\gamma_{\ell}$ have been numerically measured on figures \ref{fig: P(S_c)} \ref{fig: P(ell_c)} \ref{fig: Nc vs S} and \ref{fig: Nc vs ell} respectively. 
Table~\ref{tab: measurement vs predictions} summarizes the results and the comparison with the scaling predictions. The measurement uncertainties are summarized in table~\ref{tab: measurement with errors}.
The measurements are found to be in good agreement with the predictions.

\begin{figure}[p]
    \begin{adjustwidth}{-2cm}{-2cm}
	\centering
	\includegraphics[scale=0.25]{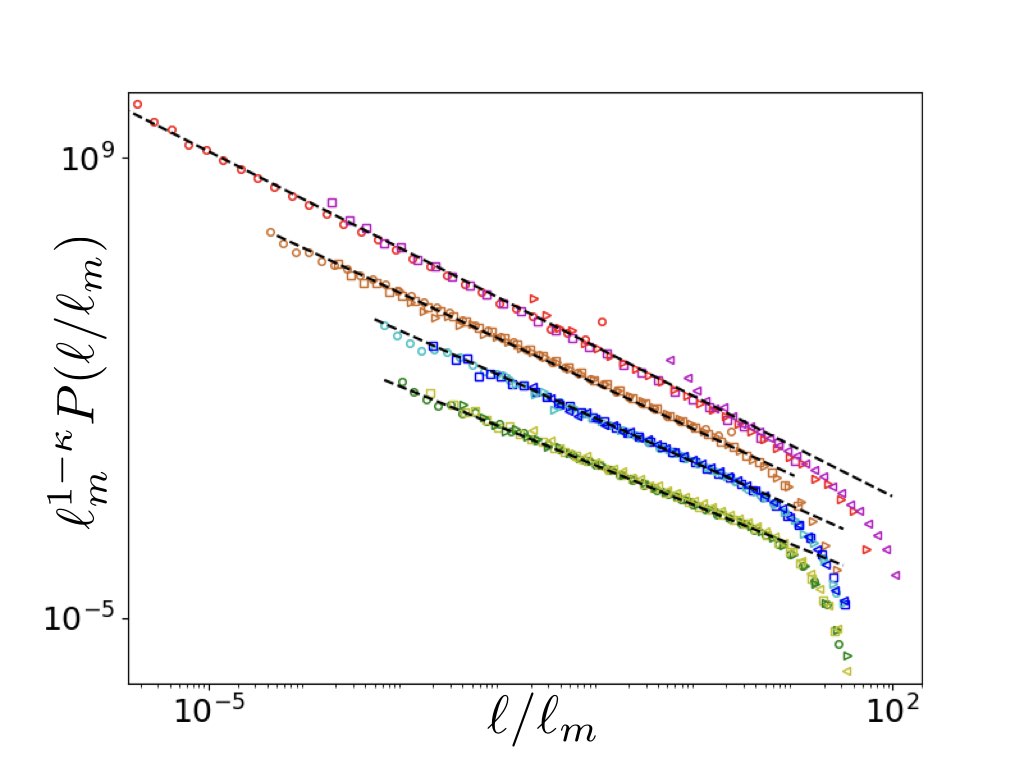}\includegraphics[scale=0.25]{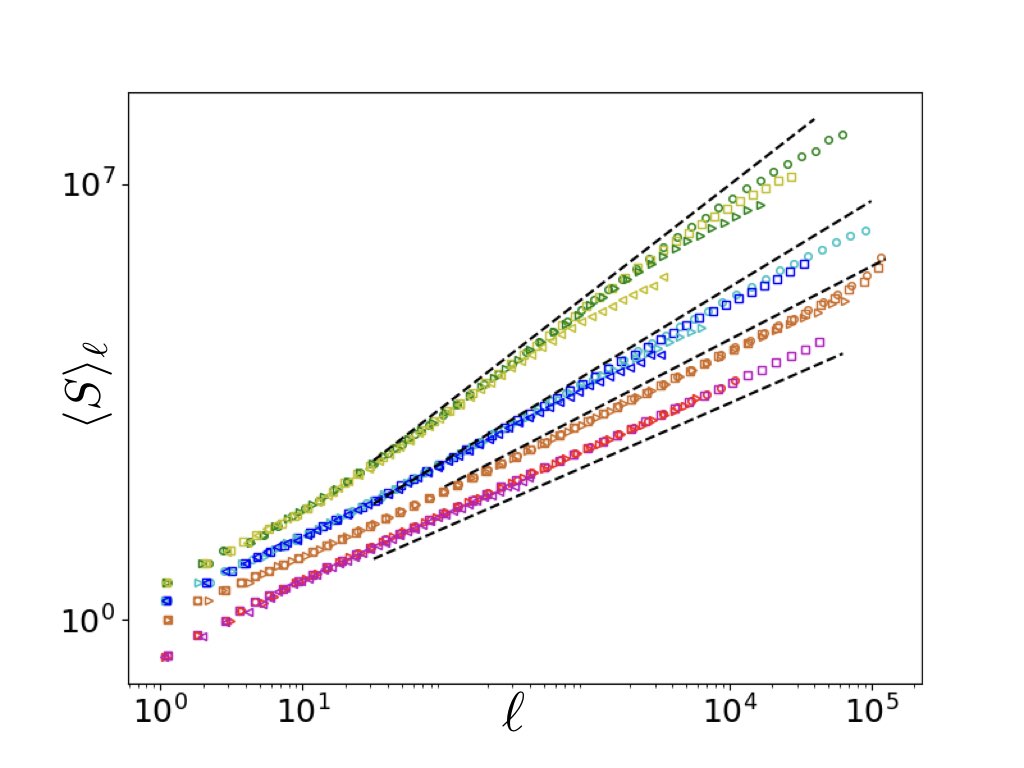}
	\caption{\emph{Left :} Collapse of the avalanche linear size distribution. The dashed lines correspond to the prediction \eqref{eq: relation kappa zeta alpha}~:
	$\kappa=1.5$ for $\alpha=0.5$ , $\kappa=1.39$ for $\alpha=1$  and $\kappa=1.26$ for $\alpha=1.5$ 
	For visibility purpose, data on the right panel have been shifted by a factor $100$ for $\alpha=0.5$ and $0.01$ for $\alpha=1.5$.
	\emph{Right :} $\langle S \rangle_{\ell}$ versus $\ell$. The dashed lines correspond to the prediction \eqref{eq: relation S vs ell} $\langle S \rangle_{\ell}\sim \ell^{d+\zeta}$. Data have been shifted vertically for visibility purpose.
	\label{fig: P(ell) and S vs ell}} 
	\end{adjustwidth}
\end{figure}
\begin{figure}[p]
	\centering
	\includegraphics[width=.7\linewidth]{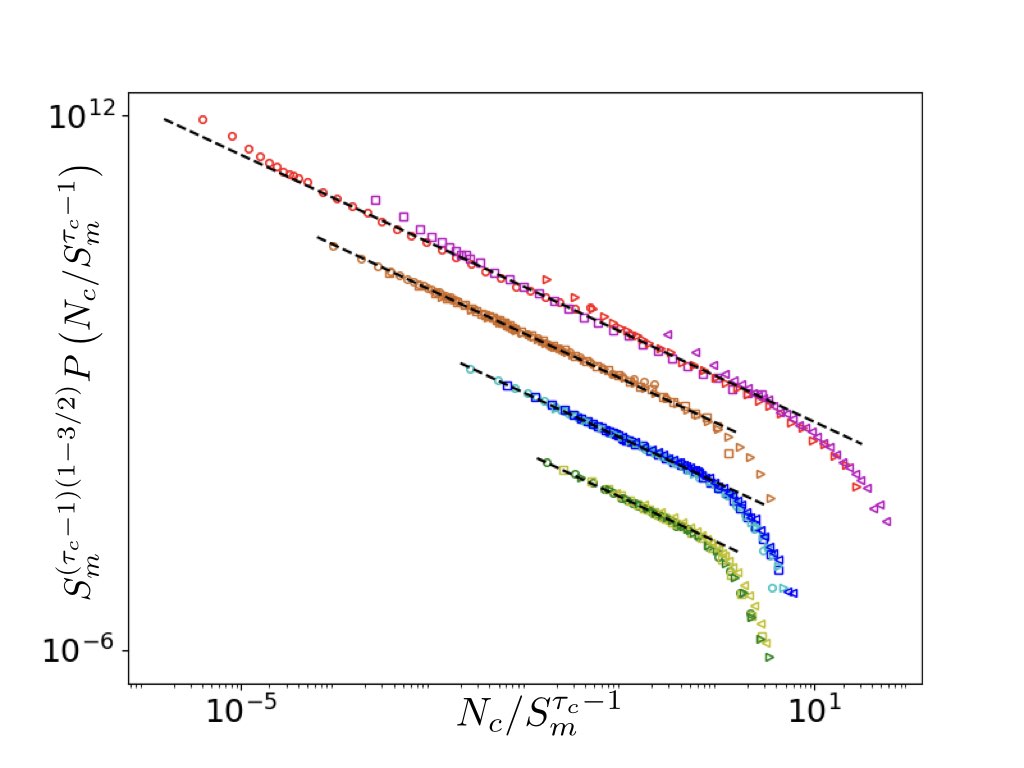}
	\caption{Collapse of the probability distribution of the number of clusters using the rescaling proposed in equation \eqref{eq: mass rescaling for Nc}.
	Data have been shifted vertically for visibility purpose.
	\label{fig: P(Nc) rescaled with the mass}} 
\end{figure} 

The scaling relations \eqref{eq: relation kappa_c tau_c zeta}, \eqref{eq: relation Nc vs S},\eqref{eq: relation Nc vs ell} and the equality \eqref{eq: P(S)dS=P(N)dN=P(ell)dell} also allow us
to recover the self-affinity between the total size and the 
linear size that is known for short-range avalanches which in turn allows to determine the exponent $\kappa$~:
\begin{align} 
S &\sim \ell^{d+\zeta} \, ,  \label{eq: relation S vs ell} \\
\kappa &= 1 + d+\zeta - \alpha \, .  \label{eq: relation kappa zeta alpha}
\end{align}
Note that the self-affinity holds as long as $P(N_c)$ is a power law, even if its exponent is different from $3/2$. The scaling relation 
\eqref{eq: relation kappa zeta alpha} can also be obtained from \eqref{eq: relation kappa_c kappa} and \eqref{eq: relation kappa_c zeta alpha}. 
The linear size distributions are plotted on figure~\ref{fig: P(ell) and S vs ell} left panel with fits of the exponent $\kappa$. The results are compared with the prediction in table~\ref{tab: measurement vs predictions} and are once again in good agreement. 
A comparison of the self-affinity prediction \eqref{eq: relation S vs ell} is made on the right panel of figure~\ref{fig: P(ell) and S vs ell}. The agreement for this relation is not as good as for the all the other ones. I also made measurements of the effective exponent which are available in tables~\ref{tab: measurement vs predictions} and \ref{tab: measurement with errors}.

As a final remark to conclude this section let us note that from \eqref{eq: relation Nc vs S} we can expect the cutoff of 
$P(N_c)$ to scale as~:
\begin{equation}
N_m \sim S_m^{\tau_c-1} \sim \ell_m^{2(d+\zeta-\alpha)} \sim m^{4\left(1 - (d+\zeta)/\alpha\right)} \, .\label{eq: mass rescaling for Nc}
\end{equation}
This scaling is checked on figure \ref{fig: P(Nc) rescaled with the mass} where I performed a data collapse of $P(N_c)$ with the prediction \eqref{eq: mass rescaling for Nc}.

\subsection{Discussion of the study of Planet et al.}

\subsubsection{Presentation of the study}

In Ref.~\cite{planet2018}, Planet et al. studied how local bursts of activity (i.e. clusters) organize into global avalanches. They used scaling arguments to derive a scaling relation between $\tau_c$ and $\tau$ and tested their relation on a stable imbibition experiment. 
Stable imbibition occurs when a fluid (e.g. water) invades a porour medium, taking the place of an other fluid that is both less wetting and less viscous than the invading fluid (e.g. air).
The mean velocity $v$ of the front is imposed by pushing water into the medium at a constant rate. 
In this case the equation of motion of the front can be written in Fourier space as~\cite{ortin2017}~:
\begin{equation}
\partial_t \hu(q,t) = v\delta(q) - v|q|\hu(q,t)-K_0 |q|q^2 \hu(q,t) + \hat{F}_{\dis}\left(q, g[u]\right) \, .
\end{equation}
The first term on the RHS, $v\delta(q)$, means that the average position of the front advances with an instantaneous velocity $v$, a property which is enforced by the incompressibility of the fluid. 
The second term corresponds to long-range elasticity, with an interaction coefficient proportional to $v$. It arises from viscous damping and is dominant at small $q$, i.e. at large scale. The third term corresponds to short-range elasticity. It arises form surface-tension damping and is dominant at large $q$, i.e. at small scales. 
Finally the fourth term corresponds to quenched disorder. 

The competition between the two elastic forces leads to a characteristic length scale
$\xi_c = \sqrt{K_0/v}$.
The interface is characterized by local roughness and dynamical exponents $\zeta_{\text{loc}}$  and $z_{\text{loc}}$ below $\xi_c$ and by global exponents $\zeta$ and $z$ above $\xi_c$.
The crossover between different exponents at short and large scales implies that the scaling of the events size with their duration is different for clusters and for global avalanches. We have two distinct exponents, $\gamma$ and $\gamma_c$ such that~:
\begin{equation}
S \sim T^{\gamma} \quad , \quad S_c \sim T_c^{\gamma_c} \, ,
\end{equation}
where $T$ and $T_c$ are respectively the global avalanche and cluster durations.

The authors distinguish experimentally three different arrangements of local events leading to three kinds of avalanches~:
\begin{itemize}
\item[(I)] small avalanches made of a single cluster ;
\item[(II)] intermediate avalanches made of several clusters among them $N(T)$ clusters have almost the same duration as the avalanche, i.e. $T_c=T$ ;
\item[(III)] large and long avalanches made of many clusters with no dominant cluster.
\end{itemize}

In the intermediate regime (II), scaling arguments are made to propose a scaling relation relating the local and global exponents~:
\begin{equation}\label{eq: Planet scaling 1}
\gamma - \frac{\zeta}{z} = \gamma_c - \frac{\zeta_{\text{loc}}}{z_{\text{loc}}} \, .
\end{equation}
Then considering global avalanches and clusters of the same duration, the authors apply the relation 
$P(S)dS = P(T) dT = P(S_c) dS_c$ which yield a new scaling relation relating $\tau_c$ and $\tau$~:
\begin{equation}\label{eq: Planet scaling 2}
\tau_c = 1 + \frac{\gamma}{\gamma_c}(\tau -1) \, .
\end{equation}
Experimentally the authors measure $\tau=1.00\pm0.15$ for which equation~\eqref{eq: Planet scaling 2} yields a prediction $\tau_c=1.00\pm0.14$.
This prediction compares well with the experimental measurement $\tau_c=1.08\pm0.05$.

\subsubsection{Discussion}

The argument of Planet et al.~\cite{planet2018} relies on the existence of two scales characterized by different roughness and dynamical exponents. 
We have seen in chapter~\ref{chap: experiments} that a crossover in the value of the roughness exponent is indeed present in crack front experiments and probably also in wetting front experiments. 
This crossover is however not explained by the model of an interface with long-range elasticity studied in this chapter. Indeed a single value of the roughness exponent is expected analytically and observed numerically. This is visible in figure~\ref{fig: Roughness} and confirmed by the fact that 
the scalings $S_c\sim \ell_c^{d+\zeta}$ and $S \sim \ell^{d+\zeta}$ are characterized by the same exponent $\zeta$. (I note however that on this point my results disagree with~\cite{laurson2010} where a crossover in the roughness exponent was observed in numerical simulations, using the same model but with a larger value of the mass, $m^2=0.0125$.) A single value of the dynamical exponent $z$ is also observed (see e.g.the numerical determination~\cite{duemmer2007}).
So for our model the scaling relation~\eqref{eq: Planet scaling 1} yields $\gamma_c=\gamma$. 
This equality can also be obtained from the scaling relation 
$\gamma=(d+\zeta)/z$~\eqref{eq: scaling relation ta tau gamma}.
With this equality the scaling relation~\eqref{eq: Planet scaling 2} then yields $\tau_c=\tau$ which is at odd with both numerical and experimental observations. 
This suggests that the relation~\eqref{eq: Planet scaling 2} requires the existence of two distinct scales characterized by different exponents and is not valid for our model. 

Finally we note that with the value $\tau=1$ measured in~\cite{planet2018}, the relation 
$\tau_c=2\tau-1$~\eqref{eq: ch4 tau_c = 2tau-1} gives the same prediction as~\eqref{eq: Planet scaling 2}, i.e. $\tau_c=1$. 
Hence it is plausible that this relation also applies to the imbibition experiment studied in~\cite{planet2018}\footnote{Note that the relation~\eqref{eq: ch4 tau_c = 2tau-1} accurately describes the experimental distribution of cluster sizes~\cite{tallakstad2011} despite the crossover in the experimental value of $\zeta$.}. 
It would therefore be interesting to study the statistics of the number of clusters$N_c$ in this experiment. A scaling $P(N_c)\sim N_c^{-3/2}$ would confirm the validity of the relation~\eqref{eq: ch4 tau_c = 2tau-1}.

\section{Statistics of gaps and avalanche diameter}\label{sec: statistics gaps and diameter}

\begin{figure}
	\centering
	\includegraphics[scale=0.3]{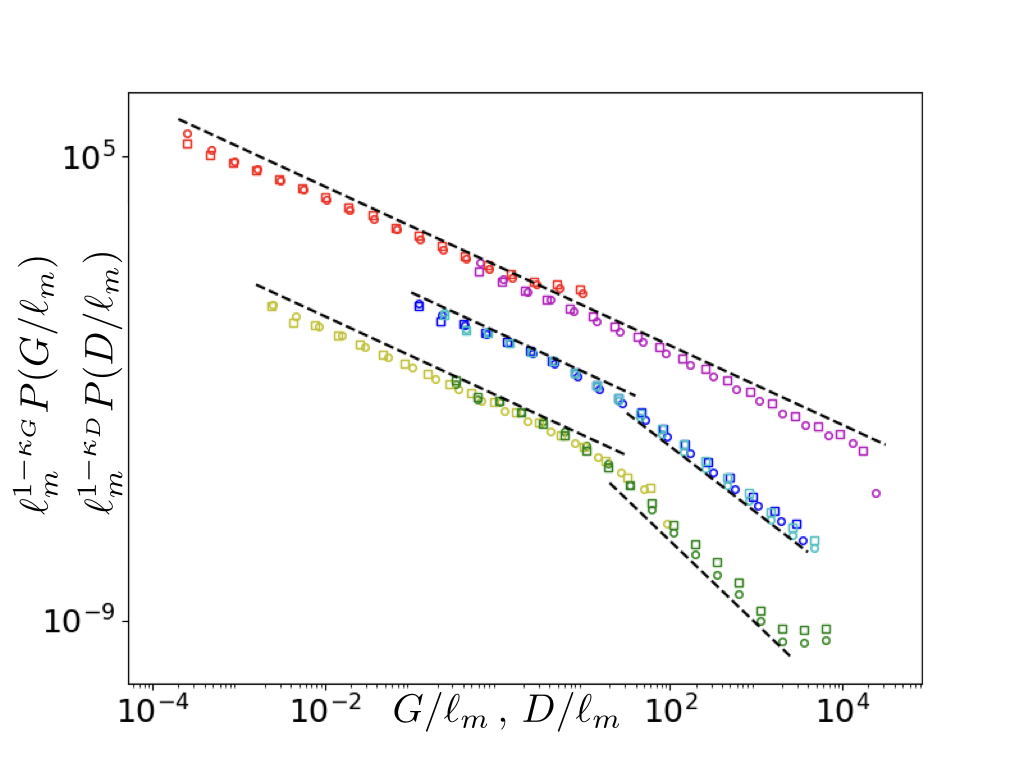}
	\caption{$P(G)$ (squares) and $P(D)$ (circles). The two distributions are indistinguishable. The dashed lines fit with exponent $1.2$ and $d+\alpha$. 
	Parameter values are : 
	$\alpha=0.5$ : $L=2^{17}$ and $m^2=0.01$ (red symbols) and $L=2^{17}$ and $m^2=0.5$ (magenta) ; 
	$\alpha=1$ : $L=2^{17}$ and $m^2=0.05$ (blue) and $L=2^{17}$and $m^2=0.1$ (cyan) ;
	$\alpha=1.5$ : $L=2^{17}$ and $m^2=3.10^{-5}$ (yellow) and $L=2^{17}$ and $m^2=0.03$ (green).
	For visibility purpose, data have been shifted by a factor $100$ for $\alpha=0.5$ and $0.01$ for $\alpha=1.5$. \label{fig: P(G), P(D)}} 
\end{figure}

Regarding the global avalanche a natural, and novel, object to study is the avalanche diameter $D$ that is the distance between the two extremal points of the avalanche. 
For SR avalanches $D=\ell$ is the avalanche extension. Its statistics was theoretically studied within a mean-field model in~\cite{delorme2016}.
In LR avalanches however $D$ is much larger than the linear size $\ell$ and exhibit a new distribution $P(D)$.
This distribution is plotted in figure \ref{fig: P(G), P(D)} together with the total gap length distribution $P(G)$ for $\alpha=0.5$, $1$ and $1.5$. 
We see that $P(D)$ has a universal exponent $\kappa_D=1.2$ at small scales that does not depend on $\alpha$. At large scales there is a crossover to an exponent $d+\alpha$ for 
$\alpha=1$ and $1.5$~:
\begin{align}
P(D) &\sim \left\lbrace 
\begin{array}{l}
D^{-\kappa_D} \, \text{ for } \, D < D_{\text{cross}} \, , \\
D^{-(d+\alpha)} \, \text{ for } \, D > D_{\text{cross}} \, , 
\end{array}\right. \\
D_{\text{cross}} &\sim \ell_m \, .
\end{align}
Once again the crossover appears to be governed by $\ell_m$. 
No crossover is visible for $\alpha=0.5$ but we can not yet exclude that it occurs at larger scale.
We notice that the statistics of the diameter is very different from the one of the linear size $\ell$. In particular there is no exponential cutoff. 
Since $D = \ell + G$, where $G = \sum g$ is the total gap length, we need to investigate the statistics of the gaps $g$ in order to understand the behavior of $P(D)$.

\begin{figure}[!p]
	\begin{subfigure}[t]{1\linewidth}%
	\centering
	\includegraphics[width=.7\linewidth]{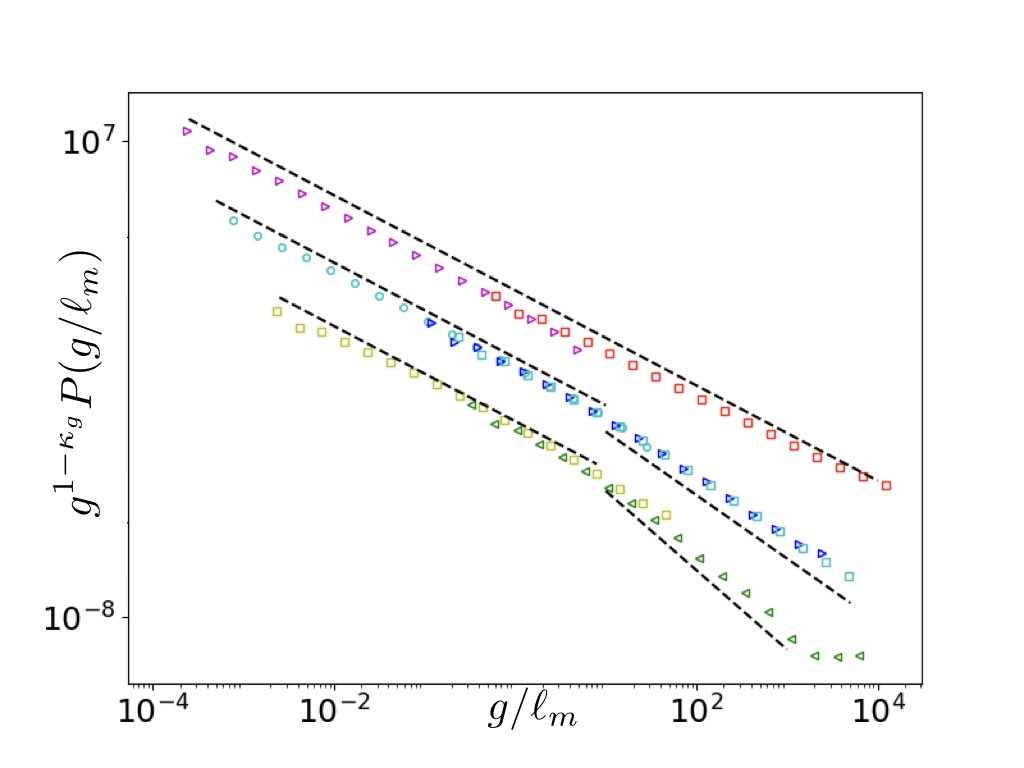}
	\caption{Collapse of the gap distribution. $P(g/\ell_m)$ is the probability distribution of the variable $g/\ell_m$, 
	$\ell_m = m^{-2/\alpha}$. 
	The dashed lines are fit with exponents $\kappa_g=3/2$ at small scale and $d+\alpha$ at large scale. Parameter values are : 
	$\alpha=0.5$ : $L=2^{17}$ and $m^2=0.01$ (magenta triangles) and $L=2^{17}$ and $m^2=0.5$ (red squares) ; 
	$\alpha=1$ : $L=2^{18}$ and $m^2=3.10^{-4}$ (cyan circles), $L=2^{17}$ and $m^2=0.05$ (blue triangles) and $L=2^{17}$and $m^2=0.1$ (cyan squares) ;
	$\alpha=1.5$ : $L=2^{17}$ and $m^2=3.10^{-5}$ (yellow squares) and $L=2^{17}$ and $m^2=0.03$ (green triangles). 
	Data have been shifted vertically for visibility purpose.
	\label{fig: P(g)}}
	\end{subfigure}
	\vspace{1cm}
    \begin{subfigure}[t]{1\linewidth}%
    \centering
	\includegraphics[width=.7\linewidth]{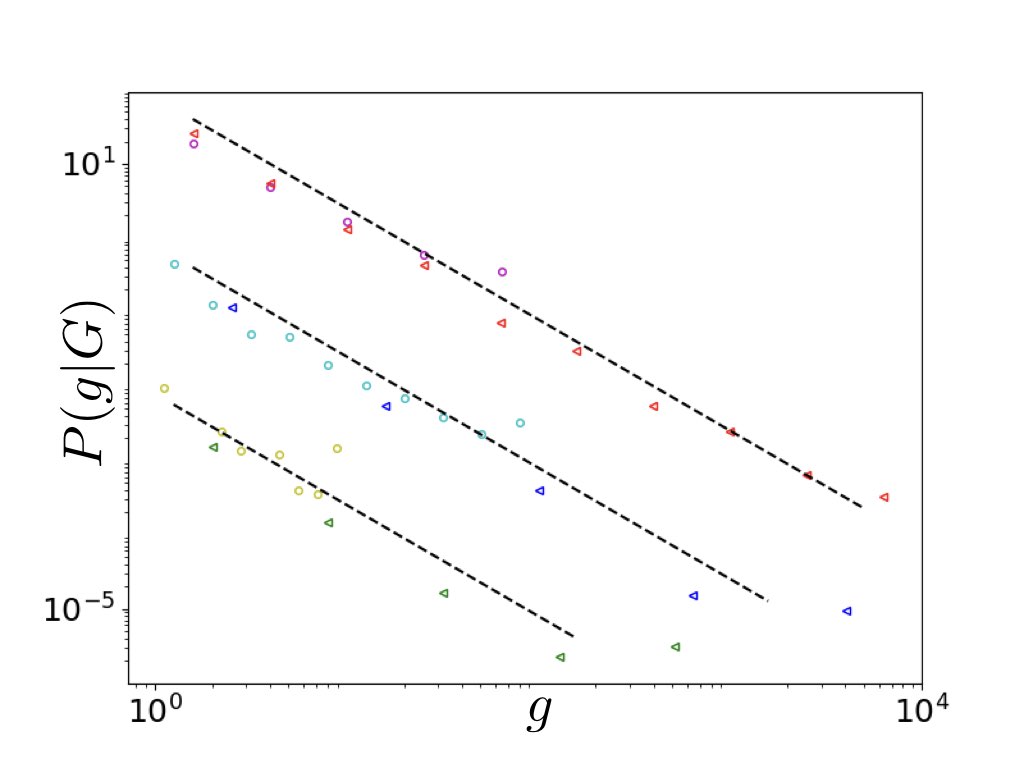}
	\caption{Distribution of gaps $g$ conditioned by the total gap length $G$. For each $\alpha$ two values of $G$ are compared. For $\alpha=0.5$ and $\alpha=1$, $G=10^2$ (circles) and $G=10^4$ (triangles). For $\alpha=1.5$, $G=10$ (circles) and $G=10^3$ (triangles). We observe that the conditional distribution $P(g|G)$ is a power law cut at $G$.
	The dashed lines are guide to the eyes with the exponent $\kappa_g=3/2$ determined in the figure above.
	Parameters values are $\alpha=0.5$, $m^2=0.01$, $L=2^{17}$ ; 
	$\alpha=1$, $m^2=0.05$, $L=2^{17}$ ; 
	$\alpha=1.5$, $m^2=0.03$, $L=2^{17}$.
	Data have been shifted vertically for visibility purpose. \label{fig: P(g|G)}}
	\end{subfigure}
	\vspace{-1cm}
	\caption{Distributions of cluster extensions. \label{fig: P(g) and P(g|G)}}
\end{figure}

The gap distribution $P(g)$ is plotted in figure \ref{fig: P(g)}. 
It presents the following universal distribution:
\begin{align}
P(g) &\sim \left\lbrace 
\begin{array}{l}
g^{-\kappa_g} \, \text{ for } \, g < g_{\text{cross}} \, , \\
g^{-(d+\alpha)} \, \text{ for } \, g > g_{\text{cross}} \, , 
\end{array}\right. \label{eq: P(g)} \\
g_{\text{cross}} &\simeq 10 \ell_m \, ,
\end{align}
with $\kappa_g=3/2$ independentlty from $\alpha$
For $\alpha=0.5$ (and $d=1$) $\kappa_g=d+\alpha$ while for $\alpha=1$ and $\alpha=1.5$ there is a crossover from a power law to a faster power law.
Note that $P(g)$ has a much slower decay than $P(\ell_c)$~: the exponent $\kappa_g=3/2$ is in general lower than the exponent $\kappa_c$ and the decay at large scales is a power law for $P(g)$ while it is faster than any power law for $P(\ell_c)$.
Therefore we expect that there are more large gaps than large cluster extensions and hence that $G \gg \ell$. 
This in turn implies that $D=\ell + G \sim G$ and the distributions of both $G$ and $D$ must be dominated by the large gaps.
This is confirmed by figure \ref{fig: P(G), P(D)} where both distributions $P(G)$ and $P(D)$ are plotted together and cannot be distinguished.

As for the clusters we expect that the gaps within an avalanche of total gap length $G$ follow the 
same distribution as $P(g)$~\eqref{eq: P(g)} but with a sharp cutoff at $G$. 
The conditional distribution $P(g|G)$ is plotted on figure~\ref{fig: P(g|G)}.
Only the first power law with exponent $\kappa_g=3/2$ is visible.
Due to a lack of statistics I was not able to explore a large $G$ regime.
A crossover should exist at larger scales as otherwise
there would be no reason to observe a crossover in $P(G)$ (visible on figure \ref{fig: P(G), P(D)}).
Base on this assumption we derive a relation between the total gap length $G$ and the number of clusters $N_c$. 
Two cases must be distinguished.

\paragraph{a) $G < G_{\text{cross}}$}
In this case we assume that all the gaps are in the $g^{-3/2}$ regime.
We can apply the same reasoning as for equations \eqref{eq: relation Nc vs S} and \eqref{eq: relation Nc vs ell}. This yields the scaling relations~:
\begin{align}
\langle g \rangle_G &\sim G^{2-\kappa_g} \, ,\\
\langle N_c \rangle_G &= \frac{G}{\langle g \rangle_G } \sim G^{\kappa_g-1} \, . \label{eq: relation Nc vs G}
\end{align}
Inverting relations \eqref{eq: relation Nc vs G} and \eqref{eq: relation Nc vs ell} we have 
$G \sim N_c^{1/(\kappa_g-1)}$ and $\ell \sim N_c^{1/(\kappa_c-1)}$. 
This confirms that $\ell \ll G$ within an avanlanche when $\kappa_c > \kappa_g=3/2$.
Then using $D = \ell+G \sim G \sim N_c^{1/(\kappa_g-1)}$ and $P(D)dD = P(N_c)dN_c$ we find
\begin{equation}
\kappa_D = (\kappa_g + 1)/2 = 1.25 \, . \label{eq: conj. kappa_D}
\end{equation}
This value is close from the exponents I measured for various $\alpha$ and which are around $\kappa_D = 1.2$.

\paragraph{b) $G > G_{\text{cross}}$}
In this case we assume that we have a gap in the $g^{-(d+\alpha)}$ regime.
Since this is a rare event we can assume there is at most one (or a very few) such gap in an avalanche. This gap dominates the avalanche diameter which then follow the same power law. 
Note that this holds only when there is a crossover in the gap distribution i.e. for $\alpha=1$ and $\alpha=1.5$. For $\alpha=0.5$ we have $d+\alpha = 3/2$ so there is no crossover in $P(g)$ and therefore we do not expect any crossover in $P(G)$ and $P(D)$. 

\vspace{0.7cm}
To conclude this section let us note that the gaps can possibly be identified with the \textit{pinning clusters} defined and studied by Tallakstad et al. in the context of the Olso experiment~\cite{tallakstad2011}. Pinning clusters are defined as regions where the local velocity $v(x,t)$ is smaller than
$v/C$ where $v$ is the mean velocity of the front and $C$ an arbitrary theshold that ranges from $2$ to $12$ in their study. 
The pinning clusters are extended in the $x$ direction along the front and are very narrow in the direction of propagation. Therefore they can be considered as lines and the variation of their sizes can be identified with the variation of their extension in the $x$ directions. 
Tallakstad et al. found that the size distribution of the pinning clusters is power law distributed with an exponent $1.56\pm0.04$ which is compatible with the exponent $\kappa_g=3/2$ that is measured in figure~\ref{fig: P(g)}.

\section{Tables of exponents}\label{sec: table of exponents}

The final results of this chapter are summarized in table~\ref{tab: measurement vs predictions}. 
All scaling predictions can be expressed in terms on only three parameters~: the internal dimension $d$ of the interface, the elasticity range parameter $\alpha$ and the roughness exponent $\zeta$.
The predictions, computed for $d=1$ with the values of $\zeta$ listed in the first line are compared with the measurements obtained by fitting the data and performing data collapse. 
The predictions and measurements are in good agreement overall. 
The uncertainties over the measurements are given in a second table~\ref{tab: measurement with errors}.
\vspace{1cm}

\begin{adjustwidth}{-2cm}{-2cm}
\begin{center}
	\begin{tabular}{c c c c c c c}
		Exponent & Expression & Relation & $\alpha=0.5$ & $\alpha=0.75$ & $\alpha=1$ & $\alpha=1.5$ \\ 
		 & & & \multicolumn{4}{c}{measured/prediction}  \\		
		\hline
		$\zeta$ & $S(q)\sim q^{-(d+2\zeta)}$ &  & 0.00(MF) & 0.18(1) & 0.39~\cite{rosso2002} & 0.77(2) \\ 
		$\gamma_S$ & $\langle N_c \rangle_S\sim S^{\gamma_S}$ & $\gamma_S=2-\frac{2\alpha}{d+\zeta}$ &
		0.89/1 & 0.73/0.73 & 0.52/0.56 & 0.29/0.30 \\ 
		$\tau$ & $P(S)\sim S^{-\tau}$ & $\tau = 2-\frac{\alpha}{d+\zeta}$ & 1.5/1.5 & 1.36/1.36 &
		1.26/1.28 & 1.14/1.15 \\ 
		$\tau_c$ & $P(S_c)\sim S^{-\tau_c}$ & $\tau_c = 3-\frac{2\alpha}{d+\zeta}$ & 2/2 &
		1.72/1.73 & 1.56/1.56 & 1.28/1.30 \\ 
		$\kappa$ & $P(\ell)\sim S^{-\kappa}$ & $\kappa = 1+d+\zeta-\alpha$ & 1.5/1.5 & 
		1.38/1.43 & 1.33/1.39 & 1.20/1.27 \\ 
		$\kappa_c$ & $P(\ell_c)\sim S^{-\kappa_c}$ & $\kappa_c = 1+2(d+\zeta)-2\alpha$ & 2.05/2 &
		1.9/1.86 & 1.80/1.78 & 1.45/1.54 \\ 
		$\kappa_D$ & $P(D)\sim S^{-\kappa_D}$ & $\kappa_D=\frac{5}{4}$ &
		1.20/1.25 & 1.15/1.25 & 1.20/1.25 & 1.21/1.25 \\ 
		$\gamma_{\ell}$ & $\langle N_c \rangle_{\ell} \sim S^{\gamma_{\ell}}$ & 
		$\gamma_{\ell}=2(d+\zeta - \alpha)$ & 0.93/1 & 0.80/0.86 & 0.70/0.78 & 0.50/0.54 \\
		$\gamma_{S/\ell}$ & $\langle S \rangle_{\ell} \sim \ell^{\gamma_{S/\ell}}$ & 
		$\gamma_{S/\ell}=d+\zeta$ & 1.02/1 & 1.14/1.18 & 1.31/1.39 & 1.71/1.77 \\ \hline		
	\end{tabular}
\end{center}
\end{adjustwidth}
\captionof{table}{Comparison between measurements and predictions. The measured values correspond to best fits of my data. The predictions correspond to the scaling relations indicated in the third column, for $d=1$, using the values of $\zeta$ listed in the first line. I used the mean-field value $\zeta=0$ for $\alpha=0.5$ and a previous more precise numerical determination $\zeta=0.39$ for $\alpha=1$. \label{tab: measurement vs predictions}}



\begin{center}
	\begin{tabular}{c c  c c c c}
		Exponent & Expression & $\alpha=0.5$ & $\alpha=0.75$ & $\alpha=1$ & $\alpha=1.5$ \\ 		
		\hline
		$\zeta$ & $S(q)\sim q^{-(d+2\zeta)}$  & $-0.03(2)$ & $0.18(1)$ & $0.37(2)$ & 
		$0.77(2)$ \\
		$\gamma_S$ & $\langle N_c \rangle_S\sim S^{\gamma_S}$ &
		0.89(1) & 0.73(1) & 0.52(1) & 0.29(1) \\ 
		$\tau$ & $P(S)\sim S^{-\tau}$ & 1.50(1) & 1.36(2) &
		1.26(2) & 1.14(2) \\ 
		$\tau_c$ & $P(S_c)\sim S^{-\tau_c}$ & 2.00(10) &
		1.72(4) & 1.56(2) & 1.28(2) \\ 
		$\kappa$ & $P(\ell)\sim S^{-\kappa}$ & 1.50(2) & 
		1.38(5) & 1.33(4) & 1.20(3) \\ 
		$\kappa_c$ & $P(\ell_c)\sim S^{-\kappa_c}$ &  2.05(5) &
		1.90(5) & 1.80(2) & 1.45(3) \\ 
		$\kappa_D$ & $P(D)\sim S^{-\kappa_D}$ &
		1.20(7) & 1.15(5) & 1.20(10) & 1.21(6) \\ 
		$\mu$ &  $P(N_c) \sim N_c^{-\mu}$ & 1.50(2) & 1.50(3) & 1.50(6) & 1.50(8) \\
		$\kappa_g$ & $P(g)\sim g^{-\kappa_g}$ & 1.50(5) & 1.50(3) & 1.50(5) & 1.50(6) \\ 
		$\gamma_{\ell}$ & $\langle N_c \rangle_{\ell} \sim S^{\gamma_{\ell}}$ &
		0.93(2) & 0.80(2) & 0.70(2) & 0.50(2) \\
		$\gamma_{S/\ell}$ & $\langle S \rangle_{\ell} \sim \ell^{\gamma_{S/\ell}}$ &
		1.02(3) & 1.14(3) & 1.31(2) & 1.71(2) \\ \hline		
	\end{tabular}
	\captionof{table}{Summary of measurements with error margins. All measurements have been made by fitting power law and confirmed by data collapse.
	\label{tab: measurement with errors}}
\end{center}

\section{Conclusion}\label{sec: conclustion ch4}

I studied the statistics of clusters in long-range avalanches with extensive numerical simulations.
The number of clusters seems to obey a BGW law, independently from $\alpha$.
This law is characterized by an exponent $\mu=3/2$ which is a key element allowing to establish the scaling relations $\tau_c=2\tau-1$~\eqref{eq: relation tau_c tau} and
$\kappa_c = 2\kappa-1$~\eqref{eq: relation kappa_c kappa} that link the cluster exponents to the ones of the global avalanches. 
This confirms on firmer ground the claim made by Laurson et al. in the abstract of~\cite{laurson2010}.
A consequence of these relations is that all the exponents characterizing the avalanches can be expressed as functions of only three parameters~:
the internal dimension of the interface $d$, the elasticity range parameter $\alpha$ and the roughness exponent $\zeta$.
We have retrieved the self-affinity relation $S\sim \ell^{d+\zeta}$ which was known for SR elasticity avalanches. This required to replace the avalanche extension by an adequate quantity, namely the linear size. 

The prediction $\mu =3/2$ deserves experimental confirmations. In Ref.~\cite{planet2018}, a new method was introduced to define simultaneously local clusters and global avalanches. With this method a cluster can be unambiguously attributed to one avalanche. 
This opens the possibility to measure experimentally the distribution  $P(N_c)$ and to test the numerical observation $P(N_c)\sim N_c^{-3/2}$ made in this chapter.

A new result was also obtained for the exponent of the avalanche diameter distribution.
For this purpose we introduced a new object, 
namely the gaps between clusters. 
Remarkably their distribution also shows a power law with an $\alpha$ independent short-scale exponent 
$\kappa_g=3/2$. This exponent is a key to obtain the short-scale exponent of the diameter distribution which is also $\alpha$ independent. 
At large scale the diameter and gap distributions crossover to a faster power law with an exponent 
$d+\alpha$ which is reminiscent of the long-range kernel. This is similar to the result obtained for the spatial velocity correlation function $C_v(x)$ in the previous chapter. 

It is very intriguing and fascinating that the number of clusters and gap distributions are characterized by mean-field exponents $3/2$ that do not depend on $\alpha$. 
At present there is no obvious explanation for that fact,
and it remains as a challenging question for future work.



%% file: Chapter_5/Chapter_5.tex


\chapter{Mean-Field models}\label{chap: chap5}


In this chapter I discuss several mean-field models relevant for the depinning transition.
I start with a discussion about the concept of mean-field in the first section. 
In the second section I introduce the fully-connected (FC) model for an elastic interface driven in a disordered medium and
show how it can be mapped onto the Alessandro-Beatrice-Bertotti-Montorsi (ABBM) model
of a single particle driven in a Brownian force landscape. This model was introduced in the context of the Barkhausen noise~\cite{alessandro1990, alessandro1990a} and was already mentioned in the corresponding section of chapter~\ref{chap: experiments} in equation~\eqref{eq: ABBM model}.
In the third section I present a spatial extension of the ABBM~: the Brownian force model (BFM). 
This model is more recent and appeared in FRG calculations to be the mean-field model for the local velocity field, taking into account the range of the interactions.
I explain how several statistical distributions have been obtained within this model. 
For global quantities such as the total avalanche size, duration or temporal shape, the distribution do not depend on the range of the interactions. 
Previous works on the BFM focused on an interface with short-range (SR) elasticity for which the distributions of spatial quantities, such as the jump of a single point, the spatial extension of avalanches and their spatial shape, have been obtained. 
I worked on the BFM incorporating long-range (LR) elasticity. Obtaining exact results for spatial quantities is more difficult than for SR elasticity. I present some results that I obtained using perturbative expansions. 

\section{What is mean-field ?}

Many models go under the name of \emph{mean-field}. Two historical examples of mean-field models are the Van-der-Waals equation of state for a gas and the Curie-Weiss model of a magnet. 
In these models a particle interacts equally with all other particles in the system. 
They are said to be fully-connected. We can also say that the range of the interaction is infinite.
The notion of neighbours and, consequently, of space is irrelevant.
The sum of the interactions of one particle with the rest of the system reduces to a single interaction with an "average particle"\footnote{If one uses a field-theoretical approach, the state of the particles is represented by some field and a particle interacts with the mean field of the system, hence the name of the models.} and
the number of degree of freedom is reduced from the particle number $N$ to a single effective one.

Mean-field models have proven to be very useful to understand phase transitions. 
But, the range of interactions being finite in real systems, they often yield predictions about the critical exponents that disagree with experiments.
Nevertheless mean-field models may yield the correct exponents if the \emph{connectivity} of a particle is high enough. 
This high connectivity can be realized either because the range of the interactions is very long or because particles have many neighbours. 
The second case is realized for systems of high dimensions. 
Formally when the dimension is $d=+\infty$ a particle interacts with infinitely many neighbours and one can expect their average to be the same as the one of the whole system.
The infinite dimension might seem far from real systems but it turns out that even in finite dimension mean-field predictions can remain true~:
there is often a finite \emph{upper critical dimension} $d_{uc}$ above which mean-field theory holds and this dimension may be not so high. 
In particular for elastic interfaces the Larkin argument presented in section \ref{sec: Ch1 Larkin length and d_uc} shows that the upper critical dimension is $d_{uc}=2\alpha$. 
For the short-range elasticity this corresponds to $d_{uc}=4$. 
We see that an increasing range of interactions (i.e. decreasing $\alpha$) corresponds to a decreasing upper critical dimension. The validity if the mean-field approximation is facilitated by the range of the interactions.  
In particular for physical cases where $\alpha=1$ the mean-field predictions are correct for two-dimensional interfaces, such as domain walls in three-dimensional ferromagnets (see section~\ref{sec: Ch2 Barkhausen noise}). This is thus a relevant theory to study.

\section{Fully-connected and ABBM models}\label{sec: FC and ABBM ch5}

In this section I introduce the fully-connected (FC) model of elastic interfaces and show how it can be mapped onto the ABBM model. This mapping was first derived in Ref.~\cite{zapperi1998}.

\subsection{Mapping of the Fully-connected onto the ABBM model}
In the FC model each point of the interface interacts equally with all the others. 
In order for the interaction energy to be extensive in the system size, the strength of the interaction must be inversely proportional to the system size.
For a one-dimensional system, the elastic interaction term defined in equations \eqref{eq: F el LR generalized} and \eqref{eq: Hamiltonian el LR} now reads 
$\mathcal{C}(x-y) = \frac{c}{L}$ with $L$ the system size. 
The elastic force felt at position $x$ is~:
\begin{equation}
F_{\el}^{\LR} (x,t) =  \int_0^L \frac{c}{L}\left( u(y,t)-u(x,t)\right) dy = c\left(u(t)- u(x,t)\right)
\end{equation}
where $u(t):=\langle u\rangle (t)=\frac{1}{L}\int_0^{L} u(y,t) dy$ is the average position of the interface at time $t$.
The motion of the interface is thus governed by~:
\begin{equation}
\eta \partial_t u(x,t) = m^2\left( w(t)-u(x,t)\right) + c\left(u(t)- u(x,t)\right) + F_{\dis} \left( x, u(x,t) \right) \label{eq: Motion FC continuous}
\end{equation}
where we use the external force \eqref{eq: F ext} discussed in chapter~\ref{chap: DES and depinning}. 

For the sake of simplicity, in the following, we will use a discretized version of \eqref{eq: Motion FC continuous}, assuming a discretization step $a=1$ along the line :
\begin{equation}
\eta \partial_t u(i,t) = m^2\left( w(t)-u(i,t)\right) + c\left(u (t)- u(i,t)\right) + F_{\dis} \left( i, u(i,t) \right) \label{eq: Motion FC discretized}
\end{equation}

To obtain an equivalence with the ABBM model we must take the average of equation \eqref{eq: Motion FC discretized} in order to obtain an equation for the center-of-mass of the interface $u(t)$~:
\begin{align}
\eta \partial_t u(t) &= m^2\left( w(t)-u(t)\right) + W \left(u(t) \right) \, ,\label{eq: ABBM} \\
W \left(u(t) \right) &=  \frac{1}{L} \sum_{i=1}^{L} F_{\dis} \left( i, u(i,t) \right) \, . \label{eq: def pinning force ABBM}
\end{align}
Note that since there exists a unique asymptotic solution for a given monotonous driving (third Middleton's theorem) there is a unique configuration $\left(u(i) \right)_i$ associated to a given average position $u$. Thus the effective pinning force $W \left(u(t) \right) $ is defined without ambiguity by equation~\eqref{eq: def pinning force ABBM}.
Note also that the internal interactions have vanished when taking the average. 
Equation \eqref{eq: ABBM} is equivalent to the ABBM model under the condition that the effective pinning force $W(u)$ has brownian correlations.
When the interface moves between two stable configurations, of average position $u$ and $u'$, the increment in $W(u)$ is~:
\begin{equation}
W(u')-W(u) = \frac{1}{L} \sum_{i=1}^{n} \Delta F_{\dis}^{i}
\end{equation}
where the sum is restricted to the $n$ sites that have moved during the avalanche.
Since the points of the interface interact equally alltogether, the interface has no spatial structure. In particular the roughness exponent is $\zeta=0$. 
In this case the number of sites which move during an avalanche is proportional to the avalanche size : $n\sim S$ with $S=L|u'-u|$. 
For each site the increment in the disorder force, $\Delta F_{\dis}^{i}$, is a random variable with mean $0$ and variance\footnote{One usually uses $\sigma^2$ for the variance but I used $\sigma$ for the sake of coherence with the notations used in section~\ref{Sec: definition BFM} and after.}
$\tilde{\sigma}$. 
All increments are independent so the sum of the increments follows a normal law~:
\begin{equation}
W(u')-W(u) = \frac{1}{L} \sum_{i=1}^{n} \Delta F_{\dis}^{i}\underrel{\text{law}} {\sim} \frac{1}{L} \mathcal{N}\left(0, n \tilde{\sigma}\right) = \mathcal{N}\left(0, \frac{n \tilde{\sigma}}{L^2}\right) \, ,
\end{equation}
where the symbol $\underrel{\text{law}} {\sim}$ means that the random variable on the left follows the probability law on the right.
Since $n\sim S$ the variance of the normal law is proportional to $S/L^2=|u'-u|/L$. Absorbing all proportionality constants into a new variance $2\sigma$ we have~:
\begin{equation}\label{eq: Brownian correlations ABBM}
W(u')-W(u) \underrel{\text{law}} {\sim} \mathcal{N}\left(0, 2 \sigma |u'-u|\right)\, ,
\end{equation}
which is indeed the probability law for the increments of a Brownian motion.

\subsection{Statistics of the ABBM model}\label{sec: statistics of the ABBM ch5}

The ABBM model corresponds to \eqref{eq: ABBM} where the drive is increased at the constant rate
$w(t)=vt$~:
\begin{align}\label{eq: ABBM 2}
&\eta \partial_t u(t) = m^2\left( w(t) - u(t)\right) + W \left(u(t) \right) \ .
\end{align}
The correlations of the Brownian motion $W$ are given by~\eqref{eq: Brownian correlations ABBM}.
Avalanches starts when the left-hand side (LHS) of \eqref{eq: ABBM 2} is equal to $0^+$ and stops when it returns to $0$.
Therefore the statistics of avalanches in the ABBM corresponds to the statistics of the first return to the origin of a Brownian motion with a drift or, equivalently, to the first-crossing of a Brownian motion with a line. This problem has been studied by Sinaï~\cite{sinai1992} in the quasi-static limit 
$v=0^+$.
For a drive at finite velocity $v>0$, scaling exponents have been obtained by Bertotti et al.~\cite{bertotti1994} and a detailed solution was given in~\cite{ledoussal2009c}.

\subsubsection{Quasi-static drive}

\begin{figure}
	\centering
	\includegraphics[scale=0.3]{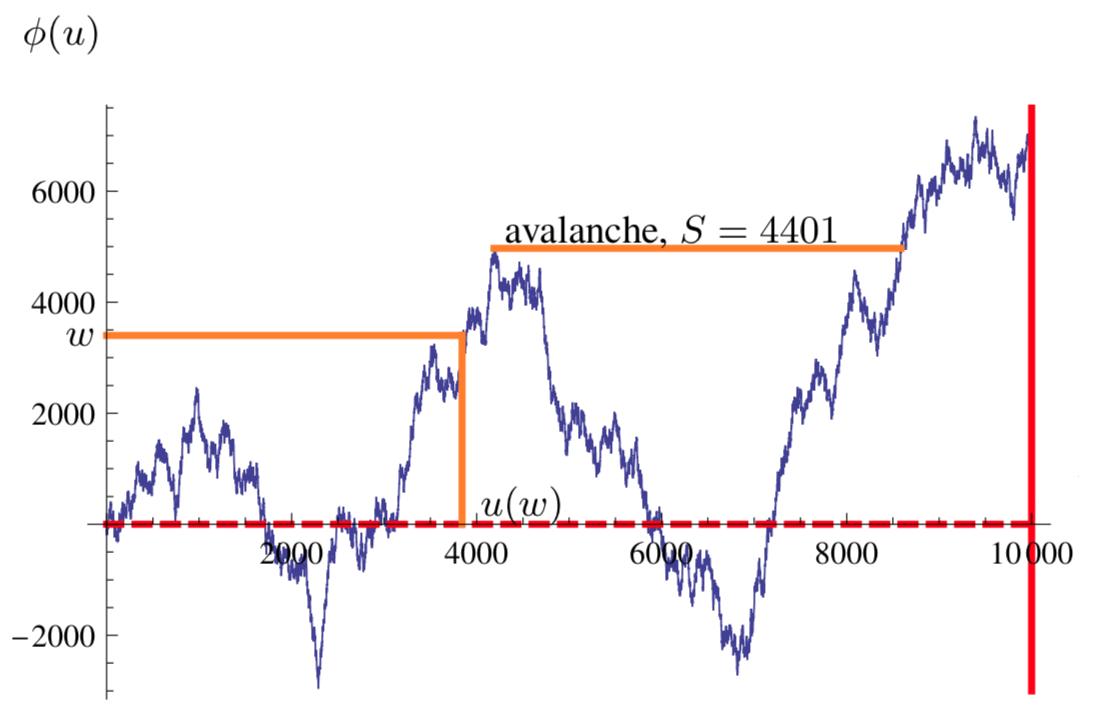}
	\caption{Illustration of the process $\phi(u)$. $u(w)$ is the first time that the process $\phi$ reaches the altitude $w$. An avalanche occurs when there is a discontinuity in $u(w)$. Image from~\cite{ledoussal2009c}. \label{fig: Maxwell construction}} 
\end{figure}

We define the process $\phi(u) = u-m^{-2}W(u)$. Equation~\eqref{eq: ABBM 2} can be rewritten as~
\begin{equation}
m^{-2}\eta \partial_t u(t) = w(t) - \phi\left(u(t)\right) \, . 
\end{equation}
The position of the particle
$u=u(w)$ corresponds to the first "time" that the process $\phi(u)$ reaches altitude $w$ (with $u$ being the "time" of the process $\phi(u)$).
This process is illustrated in figure~\ref{fig: Maxwell construction}.
An avalanche corresponds to a discontinuity in the process with a jump forward $u(w^+) > u(w)$.
The size of the avalanche is $S=u(w^+) - u(w)$.
During an avalanche triggered at $w$, the process makes an excursion below $\phi\left(u(w)\right)$ 
(i.e. $\phi\left(u \right) < \phi\left(u(w)\right)$). The avalanche stops when the process returns to 
$\phi\left(u(w)\right)$.
It corresponds to the first crossing of the Brownian motion $W$ with a line of slope $m^2$.
The probability of the first return to the origin of a Brownian motion $W(S)$\footnote{I now use $S$ because it corresponds to the avalanche size.} decays as $S^{-3/2}$.
The probability of the first crossing with a line is similar but is exponentially cut above a scale $S_m$ which can be obtained by scaling arguments~:
when the avalanche starts the Brownian $W(S)$ grows as $(\sigma S)^{1/2}$. 
The probability of the Brownian staying above the line is overwhelmingly suppressed when the typical fluctuations of the Brownian becomes smaller than  the linear term, i.e. when $(\sigma S)^{1/2} \sim m^2 S$. This corresponds to a cutoff scale of the order
\begin{equation}\label{eq: scaling Sm ch5}
S_m = \sigma /m^{4} \, .
\end{equation}
To define the avalanche probability density we need to introduce a small scale cutoff $S_0$. 
In the limit of small $m$ we have $S_0 \ll S_m$ and the probability in the limit 
$S \gg S_0$ is given by~\cite{ledoussal2009c}~:
\begin{equation}
P(S) =  \frac{\langle S \rangle}{2\sqrt{\pi S_m}} S^{-3/2} e^{-S/(4S_m)}
\end{equation}
where $\langle S \rangle = \sqrt{S_0 S_m}$.

\subsubsection{Finite velocity drive}

In the case where the driving velocity is finite $v>0$, the equation~\eqref{eq: ABBM 2} is a Langevin equation and one can derive from it a Fokker-Planck equation for the probability of observing a velocity $V=\partial_t u$ at position $u$ given an initial velocity $V_0$ at position $u_0$.
From this equation one can then deduce the stationary probability distribution~\cite{alessandro1990, bertotti1994, ledoussal2009c}~:
\begin{equation}\label{eq: stat Prob position ABBM}
P_{u}(V) = \frac{1}{v \Gamma(\frac{v}{v_m})} \left(\frac{V}{v_m}\right)^{\frac{v}{v_m}} e^{-\frac{V}{v_m}} \quad \text{ with}
\quad v_m = \frac{\sigma}{\eta m^2} \, .
\end{equation}
Similarly a Fokker-Planck equation for the probability of observing a velocity $V$ at a time $t$ given an initial velocity $V_0$ at time $t_0$ can be derived. Its stationary solution is~\cite{alessandro1990, bertotti1994, ledoussal2009c}~:
\begin{equation}\label{eq: stat Prob time ABBM}
P_{t}(V) = \frac{1}{v_m \Gamma(\frac{v}{v_m})} \left(\frac{V}{v_m}\right)^{\frac{v}{v_m}-1} e^{-\frac{V}{v_m}} \, .
\end{equation}

\paragraph{Transition at $v=v_m$ and scaling exponents}
The behavior of the distribution $P_{t}(V)$ near $V=0$ presents a transition at $v= v_m = \frac{\sigma}{\eta m^2}$ between an intermittent motion and a continuous motion.
On the one hand when $ v/v_m  > 1$, $P_{t}(V)$ vanishes when $V\to 0$  and one can show that after a finite time there is a single infinite avalanche 
and the velocity never vanishes again.  
On the other hand when  $ v/v_m < 1$, $P_{t}(V)$ diverges when $V\to 0$. 
One can show that indeed the instantaneous total velocity becomes zero infinitely often and that there are well-separated avalanches.
In this case the avalanche size and duration corresponds to the first return at the origin of the process defined by equations~\eqref{eq: stat Prob position ABBM} and \eqref{eq: stat Prob time ABBM} respectively. Scaling arguments~\cite{bertotti1994} or a complete solution~\cite{ledoussal2009c} gives the avalanche size and duration exponent~:
\begin{align}
\tau = \frac{3}{2} - \frac{v}{2 v_m} \quad , \quad \ta = 2 - \frac{v}{v_m} \quad  \text{with} \quad \quad v_m = \frac{\sigma}{\eta m^2}  \label{eq: ABBM exponents ch5} \, .
\end{align}
These exponents were already given in section~\ref{sec: Ch2 Barkhausen noise} about the Barkhausen effect where $m^2v=c$ was the magnetization rate and $\eta$ was set equal to 1.

\section{Introduction of the Brownian force model}

I now present a generalization of the ABBM model, the Brownian force model (BFM), which describes a multidimensional elastic interface driven in a Brownian random force landscape. This model introduced in \cite{ledoussal2012, ledoussal2012b, dobrinevski2012, ledoussal2013} provides the correct mean-field theory describing the full space-time statistics of the velocity in a single avalanche for $d$-dimensional realistic interfaces close to the depinning transition. Restricted to the center of mass it reproduces the ABBM model. 

Until now statistics of the spatial structure of avalanches have been calculated in the case of the short-range elasticity \cite{dobrinevski2012, ledoussal2013, delorme2016, thiery2015, thiery2016a}. Yet domain walls, fluid contact lines or crack fronts are described by long-ranged elasticity (see chapter~\ref{chap: experiments}). 
In this section I present first results that have been obtained for spatial observables of the BFM with long-range elasticity.  

\subsection{Definition of the model}\label{Sec: definition BFM}

The starting point is the equation of motion for a $d$-dimensional interface introduced in chapter~\ref{chap: DES and depinning}~:
\begin{equation}
\eta \partial_t u(x,t) = m^2\left( w(x,t)-u(x,t)\right) + \int \mathcal{C}(y-x)\left(u(y,t)- u(x,t)\right) d^dy + F_{\dis} \left( x, u(x,t) \right) \label{eq: motion general}
\end{equation}
Drawing on the ABBM model, we assume the random force field to be a collection of independant one-dimensional Brownian motion of amplitude $\sigma$~:
\begin{align}
\overline{\left[ F_{\dis}(x,u) - F_{\dis}(x',u') \right]^2} &= 2\sigma \delta^d(x-x') |u-u'| \label{eq: force correlator 0} \\
\overline{\partial_u F_{\dis}(x,u) \partial_{u'} F_{\dis}(x',u')}
	&= 2\sigma \delta^d(x-x') \delta(u-u') \, .\label{eq: force correlator}
\end{align}
Note that for the disorder correlator $\Delta$ defined in section~\ref{sec: Ch1 disorder force}, 
equation~\eqref{eq: force correlator} implies~:
\begin{equation}\label{eq: force correlator detail}
\partial_u \partial_{u'} \Delta(u-u')  = -\sigma \partial_{u'} \text{sign} \left( u-u' \right)
  = 2\sigma \delta(u-u')
\end{equation}
Let us compute the time derivative of the disorder correlations.
Writing $\du (x,t) = \partial_t u(x,t)$ we have~:
\begin{align}\label{eq: force correlator time derivative intermediate}
\overline{\partial_t F_{\dis}\left(x,u(x,t)\right) \partial_{t'} F_{\dis}\left(x',u(x',t')\right)} &=
\delta^d(x-x') \partial_t \partial_{t'} \Delta\left(u(x,t)-u(x',t')\right) \notag \\
&= -\sigma \delta^d(x-x') \du (x,t) \partial_{t'} \text{sign} \left(u(x,t)-u(x',t')\right) \notag  \\
& = 2 \sigma \du (x,t) \delta^d(x-x') \delta(t-t') \, .
\end{align}
When deriving the last equality we assumed a strictly monotonous motion so that \newline
$\text{sign} \left(u(x,t)-u(x,t')\right) = \text{sign} \left(t-t'\right)$.
We can rewrite this equation using a Gaussian white noise $\xi$ of variance $2\sigma$~:
\begin{align}
\overline{\partial_t F_{\dis}\left(x,u(x,t)\right) \partial_{t'} F_{\dis}\left(x',u(x',t')\right)} &=
\overline{\sqrt{\du(x,t)}\xi(x,t)\sqrt{\du(x',t')}\xi(x',t')} \, , \label{eq: force correlator time derivative} \\
\overline{\xi(x,t)\xi(x',t')} & = 2 \sigma \delta^d(x-x') \delta(t-t') \, . \label{eq: white noise def}
\end{align}
Equation \eqref{eq: force correlator time derivative} shows that the time derivative of the disorder is a white noise multiplied by the square root of the local velocity.
Using this result we can now take the time derivative of equation \eqref{eq: motion general}. 
We obtain a close stochastic differential equation for the velocity field~:
\begin{equation}\label{eq: definition BFM}
\eta \partial_t \du(x,t) = m^2\left(\dot{w}(x,t)-\du(x,t)\right) + \int \mathcal{C}(y-x)\left(\du(y,t)- \du(x,t)\right) d^dy + \sqrt{\du (x,t)} \xi (x,t) \, .
\end{equation}
This equation defines the Brownian force model. It is valid for any monotonous driving. As such it allows for instance to study the response of the interface to a constant velocity driving $\dot{w}(x,t)=v>0$.

\paragraph{Monotonocity condition}
Note that the derivation of equation \eqref{eq: definition BFM} holds under the requirement that the motion of the interface is strictly monotonous. 
This is ensured by Middleton's second theorem under the assumptions that~: (i) the drive is monotonous $\dot{w}(x,t) \geq 0$, $\forall x, \, t$ ;
(ii) the initial velocity at some time $t_0$ is positive everywhere $\du(x,t_0) \geq 0$ $\forall x$ and (iii) there exists at least one point $x_0$ which has a strictly positive initial velocity $\du(x_0, t_0) >0$   (see section~\ref{sec: Ch1 Middleton thm}). 

\paragraph{Loose definition of avalanches}
We define avalanches as the response of the interface to any finite total drive~: $ w^{\tot} = \int \dot{w}(x,t) d^d xdt < +\infty$.
Under such a drive the interface, starting from a stable configuration will eventually reach a new stable configuration.
Middleton's third theorem ensures that the final configuration reached by the line is independant from how the driving is performed. Hence if we are interested in integral quantities depending only on the difference between the final and the inital state (e.g. the avalanche size) we  can assume the driving to be a kick at $t=0$ : $\dot{w}(x,t)=w(x) \delta(t)$.
We will mainly use two type of kicks~: a homogeneous kick $w(x)=w$ and a localized kick $w(x)=w\delta(x)$.

\subsection{Generating functional and instanton equation}\label{subsec: Generating functional and instanton}

\paragraph{Generating functional of a random variable}
Let $Y$ be a random variable of probability distribution $P(y)$. All the information about the statistics of $Y$ are encoded in the generating functional~:
\begin{equation}\label{def: generating functional general}
G_Y(\lambda) := \int e^{\lambda y} P(y) dy \, . 
\end{equation}
Note that since $P(y)$ is a probability distribution the integral converges for $\lambda = 0$.
We will only consider non-positive values of $\lambda$, for which the integral always converges~: $\lambda \leq 0$.
The generating functional $G_Y$ encodes all the informations about the statistics of the random variable $Y$. In particular, since $G_Y$ is the Laplace transform  of the probability distribution $P(y)$, this latter can be recovered by taking the inverse Laplace transform from $G_Y$~:
\begin{equation}\label{eq: P is inverse Laplace transform from G_Y}
P(y) = \mathcal{L}^{-1} \left( G_Y \right)
\end{equation}
where $\mathcal{L}$ denotes the Laplace transform operator and $\mathcal{L}^{-1}$ is the inverse Laplace transform operator.

\subsubsection{Generating functional of the velocity field}

The notion of generating functional can be extended to the stochastic field $\dot{u}(x,t)$ as follows~:
\begin{equation}\label{eq: Generating functional}
G[\lambda(x,t), \dot{w}(x,t)] := \overline{\exp \left( \int\lambda(x,t)\dot{u}(x,t) d^d x dt \right)}^{\dot{w}(x,t)}
\end{equation}
where $\overline{\mbox{...\rule{0pt}{2.5mm}}}^{\dot{w}(x,t)}$ denotes the disorder average for a given driving $\dot{w}(x,t)$.
The term $\lambda(x,t)$, called the \emph{source}, is arbitrary, 
we are free to make any choice for the source. 
All the informations about the distribution of the interface local velocity field $\dot{u}(x,t)$ 
for a given driving $\dot{w}(x,t)$ are encoded in $G[\lambda(x,t), \dot{w}(x,t)]$.
In particular, when choosing a source $\lambda(x,t) = \lambda \delta^{d}(x-x_0) \delta (t-t_0)$, 
equation \eqref{eq: Generating functional} becomes the generating functional of the random variable 
$\du (x_0, t_0)$. 
This can be extended to any joint probability distribution of the velocity field.
The statistics of different observables can be accessed by choosing an appropriate source $\lambda(x,t)$.
I give a few examples below, before diving into the details of the computation of $G$.
We will consider mainly two cases.
The first one is a finite total increment in the drive~:
\begin{equation}\label{eq: def w^tot}
w^{\tot} = \int \dot{w}(x,t) d^d x dt < +\infty
\end{equation}
The second one is a drive at constant velocity~: $\dot{w}(x,t)=\dot{w}(x)$ is independent of time.
In this case we can define the total driving velocity~:
\begin{equation}\label{eq: def v^tot}
v^{\tot} = \int \dot{w}(x) d^d x
\end{equation}

\paragraph{Total velocity}
The total velocity of the interface $V(t_0)$ at time $t_0$ is defined as $V(t_0) = \int \du(x,t_0) d^d x$. It corresponds to the choice of the source $\lambda(x,t) = \lambda \delta (t-t_0)$.
Under this choice \eqref{eq: Generating functional} reads~:
\begin{equation}\label{eq: generating functional global velocity}
G[\lambda \delta (t-t_0), \dot{w}(x,t)] = \overline{\exp \left( \lambda V(t_0) \right)}^{\dot{w}(x,t)}
\end{equation}
which is the generating functional for the random variable $V(t_0)$. 

\paragraph{Avalanche size}
For a finite drive, $ w^{\tot} < +\infty$, 
the avalanche size is defined as $S = \int \du(x,t) d^d x dt$. It corresponds to a constant and homogeneous source $\lambda(x,t) = \lambda$. 
Under this choice \eqref{eq: Generating functional} becomes~:
\begin{equation}\label{eq: generating functional avalanche size}
G[\lambda, \dot{w}(x,t)] = \overline{\exp \left( \lambda S \right)}^{\dot{w}(x,t)}
\end{equation}

\paragraph{Local avalanche size}
The local avalanche size is defined as the jump forward of a single point $x_0$ during the avalanche. It is formally defined as $S(x_0) = \int \du(x_0,t) dt$. It corresponds to a choice of the source 
$\lambda(x,t) = \lambda^d \delta(x-x_0) $. 
Under this choice \eqref{eq: Generating functional} becomes~:
\begin{equation}\label{eq: generating functional local avalanche size}
G[\lambda^d \delta(x-x_0), \dot{w}(x,t)] = \overline{\exp \left( \lambda S(x_0) \right)}^{\dot{w}(x,t)}
\end{equation}

\subsubsection{Computation of the generating functional}

Using Martin-Siggia-Rose formalism the generating functional can be rewritten as~\cite{dobrinevski2012, ledoussal2013}
\begin{equation}\label{eq: Generating functional from instanton}
G[\lambda(x,t), \dot{w}(x,t)] = \exp \left (\int m^2\dot{w}(x,t) \tu^{[\lambda]}(x,t)  d^d xdt\right)
\end{equation}
where $\tu^{[\lambda]}(x,t)$ is the solution of the following ordinary non-linear differential equation, called \emph{instanton} equation~:
\begin{equation}
\eta \partial_t \tu(x,t) + \int \mathcal{C}(y-x) \left(\tu(y,t) - \tu(x,t) \right) d^d y - m^2\tu(x,t) 
+ \sigma \left(\tu(x,t)\right)^2 = -\lambda(x,t) \, . \label{eq: Instanton real space} 
\end{equation}
The derivation of this result is sketched in appendix~\ref{sec:Appendix MSR}.
With this result we see that finding the statistical distribution of any observable can be reduced to the problem of solving the ordinary equation~\eqref{eq: Instanton real space} with the corresponding appropriate source $\lambda(x,t)$.

\subsection{Dimensionless equations}

\subsubsection{Units}

In the following of this chapter it will be convenient to work in dimensionless units, i.e. to have 
$\eta=m^2=\sigma=1$ in equations \eqref{eq: definition BFM} and \eqref{eq: Instanton real space}.
This can be achieved upon proper rescaling of all quantities. 
We need to specify the form of the elasticity kernel which we take to be 
\begin{equation}
\mathcal{C}(y-x) = c/|y-x|^{d+\alpha} \, . \label{eq: specification elastic kernel}
\end{equation}
We also want to set the elasticity constant $c=1$.
In order to do this we define a transversal length scale $\ell_m$, a time scale $\tau_m$ and an avalanche scale $S_m$~:
\begin{equation}\label{eq: characteristic scales Ch5}
\ell_m = \left(\frac{c}{m^2} \right)^{\frac{1}{\alpha}} \, , \quad 
\tau_m = \eta m^{-2} \, , \quad 
S_m = \frac{\sigma}{m^4} \, ,
\end{equation}
and we set~:
\begin{gather}
x = \ell_m x' \, , \quad t = \tau_m t'  \, , \quad  
\du(x,t) = \frac{S_m}{\tau_m \ell_m^d} \du'(x',t') \, , \quad  \tu(x,t) = \frac{1}{m^{2} S_m} \tu'(x',t') \, , \label{eq: dimensionless units 1/2}\\
\dot{w}(x,t) = \frac{S_m}{\tau_m \ell_m^d} \dot{w}'(x',t') \, , \quad 
\xi(x,t) = m^2\sqrt{\frac{S_m}{\tau_m \ell_m^d}} \xi'(x',t')\, , \quad
\lambda(x,t) = S_m^{-1} \lambda'(x',t') \, . \label{eq: dimensionless units 2/2}
\end{gather}

\subsubsection{Equations}

Rewriting equation~\eqref{eq: definition BFM} in function of the primed quantities, all terms become proportional to $m^2 S_m /(\tau_m \ell_m^d)$. This term can be cancelled and the dimensionless equation of motion read~:
\begin{gather}
\partial_{t'} \du'(x',t') = \left(\dot{w}'(x',t') - \du'(x',t')\right) + \int \frac{\du'(y',t')- \du'(x',t')}{|y'-x'|^{d+\alpha}} d^dy' + \sqrt{\du'(x',t')} \xi'(x',t') \, , \label{eq: BFM dimensionless} \\
\overline{\xi'(x'_1,t'_1)\xi'(x_2't_2')} = 2 \delta^d(x'_1-x'_2) \delta(t'_1-t'_2) \, . \label{eq: white noise dimensionless}
\end{gather}

Similarly, rewriting equation~\eqref{eq: Instanton real space}  in function of the primed quantities, all terms become proportional to $S_m^{-1}$ which can be cancelled.
The dimensionless instanton equation is~:
\begin{equation}
\partial_{t'} \tu'(x',t') + \int \frac{\tu'(y',t') - \tu'(x',t')}{|y'-x'|^{d+\alpha}} d^d y' - \tu'(x',t') + 
 \left(\tu'(x',t')\right)^2 = -\lambda'(x',t') \, . \label{eq: Instanton RS dimensionless}
\end{equation}

\subsection{Massless equations}
Alternatively to the external force 
$m^2\left( w(x,t)-u(x,t)\right)$, one can also use a driving force $f(x,t)$ that does not depend on the position of the interface.
In this case the stochastic equation for the velocity field reads~:
\begin{equation}\label{eq: definition BFM massless}
\eta \partial_t \du(x,t) = \dot{f}(x,t) + \int \mathcal{C}(y-x)\left(\du(y,t)- \du(x,t)\right) d^dy + \sqrt{\du (x,t)} \xi (x,t) \, 
\end{equation}
and the generating functional now depends on the ramp rate of the force $\dot{f}(x,t)$~:
\begin{equation}\label{eq:Generating functional massless}
G[\lambda(x,t), \dot{f}(x,t)] := \overline{\exp \left( \int \lambda(x,t)\dot{u}(x,t) d^d  xdt \right)}^{\dot{f}(x,t)} = \exp \left (\int \dot{f}(x,t) \tu^{[\lambda]}(x,t) d^d x dt\right) \, ,
\end{equation}
where $\tu^{[\lambda]}(x,t)$ is now solution of a \emph{massless} instanton equation~:
\begin{equation}
\eta \partial_t \tu(x,t) + \int \mathcal{C}(y-x) \left(\tu(y,t) - \tu(x,t) \right) d^d y
+ \sigma \left(\tu(x,t)\right)^2 = -\lambda(x,t) \, . \label{eq: Instanton real space massless} 
\end{equation}

Equations \eqref{eq: definition BFM massless} and \eqref{eq: Instanton real space massless} can also be
written in dimensionless units. 
The characteristic scales that we need to introduce for this purpose are similar to the ones in \eqref{eq: characteristic scales Ch5} except that they do not depend on the mass
(we keep the same notation for convenience but these scales do not contain any mass and do not result in a large scale cutoff)~:
\begin{equation}\label{eq: characteristic scales dimensionless Ch5}
\ell_m = c^{\frac{1}{\alpha}} \, , \quad 
\tau_m = \eta \, , \quad 
S_m = \sigma \, ,
\end{equation}
Then, besides introducing the rescaling of $\dot{f}(x,t)$, we can use the same rescalings as in equations~\eqref{eq: dimensionless units 1/2} and \eqref{eq: dimensionless units 2/2} except for $\tu$ and $\xi$~:
\begin{equation}\label{eq: dimensionless units massless}
\dot{f}(x,t) = \sqrt{\frac{S_m}{\tau_m \ell_m^d}} \dot{f}'(x',t') \, , \quad  \tu(x,t) = \frac{1}{ S_m} \tu'(x',t')   \, , \quad
\xi(x,t) = \sqrt{\frac{S_m}{\tau_m \ell_m^d}} \xi'(x',t') \, .
\end{equation}
With this rescaling we obtain the dimensionless massless instanton equation~:
\begin{equation}
\partial_{t'} \tu'(x',t') + \int \frac{\tu'(y',t') - \tu'(x',t')}{|y'-x'|^{d+\alpha}} d^d y' + 
\left(\tu'(x',t')\right)^2 = -\lambda'(x',t') \, . \label{eq: Instanton massless dimensionless}
\end{equation}

\smallskip
In the following we will mainly use the dimensionless equations~\eqref{eq: BFM dimensionless}, \eqref{eq: Instanton RS dimensionless} and \eqref{eq: Instanton massless dimensionless} and drop the primes for readibility.

\subsection{Previous results for global quantities : avalanche size, total velocity and duration}\label{sec: results global quantities ch5}

For obtaining the statistics of global quantities we choose sources that do not depend on space.
In consequence the solution of the instanton equation is also space-independent, the elasticity term vanishes and the instanton equations we need to solve are of the form (in dimensionless units)~:
\begin{equation}\label{eq: instanton global quantities}
\partial_t u(t) - u(t) +u(t)^2 = -\lambda(t) \, .
\end{equation}

\subsubsection{Avalanche size}
The avalanche size is $S = \int \du(x,t) d^d x dt$ and corresponds to a constant source $\lambda(x,t) = \lambda$. By symmetry the solution is also independent of time. The instanton equation reduces to~:
\begin{equation}
\tu^2 - \tu +\lambda = 0 \, .
\end{equation}
This equation has two roots. For choosing the root we notice that from~\eqref{eq: Generating functional}, we have $G[\lambda, \dot{w}(x,t)] < 1$ for $\lambda<0$ which corresponds to 
$\tu^{\lambda}<0$ in~\eqref{eq: Generating functional from instanton}.
The solution is thus $\tu^{\lambda} = \left(1-\sqrt{1-4\lambda}\right)/2$.
It gives  the solution for a total drive $w^{\tot}$~:
\begin{equation}\label{eq: P_w(S)}
\overline{e^{\lambda S}} = \exp \left( \frac{w^{\tot}}{2}\left(1-\sqrt{1-4\lambda}\right) \right) \Leftrightarrow 
P_{w^{\tot}}(S) = \frac{w^{\tot}}{\sqrt{4\pi}S^{3/2}} \exp\left( -\frac{(S-w^{\tot})^2}{4S} \right) 
\end{equation}

\subsubsection{Avalanche total velocity}
The instantaneous total velocity of the interface, also called areal velocity, is defined as $V(t_f) = \int \du(x,t_f) d^d x$. It corresponds to the choice of source $\lambda(x,t) = \lambda \delta (t-t_f)$.
The solution of the instanton equation with this source is~:
\begin{equation}
\tu_{t}^{\lambda} = \frac{\lambda}{\lambda + (1-\lambda)e^{t_f-t}} \theta(t_f-t) \, ,  
\end{equation}
with $\theta$ the Heaviside function.
The distribution obtained depends on the driving $\dot{w}_{xt}$ in~\eqref{eq: Generating functional from instanton}. 
We will first consider a constant driving $\dot{w}(x,t) = v$. We note $v^{\tot} = L^d v$ the total driving velocity.
This choice gives the stationary total velocity distribution of the interface. The result does not depend on time hence we can drop the time $t_f$ and write $V$ for the total instantaneous velocity~:
\begin{equation}\label{eq: Total velocity distribution BFM}
\overline{e^{\lambda V}} =  \left( 1-\lambda \right)^{-v^{\tot}} \Leftrightarrow 
P_{v^{\tot}}(V) = \frac{V^{-1+v^{\tot}}}{\Gamma[v^{\tot}]} \exp\left( -V \right) \, .
\end{equation}
Note that the probability distribution \eqref{eq: Total velocity distribution BFM} is exactly the same one as the probability distribution for the velocity in the ABBM model~\eqref{eq: stat Prob time ABBM}
with the prefactor $v_m = \frac{\sigma}{\eta m^2} = 1$ in dimensionless units\footnote{
If we restore the dimensions in~\eqref{eq: Total velocity distribution BFM} we 
obtain exactly the formula~\eqref{eq: stat Prob time ABBM} with $v_m=S_m/\tau_m$
}.
Here $v^{\tot}=L^d v$ plays the role of the mean velocity $v$ in~\eqref{eq: stat Prob time ABBM}.
This distribution has a transition at $v^{\tot}=1$ between a phase where the interface stops infinitely often ($v^{\tot} <1$) and a phase where the interface never stops ($v^{\tot}>1$).
In terms of the mean velocity, the transition happens at $v=L^{-d}$ which decays with the system size. 


\subsubsection{Avalanche duration}
A second choice of driving is to consider an initial kick at time $t_i=0$ : 
$w(x,t) = w \delta(t) = w^{\tot} L^{-d} \delta(t)$. 
In this case the generating functional reads~:
\begin{align}\label{eq: Gen func av. duration}
\int e^{\lambda V(t_f)} P_{w^{\tot}}(V(t_f)) dV(t_f) &= \overline{e^{\lambda V(t_f)}} = \exp \left ( w L^d \tilde{u}^{\lambda}_{t=0} \right) \notag \\
	&= \exp \left ( w^{\tot} \frac{\lambda}{\lambda(1 - e^{t_f}) + e^{t_f}} \theta(t_f) \right)
\end{align}
Note that in the case $t_f<0$, we expect the velocity to be constantly zero. This is indeed the case as the generating functional is independent of $\lambda$ and equal to $1$ which corresponds to a probability distribution $P_{w^{\tot}}(V) = \delta(V)$.
In the case $t_f > 0$ we will not intend to compute the probability distribution but only consider the probability for the total velocity to be $0$ at time $t_f$~:  
\begin{align}
\mathbb{P}_{w^{\tot}}(V(t_f)=0) &= \lim_{\lambda\to -\infty} \int e^{\lambda V(t_f)} P_{w^{\tot}}(V(t_f)) dV(t_f) \\
	&= \exp \left ( \frac{w^{\tot}}{1 - e^{t_f}} \right)
\end{align}
This gives access to the distribution of the avalanche duration $T$ by noticing that 
$ \mathbb{P}_{w^{\tot}}(V(t_f)=0) = \mathbb{P}_{w^{\tot}}(T<t_f)$. 
We then have~:
\begin{align}
P_{w^{\tot}}(T=t_f) &= \partial_{T_f} \mathbb{P}_{w^{\tot}}(T<t_f) = \frac{w^{\tot}e^{t_f}}{(1 - e^{t_f})^2} \exp \left ( \frac{w^{\tot}}{1 - e^{t_f}} \right) \\
	&= \frac{w^{\tot}}{(2 \sinh t_f/2 )^2} \exp \left ( \frac{w^{\tot}}{1 - e^{t_f}} \right)
\end{align}

\subsubsection{Single avalanche density}
The probability distribution $P_{w^{\tot}}(S)$computed above is defined for a finite increment in the drive $w^{\tot}$. 
Within this procedure several avalanches might be triggered during the same drive increment. 
To ensure that a single avalanche is triggered we must take the limit $w^{\tot} \to 0$.
We can compute the single avalanche density per unit increment $\rho(S)$ which is defined by~:
\begin{equation}\label{eq: rho(S)}
\rho (S) = \frac{\partial P_{w^{\tot}}(S)}{\partial w^{\tot}}\Bigr|_{w^{\tot}=0}
	 = \frac{1}{\sqrt{4\pi} S^{3/2}}\exp \left(- \frac{S}{4} \right) \, .
\end{equation}
Similarly we can define the single avalanche density $\rho(T)$ for the avalanche duration :
\begin{equation}\label{eq: rho(T)}
\rho (T) = \frac{\partial P_{w^{\tot}}(T)}{\partial w^{\tot}}\Bigr|_{w^{\tot}=0}
	 = \frac{1}{(2 \sinh t_f/2 )^2} \, .
\end{equation}

\paragraph{Link between the massive and massless cases}
The results above were obtained for the massive case with the dimensionless massive units. 
The results for the massless case can be obtained from the behavior for $T,S,V \ll 1$.

\subsection{Link with the ABBM model}\label{sec: link ABBM BFM}

As for the BFM, the single particle equation of the ABBM model~\eqref{eq: ABBM 2} can be turned into a stochastic equation for the particle velocity~:
\begin{equation}\label{eq: ABBM velocity theory}
\eta \partial_t \du (t) = m^2\left(\dot{w}-\du(t) \right) + \sqrt{\du(t)} \xi(t) \, ,
\end{equation}
where $\xi$ is a white noise with correlations $\overline{\xi(t) \xi(t')} = 2\sigma \delta(t-t')$.
The same field-theory formalism as in section~\ref{subsec: Generating functional and instanton} can be used to write the generating functional~\cite{dobrinevski2012}~:
\begin{equation}\label{eq: Generating functional ABBM}
G^{\text{ABBM}}[\lambda(t), \dot{w}(t)] := \overline{\exp \left( \int\lambda(t)\dot{u}(t) dt \right)}^{\dot{w}(t)} = \exp\left( m^2\int \tu^{[\lambda]}(t) dt \right)
\end{equation}
where $\tu^{[\lambda]}$ is solution of the zero-dimensional instanton equation~:
\begin{equation}\label{eq: Instanton ABBM}
\eta \partial_t \tu(t) -m^2 \tu(t) + \sigma \tu^2(t) = -\lambda(t) \, .
\end{equation}
This is exactly the same equation (with dimension) as the general instanton equation for global quantities~\eqref{eq: instanton global quantities}. This means that the statistics of the velocity in the ABBM model is exactly the one of the total velocity in the BFM and the driving velocity $v$ in the ABBM corresponds to the total drive $v^{\tot} = L^d v$ in the BFM.

Equation \eqref{eq: ABBM velocity theory} can also be derived from the BFM equation for the local velocity field~\eqref{eq: definition BFM}. Integrating over space we obtain an equation for the total velocity~:
\begin{equation}
\eta \partial_t V(t) = m^2\left(\dot{w}^{\tot}(t) - V(t)\right)+\int \sqrt{\du(x,t)} \xi(x,t)d^dx \, ,
\end{equation}
where $\dot{w}^{\tot}(t) = \int \dot{w}(x,t) d^d x$.
The computation of the variance of the disorder term gives~:
\begin{align}
\overline{\int \sqrt{\du(x,t)} \xi(x,t)d^dx \int \sqrt{\du(y,t')} \xi(y,t')d^dy} &= 
\int \int \sqrt{\du(x,t)\du(y,t')}  \overline{\xi(x,t)\xi(y,t')} d^dx d^dy \notag \\
&= \int \int \sqrt{\du(x,t)\du(y,t')} 2\sigma \delta^d(x-y) \delta(t-t') d^dx d^dy \notag \\
&= 2\sigma \int \du(x,t)  \delta(t-t') d^dx  \notag \\
&= 2\sigma V(t) \delta(t-t') \, . \label{eq: variance integrated disorder ABBM}
\end{align}
The last term is equal to the variance of the disorder term $\sqrt{\du(t)} \xi(t)$ in \eqref{eq: ABBM velocity theory} with the correspondance $V(t) \equiv \du(t)$.

\section{Insights into the long-range instanton equation with a local source}

For global quantities like the avalanche size and duration or the total instantaneous velocity of the interface, the probability distribution are obtained by solving the instanton equation with sources that are independent of $x$. The results are thus the same for LR elasticity as for SR elasticity and are given in the previous section~\ref{sec: results global quantities ch5}.
On the other hand obtaining informations about local quantities, and hence about the spatial structure of the avalanches, requires to solve the instanton equation with sources that depend on $x$. In this case we expect to obtain different results for LR elasticity than for SR elasticity. 
The spatial structure also depends on the dimension $d$. In this section I focus on the one-dimensional case, $d=1$.

The most simple spatial quantity that we can study is the distribution of local jumps $P(S(x_0))$, with 
$S(x_0):=\int \du(x_0,t) dt$.
This distribution can be obtained by solving the instanton equation~\eqref{eq: Instanton RS dimensionless} with a local source $\lambda(x) = \lambda \delta(x-x_0)$.
This equation has been solved for SR elasticity and $d=1$ in~\cite{ledoussal2013, delorme2016}.
It enabled then to obtain the joint probability distribution $P(S(x_1), S(x_2))$ which paved the way to the computation of the probability distribution of the extension of the avalanche. 
Thus solving the instanton equation for the local jumps is a key step on the way to accessing more spatial quantities and being able to compute for instance the number of clusters inside one avalanche or the extension of clusters.
The instanton equation we need to solve for LR elasticity reads in dimensionless units~:
\begin{equation}
\int \frac{\tu(y) - \tu(x)}{|y-x]^{1+\alpha}}dy - A\tu(x) + \tu^2(x) = -\lambda \delta(x) \label{eq: Instanton  LR local size dimensionless}
\end{equation}
where $A=1$ for the massive instanton equation~\eqref{eq: Instanton RS dimensionless} and $A=0$ for the massless instanton equation~\eqref{eq: Instanton massless dimensionless}.
This equation is non linear and non local. Although we were not able to solve it yet we could still obtain some information about the statistics of $S(x)$.

\subsection{Outline and main results of the section}

In section~\ref{sec: Instanton Fourier} I start by studying the Fourier transform of the instanton equation as it will be useful for the following sections.

In section~\ref{subsec: scaling argument for the critical exponent of the local size} I make the Ansatz that the solution of the massless instanton equation is scale invariant and find the scale transformation under which the scale invariance holds. An homogeneity relation involving the generating functional then allows to deduce the value of the exponent $\tau_0$ which controls the decay of 
$P\left(S(x)\right)$. I find that 
$\tau_0 = 1 + \frac{\alpha-1}{2\alpha-1}$~\eqref{eq: tau_0} 
for an homogeneous drive and 
$\tau_{0, \text{loc}} = 1 + \frac{\alpha}{2\alpha-1}$~\eqref{eq: tau_(0,loc)}
for a local drive. 
The values for $\alpha=2$, $\tau_0 = 4/3$ and $\tau_{0, \text{loc}} = 5/3$, are in agreement with the results of the calculation in~\cite{delorme2016} for SR elasticity.
For $\alpha=1$ I obtain the values $\tau_0 = 1$ and $\tau_{0, \text{loc}} = 2$.

In section~\ref{sec: power law decay massless solution} I assume that the solution of the instanton equation decays as a power law at large scale $\tu(x) \sim |x|^{-\beta}$. Matching the leading order non-analytic terms in a small $q$ expansion of the instanton equation in Fourier space, I obtain an exponent $\beta=\alpha$ for $0<\alpha<1$, $\beta=(1+\alpha)/2$ for $1< \alpha <3$ and $\beta=2$ for $\alpha>3$.
These results are summarized in table~\ref{tab:beta power law decay}.
I deduce a bound on the exponent $\kappa_c$ of the cluster extension~\eqref{eq: bound kappa_c}.

Finally in section~\ref{sec: expansion and first moments} I use a formal expansion of the solution of the instanton equation to compute the large scale behavior of the  first two moments of the local jump distribution (i.e. $\overline{S(x)}$ and $\overline{S^2(x)}$). 
Following a local kick $x=0$ both moments decay as $1/|x|^{1+\alpha}$ at large scale for any $\alpha$ between 0 and 2 
(cf. equations \eqref{eq: 1st moment S(x) local kick general alpha dimensionfull} and \eqref{eq: 2nd moment S(x) local kick general alpha dimensionfull})
The same decay holds when the moments are computed for a fixed total avalanche size $S$ (cf. equations
\eqref{eq: 1st moment S(x) conditonned by S general alpha} and \eqref{eq: 2nd moment S(x) conditonned by S general alpha}).
This suggests a kind of intermittency, i.e. that $S(x)$ is very small most of the time but that 
the moments are dominated by rare events, where $S(x)$ is large, of order 1, that arise with a probability proportional to $1/x^{1+\alpha}$. 
I deduce a bound on the exponent characterizing the large scale decay of the avalanche diameter~\eqref{eq: bound kappatilde_D}.

\subsection{Instanton equation in Fourier}\label{sec: Instanton Fourier}
The dimensionfull instanton equation ~\eqref{eq: Instanton real space} reads in Fourier space~:
\begin{gather}
\eta \partial_t \hat{u}(q,t) + \left(\hat{\mathcal{C}}(q) -\hat{\mathcal{C}}(0) \right) \hat{u}(q,t)  - m^2\hat{u}_{qt} + 
\sigma \left(\hat{u}*\hat{u}\right)(q,t) = -\hat{\lambda}(q,t) \, , \label{eq: Instanton fourier space} 
\end{gather}
where $*$ denotes the convolution product in the Fourier variable~:
\begin{equation}
\left(\hat{u}*\hat{u}\right)(q,t) = \int \hu(k,t) \hu(q-k,t) \frac{dk}{2\pi} \, .
\end{equation}
The elastic term can be explicitly computed for the elastic kernel~\eqref{eq: specification elastic kernel} with $0<\alpha<2$ and reads (see appendix~\ref{sec:Appendix Fourier Transform elastic force} for details of the calculation)~:
\begin{align}\label{eq: specification LR kernel Fourier}
\hat{\mathcal{C}}_q-\hat{\mathcal{C}}_0
	&= \left\lbrace
\begin{array}{l c}
2 c \cos\left(\frac{\alpha\pi}{2}\right) \Gamma(-\alpha) \abs{q}^{\alpha} & \text{for }
0< \alpha < 2 \text{ and } \alpha \neq 1 \, , \\
-c\pi |q| & \text{for } \alpha=1 \, .
\end{array}\right.
\end{align}
With this kernel the Fourier transform of the dimensionless instanton equation~\eqref{eq: Instanton RS dimensionless} is thus~:
\begin{gather}
\partial_t \hat{u}(q,t) -\kappa_{\alpha}|q|^{\alpha} \hat{u}(q,t)  - \hat{u}(q,t) + 
\left(\hat{u}*\hat{u}\right)(q,t) = -\hat{\lambda}(q,t) \, , \label{eq: Instanton fourier space dimensionless} 
\end{gather}
with $\kappa_{\alpha} = -2 \cos\left(\frac{\alpha\pi}{2}\right) \Gamma(-\alpha)>0$ for $\alpha\neq 1$ and 
$\kappa_{1}=\pi$.


\subsection{Scaling argument for the exponent of the local size distribution}\label{subsec: scaling argument for the critical exponent of the local size}



We can estimate the critital exponent for the distribution of local jumps via a scaling argument. 
As the exponent is the hallmark of the behaviour at the critical point we can set the mass 
$m\to 0$ (i.e. $A=0$ in \eqref{eq: Instanton  LR local size dimensionless})
We will make the assumption that the solution of the massless instanton equation has a scale-invariance.
This assumption is motived by the fact that the solution of the massless instanton equation for SR elasticity is scale invariant. Indeed this solution reads~\cite{delorme2016}~:
\begin{equation}\label{eq:solution massless instanton SRE v1}
\tu(x) = -\frac{6}{(|x|+(-24/\lambda)^{1/3})^2} = -\frac{(-\lambda)^{2/3}6}{\left((-\lambda)^{1/3}|x|+24^{1/3}\right)^2} \, \text{ with }\, -\lambda >0 \, .
\end{equation}
It can thus be written as 
$\tu(x) = (-\lambda)^{\beta} \check{u}((-\lambda)^{\gamma} x)$ with $\beta=2/3$, $\gamma=1/3$ and 
$\check{u}(\check{x}) = -6/(|\check{x}| + 24^{1/3)^2}$ independent of $\lambda$.

Therefore we consider the LR massless intanton equation~:
\begin{equation}
\int \frac{\tu (y) - \tu (x)}{|y-x|^{1+\alpha}}dy + \tu^2(x) = -\lambda \delta(x) \label{eq: Instanton local size massless}.
\end{equation}
and assume that its solution scales with $-\lambda$ (which is positive) as~:
\begin{equation}\label{eq: scaling hypothesis}
\tu(x) = (-\lambda)^{\beta} \check{u}(\check{x})  \text{ where } \check{x}=(-\lambda)^{\gamma} x.
\end{equation}
Under this rescaling the instanton equation~\eqref{eq: Instanton local size massless} becomes~:
\begin{equation}
(-\lambda)^{\beta + \alpha \gamma} \int \frac{\check{u}(\check{y}) - \check{u}(\check{x})}{|\check{y}-\check{x}|^{1+\alpha}}d\check{y} + (-\lambda)^{2\beta} \check{u}^2(\check{x}) = (-\lambda)^{1+\gamma} \delta(\check{x})  \, .
\end{equation}
The scale invariance of the equation implies that 
$\beta + \alpha \gamma = 2\beta = 1+\gamma$ and hence that~:
\begin{equation}
\beta = \frac{\alpha}{2\alpha-1} \, , \quad \gamma = \frac{1}{2\alpha-1} \, .
\end{equation}
Note that for $\alpha=2$ we recover the values $\beta=2/3$ and $\gamma=1/3$ which correspond to the scale invariance of the SR solution~\eqref{eq:solution massless instanton SRE v1}.
The constants $\beta$ and $\gamma$ diverge for $\alpha=1/2$. Therefore we assume from now on that $\alpha > 1/2$.
Now we look at the consequence of this scaling solution for the distribution of the local size.
The distribution of $S_0=S(x=0)$ is encoded in the generating functional~:
\begin{equation}\label{eq: generating functional local size}
\overline{\exp(\lambda S_0)}^{\dot{w}(x,t)} = G[\lambda\delta(x), \dot{w}(x,t)] 
	= \exp \left( \int \dot{w}(x,t) \tu(x) dx dt \right) \, .
\end{equation}
The integral over time does not depend on $\tu$ so we can consider the drive to be a kick at $t=0$~:
$\dot{w}(x,t) = w(x) \delta(t)$.
The result depends on the specific form of the function $w(x)$. We consider to case, an homogeneous 
drive, $w(x)=w$, and a local drive $w(x)=w\delta(x-x_0)$.

\paragraph{Homogeneous drive : $w(x)=w$}
In this case the generating functional reads~:
\begin{align}
\overline{\exp(\lambda S_0)}^{w\delta(t)} &= \exp \left( w \int \tu(x) dx \right)
 = \exp \left( w (-\lambda)^{\beta-\gamma} \int \check{u}(\check{x}) d\check{x} \right) \notag \\
 &=  1 + w K (-\lambda)^{\beta-\gamma} + O(w^2) \, .
\end{align}
where $K=\int \check{u}(\check{x}) d\check{x}$.
From this we deduce~:
\begin{equation}
\int P_w(S_0) \left( e^{\lambda S_0} -1 \right) dS_0 = wK (-\lambda)^{\beta-\gamma} +O(w^2) \, .
\end{equation}
For the single avalanche density $\rho (S_0) = \frac{\partial P_{w}(S_0)}{\partial w}\Bigr|_{w=0}$ we obtain~:
\begin{equation}
\int \rho(S_0) \left( e^{\lambda S_0} -1 \right) dS_0 = K (-\lambda)^{\beta-\gamma} \, .
\end{equation}
Assuming a power law $\rho(S_0) \sim S_0^{-\tau_0}$, we see that the left-hand side (LHS) has dimension
$S_0^{1-\tau_0} \sim \lambda^{-1+\tau_0}$ (because $\lambda S_0 \sim 1$).
By homogeneity this yields~:
\begin{equation}\label{eq: tau_0}
\tau_0 = 1 + \beta - \gamma = 1 + \frac{\alpha-1}{2\alpha-1} = 2-\frac{\alpha}{2\alpha - 1}
\end{equation}
Thus for $\alpha=1$ this argument predicts
\begin{equation}\label{eq: tau_0 alpha=1}
\tau_0=1 \, .
\end{equation} 
For $\alpha=2$ it predicts $\tau_0=4/3$ which is consistent with previous results for SR elasticity.
Indeed for SR elasticity, the full probability density function of local jumps has been obtained in \cite{ledoussal2013, delorme2016}. 
For an homogeneous driving $w(x)=w$, it does not depend on $x$. In this case we write the result
for $S_{0}:=S(x=0)$~:
\begin{equation}\label{eq: SRE LocalSize distribution}
P_w(S_0) = \frac{2 \times 3^{1/3}}{S_0^{4/3}} e^{6w} w 
Ai \left ( \left ( \frac{3}{S_0} \right )^{1/3}\left ( S_0+2w \right ) \right ),
\end{equation}
where $S_0$ and $w$ are expressed in units of $\sigma/m^3$.

\paragraph{Local drive : $w(x) = w\delta(x-x_0)$}
We assume a kick at $x=x_0$.
The generating functional for the distribution of jumps in $x=0$ is~:
\begin{align}
\overline{\exp(\lambda S_0)}^{w\delta(x-x_0)\delta(t)} &= \exp \left( w \tu(x_0)\right)
 = \exp \left( w (-\lambda)^{\beta} \check{u}(\check{x}_0) \right) \notag \\
 &= 1 + w \check{u}(\check{x}_0) (-\lambda)^{\beta} +O(w^2) \, .
\end{align}
From which we deduce~:
\begin{equation}
\int \rho(S_0) \left( e^{\lambda S_0} -1 \right) dS_0 = \check{u}(\check{x}_0) (-\lambda)^{\beta}\, .
\end{equation}
Assuming a power law decay $\rho(S_0) \sim S_0^{-\tau_{0, \text{loc}}}$,  the LHS has dimension
$S_0^{1-\tau_{0, \text{loc}}} \sim \lambda^{-1+\tau_{0, \text{loc}}}$.
The homogeneity now imposes the relations~:
\begin{equation}\label{eq: tau_(0,loc)}
\tau_{0, \text{loc}} = 1 + \beta = 1 + \frac{\alpha}{2\alpha-1} = 2-\frac{\alpha-1}{2\alpha - 1}
\end{equation}
For $\alpha=1$ this argument now predicts~:
\begin{equation}\label{eq: tau_(0,loc) aplha=1}
\tau_{0, \text{loc}} = 2 \, .
\end{equation}
For $\alpha = 2$ it predicts $\tau_{0, \text{loc}} = 5/3$ which is once again in agreement with the result of the calculation in~\cite{delorme2016} for SR elasticity.

Table~\ref{tab: predictions tau_0} gives the numerical values according to the predictions~\eqref{eq: tau_0} and~\eqref{eq: tau_(0,loc)} for some values of $\alpha$.

\subsection{Power law decay of the massless solution}\label{sec: power law decay massless solution}

For SR elasticity the solution of the massless instanton equation is~\cite{delorme2016}~:
\begin{equation}\label{eq:solution massless instanton SRE}
\tu(x) = -\frac{6}{(|x|+x_{\lambda})^2} \, , \quad \text{ with } \, x_{\lambda}^3 = -\frac{24}{\lambda} >0 \, .
\end{equation}
Taking the limit $\lambda\to -\infty$ allows to obtain the probability that the interface does not move in $x=0$ (i.e. $S_0=0$) when a kick 
$\dot{f}(x,t) = f \delta(x-x_0)\delta(t)$ is applied in $x_0$. 
In this limit the solution of the instanton equation becomes\footnote{Note that this function is also solution of $\partial^2_x \tu(x) + \tu^2(x) =0$ for $x\neq 0$.}
$\tu(x)=-6/x^2$
and the probability reads 
\begin{equation}\label{eq:P(S_0=0) SRE}
\mathbb{P}_{f \delta(x-x_0)\delta(t)} (S_0 = 0) = \exp \left( f \tu(x_0)\right) 
= \exp \left(-\frac{6f}{x_0^2} \right) \, .
\end{equation}
This probability is finite for $x_0\neq 0$ and tends to $1$ when $|x_0| \to \infty$. 
For SR elasticity, \eqref{eq:P(S_0=0) SRE} is the probability that an avalanche starting from a seed at $x=x_0$ does not reach $x=0$. 
By taking the derivative with respect to $|x_0|$ 
of~\eqref{eq:P(S_0=0) SRE} one
obtains the probability that the avalanche
reaches a distance $|x_0|$ from the seed.
This probability decays as $1/|x_0|^3$, hence the exponent of the avalanche extension
is $\kappa=3$, in agreement with the general scaling relation 
$\kappa=d+\zeta-1$ (with $d+\zeta=4$ for the BFM with SR elasticity).

\vspace{0.3cm}
By analogy let us assume that the solution of the massless LR instanton equation also has a power law decay when $|x| \to \infty$~:
\begin{equation}
\tu(x) \sim_{|x|\to \infty} \frac{B}{|x|^{\beta}}
\end{equation}
where $\beta$ is an unknown exponent that depends on $\alpha$ and that we want to determine. 
In order to determine $\beta$ we can make two arguments. 
The first one is in real space.
It consists in making an Ansatz $\beta=\alpha$ and checking that it is consistent. 
It leads to an exact solution of the LR instanton equation for all $x \neq 0$ valid for $\alpha<1$. 
This solution was noticed earlier in~\cite{delorme}.
The second one consists in matching the non-analytic exponents of the equation in Fourier space.

\subsubsection{Real space argument}

We make the Ansatz for $x\neq 0$~:
\begin{equation}\label{eq: Ansatz power law decay real space}
\tu(x) = \frac{B(\alpha)}{|x|^{\alpha}}
\end{equation}
and plug this Ansatz into \eqref{eq: Instanton  LR local size dimensionless} with $A=0$.
For $x \neq 0$ the elastic term becomes~:
\begin{align}
\int \frac{\tu(y) - \tu(x)}{|y-x|^{1+\alpha}}dy =
	\frac{B(\alpha)}{|x|^{2\alpha}} \int \frac{1}{|s-1|^{1+\alpha}}\left(\frac{1}{|s|^{\alpha}}-1\right)ds =  - \frac{B(\alpha)Z(\alpha)}{|x|^{2\alpha}}
\end{align}
with 
\begin{equation}\label{eq: Z(alpha)}
Z(\alpha) = -\int \frac{1}{|s-1|^{1+\alpha}}\left(\frac{1}{|s|^{\alpha}}-1\right)ds = \left\lbrace
\begin{array}{l c}
\frac{\Gamma(1 - \alpha)^2}{\alpha \Gamma[1 - 2\alpha]} & \text{if } \, \alpha < 1 \\
- \infty & \text{if } \, \alpha \geq 1
\end{array}\right.
\end{equation}
Since $\tu^2(x) = B^2(\alpha)/|x|^{2\alpha}$, we must have for $x\neq 0$~:
\begin{equation}
\frac{B^2(\alpha) - B(\alpha)Z(\alpha)}{|x|^{2\alpha}} = 0
\end{equation}
which imposes $B(\alpha) = Z(\alpha)$. 
For $\alpha<1$ this gives a solution for the instanton equation for all $x\neq 0$.
Taking into account the source $\lambda \delta(x)$ would require to modify the Ansatz.
By analogy with the SR case it is reasonable to assume that this gives the solution for $\lambda \to -\infty$.

The Ansatz \eqref{eq: Ansatz power law decay real space} does not work for $\alpha \geq 1$ as the integral $Z(\alpha)$ diverges in this case. 
Let us turn to the Fourier space to obtain further information about the power law decay of $\tu(x)$.

\subsubsection{Fourier space argument}

We assume that $\tu (x) \sim |x|^{-\beta}$ at large distance.
The Fourier transform of the elastic kernel is given in~\eqref{eq: specification LR kernel Fourier} for $\alpha < 2$ and is proportional to $|q|^{\alpha}$. 
If $\alpha > 2$ there are singularities and even power of $q$ arises (see Appendix~\ref{sec:Appendix Fourier Transform elastic force} for details). 
In the most general setting we can write the small $q$ expansion of the elastic kernel as~:
\begin{equation}
\hat{\mathcal{C}}_q-\hat{\mathcal{C}}_0 = -\left(\kappa_{\alpha} |q|^{\alpha} + c_2 q^2 + O(q^4)\right)\, .
\end{equation}
The Fourier transforms of $\tu(x)$ and $\tu^2(x)$ can be expanded near $q=0$ as~:
\begin{align}
\hu(q) &= u_0 + u_2 q^2 + u_{\beta} |q|^{\beta-1} + O(q^4) \, , \\
\widehat{u^2}(q) &= U_0 + U_2 q^2  + U_{\beta} |q|^{2\beta-1} + O(q^4) \, .
\end{align}
For the source we only assume a fast decay so that its Fourier transform is analytic~:
\begin{equation}
\hat{\lambda}(q) = \lambda_0 + \lambda_2 q^2 + O(q^4)
\end{equation}
with $\hat{\lambda}(q) \equiv \lambda_0 = \lambda$ corresponding to $\lambda(x) = \lambda \delta (x)$.
A necessary condition for our ansatz to be solution of the instanton equation is that the leading order non-analytic term $U_{\beta} |q|^{2\beta-1}$ of $\widehat{u^2}(q)$ cancels against the leading order non-analytic term of $\left(\hat{\mathcal{C}}_q-\hat{\mathcal{C}}_0\right)\hu(q)$.
To proceed let us distinguish three cases.

\paragraph{Case 1}
For $\alpha<2$ and $\beta<1$ the leading order of the elastic term is 
$\kappa_{\alpha} u_{\beta} |q|^{\alpha+\beta-1}$. 
This yields the solution $\beta=\alpha$ with the condition $\alpha <1$.
We retrieve the result that we obtained in real space.
\paragraph{Case 2}
If $\alpha<2$ and $\beta>1$ the leading order term is now
$\kappa_{\alpha} u_{0} |q|^{\alpha}$. 
We must have $\alpha = 2\beta-1 $ which implies $\beta = (1+\alpha)/2$.

\paragraph{Case 3}
If $\alpha>2$ we have a competition between two plausible leading terms : $|q|^{\alpha}$ or $|q|^{2+\beta -1}$.
If the latter is the leading term it implies 
\begin{align}
\left\lbrace
\begin{array}{lll}
2+\beta -1  = 2\beta-1 & \Rightarrow & \beta = 2 \\
2+\beta -1 < \alpha & \Rightarrow & \alpha > 3 
\end{array}\right .
\end{align}
So $|q|^{2+\beta -1}$ can be the leading term only if $\alpha>3$. For $2< \alpha <3$, the leading term must be $|q|^{\alpha}$ and we have again $\beta = (1+\alpha)/2$.

The results are summarized in table~\ref{tab:beta power law decay}.

\begin{center}
\begin{table}[hbt]
\centering
\begin{tabular}{c c c}\hline
    $0< \alpha <1$ & $1< \alpha <3$ & $ 3 < \alpha $ \\ \hline
	$\beta=\alpha$ & $\beta=(1+\alpha)/2$ & $\beta=2$ \\ \hline
\end{tabular}
\caption{Table of the decay exponent $\beta$ of the instanton equation solution
tilde $\tu(x) \sim |x|^{-\beta}$ as a function of the elastic kernel exponent $\alpha$. \label{tab:beta power law decay}}
\end{table}
\end{center}

\subsubsection{Interpretation}

We would like to make the same interpretation as for SR elasticity and conclude that the instanton equation solution
that we obtained above, which is infinite at $x=0$, gives the probability that the line does not move in $x=0$ when a kick is given in $x_0 \neq 0$, as in \eqref{eq:P(S_0=0) SRE}. 
However for LR interactions with the elastic kernel~\eqref{eq: specification elastic kernel}, the equations~\eqref{eq: motion general} and~\eqref{eq: definition BFM} show that, as soon as a small portion of the interface moves, all points must move from a finite amount. 
Thus we will interpret the result as the probability that the local jump $S_0=S(0)$ is "small"\footnote{We have not given a precise meaning of what "small" means. This would require further investigations.}   
when a local kick $\dot{f}(x,t)=f\delta(x-x_0)\delta(t)$ is applied at $x_0$~:
\begin{equation}\label{eq:P(S_0 simeq 0) LRE}
\mathbb{P}_{f\delta(x-x_0)\delta(t)} (S_0 \simeq 0) = \exp \left( f \tu(x_0)\right)
	\sim \exp \left(-\frac{B}{|x_0|^{\beta}} f \right) \, ,
\end{equation}
where the value of $\beta$ depends on $\alpha$ and is given in table~\ref{tab:beta power law decay}.
For $\alpha < 1$ the scaling \eqref{eq:P(S_0 simeq 0) LRE}  holds for all $x_0 \neq 0$ with $B=B(\alpha)=Z(\alpha)$ calculated in~\eqref{eq: Z(alpha)} while for $\alpha>1$ it is true only for $x_0$ large.

A second complication is that for LR elasticity we expect the avalanche to be made of several disconnected clusters, so we cannot infer immediately information about the avalanche extension and its exponent $\kappa$. Instead we can infer information about the extension of the clusters.
We consider that two clusters are disconnected if they are separated by a portion 
$\Omega$ of the interface where $S(x)\simeq 0$ for all $x \in \Omega$.
For a local kick at $x_0$, having $S(0) \simeq 0 $ implies that the distance $\ell_c(x_0)$
reached by the cluster which contains $x_0$  is smaller than $|x_0|$~:
\begin{equation}
S(0) \simeq 0   =>   \ell_c(x_0) < |x_0|  \, .
\end{equation}
Hence we can deduce a bound on the probability that $\ell_c(x_0) < |x_0|$. Writing $\mathbb{P}_{f}$ instead of $\mathbb{P}_{f\delta(x-x_0)\delta(t)}$ we have~:
\begin{align}
&\mathbb{P}_{f} \left( \ell_c(x_0) < |x_0| \right) \geq \mathbb{P}_{f} \left( S(0)\simeq 0 \right) \\
&\mathbb{P}_{f} \left(\ell_c(x_0) > |x_0| \right) \leq 1- \mathbb{P}_{f} \left( S(0)\simeq 0 \right) 
\underrel{|x_0| \gg 1}{\sim} f \frac{B}{|x_0|^{\beta}} \label{eq: ijkl}
\end{align}
Assuming that all clusters are described by the same decay exponent $\kappa_c$, the second probability scales as~:
\begin{equation}
\mathbb{P}_{f} \left(\ell_c(x_0) > |x_0| \right) \sim f\int_{|x_0|}^{+\infty} \frac{d\ell_c}{\ell_c^{\kappa_c}} \sim \frac{f}{|x_0|^{\kappa_c-1}} \, .
\end{equation}
Then the inequality in~\eqref{eq: ijkl} implies that~:
\begin{equation}\label{eq: bound kappa_c}
\kappa_c \geq \beta+1 = \left\lbrace
\begin{array}{l l}
\alpha+1 \; & \text{ if }  \alpha \leq 1 \\
\frac{3+\alpha}{2} \; & \text{ if } 1\leq \alpha \leq 3
\end{array}\right. 
\end{equation}
This is consistent with the scaling relation for $\kappa_c$
derived in chapter~\ref{chap: cluster statistics}.
Indeed for the BFM one has $d+\zeta = 2\alpha$ so this scaling relation gives
$\kappa_c = 1 + 2(d + \zeta) - 2\alpha = 1+2\alpha$ which satisfies the bound~\eqref{eq: bound kappa_c}.
We recall also that
$\kappa =1 + d + \zeta - \alpha =1 + \alpha$ for the BFM.
Hence one can note that $\kappa$ seems to obey the same bound and that
it saturates it for $\alpha \leq 1$.

\subsection{Expansion of the solution and first moments of local jumps}\label{sec: expansion and first moments}

Although we are not able to solve the instanton equation~\eqref{eq: Instanton  LR local size dimensionless} which would give the full probability distribution of the local jumps $S(x)$, 
a formal expansion of the solution allows to compute the first moments of $S(x)$.
For this purpose we start once again from the generating functional.
Since $\tu(x)$ does not depend on time we can consider the drive to be a kick at $t=0$:
$\dot{f}(x,t) = f(x) \delta(t) = m^2 w(x) \delta(t)$.
With this convention the generating functional \eqref{eq: Generating functional}, \eqref{eq: Generating functional from instanton}  reads~:
\begin{equation}\label{eq: generating functional for expansion}
G[\lambda(x), f(x)] := \overline{\exp \left( \int\lambda(x) S(x) dx \right)}^{f(x)} = \exp \left (\int f(x) \tu^{\lambda}(x) dx \right) \, .
\end{equation}
One can see that expanding the first exponential in~\eqref{eq: generating functional for expansion}
with a source $\lambda(x)=\lambda \delta(x-x_0)$ would give the first moments of $S(x_0)$.
The first terms of the expansion of the second exponential can be computed by writting a
formal expansion of the solution of the instanton equation
as~:
\begin{equation}\label{eq: expansion solution instanton}
\tu^{\lambda}(x) = \lambda \tu_1(x) + \frac{\lambda^2}{2} \tu_2(x) + ... 
= \sum_{n=1}^{\infty} \frac{\lambda^n}{n!} \tu_n(x)
\end{equation}
Note that we can perform this expansion for any source $\lambda(x)$ that can be written as
$\lambda(x) = \lambda g(x)$ with $g$ some function.
Expanding both sides of~\eqref{eq: generating functional for expansion} in power of $\lambda$ we have~:
\begin{align}
G[\lambda(x), f(x)] &=  1 + \overline{\int\lambda(x) S(x) dx} + \frac{1}{2}\overline{\left( \int\lambda(x) S(x) dx \right)^2} + ... \\
	&= 1 + \lambda \int f(x) \tu_1(x) dx +  \frac{\lambda^2}{2} \int f(x) \tu_2(x) + \frac{\lambda^2}{2} \left(\int f(x) \tu_1(x) dx\right)^2 + ...
\end{align}
Matching the same orders in $\lambda$ yields~:
\begin{align}
\overline{\int\lambda(x) S(x) dx} &=  \lambda \int f(x) \tu_1(x) dx \label{eq: first moment formal formula} \\
\frac{1}{2}\overline{\left( \int\lambda(x) S(x) dx \right)^2} &= \frac{\lambda^2}{2} \int f(x) \tu_2(x) + \frac{\lambda^2}{2} \left(\int f(x) \tu_1(x) dx\right)^2
\end{align}
The last term in the second line is the square of the first line. Thus we obtain a formula for the second connected moment~:
\begin{equation}\label{eq: second connected moment formal formula}
\overline{\left( \int\lambda(x) S(x) dx \right)^2}^c =
\overline{\left( \int\lambda(x) S(x) dx \right)^2} - \left( \overline{\int\lambda(x) S(x) dx} \right)^2 = \lambda^2 \int f(x) \tu_2(x)  dx
\end{equation}
More generally the n-th connected moment (i.e. the n-th cumulant) is, 
for $n \geq 1$~:
\begin{equation}\label{eq: n-th connected moment formal formula}
\overline{\left( \int\lambda(x) S(x) dx \right)^n}^c = \lambda^n \int f(x) \tu_n(x)  dx
\end{equation}

\subsubsection{Computation of the first terms of the expansion and first moments for a local kick}
To compute the first terms of the expansion of $\tu^{\lambda}(x)$ we turn to Fourier space.
The Fourier transform of the dimensionless instanton equation with a source $\lambda(x) = \lambda g(x)$ is~:
\begin{equation}
-\left(\kappa_{\alpha}|q|^{\alpha} + A\right) \hat{u}(q) + 
\left(\hat{u}*\hat{u}\right)(q) = -\hat{\lambda}(q) = -\lambda \hat{g}(q)\, , \label{eq: Instanton fourier for expansion} 
\end{equation}
with $A=1$ for the massive equation and $A=0$ for the massless equation.
Plugging in the expansion~\eqref{eq: expansion solution instanton} into~\eqref{eq: Instanton fourier for expansion}, we see that the first order term matches with $\hat{g}(q)$ while the highest order terms can be expressed as convolution products of lowest order terms. 
Hence the terms of the expansion can be computed successively by iteration.
Restricting to the two lowest order terms, we have~:
\begin{align}
\hat{u}_1(q) &= \frac{\hat{g}(q)}{\kappa_{\alpha}|q|^{\alpha} + A} \label{eq: hu_1(q)}\\
\hat{u}_2(q) &= \frac{2}{\kappa_{\alpha}|q|^{\alpha} + A}\left(\hat{u}_1*\hat{u}_1\right)(q) \label{eq: hu_2(q) is convolution product of hu_1(q)}
\end{align}

To proceed we first focus on $\alpha=1$, which is the case of most relevant interest, and then extend the result for $0<\alpha<2$..
We consider a source $\lambda(x) = \lambda \delta(x)$.
In order to compute the convolution product in~\eqref{eq: hu_2(q) is convolution product of hu_1(q)}
we must set $A$. We choose $A=1$ which corresponds to the massive instanton equation.
With this convention we have, for $\alpha=1$ and using $\kappa_{1}=\pi$~:
\begin{align}
\hat{u}_1(q) &= \frac{1}{\pi|q| + 1}  \label{eq: hu_1(q) massive kernel alpha=1}\\
\hat{u}_2(q) &= \left\lbrace
\begin{array}{l l}
\frac{4}{2+\pi |q|}\times \frac{\ln(1+\pi |q|)}{\pi^3 |q|} & \text{if } q\neq 0 \\
\frac{2}{\pi^2} & \text{if } q=0
\end{array}\right. \label{eq: hu_2(q) massive kernel alpha=1}
\end{align}
To obtain the moments of $S(x)$ following  a local kick we must take the inverse Fourier transform.
The inverse Fourier transform of $\hu_1(q)$ reads\footnote{Note that for a source $\lambda(x) = \lambda \delta(x-x_0)$ one has that $\tu(x)$ and all the $\tu_n(x)$ depend only on $|x-x_0|$. We have used that below in several places to simplify the equations.}~:
\begin{equation}\label{eq: tu_1 massive kernel alpha=1}
\tu_1(x) = \frac{-2 \text{Ci}\left(\frac{x}{\pi }\right) \cos \left(\frac{x}{\pi }\right)-2
   \text{Si}\left(\frac{x}{\pi }\right) \sin \left(\frac{x}{\pi }\right)+\pi  \sin
   \left(\frac{x}{\pi }\right)}{2 \pi ^2}
\end{equation}
where Ci and Si are the cosine and sine integral functions defined as~:
\begin{equation}
\text{Ci}(z) = -\int_{z}^{+\infty} \frac{\cos(t)}{t} dt \quad , \quad 
\text{Si}(z) = \int_0^{z} \frac{\sin(t)}{t} dt
\end{equation}
Equation~\eqref{eq: tu_1 massive kernel alpha=1}  can be expanded near $x=0$ and $x=+\infty$ yielding~:
\begin{align}\label{eq: expansion tu_1 massive kernel alpha=1}
\begin{array}{l l }
\tu_1(x) = \frac{-\ln(|x|)}{\pi^2} + O(1)  & \text{ when } x\to 0 \\
\tu_1(x) = \frac{1}{x^2} + O\left( \frac{1}{x^4} \right) & \text{ when } x\to \pm\infty
\end{array}
\end{align}
With this result we obtain the first moment of $S(x)$ following a local kick at $x=0$ (i.e. $f(x)=f\delta(x)$)~: 
\begin{align}\label{eq: 1st moment S(x) local kick}
\overline{S(x)} = f \tu_1(x) \simeq \left\lbrace
\begin{array}{l l }
-f\frac{\ln(|x|)}{\pi^2}  & \text{ when } x\to 0 \\
 f\frac{1}{x^2}  & \text{ when } x\to \pm\infty
\end{array}\right.
\end{align}
Regarding the second moment $\overline{S^2(x)}^{c}:=\overline{\left(S(x)\right)^2}^{c}$, we are not able to compute the inverse Fourier transform of $\hu_2(q)$ but
we note that $\tu_2(0) = \int \hu_2(q) \frac{dq}{2\pi} = \frac{1}{2\pi^2}$ is finite.
Hence the second connected moment of the jump in $x=0$ following a kick at the same location is finite and reads~:
\begin{equation}
\overline{S^2(0)}^{c} = f\tu_2(0) = \frac{f}{2\pi^2} \, .
\end{equation}
This results is remarkable. Indeed although the expectation of $S(0)$ is infinite (as can be seen by considering $x\to 0$ in~\eqref{eq: 1st moment S(x) local kick}), the fluctuations around this expectation are finite.

We can also compute the large $x$ behavior of $\overline{S^2(x)}^{c}$ by considering the expansion of $\hu_2(q)$ near $q=0$~:
\begin{equation}\label{eq: hu_2(q) expanded near q=0}
\hu_2(q) = \frac{2}{\pi^2}\frac{1}{1+\pi |q|} + O(q^2)=\frac{2}{\pi^2} \hu_1\left( q \right) + O(q^2)
\end{equation}
Since the large $x$ behavior is controlled by the leading non-analytic 
term $|q|$ in Fourier, we can use the above result for $\hu_1$ so we have
(for $\alpha=1$)~:
\begin{align}
\overline{S^2(x)}^{c} &= f\tu_2(x) \underrel{x\to \pm \infty}{\simeq} f\frac{2}{(\pi x)^2}  \\
\overline{S^2(x)} &= \overline{S^2(x)}^{c} + \left(\overline{S(x)}\right)^2 \underrel{x\to \pm \infty}{\simeq} f\frac{2}{(\pi x)^2} \label{eq: 2nd moment S(x) local kick}
\end{align}
The computation of the large $x$ behavior for the first and second moments can be extended to $0< \alpha <2$. We have~:
\begin{align}
\overline{S(x)} &= f \tu_1(x) \underrel{x\to \pm \infty}{\simeq} \frac{f}{ |x|^{1+\alpha}}  \label{eq: 1st moment S(x) local kick general alpha} \\
\overline{S^2(x)} &= f \tu_2(x) + \left(\overline{S(x)}\right)^2  \underrel{x\to \pm \infty}{\simeq} \frac{f K_{\alpha}}{ |x|^{1+\alpha}}  \label{eq: 2nd moment S(x) local kick general alpha}
\end{align}
with $K_{\alpha} = 2\left(\hat{u}_1*\hat{u}_1\right)(0) = 2 (\kappa_{\alpha})^{-1/\alpha} \frac{\alpha-1}{\alpha^2 \sin(\pi/\alpha)} $ and $K_1 = 2/\pi^2$.
This result can be obtained by expanding equations~\eqref{eq: hu_1(q)} and~\eqref{eq: hu_2(q) is convolution product of hu_1(q)}
near $q=0$~:
\begin{align}
\hat{u}_1(q) &= \frac{1}{\kappa_{\alpha}|q|^{\alpha} + 1} = 1-\kappa_{\alpha}|q|^{\alpha} + O(|q|^{2\alpha}) \label{eq: hu_1(q) expanded near q=0}\\
\hat{u}_2(q) &= \frac{2}{\kappa_{\alpha}|q|^{\alpha} + 1}\left[\left(\hat{u}_1*\hat{u}_1\right)(0) + O(q^2) \right] \label{eq: hu_2(q) expanded near q=0}
\end{align}
One can check that the expansion~\eqref{eq: hu_2(q) expanded near q=0} is analytic in $q$ at least to order (and including) $q^2$.
The term of order $q$ in the expansion of the convolution product must be zero because $\left(\hat{u}_1*\hat{u}_1\right)(q)$ is even. The term of order $q^2$ is a convergent integral.
We see in~\eqref{eq: hu_2(q) expanded near q=0} that $\hu_2(q)$ is proportional to $\hu_1(q)$ to first order in $q$ which means that the large scale behaviors of $\tu_2(x)$ and $\tu_1(x)$ are similar.
Taking the inverse Fourier transform of~\eqref{eq: hu_1(q) expanded near q=0}, the first term yields a $\delta(x)$, which is irrelevant for the large scale behavior. The second term $-\kappa_{\alpha}|q|^{\alpha}$ is the Fourier transform of our favorite elastic kernel $\mathcal{C}(x) = 1/|x|^{1+\alpha}$ so it gives the scaling\footnote{More precisely the inverse Fourier transform of 
$\kappa_{\alpha}|q|^{\alpha}$ is $\hat{\mathcal{C}}(0)\delta(x) - 1/|x|^{1+\alpha}$, with 
$\hat{\mathcal{C}}(0)=+\infty$, but the $\delta(x)$  is irrelevant at large 
scale.}~\eqref{eq: 1st moment S(x) local kick general alpha}.

\paragraph{Dimensionfull results}
We can set back the dimension by using the scales defined in equations~\eqref{eq: dimensionless units 1/2} and~\eqref{eq: dimensionless units 2/2} for the massive case.
One can see that $S(x)=\int \du(x,t) dt$ has dimension $S_m/\ell_m^d$ and $f=m^2\int \dot{w}(x,t)dx dt$ has dimension $m^2 S_m$.
Hence, in dimensionfull units the results~\eqref{eq: 1st moment S(x) local kick general alpha} and ~\eqref{eq: 2nd moment S(x) local kick general alpha} now read~:
\begin{align}
\overline{S(x)} &\underrel{x\to \pm \infty}{\simeq} \frac{1}{m^2 \ell_m} f \left( \frac{\ell_m}{|x|}\right)^{1+\alpha} \label{eq: 1st moment S(x) local kick general alpha dimensionfull} \\
\overline{S^2(x)} &\underrel{x\to \pm \infty}{\simeq} \frac{S_m}{m^2 \ell_m^2} f K_{\alpha} \left( \frac{\ell_m}{|x|}\right)^{1+\alpha} \label{eq: 2nd moment S(x) local kick general alpha dimensionfull}
\end{align}


\paragraph{Interpretation and consequence for the large scale decay of the avalanche diameter}
We see in equations~\eqref{eq: 1st moment S(x) local kick general alpha} and~\eqref{eq: 2nd moment S(x) local kick general alpha}
that the first and second moment of $S(x)$ are of the same order of magnitude at large $x$.
This property suggests intermittency at large $x$, i.e. $S(x)$ is almost zero most of the time but there are rare events, that arise with a small probability $p(x)  \sim f/x^{1+\alpha}$, where $S(x)$ is of order one: $S(x) \sim 1$.
This property allows to set a bound on the probability of the avalanche diameter.
Indeed $S(x) \sim 1 \Rightarrow D \geq |x|$ so it follows that~:
\begin{equation}\label{eq: bound proba diameter}
\mathbb{P}_f\left( D \geq |x| \right) \geq \mathbb{P}_f\left( S(x) \sim 1 \right) \sim \frac{f}{x^{1+\alpha}}
\end{equation}
Assuming that  at large scale the probability density function of the diameter decays as a power law with an exponent $\tilde{\kappa}_D$, the probability of $D$ being larger that $x$ scales as~:
\begin{equation}
\mathbb{P}_f\left( D \geq |x| \right) \sim  f\int_{|x|}^{+\infty} \frac{dD}{D^{\tilde{\kappa}_D}} \sim \frac{f}{|x|^{\tilde{\kappa}_D-1}} \, .
\end{equation}
Then the inequality in~\eqref{eq: bound proba diameter} implies that~:
\begin{equation}\label{eq: bound kappatilde_D}
\tilde{\kappa}_D \leq 2+\alpha  \, .
\end{equation}
This bound is consistent with the result obtained in chapter~\ref{chap: cluster statistics} for a one-dimensional interface which is that the large scale exponent of the diameter is 
$\tilde{\kappa}_D =  1+\alpha$.

\paragraph{Remark about the massless case}
In the massless case, $A=0$, we know that the moments $\overline{S}$ (resp. $\overline{S^2}$) of the total avalanche sizes are infinite since there
is no more cutoff for large avalanches, and $\tau=3/2 < 2$ (resp. $\tau<3$).
Note however that for $\alpha < 1$ the first moment of the local 
avalanche size $\overline{S(x)}$ following a kick in $x=0$ is finite. Indeed one has
\begin{equation}
\overline{S(x)} = f \tu_1(x) = f \int \frac{dq}{2\pi} \frac{e^{i q x}}{\kappa_{\alpha} |q|^{\alpha}} 
= f \frac{\sin \left(\frac{\pi  \alpha}{2}\right) \Gamma (1-\alpha)}{ \kappa_{\alpha}\pi |x|^{1-\alpha}} \, .
\end{equation}
The finiteness of the moment is consistent with the fact that from~\eqref{eq: tau_(0,loc)}
one has $\tau_{0, \text{loc}} = 1 + \frac{\alpha}{2\alpha-1} > 2$ for $\alpha<1$.
For $\alpha \geq 1$ however even the first moment is infinite (which is consistent with 
$\tau_{0, \text{loc}}<2$).
This is why here we focus on the massive case, where
some information can be gained from the moments.


\subsubsection{Computation of the first moments for an homogeneous driving $f(x)=f=m^2w$}

For an homogeneous driving $f(x)=f=m^2w$ the distribution of $S(x)$ does not depend on $x$ and is the same as the one of $S(0)=S_0$.
We see from~\eqref{eq: n-th connected moment formal formula} that~:
\begin{equation}
\overline{S^n_0}^c = f\int \tu_n(x) dx = f\hu_n(0) \, .
\end{equation}
So the first two moments of $S_0$ for $\alpha=1$ are obtained directly by setting $q=0$ in equations
\eqref{eq: hu_1(q) massive kernel alpha=1} and~\eqref{eq: hu_2(q) massive kernel alpha=1}~:
\begin{align}
\overline{S_0} &= f\hu_1(0) =  f\\
\overline{S^2_0}^c &= f\hu_2(0) = \frac{2}{\pi^2} f
\end{align}

We can also obtain the two-point correlation function by considering a source 
$\lambda(x) = \lambda g(x)$ with
$g(x) = \delta(x+r/2) - \delta(x-r/2)$ which in Fourier space reads
$\hat{g}(q) = 2i \sin (qr/2)$.
Using \eqref{eq: hu_1(q)} and \eqref{eq: hu_2(q) is convolution product of hu_1(q)},
we have for $0<\alpha<2$~:
\begin{align}
\overline{(S(x)-S(x+r))^2}^c &= \overline{(S(-r/2)-S(r/2))^2}^c=f \hu_2(0) = 2 f \left(\hu_1*\hu_1\right)(0) \\
	&= 4f \int \frac{dk}{2\pi} \frac{1-\cos(kr)}{\left(1+\kappa_{\alpha} |k|^{\alpha}\right)^2} 
\end{align}
For $\alpha=1$ the integral can be computed exactly and yields (using $\kappa_{1}=\pi$)~:
\begin{equation}
\overline{(S(x)-S(x+r))^2}^c = f \frac{2r}{\pi^3}
 \left(2 \text{Ci}\left(\frac{r}{\pi}\right) \sin \left(\frac{r}{\pi}\right)+\left(\pi
   -2 \text{Si}\left(\frac{r}{\pi}\right)\right) \cos \left(\frac{r}{\pi}\right)\right)
\end{equation}
We have the small and large $r$ expansions~:
\begin{equation}
\overline{(S(x)-S(x+r))^2}^c = \left\lbrace
\begin{array}{l l}
f\left(\frac{2r}{\pi^2} + O(r^2 \ln (r))\right) \; & \text{ when } r \to 0 \\
f\left(\frac{4}{\pi^2} - \frac{8}{r^2} + O\left(\frac{1}{r^3}\right) \right) \; & \text{ when } r \to \infty
\end{array}\right.
\end{equation}
For $\alpha \neq 1$ we are not able to compute the integral but the small and large $r$ behaviors can be obtained from the large and small $k$ behaviors of the integrand respectively. 
The scalings are~:
\begin{equation}
\overline{(S(x)-S(x+r))^2}^c \sim \left\lbrace
\begin{array}{l l}
f r^{\min(2, 2\alpha-1)} \; & \text{ when } r \to 0 \\
f\left(\beta_0- \frac{\beta_{\alpha}}{r^{1+\alpha}}  \right) \; & \text{ when } r \to \infty
\end{array}\right.
\end{equation}
where $\beta_{0}$ and $\beta_{\alpha}$ are some undetermined constants.

\subsubsection{Spatial shape at fixed avalanche size}

We now focus on computing the first moments conditioned by the total size of the avalanche to be fixed.
The average shape of the avalanche for a fixed total size S and following a kick at $x=0$, 
$\overline{S(x)}^S$,  was calculated in~\cite{thiery2015} for SR and LR elasticity with $\alpha=1$. 
Let us present the result in our units. One has in the limit $f \to 0$~:
\begin{equation}
\overline{S(x)}^S = 2 \sqrt{\pi} S^{3/2} e^{S/4} \mathcal{L}^{-1}_{\mu \to S} \left[ \int \frac{dq}{2 \pi} \frac{e^{iqx}}{\kappa_{\alpha} |q|^{\alpha} + \sqrt{1+ 4 \mu}} \right]
\end{equation}
where $\mathcal{L}^{-1}$ is the inverse Laplace transform operator.
Note that the prefactor is the inverse of the avalanche density $\rho(S)$ (cf. equation~\eqref{eq: rho(S)}) and comes from conditioning at fixed $S$.
Performing the inverse Laplace transformed followed by the Fourier integral
one obtains for $\alpha=1$~:
\begin{equation}
\overline{S(x)}^S = \frac{1}{\pi^2} \left( \sqrt{\pi S} - |x| e^{\frac{x^2}{\pi^2 S}} \text{erfc}  
 \left( \frac{|x|}{\pi\sqrt{S}} \right) \right)  
   \simeq \left\lbrace
   \begin{array}{l l}
   \frac{\sqrt{S}}{\pi^{3/2}} - \frac{|x|}{\pi^2} \; & \text{ when } x\to 0 \\
   \frac{\sqrt{\pi} S^{3/2}}{2 x^2} \; & \text{ when } x\to \pm\infty
   \end{array}\right.
\end{equation}
with erfc the complementary error function defined by
$\text{erfc}(z)= \frac{2}{\sqrt{\pi}} \int_z^{+\infty} e^{-t^2} dt$.
Note that there is a cusp at the origin and that the power of the
decay at large $x$ (i.e. $1/x^2$) is the same as the one we obtained above for $\overline{S(x)}$
(but the prefactor is different). 

\vspace{0.3cm}
We will use our formal expansion, with a different source, to compute the large $x$ behaviors of the first moments for any $\alpha$ between 0 and 2.

\paragraph{Strategy for computing $\overline{S(x)}^S$ and $\overline{S^2(x)}^S$}

Consider the instanton equation with the source $\Lambda(x) = -\mu + \lambda \delta(x-x_0)$ with 
$\mu>0$ and $\lambda<0$. With this source the generating functional reads~: 
\begin{align}
G[\Lambda(x),f(x)] &= \overline{\exp\left(\int \lambda(x) S(x) dx \right)}
	=\overline{\exp\left(-\mu S + \lambda S(x_0)\right)} \\
	&= \int \int e^{-\mu S + \lambda S(x_0)} P_f(S, S(x_0)) dS dS(x_0)
\end{align}
We see that the total size $S$ appears in the generating functional. 
Writting $P_f(S, S(x_0))=P_f(S(x_0)|S) P_f(S)$ and taking derivatives with respect to $\lambda$ evaluated in $\lambda=0$ we obtain the Laplace transforms of the first moments conditioned by $S$~:
\begin{align}
\partial_{\lambda} G _{|\lambda=0} &=  \int e^{-\mu S} P_f(S) \int S(x_0)  P_f(S(x_0)|S) dS(x_0)  dS  \notag\\
			&= \mathcal{L}_{S\to \mu} \left( \overline{S(x_0)}^S P_{f}(S) \right) \\
\partial^2_{\lambda} G _{|\lambda=0} &=  \int e^{-\mu S} P_f(S) \int S^2(x_0)  P_f(S(x_0)|S) dS(x_0)  dS   \notag\\
			&= \mathcal{L}_{S\to \mu} \left( \overline{S^2(x_0)}^S P_{f}(S) \right)		
\end{align}
On the other hand by expanding the solution of the instanton equation as 
\begin{equation}\label{eq: expansion for spatial shape}
\tu^{\Lambda}(x) = \tu_0(x) + \lambda \tu_1(x) + \frac{\lambda^2}{2} \tu_2(x) + ...
\end{equation} 
we have~:
\begin{align}
G[\Lambda(x),f(x)] &= \exp\left( \int f(x) \tu(x) dx \right) \\
\partial_{\lambda} G _{|\lambda=0} &=  \left(\int f(x) \tu_1(x) dx \right) e^{\int f(x) \tu_0(x) dx} \\
\partial^2_{\lambda} G _{|\lambda=0} &= \left(\int f(x) \tu_2(x) dx + 
\left( \int f(x) \tu_1(x) dx\right)^2 \right)  e^{\int f(x) \tu_0(x) dx}
\end{align}
Matching the two expressions for $\partial_{\lambda} G _{|\lambda=0}$ and $\partial^2_{\lambda} G _{|\lambda=0}$ we obtain the formulae~:
\begin{align}
\overline{S(x_0)}^S  &= \frac{1}{P_{f}(S)} \mathcal{L}^{-1}_{\mu \to S} \left(\left(\int f(x) \tu_1(x) dx \right) e^{\int f(x) \tu_0(x) dx}  \right) \\
\overline{S^2(x_0)}^S &= \frac{1}{P_{f}(S)}  \mathcal{L}^{-1}_{\mu \to S} \left( \left(\int f(x) \tu_2(x) dx + 
\left( \int f(x) \tu_1(x) dx\right)^2 \right)  e^{\int f(x) \tu_0(x) dx} \right)
\end{align}

\paragraph{Sketch of the calculation and results}
The first term of the expansion~\eqref{eq: expansion for spatial shape} is actually independent of $x$ and solution of the equation $\tu^2_0-\tu_0 = \mu$.
It reads $\tu_0 = (1-\sqrt{1+4\mu})/2$.
The expansion can be perfomed as above (starting from~\eqref{eq: Instanton fourier for expansion}) by making the change~:
\begin{equation}
(\kappa_{\alpha} |q|^{\alpha} +1) \; \rightarrow \; (\kappa_{\alpha} |q|^{\alpha} +1)- 2 \tu_0 = (\kappa_{\alpha} |q|^{\alpha} +\sqrt{1+4\mu}) \, .
\end{equation}
The following terms of the expansion are thus~:
\begin{align}
\hu_1(q) &= \frac{e^{iqx_0}}{\kappa_{\alpha} |q|^{\alpha} +\sqrt{1+4\mu}} \\
\hu_2(q) &= \frac{2}{\kappa_{\alpha} |q|^{\alpha} +\sqrt{1+4\mu}} (\hu_1*\hu_1)(q)
\end{align}
To obtain the large $x$ behaviors, we perform a small $q$ expansion followed by an inverse Fourier transform and finally take the inverse Laplace transform.
Skipping the details of the calculation, the large $x$ behaviors for $0< \alpha <2$, following a local kick $f(x)=f\delta(x)$ and in the limit $f\to 0$ are~:
\begin{align}
\overline{S(x)}^S  &\underrel{x\to\pm\infty}{\simeq} \frac{\sqrt{\pi}}{2} \frac{S^{\frac{3}{2}}}{|x|^{1+\alpha}}\\
\overline{S^2(x)}^S &\underrel{x\to\pm\infty}{\simeq} \gamma_{\alpha} \frac{S^{\frac{5\alpha-1}{2\alpha}}}{|x|^{1+\alpha}}
\end{align}
where $\gamma_{\alpha}$ is an undetermined constant.
We also used that $\underrel{f\to 0}{\lim} P_f(S) = \rho(S)$ defined in equation~\eqref{eq: rho(S)}.
In dimensionfull units these results become~:
\begin{align}
\overline{S(x)}^S  &\underrel{x\to\pm\infty}{\simeq} \frac{\sqrt{\pi}}{2} \frac{\ell_m^{\alpha}}{\sqrt{S_m}} \frac{S^{\frac{3}{2}}}{|x|^{1+\alpha}} = \frac{\sqrt{\pi}}{2\sqrt{\sigma}} \frac{S^{\frac{3}{2}}}{|x|^{1+\alpha}} \label{eq: 1st moment S(x) conditonned by S general alpha}\\
\overline{S^2(x)}^S &\underrel{x\to\pm\infty}{\simeq} \gamma_{\alpha} 
 \frac{\ell_m^{\alpha-1}}{\sqrt{S_m^{1-1/\alpha}}} \frac{S^{\frac{5\alpha-1}{2\alpha}}}{|x|^{1+\alpha}} = \gamma_{\alpha} \sigma^{\frac{1-\alpha}{2\alpha}} \frac{S^{\frac{5\alpha-1}{2\alpha}}}{|x|^{1+\alpha}} \label{eq: 2nd moment S(x) conditonned by S general alpha}
\end{align}
The dimensionfull results only depend on the amplitude $\sigma$ of the Brownian force field. 
They do not depend on the mass $m$ so they are also valid in the limit $m\to 0$.
Remark also that the moments conditioned by the avalanche size $S$ have the same large $x$ scaling as the unconditioned moments obtained above.
Hence the same discussion about intermittency can be made and the same conclusions hold. 

\subsubsection{Alternative choice of elasticity}

\begin{figure}
	\centering
	\includegraphics[scale=0.6]{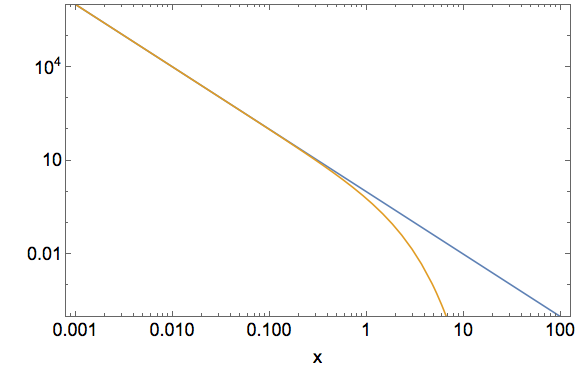}
	\caption{Log-log plot of $\mathcal{C}(x)$ (blue line) and $\mathcal{K}(x)$ (yellow line, bottom) with $\mu=1$ for $\alpha=1$. $\mathcal{K}(x)$ shows the same power-law behavior as $\mathcal{C}(x)$ at small $x$ but it is cut at $x=\frac{1}{\mu}$.\label{fig: twoDifferentElasticities}} 
\end{figure}

In this thesis I mainly worked with a long-range kernel that in Fourier space gives a term
$-(m^2 +|q|^{\alpha})$ in front of $\hu(q)$.
However the wetting line is actually described by another elastic kernel which in Fourier space gives a term $-\sqrt{q^2+\mu^2}$ in front of $\hu(q)$~\cite{ledoussal2010}. 
This kernel corresponds to $\alpha=1$. An essential difference with 
$-(m^2+|q|)$ is that it is analytic in $q$. We can generalize it to arbitrary value of $\alpha$
by making the change:
\begin{align}\label{eq: kernel change}
-\left(m^2 + \abs{q}^{\alpha}\right) \ \rightarrow \ -(q^2+\mu^2)^{\frac{\alpha}{2}} \quad \text{with} \quad \mu^{\alpha} = m^2 \, .
\end{align}
The last equality ensures that the two kernels are equal in $q=0$.
This defines an alternative elasticity. Its kernel, denoted by $\mathcal{K}$, reads, in Fourier and real space~:
\begin{gather}
\hat{\mathcal{K}}(q) = -(q^2+\mu^2)^{\frac{\alpha}{2}} \label{eq: def elasticity analytic fourier space}\\
\mathcal{K}(x) = \frac{-1}{\sqrt{\pi} \Gamma\left(-\frac{\alpha}{2}\right)}
	\left(\frac{2 \mu}{|x|} \right) ^{\frac{1+\alpha}{2}} K_{\frac{\alpha+1}{2}}(\mu |x|) \label{eq: def elasticity analytic real space}
\end{gather}
where $K$ is the modified Bessel function of the second kind.
Indeed one can check that~:
\begin{equation}
\int {\cal K}(x-y) \left(u(y)-u(x)\right) dy - \mu^{\alpha} u(x) = \int e^{iqx} \left(- (q^2 + \mu^2)^{\frac{\alpha}{2}} \right) \hu(q) \frac{dq}{2\pi} \, .
\end{equation}
The elasticity kernel $\mathcal{K}(x)$ is plotted along $\mathcal{C}(x)$ for comparison in figure~\ref{fig: twoDifferentElasticities}. It is exactly the same power law at small $x$ but $\mathcal{K}(x)$ has an exponential cutoff at $x=1/\mu$.
This cutoff induces a different physics where all power laws now have some cutoffs controlled by $\mu$.
With this kernel we have $\hu_1(q) = (q^2+\mu^2)^{\frac{-\alpha}{2}}$ 
for a local source $\lambda(x)=\lambda \delta(x)$ 
and the first moment 
following a local kick $f(x)=f\delta(x)$ is~:
\begin{align}
\overline{S(x)} &= f\tu_1(x) = \frac{f}{\sqrt{\pi} \Gamma\left(\frac{\alpha}{2}\right)}
	\left(\frac{2 \mu}{|x|} \right) ^{\frac{1-\alpha}{2}} K_{\frac{\alpha-1}{2}}(\mu |x|)
\end{align}
which becomes for $\alpha=1$~:
\begin{equation}
\overline{S(x)}= \frac{f}{\pi} K_0(\mu |x|) \, .
\end{equation}
Due to the analyticity in $q$ of this elastic kernel, the large distance
decay in $x$ of the observables computed above 
$\overline{S(x)}\,$, $\overline{S^2(x)}\,$, $\overline{S(x)}^S\,$, $\overline{S^2(x)}^S\,$ 
are all exponentials. 
Note however that, since both kernel are equal in the limit $m^2=\mu^{\alpha}\to 0$
and since the results~\eqref{eq: 1st moment S(x) conditonned by S general alpha}, \eqref{eq: 2nd moment S(x) conditonned by S general alpha} derived for the first kernel, do not depend on the mass and are valid when $m\to 0$,
we can expect these results to hold also for this kernel in a broad
range of $|x| < \ell_m$.

\section{Conclusion}

In this chapter I presented two classical mean-field models of the depinning transition, namely the fully-connected model and the ABBM model, and showed how the first one can be mapped onto the second one.
Then I introduced a more recent model, the Brownian force model (BFM), which 
captures the full space-time statistics of the velocity field. This allows to study the spatial structure of the avalanches, whereas spatial structure is absent from the fully-connected and ABBM models. 
I explained the method to obtain the statistics of different observables in the BFM. 
It relies on calculating the generating functional of the velocity field and requires to solve 
the instanton equation with a source term that is adequate for the observable we want to compute. 
Global observables are independent of the elasticity and had been obtained before. 
Regarding spatial observables previous results had been derived mainly for SR elasticity, for which the full distributions of local jumps and of the extension of the avalanches had been computed for an homogeneous driving~\cite{delorme2016}.
For LR elasticity, only the mean spatial shape at fixed size for $\alpha=1$ had previously been obtained~\cite{thiery2015}.
Solving the instanton equation which encodes for the statistics of the local jumps with LR interactions 
is very difficult, as this equation is non-local and non-linear. 
Nevertheless I made some progress in three directions. 
First I proposed a scaling Ansatz for the solution
which allows to predict the exponent of the local avalanche sizes
for homogeneous driving ($\tau_0=1$ for $\alpha=1$) and 
local driving ($\tau_{0,\text{loc}}=2$ for $\alpha=1$). 
Second, I obtained the generic large distance decay of the solution of this equation, 
which allowed to set a bound on the exponent of the clusters extensions. 
Finally, a formal expansion of the solution in powers of $\lambda$ allowed to
capture the large $x$ behavior of the two first moments of the local jumps, $\overline{S(x)}$ and $\overline{S^2(x)}$, for any $\alpha$ between 0 and 2. 
The two first moments of $S(x)$ conditioned by the total size $S$ of the avalanche, $\overline{S(x)}^S$ and $\overline{S^2(x)}^S$, were obtained similarly.
It is interesting to note that, following a kick in $x=0$, the two first moments have the same decay in $1/x^{1+\alpha}$ at large $x$ and this with or without conditioning by the avalanche size.
This suggests a kind of intermittency, i.e. that $S(x)$ is very small most of the time but that 
the moments are dominated by rare events, where $S(x)$ is large, of order 1, that arise with a probability proportional to $1/x^{1+\alpha}$. 
Thus a cluster is triggered at distance $x$ from the seed of the avalanche with a probability 
$\sim 1/x^{1+\alpha}$.
This scaling corresponds exactly to the kernel that mediates the elastic interactions.
We remark that this scaling is also the one of the large scale decay of the spatial correlation function of the velocity field $C_v(x)$ predicted in chapter~\ref{chap: chap3} and of the gaps and diameter distributions, $P(g)$ and $P(D)$, observed in chapter~\ref{chap: cluster statistics}. 
This connection strongly suggests that the scalings derived for the BFM are actually valid beyond 
mean-field.
Note that the results we obtained here for $d=1$ also describe the size of the avalanche restricted to a plane in $d=2$ which is defined as $S(x)=\int S(x,y) dy$. These results can thus directly describe 
two dimensional magnetic domain walls with LR interactions.

Regarding future works on the BFM, obtaining moments of higher order is doable in principle but becomes increasingly difficult.
Calculating the n-point correlation function $\overline{S(x_1)...S(x_n)}$ with various conditioning
could unravel information about the statistics of the number of clusters $N_c$ inside the avalanche. 
An important progress would be to solve the instanton equation with a number of local sources. 
This remains a difficult challenge.


%% file: Conclusion/Conclusion.tex


\chapter*{Summary and perspectives}\label{chap: conclusion}
\addcontentsline{toc}{chapter}{Summary and perspectives}%

I briefly summarise the main results of this thesis and some perspectives for future works.
Detailed conclusions can be found at the end of each chapter.

I started by introducing the main concepts of the theory of depinning of elastic interfaces. 
I presented three experimental systems that are well described by this theory, namely the magnetic domain walls, the crack fronts and the wetting fronts, and two others, earthquakes and avalanches of plastic events, that are different but for which connections with the depinning have been pointed out.

The first result that I obtained is the universal scaling form of the two-point correlation functions of the local velocity field. 
This result is important for several reasons.
First it is directly relevant for experimental applications 
since these functions had been previously observed in the Oslo crack experiment but were not fitted with the correct scaling forms~\cite{tallakstad2011}. 
It also provides a novel method, simple and robust, to assess the universality class of the transition. 
Finally characterizing the correlation functions of the order parameter of a phase transition is in itself of theoretical interest.
Besides testing it on different experiments, an interesting perspective would be to extend this scaling to other types of avalanches. 

The second work was to understand the statistics of the clusters that form in LR avalanches. This study was performed in the quasi-static limit for various values of the parameter 
$\alpha$ which describes the range of the elastic interactions.
I established the full set of scaling relations linking the statistics of the clusters to the one of the global avalanches.
These relations can be tested on fracture or wetting-line experiments.
The origin of these relations is even more interesting.
Numerical simulations unraveled fascinating scalings for the statistics of the number of clusters and of the gaps with an exponent $3/2$ that does not seem to depend on $\alpha$.
Regarding the number of clusters, a mysterious connection with the Bienaym{\'e}-Galton-Watson process seems to exist.
Explaining these results poses a first challenge for future works. 
Another perspective would be to find the connection with different scaling relations relating clusters and global avalanches that have been obtained for a finite velocity driving~\cite{planet2018}.
A third one would be to investigate if clusters of events can be defined and analyzed in other kinds of avalanches, such as in plasticity, and see whether similar scaling relations could be derived. 

The third part of my work was to investigate the spatial structure of avalanches at the mean-field level using the Brownian force model. Solving the non-local non-linear instanton equation that would give access to the full probability distribution of the local jumps remained out of my reach. 
I could nevertheless obtain the large distance behaviors of the first two moments of this distribution thanks to a formal expansion of the solution of the equation.
For a one-dimensional interface the same decay in $1/x^{1+\alpha}$ is shared by the first and second moments, with and without conditioning by the avalanche size. This suggests that triggering a cluster at distance $x$ far from the seed of the avalanche happens with a probability of order $1/x^{1+\alpha}$.

The emergence of the $1/x^{1+\alpha}$ decay of the elastic interactions in many different observables
is a common feature that stands out from chapters 3, 4 and 5.
This result is new and had, to the best of my knowledge, not been expected before.
It offers various possibilities for probing the range of the interactions in experimental systems.


%% file: Appendix/Appendix.tex


\chapter*{Appendix}\label{chap: Appendix}
\addcontentsline{toc}{chapter}{Appendix}
\renewcommand{\thesection}{\Alph{section}}

\setcounter{figure}{0}
\makeatletter 
\renewcommand{\thefigure}{C.\@arabic\c@figure}
\makeatother

\section{Fourier transform of the elastic force}\label{sec:Appendix Fourier Transform elastic force}

In this section I give details about the computation of the Fourier transform of the LR elastic force in the one-dimensional case ($d=1$)~:
\begin{equation}\label{eq:def elasticity appendix}
F_{\el}^{\LR} (x,t) =  \int_{-\infty}^{\infty} \mathcal{C}(x-y)\left( u(y,t)-u(x,t)\right) dy.
\end{equation}
I use the following convention for the Fourier Transform and its inversion~:
\begin{equation}\label{eq: def Fourier Transform appendix}
\hat{h}(q) = \int_{-\infty}^{\infty} e^{-iqx} h(x) dx \, , \quad
h(x) = \int_{-\infty}^{\infty} e^{iqx} \hat{h}(q)  \frac{dq}{2\pi}
\end{equation}
where $h$ is an arbitrary (integrable) function.
In the following I will write $\int$ instead of $\int_{-\infty}^{\infty}$.
Dropping the time $t$ the elastic force can be rewritten as follows~:
\begin{align*}
\int  \mathcal{C}(y-x)\left(u(y)-u(x)\right) dy &= 
	\int \frac{dqdk}{(2\pi)^2} \hat{ \mathcal{C}}(q)\hu(k) \int \left[ e^{-iqx}e^{i(q+k)y}- e^{i(k-q)x}e^{iqy}\right] dy\notag \\
	&= \int \frac{dqdk}{(2\pi)^2} \hat{ \mathcal{C}}(q)\hu(k) 2\pi \left[ e^{-iqx} \delta(q+k) - e^{i(k-q)x} \delta(q)\right] \notag \\
	&= \int \frac{dk}{2\pi} \left(\hat{ \mathcal{C}}(q) - \hat{ \mathcal{C}}(0)\right) \hu(k) e^{ikx}	
\end{align*}
where I used that $\hat{ \mathcal{C}}(-q) = \hat{ \mathcal{C}}(q)$ because $ \mathcal{C}$ is even.
This shows that 
\begin{equation}
\hat{F}_{\el}^{\LR} (q,t) = \left( \hat{\mathcal{C}}(q) - \hat{\mathcal{C}}(0)\right)\hat{u}(q,t).
\end{equation}

\paragraph{Computation for the specific kernel}
I now focus on the computation for the specific kernel $\mathcal{C}(x) = c/|x|^{1+\alpha}$.
First note that when computing the Fourier transform there is a divergence for $x\to 0$ so that
$\hat{\mathcal{C}}(q) = +\infty$.
However we are interested in the term $\left( \hat{\mathcal{C}}(q) - \hat{\mathcal{C}}(0)\right)$ which remains finite for $0< \alpha < 2$.
In this case, using that $\hat{ \mathcal{C}}(-q) = \hat{ \mathcal{C}}(q)$, we have~:
\begin{align}\label{eq: TF(C(q)-C(0)), continuum}
\hat{\mathcal{C}}(q)-\hat{\mathcal{C}}(0) &= c\int \frac{\cos(qx) - 1}{\abs{x}^{1+\alpha}}dx = \abs{q}^{\alpha} c \int \frac{\cos(s)-1}{\abs{s}^{1+\alpha}}ds \notag \\
	&= \left\lbrace
\begin{array}{l c}
2 c \cos\left(\frac{\alpha\pi}{2}\right) \Gamma(-\alpha) \abs{q}^{\alpha} & \text{for }
0< \alpha < 2 \text{ and } \alpha \neq 1 \, , \\
-c\pi |q| & \text{for } \alpha=1 \, .
\end{array}\right.
\end{align}
Note that $\Gamma(-\alpha) \cos\left(\frac{\alpha \pi}{2}\right) <0$ and hence the prefactor in front of $|q|^{\alpha}$ is always negative.
If $\alpha>2$ there is a divergence in the integral for $s\to 0$ because $\cos(s)-1 \sim s^2$.
In this case we need to regularize the integral by introducing a small scale cutoff.
Let us denote $m$ the integer part of $\alpha/2$, i.e. we assume $2m < \alpha < 2(m+1)$.
We write 
\begin{equation*}
\cos(qx)-1 = \sum_{n=1}^{m} (-1)^n \frac{(qx)^{2n}}{(2n)!} + g(qx) \, .
\end{equation*}
$\int \frac{g(qx)}{|x|^{1+\alpha}} dx$ converges and a change of variable $s:=qx$ makes appear 
$|q|^{\alpha} \int \frac{g(s)}{|s|^{1+\alpha}} ds$.
For the other terms we regularize the integral with a small scale cutoff $x_0$. We obtain~:
\begin{align*}
\int_{-\infty}^{\infty} \frac{(qx)^{2n}}{|x|^{1+\alpha}}dx \rightarrow 
	2 \int_{x_0}^{\infty} \frac{(qx)^{2n}}{|x|^{1+\alpha}}dx = 2q^{2n} \int_{x_0}^{\infty}\frac{x^{2n}}{|x|^{1+\alpha}} dx
	= \frac{2 x_0^{2n-\alpha}}{\alpha-2n} q^{2n}
\end{align*}
The Fourier Transform can thus be formally written as~:
\begin{equation}\label{eq: LR elasticity kernel Fourier 2}
\hat{\mathcal{C}}(q) - \hat{\mathcal{C}}(0) =  \sum_{n=1}^{\left\lfloor \frac{\alpha}{2} \right\rfloor} c_n q^{2n} + c_{\alpha} |q|^{\alpha} \, ,
\end{equation}
where $\left\lfloor \frac{\alpha}{2} \right\rfloor$ denotes the integer part of $\frac{\alpha}{2}$, 
$c_n = (-1)^n\frac{2 x_0^{2n-\alpha}}{\alpha-2n}$ and $c_{\alpha} = \int \frac{g(s)}{|s|^{1+\alpha}} ds$.

\section{Computation of the elastic coefficients for numerical simulations}\label{sec: appendix computation elasticity FFT}


In the numerics I use a slightly different convention for the elasticity coefficient than in the continuous setting~\eqref{eq:def elasticity appendix}. 
Dropping the time $t$ the elastic force is defined as~:
\begin{equation}\label{eq: appendix Fel 1}
F_{\el}(x) = \sum_{y=0}^{L-1} \tilde{\mathcal{C}}(y-x) u(y)
\end{equation}
There are two methods to compute the elastic coefficients $\tilde{\mathcal{C}}(y)$. 
One, called the \emph{image method}, computes the coefficient in real space. This costs $O(L^2)$ operations. 
The other uses a Fast Fourier Transform and costs $O\left(L\ln (L)\right)$ operations. It is therefore much faster when working on large systems.

\subsubsection{The image method}
In the image method the elastic coefficients $\tilde{\mathcal{C}}(y)$ are computed in real space.
The periodic boundary conditions are taken into account by summing the coefficients over several "images" of the line.
Fo $y\neq 0$ we compute~:
\begin{equation}\label{eq:image method appendix}
\tilde{\mathcal{C}}(y) = \frac{1}{\mathcal{N}} \sum_{n=-N}^{n=N} \frac{1}{|y-nL|^{2}} \, ,
\end{equation}
where $N$ must be high enough so that the remaining sum 
$\sum_{n=N+1}^{+\infty} \frac{1}{|y-nL|^{2}}$ is much smaller than $\mathcal{N} \tilde{\mathcal{C}}(y)$.
I used this method in the early time of my PhD with $N=7$. 
Note that we have the symmetry $\tmC(-y) = \tmC(L-y)=\tmC(y)$.
So in practice it is sufficient to compute $\tmC(y)$ for $y$ ranging from $1$ to $L/2$.
The choice of the normalization constant $\mathcal{N}$ is determined by the condition~:
\begin{equation}
\sum_{y=1}^{L-1} \tilde{\mathcal{C}}(y) + \tilde{\mathcal{C}}(0) = 0 \, , \quad \text{with } \tilde{\mathcal{C}}(0)=-2 \, ,
\end{equation}
which ensures that $F_{\el}(x)=0$ when the line is flat.
The choice of $\tilde{\mathcal{C}}(0)=-2$ is arbitrary but ensures the agreement with a standard choice when discretizing the laplacian ($\Delta_x u(x) \rightarrow u(x+1)+u(x-1)-2u(x)$).
Computing the elastic force with formula \eqref{eq: appendix Fel 1} requires $O(L)$ operation per site 
hence $O(L^2)$ operations to update the elastic force on the whole line. 

\subsubsection{The Fast Fourier Transform method}
The Fast Fourier Transform (FFT)  computes the Fourier transform of an array of size $L$ in $O\left(L\ln (L)\right)$ operations.
The use of the FFT to update the elasticity relies on~:
\begin{align*}
F_{\el}(x) &= \sum_{y=0}^{L-1} \tmC(y-x) u(y) \\
		   &= \sum_{y=0}^{L-1} \frac{1}{L} \sum_{q_1} e^{iq_1(y-x)} \hmC(q_1) \frac{1}{L} \sum_{q_2} \hat{u}(q_2) e^{iq_2y} \\
		   &= \frac{1}{L^2} \sum_{q_1,q_2} \hmC(q_1)\hat{u}(q_2) e^{-iq_1 x} \sum_{y=0}^{L-1} e^{i(q_1+q_2)y} \\
		   &= \frac{1}{L} \sum_{q_2} \hmC^{\dagger}(q_2)\hat{u}(q_2) e^{iq_2 x}
\end{align*}
where in the third from fourth line we have used $\sum_{y=0}^{L-1} e^{i(q_1+q_2)y} = L \delta(q_1+q_2)$ and $\hmC(-q_2) = \hmC^{\dagger}(q_2)$.
We see that $F_{\el}$ is the inverse Fourier transform of $\hmC^{\dagger}\hat{u}$.
So at each time step the elasticity force is updated by 
\begin{itemize}
\item[(i)] taking the FT of $u$ using an FFT routine, 
\item[(ii)] multiplying by$ \hmC^{\dagger}$ and
\item[(iii)] taking the inverse FT of the product.
\end{itemize}
Steps (i) and (iii) each cost $O\left(L\ln (L)\right)$ operations while step (ii) only costs $O(L)$ operations.

\section{Analysis of the experimental data}\label{sec: appendix experiment analysis}

\begin{figure}[t!]
	\centering
	\includegraphics[width=1\linewidth]{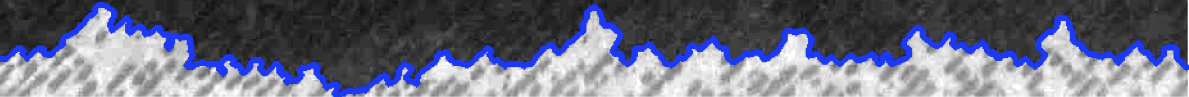}
	\caption{Picture of the d'Alembert experiment. The crack front, colored in blue, is identified as the interface between a light phase and a dark phase corresponding respectively to the broken and non-broken part of the sample. Image courtesy L. Ponson. \label{Fig: crack front image appendix}}
\end{figure}

On the raw pictures of the fracture plane, the crack front appears as the interface between a light region and a dark one (see figure \ref{Fig: crack front image appendix}).
The front has many overhangs which are not accounted for by the standard elastic interface model where the front is described by a single-valued function $u(x,t)$.
This first image of the front is then processed to obtain a single-valued function $u(x,t)$, defined as the maximum value of the front position for a given $x$.
200 000 frames are taken at a rate of one frame per second. The pixel size is $p=34.7$ $\mu$m.
The observed region has a lateral extension of $1800$ pixels, i.e. of $62.46$ mm.
I analyzed data encoding the position of the front for each of the 200 000 successive frames.

\begin{figure}[tb!]
	\centering
	\textbf{\includegraphics[width=0.45\linewidth]{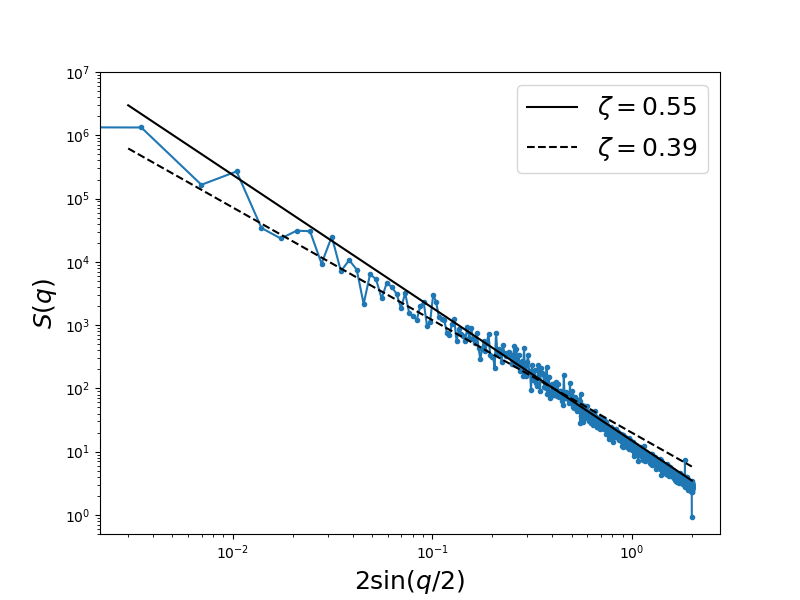}}\includegraphics[width=0.45\linewidth]{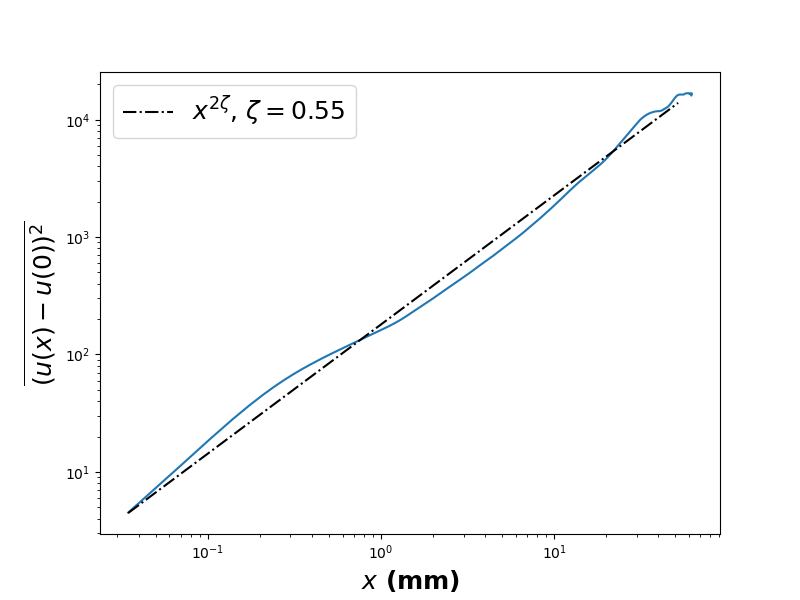}
	\caption{\textit{Left }: Structure factor of the d'Alembert experiment. 
	\textit{Right }: Autocorrelation function of the d'Alembert experiment. . \label{Fig: roughness appendix}}
\end{figure}

\paragraph{Roughness}
I studied the roughness of the interface using both the structure factor 
$S(q)=\overline{u(q)u(-q)}\sim q^{-(1+2\zeta)}$ and the autocorrelation function
$\tilde{C}(x)=\overline{\left\langle \left(u(x)-u(0)\right)\right\rangle \sim x^{2\tilde{\zeta}}}$.
For both function the average over disorder is computed by averaging over all configurations visited by the front during the experiment.
They are shown respectively in the left and right panel of figure~\ref{Fig: roughness appendix}.
The structure factor is noisy, which means that few independent configurations were visited, but two different trends $\zeta\simeq 0.6$ at large $q$ and $\zeta\simeq 0.4$ at small $q$ can be clearly distinguished.
However, the autocorrelation function can be entirely fitted with 
$x^{2\tilde{\zeta}}$ with $\tilde{\zeta}=0.55\pm 0.05$ while the structure factor presents two different trends with a crossover between $\zeta\simeq 0.55$ at small scales and $\zeta \simeq 0.39$ at large scale. 
Note that a priori we expect the LR depinning roughness exponent to be valid above the Larkin length $L_c$~\cite{santucci2010, laurson2010} which is given by
$L_c=d_0 /\sigma ^2$ where $\sigma$ is the variance of the disorder~\cite{Demery2014}.
In the d'Alembert experiment the disorder is controlled and we have $\sigma \simeq 1/2$ hence
$L_c \simeq 4 d_0 \simeq 400 \mu$m which is well below the system size.
(In the Oslo experiment the crossover was observed to be at $100 \mu$m for a disorder characteristic length $d_0=200\mu$m~\cite{santucci2010}.)


\begin{figure}[tb!]
	\centering
	\includegraphics[width=0.6\linewidth]{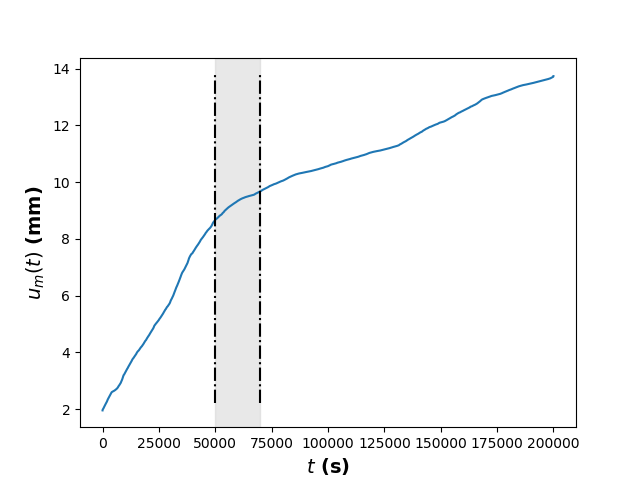}
	\caption{Mean displacement of the front versus time. Two distinct velocity regimes can be distinguished and have been analyzed separately to obtain the correlation function discussed in chapter 4. The first one for $t<50000$s corresponds to $v_1=132$ nm.s$^{-1}$ and the second one for $t>70000$s to $v_2=31$ nm.s$^{-1}$. The shaded area inbetween has not been used. \label{Fig: um(t) experiment appendix}}
\end{figure}

\paragraph{Velocity}
The data also enables to reconstruct the mean position of the front 
$u_m(t) = \sum_{x=1}^{L} u(x,t)/L$ with $L=1800$ pixels. 
I plot this mean position in figure~\ref{Fig: um(t) experiment appendix}. The slope of the curve gives the mean velocity of the front. Two distinct velocity regimes are visible and have been analyzed separately to obtain the correlation functions presented in chapter~\ref{chap: chap3}. The first one for $t<50000$s corresponds to $v_1=129$ nm.s$^{-1}$ and the second one for $t>70000$s to $v_2=31$ nm.s$^{-1}$.

\begin{figure}[tb!]
	\centering
	\includegraphics[width=0.7\linewidth]{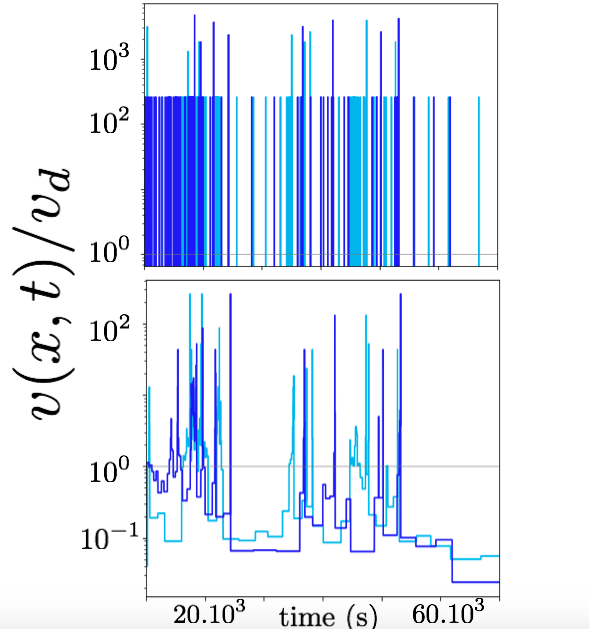}
	\caption{Velocities of two point close along the front ($x=1200$ and $x=1209$). \emph{Top~:} Direct reconstruction from the front position~: $v(x,t)=u(x,t+1)-u(x,t)$. \emph{Bottom~:} Reconstruction with the WTM using equation~\eqref{eq: velocity with WTM}. \label{Fig: 2 point velocities appendix}}
\end{figure}

There are two difficulties with the raw data. First the experimental signal is noisy with the front sometimes going backwards locally, a possibility that is ruled out by the second Middleton theorem in the model (see section~\ref{sec: Ch1 Middleton thm}) and is due to the overhangs. 
Second the local velocity of a point appears to be zero most of the time and spikes at discrete time when the front moves from one pixel to the next (see top panel of figure~\ref{Fig: 2 point velocities appendix}).

\begin{figure}[tb!]
	\centering
	\includegraphics[width=1\linewidth]{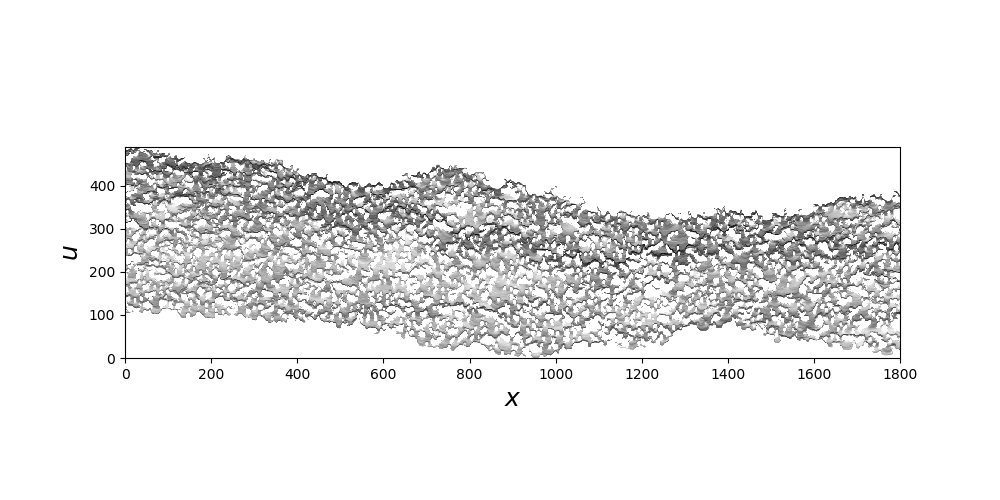}
	\caption{Waiting time matrix of the d'Alembert experiment. Units are in pixel. The greyscale map indicates the time spent by the front in each pixel. A darker shade corresponds to a longer time. There are a lot of white zones corresponding to wt=0. This zones have been "jumped" by the front. \label{Fig: WTM appendix}}
\end{figure}

The waiting time matrix, introduced in Ref.~\cite{maloy2006} 
permits to overcome both difficulties.
For each pixel we define its \emph{waiting time} as the time spent by the front inside the pixel.
Counting the time spent by the front in every pixel, we obtain a waiting time matrix (WTM).
The WTM obtained with our data is plotted in figure~\ref{Fig: WTM appendix}.
The position of the front of an abscisse $x$ at time $t$ is obtained by summing the waiting times along $x$ until the sum reaches $t$, $u(x,t) = y_0$ where $y_0$ is such that~:
\begin{equation}
\sum_{y=0}^{y_0-1}WTM(x,y) < t \; \text{  and  }\;  \sum_{y=0}^{y_0}WTM(x,y) \geq t \, .
\end{equation}
The fronts reconstructed with this method are by construction ordered in time~: $u(x,t) \leq u(x,t+1)$ for all $x$ and $t$. 

The local velocity can be estimated by dividing the pixel size by the time spent inside the pixel~:
\begin{equation}\label{eq: velocity with WTM}
v(x,t) = \frac{p}{WTM(x,y_0)} \; \text{ with } \; y_0=u(x,t)  \, .
\end{equation}
In many pixels, the waiting time is zero (see below). In this case we can consider that the front advances over many pixels between to consecutive frame. When this happens the velocity is defined as~:
\begin{equation}\label{eq: velocity with WTM 2}
v(x,t) = p N_{\text{jumped}} + \frac{p}{WTM(x,y_0)}
\end{equation}
where $N_{\text{jumped}}$ is the number of pixel jumped by the front between the frames taken at time $t-1$ and $t$ and $y_0=u(x,t)$.
With this method the local velocity is always resolved at the time step of $1$ second (which is the time interval between two frames in our experiment).
I used equations~\eqref{eq: velocity with WTM} and~\eqref{eq: velocity with WTM 2} to compute the local velocity field whose correlation functions are shown in chapter~\ref{chap: chap3}. Two local velocities are shown in figure~\ref{Fig: 2 point velocities appendix} bottom panel. This velocities are never $0$ but low velocities are characterized by large plateau corresponding to the time spent inside one pixel. Note the correspondance between the spikes on the top panel and the change of level on the bottom panel.
Both correspond to the front moving to a new pixel. 

\paragraph{Holes in the WTM}
On the plot of the WTM (figure~\ref{Fig: WTM appendix}) one can see many white zones corresponding to zero waiting time 
($WTM(x,y)=0$). 
$26.6 \%$ of the pixels localized between the initial and final configuration of the line belong to such zones. 
This could be due to the local velocity being very fast with the front advancing from many pixels in one second. 
However the sizes of the zones as well as the total number of pixel jumped by the front are surprisingly large.
Indeed the highest driving velocity being $v_1=132$ nm.s$^{-1}$ and the pixel size $p=34.7\, \mu$m, 
the average time spent inside one pixel is 265 seconds equivalent to 265 frames. Many pixels being skipped together corresponds to a velocity a few thousands time higher than the driving velocity.

\begin{figure}[tb!]
	\centering
	\includegraphics[width=0.7\linewidth]{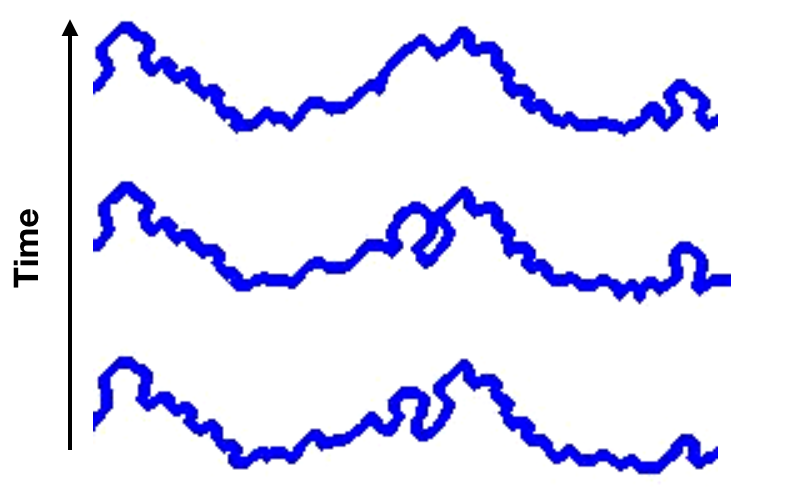}
	\caption{Zoom on a portion of the line. When two overhangs meet the new configuration of the front is defined by merging them. This leaves behind them a zone that is skipped by the reconstructed front. The waiting time for the pixels inside this zone is zero, corresponding to white regions on the WTM image~\ref{Fig: WTM appendix}. Image courtesy L. Ponson. \label{Fig: MergingOverhangs appendix}}
\end{figure}

These zero waiting time zones may originate from the many large overhangs of the front.
The reconstructed single-valued position of the front $u(x,t)$ being defined as the uppermost position of the real front for a given $x$, when a overhang developps an artificial jump is introduced in the reconstruced front.
When two overhangs meet, the new configuration of the front is defined by merging them. This leaves behind them a hole that has not been traveled by the front and where the waiting time is thereby zero.
This process is illustrated in figure~\ref{Fig: MergingOverhangs appendix}.

\section{Computation of the generating functional using the Martin-Siggia-Rose formalism}\label{sec:Appendix MSR}

In this section I sketch the computation that allows to express the generating functional
defined in chapter~\ref{chap: chap5} as a function of the instanton solution.
This computation was originally performed in~\cite{dobrinevski2012, ledoussal2013, dobrinevski2013}.
To reduce the size of equations I use subscripts notations to denote variables~:
\begin{equation}
f_{xt}:=f(x,t) \, , \quad \int_{xt} := \int d^dx dt
\end{equation}

In the Brownian Force Model, the velocity field is solution of the stochastic differential equation
\begin{equation}\label{eq:definition BFM Appendix}
\eta \partial_t \du_{xt} = m^2\left(\dot{w}_{xt}-\du_{xt}\right) + \int_y \mathcal{C}_{yx} \left(\du_{yt}- \du_{xt}\right) + \sqrt{\du_{xt}} \xi_{xt} \, ,
\end{equation}
where $\xi_{xt}$ is a Gaussian white noise with correlations
$\overline{\xi_{x t}\xi_{x' t'}}  = 2 \sigma \delta^d(x-x') \delta(t-t')$.
The statistics of the velocity field is encoded in the generating functional
\begin{equation}\label{eq:Generating functional Appendix}
G[\lambda_{xt}, \dot{w}_{xt}] := \overline{\exp \left( \int_{xt} \lambda_{xt}\dot{u}_{xt} \right)}^{\dot{w}_{xt}} \, ,
\end{equation}
where $\lambda_{xt}$ is an arbitrary function called the source.
Following the approach from Martin-Siggia-Rose (MSR), the generating functional can be expressed as an average over all solutions of the stochastic equation~\eqref{eq:definition BFM Appendix} using a path integral formalism~:
\begin{align*}
&G\left[\lambda_{x t}, \dot{w}_{x t}\right] =\int \mathcal{D}[\xi] \int \mathcal{D}[\dot{u}] e^{\int_{x t} \lambda_{x t} \dot{u}_{x t}} \prod_{x, t} \delta\left[-\eta \partial_{t} \dot{u}_{x t}+m^{2}\left(\dot{w}_{x t}-\dot{u}_{x t}\right)+\int_y \mathcal{C}_{yx}(\du_{yt}-\du_{x t})+\sqrt{\dot{u}_{x t}} \xi_{x t}\right] \\
   &\,\; =\int \mathcal{D}[\xi] \int \mathcal{D}[\dot{u}, \tilde{u}] e^{\int_{x t} \lambda_{x t} \dot{u}_{x t}} \exp \int_{x t} \tilde{u}_{x t}\left[-\eta \partial_{t} \dot{u}_{x t}+m^{2}\left(\dot{w}_{x t}-\dot{u}_{x t}\right)+\int_y \mathcal{C}_{yx}(\du_{yt}-\du_{x t})+\sqrt{\dot{u}_{x t}} \xi_{x t}\right]
\end{align*}
The $\delta$-functional in the first line ensures that $\du_{x,t}$ is solution of the equation of motion~\eqref{eq:definition BFM Appendix}. It is reexpressed in the second line using an auxiliary field
$\tu_{x,t}$. 
This allows to perform the average over the noise. Using that $\xi$ is a Gaussian white noise with the correlations 
$\overline{\xi_{x t}\xi_{x' t'}}  = 2 \sigma \delta^d(x-x') \delta(t-t')$
we have for any function $Q_{xt}$~:
$\int \mathcal{D}[\xi] \exp \left(\int_{xt} Q_{xt} \xi_{xt} \right) = \exp \left(\sigma \int_{xt} Q_{xt}^2 \right)$. The generating functional thus reads~:
\begin{equation}
G\left[\lambda_{x t}, \dot{w}_{x t}\right] = \int \mathcal{D}[\du, \tu] \exp \int_{x t}\left(\tilde{u}_{x t}\left[-\eta \partial_{t} \dot{u}_{x t}+m^{2}\left(\dot{w}_{x t}-\dot{u}_{x t}\right)+\int_y \mathcal{C}_{yx}(\du_{yt}-\du_{x t})\right]+\lambda_{x t} \dot{u}_{x t}+\sigma \tilde{u}_{x t}^{2} \dot{u}_{x t}\right) \notag 
\end{equation}
Note that the exponential is linear in $\du$. This property allows to perform the path integral over $\du$, which yields a new $\delta$-functional~:
\begin{equation}
G\left[\lambda_{x t}, \dot{w}_{x t}\right] = \int \mathcal{D}[\tu] e^{\int_{xt}m^2\dot{w}_{xt}\tu_{xt}} \prod_{xt}\delta \left( \eta \partial_t \tu_{xt} - m^2\tu_{xt} + \int_y \mathcal{C}_{yx}(\tu_{yt}-\tu_{x t})+  \sigma^2 \tu_{xt} + \lambda_{x t} \right)
\end{equation}
The change of sign $-\eta \partial_t \du_{xt} \rightarrow \eta \partial_t \tu_{xt}$ is due to an integration by part over the time axis, using that $\du_{xt}=0$ at $t=-\infty$ and $t=+\infty$.
The new $\delta$-functional enforces $\tu_{xt}$ to be solution of the instanton equation~: 
\begin{equation}
\eta \partial_t \tu_{xt} + \int_y \mathcal{C}_{yx} \left(\tu_{yt} - \tu_{xt} \right) - m^2\tu_{xt} + 
\sigma \tu^2_{xt} = -\lambda_{xt} \, . \label{eq:Instanton real space Appendix} 
\end{equation}
Then performing the last path integral, the generating functional becomes 
\begin{equation}\label{eq:Generating functional from instanton Appendix}
G[\lambda_{xt}, \dot{w}_{xt}] = \exp \left (\int_{xt} m^2\dot{w}_{xt} \tu^{[\lambda]}_{xt}\right)
\end{equation}
where $\tu^{[\lambda]}_{xt}$ is solution of~\eqref{eq:Instanton real space Appendix}.

\paragraph{Massless instanton}
We can follow the same derivation when the drive $m^2\left(w_{xt}-u_{xt}\right)$ is replaced by a general driving force $f_{xt}$ that does not depend on the interface position. 
In this case the generating functional depends on the ramp rate of the force $\dot{f}_{xt}$ and can be expressed as~:
\begin{equation}\label{eq:Generating functional massless Appendix}
G[\lambda_{xt}, \dot{f}_{xt}] := \overline{\exp \left( \int_{xt} \lambda_{xt}\dot{u}_{xt} \right)}^{\dot{f}_{xt}} = \exp \left (\int_{xt} \dot{f}_{xt} \tu^{[\lambda]}_{xt}\right) \, ,
\end{equation}
where $\tu^{[\lambda]}_{xt}$ is now solution of the \emph{massless instanton}~:
\begin{equation}
\eta \partial_t \tu_{xt} + \int_y \mathcal{C}_{yx} \left(\tu_{yt} - \tu_{xt} \right) + 
\sigma \tu^2_{xt} = -\lambda_{xt} \, . \label{eq:Instanton massless Appendix} 
\end{equation}
